\documentclass[11pt, a4paper, oneside]{Thesis} 
\usepackage{charter}
\usepackage{pdfpages}
\usepackage[square, numbers, comma, sort&compress]{natbib} 
\hypersetup{urlcolor=black, colorlinks=true} 
\title{\ttitle} 
\usepackage{pdflscape}
\usepackage{nomencl}
\usepackage{tabu}
\usepackage{tabto}
\usepackage{textcomp}
\usepackage{graphicx}
\usepackage{minitoc}
\usepackage{epstopdf}
\usepackage{amsmath}
\usepackage{verbatim}
\usepackage{amsfonts}
\usepackage{amssymb}
\usepackage{floatrow}
\usepackage{subfigure}
\usepackage{mathtools}
\usepackage[tableposition=top,justification=centering]{caption}
\usepackage[linesnumbered,lined,commentsnumbered]{algorithm2e}
\usepackage{textcomp}
\usepackage{multirow}
\usepackage{adjustbox,lipsum}
\usepackage{xcolor}

\usepackage[flushleft]{threeparttable}
\usepackage{pifont}

\usepackage{rotating}

\usepackage{color}

\begin{document}

\setstretch{1.3} 

\fancyhead{} 
\rhead{} 
\lhead{} 
\cfoot{\thepage}

\pagestyle{fancy} 

\newcommand{\HRule}{\rule{\linewidth}{0.5mm}} 

\hypersetup{pdftitle={\ttitle}}
\hypersetup{pdfsubject=\subjectname}
\hypersetup{pdfauthor=\authornames}
\hypersetup{pdfkeywords=\keywordnames}


\begin{titlepage}
\begin{center}
{\LARGE \bfseries Maximum Likelihood Coordinate Systems for Wireless Sensor Networks:\\}
{\LARGE \bfseries from physical coordinates to topology coordinates\\[2cm] }

\begin{center} \large

\Large{ Ashanie  Gunathillake}\\ [2cm]
\end{center}

{\bfseries \normalsize { 2018}} 

\vfill
\end{center}

\end{titlepage}
-------------

\clearpage 

\frontmatter 
\addtotoc{Abstract} 

\abstract{\addtocontents{toc}{\vspace{1em}} 
Many Wireless Sensor Network (WSN) protocols require the location coordinates or a map of the sensor nodes, as it is useful to consider the data collected by the sensors in the context of the location from which they were collected. However, cost constraints prevent the incorporation of expensive hardware components such as GPS, in large-scale deployments. GPS-based localization is also not feasible in many environments. Thus, one of the major challenges in WSNs is to determine the coordinates of sensors while minimizing the hardware cost. To address this, numerous localization algorithms have been proposed in the literature. However, outcomes of these algorithms are affected by noise, fading, and interference. As a result, their levels of accuracy may become unacceptable in complex environments that contain obstacles and reflecting surfaces. The alternative is to use topological maps based only on connectivity information. Since they do not contain information about physical distances, however, they are not faithful representatives of the physical layout.
 
Thus, the primary goal of this research is to discover a topology map that provides more accurate information about physical layouts such as network shapes and voids/obstacles. In doing so, this research has resulted in four main contributions. First, a novel concept Maximum-Likelihood Topology Map for radio frequency WSNs is presented. This topology map provides a more accurate physical representation, by using the probability of packet reception, an easily measurable parameter that is sensitive to the distance. The second contribution is Millimetre wave Topology Map calculation, which is a novel topology mapping algorithm based on maximum likelihood estimation for millimetre wave WSNs. It utilises the narrow beam multi-sector antenna characteristics of millimetre wave transceivers to help achieve localization. The third contribution is a distributed algorithm being proposed to calculate the topology coordinates of sensors by themselves as two algorithms above calculate centrally, which requires time.  Since a topology map contains significant non-linear distortions when compared to physical distances, two WSN applications i.e. target searching and extremum seeking, which use a proposed topology map to localize the sensors and perform its specified task are presented as the final contribution of this dissertation.

}

\clearpage 


\pagestyle{fancy} 

\lhead{\emph{Contents}} 
\tableofcontents 


\mainmatter 

\pagestyle{fancy} 
\lhead[\rm\thepage]{\fancyplain{}{\sl{\rightmark}}}
\rhead[\fancyplain{}{\sl{\leftmark}}]{\rm}
\chead{}\lfoot{}\rfoot{}\cfoot{\thepage}

\chapter{Introduction}\label{chapter:intro}

\begin{figure}[b]
  \centering
    \includegraphics[width=0.8\textwidth]{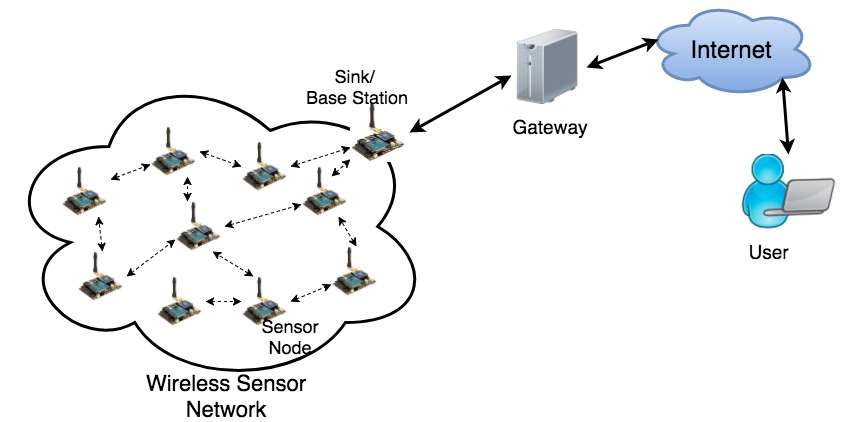}
    \caption{An illustration of WSN architecture}\label{fig::intro:architecture}
\end{figure}
A collection of sensor nodes that are able to sense, process, and transmit information about the environment in which they are deployed is called a Wireless Sensor Network (WSN). Figure \ref{fig::intro:architecture} is an illustration of a WSN architecture \cite{wsnarchi1, wsnarchi2}. Sensors are usually distributed around an environment to sense parameters such as temperature, humidity, and smoke and transmit these sensed data to a sink node using a routing protocol \cite{routingsurvey, routingsurvey2}. The sink node then transmits this information to the end user via a gateway node and the Internet. Depending on the environment in which the sensors are deployed, WSNs can be categorised into different groups- namely, terrestrial, underground, underwater, multimedia and mobile \cite{wsnTypes}. 
\begin{figure}
  \centering
    \includegraphics[width=0.99\textwidth,trim={5cm 0cm 3.5cm 0cm}]{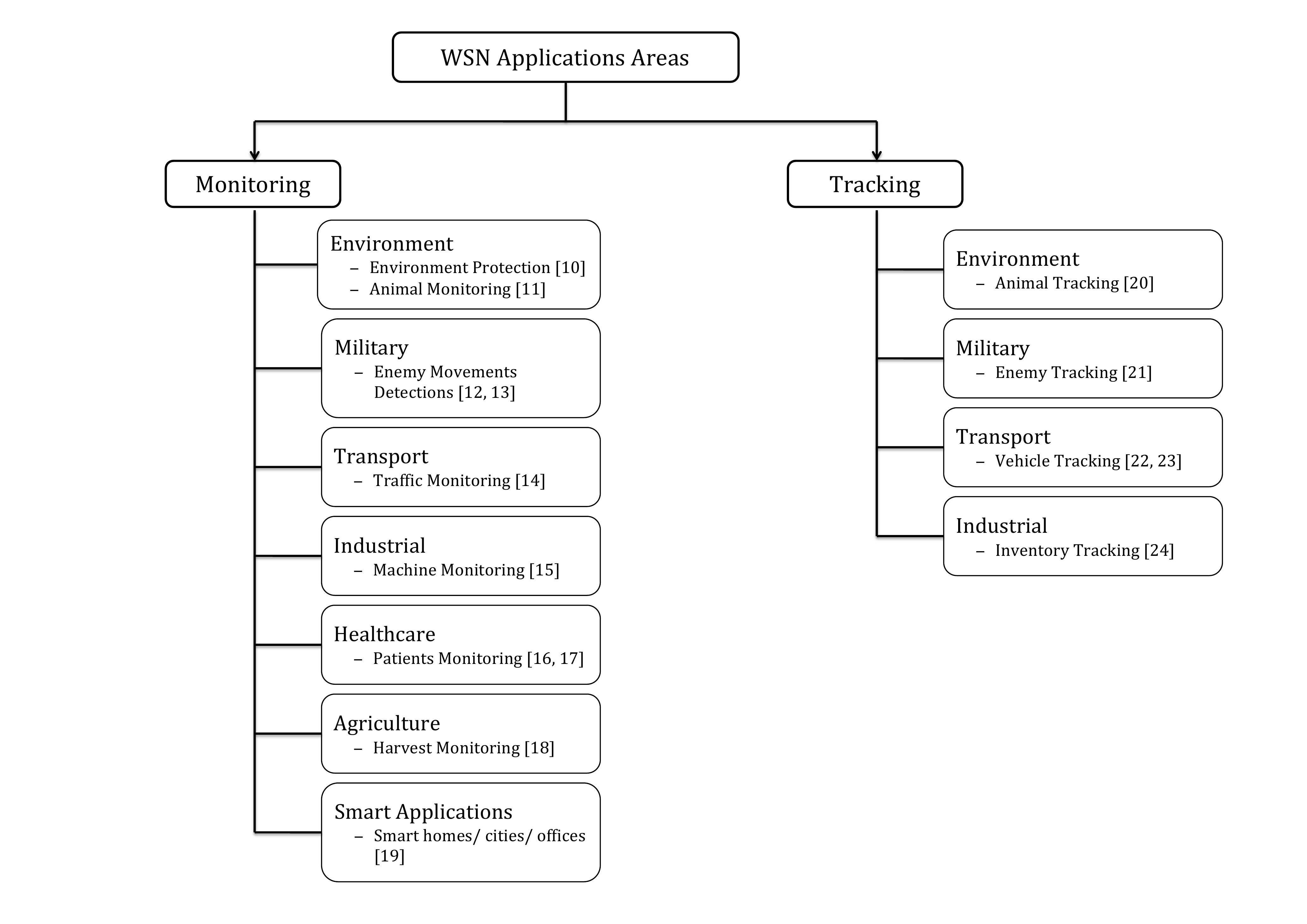}
    \caption{Some of WSN applications}\label{fig::intro:applications}
\end{figure}

One of the main advantages of WSNs is their ad-hoc nature. WSNs do not rely on any hardwired communication links. Thus, nodes can be deployed without a pre-deployed communication infrastructure \cite{vcthesis, introtoWSN}. This reduces the deployment and maintenance costs of WSNs \cite{wsnadvantages}. Moreover, WSNs can be deployed in inaccessible places such as mountains, deep forests, and rural areas, as they don not depend on pre-existing infrastructure \cite{introtoWSN}. Due to these unique features, WSNs have found their way into a large number of domains as shown in Figure \ref{fig::intro:applications}. WSN applications can be divided into two categories: monitoring and tracking \cite{wsnTypes, wsnApplicationTypes}. In monitoring applications, sensor nodes monitor environmental parameters such as temperature, humidity, and smoke and inform the end user if the sensed value exceeds a predefined value \cite{volcano, duckisland, military, militarymonitoring, trafficmonitoring, industrymonitoing, codeblue, encapsule, agreemonitoring, smartapp}. In tracking applications, sensors use infrared or ultrasound sensors to detect an intruder entering the network and track it until it leaves the network \cite{zebranet, militarytracking, vehicletracking, schoolvehicle, industrytracking}. At present, a large number of monitoring and tracking applications have been implemented to serve the public and industry(see Figure \ref{fig::intro:applications}).

\begin{figure}
 \centering
 \subfigure[Libelium waspmote (73.5 x 51 x 13 mm) \cite{Libelium} ]{
  \includegraphics[width=0.33\textwidth]{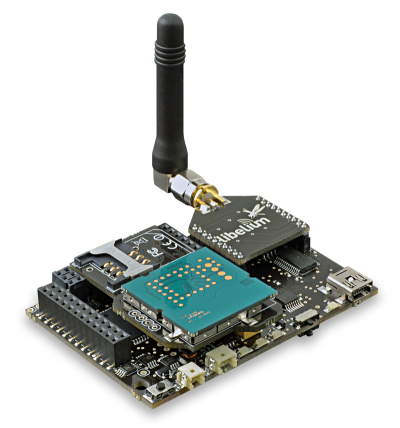}
   \label{fig::intro:waspmote}
   }
 \subfigure[Block diagram of sensor node]{
  \includegraphics[width=0.53\textwidth]{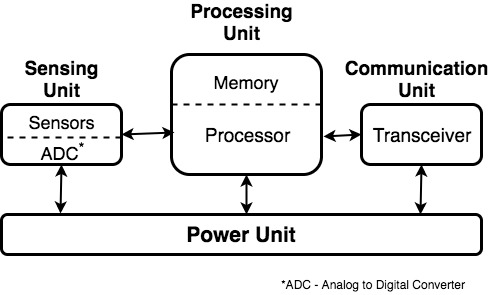}
   \label{fig::intro:blockdiagram}
   }  
  \caption{Sensor node} \label{fig::intro:sensornode}
\end{figure}

The primary of a WSN is its sensor node, which is a small device that can be as small as a millimetre scale object. This device consists of a small scale processor, memory, and a radio \cite{motecomparison}. A block diagram of a sensor node along with a Libelium Waspmote sensor node \cite{Libelium} is shown in Figure \ref{fig::intro:sensornode}. As seen in Figure \ref{fig::intro:blockdiagram}, a sensor unit comprises one or more sensors to measure environmental parameters and an Analogue to Digital Converter (ADC) to convert the analogue signals generated by sensors to a digital signals before sending them to the processor. Subsequently, the processor uses the data received from sensors and information stored in its memory to perform the tasks assigned to it. It also stores required information back in the memory for future use. The transceiver is used to transmit the processed information to the next sensor node or to receive information from a neighbouring node. Most sensor nodes available in the market use frequencies in the Radio Frequency (RF) band; however, millimetre wave (MmWave) transceivers are now being introduced into the market due to the high bandwidth demand of WSN applications \cite{mmwaveUWB}. The power unit supplies power to all the components of the sensor node. Besides this, a sensor node may have add-on components such as a Global Positioning System (GPS). 

\section{Limitations and Challenges in Sensor Networks}
Despite the numerous unique advantages of WSNs, there are some limitations and challenges that have to be considered while designning them. While WSNs share some common features with other network systems such as computer networks, the protocols used in those networks cannot be directly used in sensor networks due to the extremely limited capabilities and resources of sensor nodes  \cite{vcthesis}. Although with micro and nano technology, sensors are expected to soon become as small as a dust particle or a grain of sand \citep{smartdustintro}, sensor nodes at present are tiny devices with limited processing power and low memory. Thus, protocols need to be developed for WSNs to address the restricted resources and limitations of sensor networks. Moreover, sensor nodes are usually powered by AA or AAA batteries, which limits the amount of energy available for processing and transmission. Hence, the energy consumption for communication, sensing, and data processing in WSN protocols should be optimized to extend the network's lifespan.

Subject to technology trends, the number of nodes in a sensor network may vary from hundreds to millions of nodes depending on application requirements. Therefore, extending the lifespan of a large-scale sensor network with limited resources is challenging. Another key challenge that arises in large-scale WSNs due to their ad-hoc nature is the ability to self-organize the network. Self-organizing implies that devices cooperate and communicate with each other, form topologies, and monitor and adapt to environmental changes without human intervention \cite{selforgsurvey}. This requires additional informations such as the location of sensor nodes, providing which is another challenging task.  

\section{Challenges in Sensor Localization}
Self-organizing and other algorithms, such as those for target tracking and sensor fusion, require  locations of sensor nodes. Sensor locations must be included in data packets not only for the algorithm's decision making purposes but also to extract meaningful information from sensed data. For an example, in an application used to detect forest fires, if a user gets an alert of a fire in the forest but does not know the origin of this alert, required action cannot immediately be taken. Thus, sensor localization algorithms play an important role in WSN automation. 

Either a GPS or a localization algorithm can obtain sensor location.  However, use of GPS in large-scale networks is expensive and unfeasible for many applications. Also, GPS readings are less accurate in certain environments such as those indoor, underground, or underwater. Therefore, accurate localization algorithms are needed for WSNs. In some existing localization algorithms, analogue or range-based measurements such as Received Signal Strength Indicator (RSSI), Time of Arrival (ToA) or Angle of Arrival (AoA), are used. This increases the complexity of hardware as well as the costs involved in WSN implementation. Moreover, these measurements are highly sensitive to noise, fading, and interference, which causes erroneous calculation of location \cite{rssiemperical}. On the other hand, there are some localization algorithms based on the hop count matrix- that is the number of packet transmission hops between two nodes. Even though these algorithms do not depend on any special hardware devices to measure range-based parameters, they assume that the hop distance is a constant value. Thus, a node can move though different hop distances without making any change to the hop matrix. This causes erroneous calculation of sensor locations.

\subsection{Topology Maps}
As demonstrated, calculating the physical location of sensor nodes is challenging. It also increases the cost of WSN deployment. Researchers have therefore started to focus on locating sensor nodes using a topology map of a WSN. The topology map is an attractive alternative to the physical or geographical map of the network. It represents the arrangement of nodes while preserving node connectivity. For this reason, topology mapping algorithms deviate from geographical localization algorithms, as they are concerned with the arrangement of nodes and do not consider the actual physical location of the nodes. In other words, the mapping schemes expect the relative distances to be accurate but not the physical distances. Therefore, the expense on external hardware devices to take range-based measurements can be eliminated and the erroneous calculation due to communication effects such as noise, fading, and multipath can be minimized. 

However, topology map calculation is challenging because the calculated map must be isomorphic to the physical map of the network \citep{VC}. Moreover, the proposed topology mapping algorithms do not accurately present physical layout information such as shape, voids, and obstacles. Notwithstanding the accurate of topology map calculation, the use of topology maps in WSN applications is a challenging task. Most WSN tracking applications are based on physical distances between nodes or between nodes and the target. Thus, when moving from a physical map to a topology map of sensor networks, physical coordinate based algorithms need to be modified to use topology coordinates.

\chapter{Related Work}\label{chapter:rw}
Sensor localization is an important and crucial aspect that has generated significant interest among the research community, as it provides fundamental support to WSN location-aware protocols and applications. The use of GPS in sensor units has become infeasible due to economic constraints and power consumption, especially in large-scale WSNs. Thus, researchers have developed self-localization algorithms to determine the location of sensor nodes. Some of these algorithms are discussed in this chapter. 

Prior work in sensor localization can be divided into two categories based on the communication protocol used in the WSN: RF WSN localization and MmWave WSN localization. In RF WSNs, the sensor node is equipped with an omni directional antenna that uses the IEEE 802.15.4 standard protocol to communicate with other sensor nodes. The frequency band with highest availablility globally that uses this protocol is 2.4 GHz, however, European countries and the USA, use the frequency bands 868-868.6 MHZ and 902-928 MHz, respectively \cite{802154}. In MmWave WSNs, narrow beamwidth antenna arrays are used with the IEEE 802.11ad standard protocol. It uses the frequency band 30-300 GHz, which allows multi-Gbps data rate communication \cite{80211ad}. However, MmWave communication suffers significantly more from adverse signal propagation characteristics compared to RF communication due to its extremely high frequency use. This restricts the use of localization algorithms proposed for RF WSNs. In contrast, as MmWave communication uses narrow beamwidth antenna arrays, localization algorithms receive additional information to determine the location of sensors accurately. 

Sensor nodes in a network can be located in two ways: physical localization and topology mapping. In physical localization, the actual geographical coordinates of sensor nodes are calculated. In topology mapping, sensors are mapped to a different coordinate system based on the connectivity of nodes. Both methods include some common techniques- namely, triangulation, trilateration and maximum likelihood estimation- to calculate the coordinates of sensors. These techniques are described in the Appendix \ref{LT}. Section \ref{sec::rw:RF} discusses the RF localization schemes under physical localization and topology mapping. Approaches of MmWave physical localization approaches are analysed in Section \ref{sec::rw:MmWave}. Available information indicates that, none of the existing work has addressed MmWave WSN topology mapping, as MmWave communication is an emerging technology and topology mapping schemes are new to WSNs. 

\section{Approaches to RF WSN Localization}\label{sec::rw:RF}
RF WSN localization can be executed in two ways: physical localization and topology mapping. Prior work in the area of physical localization can be grouped into two categories- range-based localization and range-free localization \cite{locSurvey, locSurvey2, locSurvey3, locSurvey4, locSurvey5, locSurvey6, locSurvey7}. The following subsections provide a detailed background analysis of these two categories and their topology mapping algorithms.

\subsection{Range-based Localization}
Localization techniques that calculate the actual positions of nodes using anchor nodes and physical properties of communication signals are called range-based localization algorithms. Anchor nodes are the nodes that know their deployed locations. The physical properties- such as Received Signal Strength Indicator (RSSI), Time Difference of Arrival (TDoA), Time of Arrival (ToA) and Angle of Arrival (AoA)- along with the location of anchors are used to calculate the distance between two nodes. In the literature, several algorithms have been proposed using each property. Some of these are discussed below.

\subsubsection{RSSI-based Localization}
In RSSI based localization algorithms, signal strength of the received packet is used to estimate the distance between nodes \cite{rssisurvey}. The distance calculation of these algorithms uses theoretical and empirical models \cite{rssiadaptive}. In theoretical models, RF signal transmission loss is used to directly estimate the distance between two nodes \cite{rssidirect, rssidirct2}. Empirical models use a two-step process to obtain the location. First, they create an offline RSSI database using anchor nodes. Second the coordinates of non-anchor nodes are determined by matching the received signal strength to a record in the database \cite{rssiprbabilistic, rssiemperical}. The theoretical models are simple and result in lower energy consumption. However, RF communication effects such as noise, fading, and interference affect their accuracy. The empirical methods tend to be more accurate, as  a more accurate database can be created for the specific network and its environment. However, the creation of the database is time consuming, and it needs to be updated every time the operating environment changes. 

The log normal shadowing model is used in \cite{LSmethod} to improve the accuracy of the calculation, however, a larger number of sample data needs to be collected to estimate the coefficients in the model. Deieng et al. \cite{GMMmethod} proposed a two-mode Gaussian mixture model with trilateration and biased-maximum likelihood calculation to reduce localization errors in an indoor environment. An empirical model is used in \cite{empericalindoor} to estimate the distance between two nodes. This scheme has been evaluated using Crossbow IRIS motes and a conclusion drawn that the path-loss exponent of an indoor environment is larger than that of an outdoor environment. As the propagation parameters change with time, Xiao et al. \cite{gaussianfilter} proposed a method based on off-the-shelf Radio-Frequency Identification (RFID) technology that uses a low complexity Gaussian filter and a Bayesian interface to improve localization accuracy. Although this improves the accuracy of indoor localization, interference may affect the output of the algorithm.
  
To overcome the problems due to wide variations in signal levels associated with RF communication, Chengdong et al. \cite{rssiprbabilistic}  proposed a method based on a probabilistic model. The anchor nodes located closer to the unknown nodes are considered as reference nodes, and the radio coverage area of the reference nodes and unknown node is divided into a number of spatial grids. Then according to a probability model of radio signal space transmission, the degree of confidence of each grid point is calculated and proposed as the coordinates of the unknown node. The accuracy of this method depends on number of anchor nodes and the existence of obstacles. To mitigate the shadowing effects of obstacles, Chuku et al. \cite{shadowingmitigate} proposed a multilateration technique with clusterization and Hamdoun et al. \cite{mimolocalization} proposed a multilateration algorithm using multiple antennas in a node to measure accurate positions. Additionally, \cite{indoortri1, indoortri2, indoortri3} have proposed trilateration localization methods and \cite{indoorml1, indoorml2} have proposed maximum likelihood methods to localize sensor nodes accurately in an indoor environment. 

To reduce the complexity of the probabilistic method, Wang et al.\cite{rssipowerdecay} proposed an algorithm based on a transmission power adjustment strategy. The proposed method uses power decay curves of the operating environment to accurately estimate the coordinates of the unknown nodes.  Even though it reduces the complexity of determining the coordinates of the unknown nodes, power decay curves need to be update to reflect any changes to the operating environment. Ruxandra et al. \cite{MLRSSI} propose a maximum likelihood estimation algorithm derived from a triangulation technique. This algorithm takes into account the imperfections of distance measurements. However, to reduce the error in distance measurements, it requires more than three anchor nodes in the unknown node's neighbourhood. To reduce the distance error measurements in noisy environments, Marko \cite{convexRSSI} proposed an energy based localization algorithm that uses a modified semi definite relaxation (SDR) method, which is reformulated as an optimization problem. Yao et al. \cite{CN2} proposed a distributed localization scheme based on a weighted search, which includes a weighted search-based localization algorithm and a weighted search-based refinement algorithm. Authors claim this method to have a lower complexity. 

To reduce the dependency on anchor selection in RSSI localization, many researchers have proposed localization algorithms using a mobile anchor node equipped with GPS \cite{rssimobile, rssimobile2, rssimobile3, mobileanchor1, locSurvey7}. In these schemes a robot moves within the network deployment and broadcasts its position periodically. Then sensor nodes that receive the broadcasts, compute their locations using broadcasts from two different positions of the mobile anchor node. After this process, if there are still un-located sensor nodes, these sensor nodes compute their locations using the nearby localized stationary sensor nodes. These works do not address the localization of nodes in networks with obstacles. 

\subsubsection{TDoA-based Localization}
In TDoA based localization techniques, the sensor node is equipped with an RF and ultrasonic transceiver \cite{tdoasurvey, tdoasurvey2}. The transceiver simultaneously sends two signals, an RF signal and an ultrasonic signal, and the receiver records the time difference between the two signals. The receiver then turns this time difference into a distance. After calculating distance to anchor nodes, the sensor node estimates its coordinates using the location of anchor nodes and calculated distances from the TDoA method. For the calculation, the sensor needs at least four anchors to be located in its neighbourhood. 

Savarese et al.\cite{tdoa1} proposed an iterative multilateral positioning method to reduce the necissity of four anchors in each node's neighbourhood. When a sensor is localized, that node acts as an anchor node. However, if a node is localized with an error, that error is propagated to the other sensors that use this particular node for their localization. To overcome that, Luo et al. \cite{tdoa2} proposed a method which calculates the geometric distance error and decides whether a node can be upgraded to an anchor node or not.  Liu et al. \cite{tdoa3} presented a cross-correlation method for TDoA measurement and sensor localization. However TDoA requires the sensors to be synchronized when sampling the signals to locate a sensor node \cite{tdoatimesync, tdoatimesync2}. This is not feasible when there are many sensors spread in a large area.  Wang et al. \cite{tdoa4} answered this by proposing a method to obtain a sequentially algebraic solution of the source location, the sensor positions and the synchronization offsets. Furthermore, TDoA-based localization assumes that measurement noise is independent of actual source-to-sensor distance. Huang et al. \cite{tdoa5} proposed a distance-dependent noise model for TDoA measurements and obtained the location of nodes. 

Besides the time synchronization required for TDoA measurements, this method requires nodes to have both an RF and an ultrasonic radio. As most nodes are equipped with only an RF radio, the TDoA method requires additional hardware, which increases the pre-deployed hardware cost \cite{tdoasurvey2}. In addition, it was found that the accuracy of TDoA measurements  improves when the physical distance between the two nodes is increased. 

\subsubsection{ToA-based Localization}
In ToA based localization, distance is estimated based on signal propagation time. The transmitting node (that is an an anchor node) time-stamps a packet and broadcasts the packet to its neighbours. The receiving nodes calculate the distance to the transmitter by calculating the traverse time of the signal \cite{tdoasurvey, tdoasurvey2}. However, ToA distance measurement techniques are sensitive to time synchronization errors between the sender and the receiver \cite{tdoasurvey2}. To overcome this requirement, researchers have proposed a method called the Round-Trip ToA (RTToA) \cite{toa1} or Two Way ToA (TW-ToA) \cite{toa2, toa3}.  

In RTToA/TW-ToA, the sender time-stamps a packet and sends it to the receiver. The receiving node returns the packet immediately to the sender. The distance between the two nodes is then calculated at the sender using the total traverse time. As the calculation is done at the sender, time synchronization is not required. However, a major source of error in this method is the delay at the receiving node in handling the packet, procesings it, and sending it back \cite{locSurvey, toasurvey}. Moreover, these schemes are reliant on connectivity within the network. 

Besides time synchronization, time measurements are affected by noise, line of sight and a multipath environment. To overcome this, Ultra Wide Band (UWB) signals have been used in \cite{toas6}. In \cite{toas3}, an accurate measuring scheme for RF signals is proposed that involves low cost and energy consumption. Furthermore, this work claims that it can achieve similar accuracy to previously published wideband, high power indoor localization systems while using simplified hardware and low bandwidth. Yeredor et al. \cite{toa4} proposed a method, which does not require synchronization and does not involve any "hand-shaking" procedures.  It uses a single initial transmission and through an iterative procedure, each sensor estimates its own timing offset and position. 

In addition, ToA based sensor localization in underwater and underground WSNs are proposed in \cite{toas1, toas4, toas5}. A mobility assisted node localization based on ToA measurements without synchronization is proposed by Chen et al. in \cite{toas2}.

\subsubsection{AoA-based Localization}
In AoA localization, the coordinates are calculated based on the signal's direction of arrival. Direction measurements are obtained by using antenna arrays in sensor nodes \cite{aoa4}. Iterative \cite{aoa1} and non-iterative \cite{aoa2, aoa3} localization methods are used in this technique. In \cite{aoa5}, two algorithms were proposed based on the radial and the bearing. A radial is the angle at which an object is seen from another point, or more simply a radial is a reverse bearing. Kułakowski et al. \cite{aoa7} proposed an AoA localization based on antenna arrays. The AoA measurements are derived from the measurements of the phase differences in the arrival of a wave front. In \cite{aoa6}, a technique has been proposed for determining node bearings based on radio interferometric AoA measurements from multiple anchor nodes to any number of target nodes at unknown positions. Least squares triangulation is then used to estimate node position. 

AoA closed-form location estimators are studied in \cite{aoa8, aoa9} for the 2-D scenario and in \cite{aoa10, aoa11} for the 3-D scenario. In \cite{aoa14}, a closed-form AoA 3D localization method has been proposed. This method focuses on two aspects. The first is to improve the AoA localization accuracy when the sensor positions have errors. The second is to reduce the amount of estimation bias caused by the measurement noise and sensor position errors when the pseudo-linear formulation is used. Moreover, to reduce the effect of sensor position errors, array shape calibration or steering vector refinement is used in \cite{aoa12} to improve the accuracy of the AoA method. A Toeplitz Approximation Method (TAM) was proposed in \cite{aoa13}, which focuses on improving the accuracy of AoA rather than that of the point source location when sensor position errors are present.

Although AoA techniques have high accuracy, they often require the LOS between the anchor node and the receiver. As AoA techniques calculate distance using signal propagation, these localization techniques are sensitive to both shadowing and multipath effects. Also, these techniques entail an additional hardware cost \cite{tdoasurvey2}.

\subsection{Range-free Localization}
In range-free localization algorithms, the locations of nodes are obtained without using any special hardware. The location calculation relys on the connectivity information of nodes. First, range-free algorithms get the distance in hops and then map the hop distance to geometric distance \cite{rangefree, dv, rangefree2} using anchor node locations. Therefore, the accuracy of these algorithms depends greatly on the number of anchor nodes and their deployment. The key issue of range-free algorithms is distance estimation- that is the mapping of hop distance to geometric distance. 

To improve the hop distance calculation, authors in \cite{rfa1}, add a correction factor to the hop distance when computing the distance between the anchor nodes and unknown nodes. Wang et al. \cite{rangefree3} proposed an algorithm based on an accurate analysis of hop progress (i.e., quantify the relationship between the path distance and the network parameters such as the communication range and node density) in a WSN with randomly deployed sensors. However, in an anisotropic network the hop count from an anchor to sensor exhibits multiple patterns, due to the interference of multiple anisotropic factors. To compensate for that, Xiao et al. \cite{rangefree4} proposed a pattern-driven localization scheme. This adopts different anchor-sensor distance estimation algorithms for different patterns. Wu et al. \cite{CN3} proposed a regulated neighbour distance based localization algorithm to address the hop-distance ambiguity i.e., node have no ability to measure distance to their neighbours. Regulated neighbour distance is a new proximity measure for two neighbouring nodes based on their neighbour partitions.  

Wang et al. \cite{rfa2} improved the traditional Approximate Point-In-Triangulation (APIT) scheme by decreasing the probabilities of In-To-Out error and Out-To-In error. The traditional APIT algorithm is based on dividing the whole network into triangular regions that are made up of vertices formed by all the possible sets of three connected neighbouring anchor nodes. The unknown node then determines whether it is inside or outside the triangle formed and the location is estimated as the centre of gravity of the triangles overlapping region. An improved APIT-3D scheme named as Volume Test Approximate Point-In-Triangulation 3-dimension (VTAPIT-3D) is proposed in \cite{rfa3} as the practical environment is always in 3D and APIT has poor accuracy in 3D environments.

To overcome the requirements of a large number of anchor nodes to achieve the localization accuracy, researches have proposed methods utilizing the anchor mobility \cite{rangefreemobile, rangefreemobile2, rfa7}. Kuo et al. \cite{rangefreemobile} proposed a method using geometry conjecture. Each anchor moves in the network and broadcasts its current position. The sensor node that receives these broadcasts, use the location information in the broadcats and computes their own locations. Chia-Ho \cite{rangefreemobile3} extended \cite{rangefreemobile} by considering mobile sensor network localization. It assumes that the mobile sensors know their moving velocity. Therefore they calibrate their beacon points (i.e. anchor positions) using the geometric corollary. The algorithm proposed in \cite{rfa4}, uses a mobile beacon with a rotary directional antenna. This is called the Azimuthally Defined Area Localization (ADAL) method. 

In addition, a geometric constraint based range-free localization scheme is proposed in \cite{rfa8}. The constraint area of the sensor node is first determined by the intersection of two selected anchor coordinate points and then repeated with another set to narrow down the constraint area. Finally, the average of all intersection points provides the position estimation of the sensor node. In \cite{rfa9}, constraint area based localization is proposed with a mobile anchor with a specific moving trajectory. However, this scheme shows high localization error when the random mobility model is used.

Another way to calculate the hop distance is by using analytical geometry based calculation \cite{locSurvey}. In algorithms that uses this method, the average hop distance is calculated using the statistical characteristics of the network. In \cite{rfa5}, a pattern driven localization scheme for anisotropy networks is proposed. To calculate the distance, the algorithm first checks whether the anchor is slightly detoured or strongly detoured from the sensor node, and discards the strongly detoured anchors before calculating distance. Zaidi et al. \cite{rfa6} proposed another analytical algorithm based on both, the number of hops between two nodes and the number of forwarding nodes, to further improve the accuracy. However, these algorithms do not consider obstacles in the operating environment. In addition, in a noisy environment, the accuracy of range free algorithms decreases due to packet loss. 

\subsection{Topology Mapping}
Topological mapping techniques are fundamentally different to localization techniques because the mapping algorithms are concerned with the arrangement of the nodes.  They are not concerned about the actual location of the nodes. In other words, the mapping schemes expect the relative distances to be accurate, not the physical distances. Thus, given the absolute position of a subset of nodes, global localization is realizable \cite{thinplate}. However, to achieve this, the topology map should be isomorphic to the physical layout of the sensor network \cite{VC}. 

In \cite{allmap}, several unsupervised learning algorithms have been proposed that use eigenvalue decomposition for obtaining a lower dimensional embedding of the data. It provides a unified framework for extending Multi-Dimensional Scaling (MDS)\cite{mds} , Isomap\cite{isomap}, Local Linear Embedding (LLE), and Laplacian Eigenmaps (LE) \cite{allmap}. MDS \cite{mds} is a commonly used statistical technique in information visualization for exploring similarities or dissimilarities in higher-dimensional data from the complete distance matrix (similarity matrix), which is defined as the matrix of all the pairwise distances between points/nodes. A centralized range-free algorithm based on MDS is proposed in \cite{mds2}, which estimates the nodes location using the connectivity information of nodes. When the positions of adequate number of anchor nodes are known, the absolute coordinates of all nodes in the map can be estimated \cite{mds3}. However, this method suffers from high time complexity and computational power. As a solution for that Junfeng et al. \cite{mds4} proposed a method that estimates the position of nodes in a distributed way using clustering technique. 

Isomaps \cite{isomap} is an extension of MDS to geodesic distance-based topology map generation. Again, the geodesic distances are actual distances between nodes, which require expensive error prone distance estimators such as RSSI or ToA. Moreover, LLE and LE both use an iterative approach to preserve the neighbourhood distances, the realization of which is infeasible in energy-limited WSNs and also it requires more time to generate the map \cite{VC}. As a solution for the time consumption in sensor localization, several algorithms are using kernel-based machine learning technique to estimate the position of sensors. In \cite{kernal1}, a graph embedded mapping algorithm that employs an appropriate kernel function to measure the dissimilarity between sensor nodes is proposed. They have considered the sensor nodes as group of devices that construct a graph to preserve the topological structure of the network. Moreover, Wang et al. \cite{kernal2} proposed a kernel isometric mapping(KIsomap) algorithm that determine the relative locations of sensors based on geodesic distance. A semi-supervised Laplacian regularized least squares algorithm that uses the alignment criterion to learn an appropriate kernel function is presented in \cite{kernal3}. 

Dhanapala et al. \cite{VC} presented a method to obtain topology-preserving maps of WSNs using Virtual Coordinates (VCs) of sensor nodes. In Virtual Coordinate System (VCS), the layout information such as physical voids, shapes etc. are absent. To overcome that, Singular Value Decomposition (SVD) of VCs is used in this method. Furthermore, \cite{VC} shows that transformation for topological map from virtual coordinates can be generated using a subset of nodes. However, when the number of nodes increases, the time required to generate the virtual coordinate matrix also increases. Xu et al. \cite{CN4} proposed an algorithm for locating new coming nodes to the network based on polynomial mapping. In this algorithm, the pair-wise distance is obtained by geodesic distance measurement. Then a graph is constructed to represent the topological structure of the sensor networks and calculate the weight matrix and the sparse preserving matrix. Finally, physical locations of all unknown nodes are calculated by coordinate transformation.

\section{Approaches to Mmwave Physical Localization}\label{sec::rw:MmWave}
Despite its promise of enabling multi-Gbps data rates, MmWave communication significantly suffers from adverse signal propagation characteristics due to extremely high frequency band (30 to 300 GHz) communication \cite{nit14}. As the frequency is high, the transmitter and receiver need to have a clear LOS, which requires narrow beamwidth antenna arrays in sensor nodes. For this reason, the RF localization algorithms discussed above cannot directly be used in MmWave WSNs. On the other hand, very limited work has been done in MmWave localization by leveraging the communication features of MmWave systems \cite{mmwa6}.

El-Sayed et al. \cite{MMWlocalization} compared the performance of RSSI, TDoA, and AoA techniques in relation to MmWave communications. For comparative evaluation, they considered the characteristics of the MmWave wireless channel. They concluded that localization techniques based on AoA are the most promising for MmWave communication as AoA fits better with the typical properties of the MmWave systems. In \cite{mmwa6}, a set of potential approaches for localization in MmWave systems are discussed. The researchers conclude that using an RSSI-based trilateration approach results in poor performance and promising localization accuracies have been achieved by AoA triangulation and ToA trilateration when assuming two nodes have clear LOS connectivity. Moreover, they claimed that a combination of ToA and AoA signal features could yield an even better accuracy in certain scenarios. However, environment size and expected variability of signal features have to be considered for achieving optimal performance. 

An MmWave based localization algorithm named as mTrack is proposed in \cite{mmwa7}. This algorithm uses RSSI and phase of the signals to estimate the position of nodes. In \cite{mmwa8}, a mobile node is used for localization in an MmWave Multiple Input Multiple Output (MIMO) system that does not require LOS connectivity as it exploits changes in the statistics of a sparse beam-space channel matrix.

In \cite{mmLocalization1}, they have designed a lightweight algorithm that targets a single-anchor localization scheme for MmWave systems. After identifying the main propagation properties of MmWave signals that have an impact on localization, they have designed three algorithms that exploit these- namely, a triangulation validation procedure, an angle difference of arrival approach, and a scheme based on location fingerprinting. Joan et al. \cite{mmLocalization3} proposed an algorithm named JADE, which estimates the location of mobile nodes in an indoor space. This algorithm does not need any prior information about the deployed environment. It estimates the location of nodes using AoA of multipath components of the signal sent by visible anchors. In addition, a localization scheme based on time of flight and RF chain infrastructure is proposed in \cite{mmLocalization2}. 

As the MmWave communication is affected by multipath propagation, Bocquet et al. \cite{mmwa1} proposed a scheme using a focusing technique to reduce the multipath effect. An indoor localization scheme based on information gathered by multipath is proposed in \cite{mmwa4}. An algorithm is proposed in \cite{mmwa3} for an indoor positioning system that is based on location fingerprinting. In this method, the effects of grid spacing, the number of reference points and the effect of radio propagation are considered. In \cite{mmwa5}, a joint design of axis alignment and positioning with directional antenna under Non Line of Sight (NLOS) indoor conditions is proposed. However, an axis alignment scheme depending on the rotation vector method may require additional hardware to be installed in the nodes.

\chapter{Maximum Likelihood Topology Maps for Radio Frequency Wireless Sensor Networks}\label{chapter:mltm}
Many sensor network protocols require maps indicating sensor locations for automation. Physical coordinate based maps capture the physical layout including voids and shapes, but obtaining the distance values required is often not feasible or economical. The alternative is to use topological maps based only on connectivity Information. Since they do not contain physical distances, they are not faithful representations of the physical layout. In this chapter a \textbf{M}aximum \textbf{L}ikelihood-\textbf{T}opology \textbf{M}ap (ML-TM) for RF WSNs is presented. ML-TM provides a more accurate physical representation by using the probability of signal reception, an easily measurable parameter that is sensitive to distance. This approach is illustrated using a mobile robot that listens to signals transmitted by sensor nodes and maps the packet reception probability to a coordinate system using a packet receiving probability function. ML-TM is an intermediate map between exact physical maps and hop-based topology maps. 

The chapter is structured as follows and the main results of the chapter were originally published in \cite{mltm, Jmltm}. Section \ref{sec::mltm:Introduction} offers an introduction and motivation for the research presented in this chapter. Section \ref{sec::mltm:algo} discusses the details of proposed ML-TM algorithm. Section \ref{sec::mltm:error} presents a novel parameter one-hop connectivity error ($E_{total}$) that captures the connectivity error of topology maps, which accounts for node connectivity, and distance correlation between physical and topological maps. Section \ref{sec::mltm:robottrajectory} explains a robot trajectory algorithm to cover a network with least possible time and generate accurate topology map. Section \ref{sec::mltm:result} presents the performance evaluation and comparison of the algorithm. Section \ref{sec::mltm:copm} examines and compares the limitations, energy usage and complexity of the proposed algorithm. Finally, Section \ref{sec::mltm:conclusion} provides a conclusion of the chapter.

\section{Introduction}\label{sec::mltm:Introduction}
The use of WSNs have increased due to their low-cost, and characteristics such as the distributed nature and ease of deployment \cite{SNadvantages, CN5}. However, the low-cost sensor nodes have limited resources and as a result their communication range, computation power and memory are limited \cite{CN2}. In addition, cost constraints prevent the incorporation of  expensive hardware components such as GPS in large-scale deployments \cite{VC}. GPS based localization is also not feasible in many environments\cite{CN2}. However, most sensor network based applications of WSNs require the determination of the physical location of sensor nodes \cite{CN1}. Examples include cases where the data collected by the sensor nodes are useful only when considered in the context of the location from which the data was collected \cite{rangefree3}, or when location based routing is used. Therefore, one of the major challenges in WSNs is to determine the location of the sensor nodes even when nodes are deployed in harsh environments and minimizing the cost of hardware \cite{CN5}. 

This has been addressed by a large number of researchers, with numerous algorithms to calculate physical coordinates of sensors proposed in literature. However, outcomes of range-based algorithms are affected by noise, fading of the signals and interference \cite{rssiemperical} and as a result, their accuracy may become unacceptable in complex environments with obstacles and reflecting surfaces. In range-free algorithms, the accuracy is highly depend on the number of anchor nodes and their distribution \cite{rangefree, dv}. In general, maps generated by range-free algorithms are less accurate when compared to those from range-based algorithms \cite{locationinwild}.

Thus, topology map is an attractive alternative to the physical map of the network. As topology map is a representative of arrangement of node that preserves the connectivity, it is not an accurate representation of physical layout information such as shape, voids/obstacles, etc. A range-free algorithm with increased accuracy is the hop based topology map obtained by SVD of VCS \cite{VC}. In a VCS, a sensor node is identified by a vector that contains the distance from it, in hops, to a set of anchor nodes. As a result, the accuracy of VCS depends on the  distribution of anchor nodes and the density of the sensor nodes. Therefore, it is necessary to find maps with more accurate physical layout information such as shapes of network boundaries and voids/obstacles.

To this end, a novel concept \textbf{M}aximum \textbf{L}ikelihood-\textbf{T}opology \textbf{M}aps (ML-TM) for RF WSNs is presented. As it is based on a packet reception probability function, which is sensitive to the distance, this preserves the connections as well as physical layout information. In this algorithm a mobile robot is used to obtain the ML-TM. This reduces the dependency of the output on range-based parameters and anchor selection/distribution. The robot moves in the space occupied by the network and updates a binary matrix based on the received packets from different nodes. Then, based on this binary matrix and a proposed packet receiving probability function, maximum-likelihood topology coordinates of the sensors are calculated. The packet receiving probability function that proposed in this chapter is an intermediate model between actual physical distance measurement, e.g., using RSSI and VCS. This in turn enables the relaxation of the node density dependency of range-free algorithms, while eliminating the need for overcoming the uncertainties associated with RSSI under different environmental conditions, which may vary widely even within a single network. Moreover, RSSI based algorithms extract the distances from received power, which encounters significant errors due to RF communication effects. Evaluation of proposed scheme shows that the method is able to provide accurate topological maps of nodes, identify features such as physical voids and network boundaries and outperform the RSSI based geographical localization and hop based topology maps. 

\section{Maximum Likelihood Topology Map Algorithm : ML-TM}\label{sec::mltm:algo}
\begin{figure}[t]
  \centering
    \includegraphics[width=.98\textwidth]{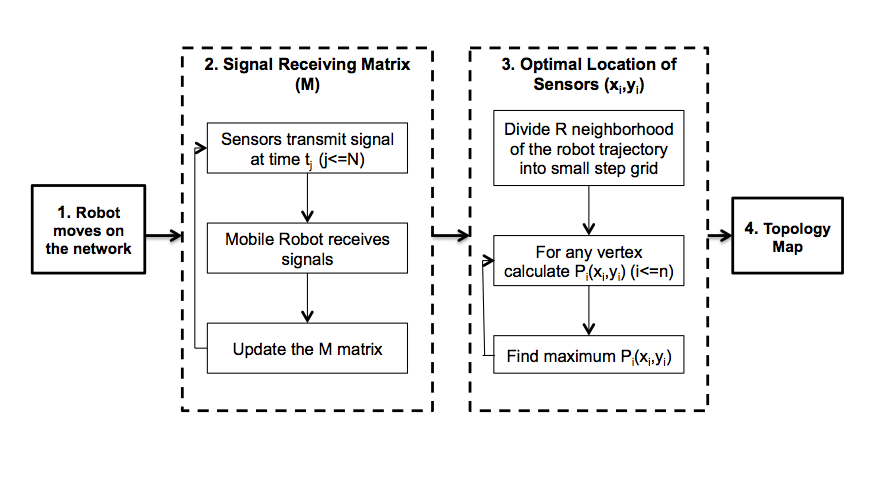} 
    \caption{Workflow of ML-TM}
    \label{fig::mltm:workflow}
\end{figure}

This section describes the ML-TM that calculates the sensor coordinates in a topological map that characterizes the sensor coordinates in such a way that it is more representative of the physical layout than TPM \cite{VC}, but still preserves connectivity. Figure \ref{fig::mltm:workflow} illustrates the work flow of the algorithm and the pseudo code of the algorithm is explained in the Appendix \ref{PC}. Sensor nodes uses the IEEE 802.15.4 standard protocol to communicate with each other and RF transceiver of the node is equipped with an omni directional antenna. Detail of the IEEE 802.15.4 protocol is discussed in the Appendix \ref{CP}. 

Topology map is achieved by using a mobile robot moving on the network for information gathering. Then the gathered information is mapped to topology coordinates of the sensors with a probability function of packet receiving, which is sensitive to the distance. The following subsections describe the probability function used and how the coordinates are calculated. 

\subsection{The Packet Receiving Probability Function}\label{subsec::mltm:pfn}
The packet receiving probability function describes the probability of receiving a packet transmitted by a sensor when robot is at a particular distance \cite{sensornets16}. Let, $S(d)$ be the probability value when robot is at distance $d$ from the sensor. Then, $S(d)$ satisfies the following constraints:
\begin{eqnarray}\nonumber
&&0\leq S(d)\leq 1  \:\:\:\:\:\:\:\: \forall  d\\ \nonumber
&&S(d_1) \leq S(d_2) \:\:\:\:\: \forall  d_1 \geq d_2\\ 
&&S(d) = 0 \:\:\:\:\:\:\:\:\:\:\:\:\:\:\:\: \forall  d > R \label{eq::mltm:sd}
\end{eqnarray}
where $R$ is some given distance.

Such a function $S(d)$ is called the packet receiving probability function. This chapter uses the following example of that function: 

\begin{eqnarray}\nonumber
S(d) &:=& p_0 \:\:\:\:\:\:\:\:\:\:\:\:\:\:\:\:\:\:\:\: \forall d\leq r \\ \nonumber
S(d) &:=& 0 \:\:\:\:\:\:\:\:\:\:\:\:\:\:\:\:\:\:\:\:\:\: \forall d\geq R \\ 
S(d) &:=& \frac{p_0(R-d)}{(R-r)} \:\:\:\: \forall r<d<R \label{eq::mltm:sd2}
\end{eqnarray}
where $0 < p_0 \leq 1$, $0 < r < R \leq R_c$ are some given constants. $R_c$ is the communication range of a sensor node. It is obvious that the function (\ref{eq::mltm:sd2}) satisfies all the conditions in (\ref{eq::mltm:sd}).

\begin{figure}[ht!]
  \centering
    \includegraphics[width=0.65\textwidth]{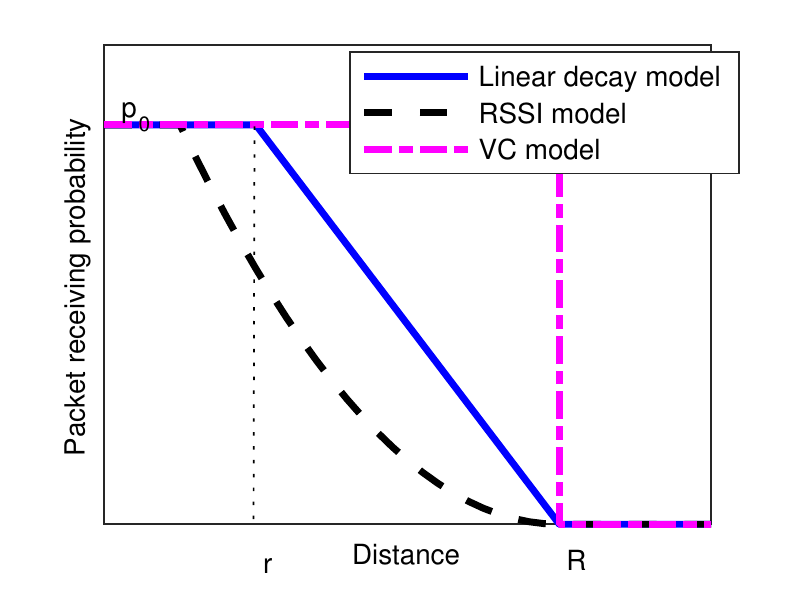}
    \caption{Packet receiving probability function for different models}\label{fig::mltm:sdmodel}
\end{figure}

The packet receiving probability function that considered in this chapter is an intermediate model of RSSI and VC. As shown in Figure \ref{fig::mltm:sdmodel}, VC uses the $r=R$ assumption, which is not the case in real environment. On the other hand, RSSI localization uses a polynomial function to estimate packet receiving probability, which is hard to estimate for different environmental situations. Therefore, an intermediate level between RSSI model and VC model is considered to obtain the proposed topological map. 

\subsection{Calculating Topological Coordinates}
This section describes the information gathering by mobile robot and calculating the sensor coordinates to form a topological map. In information gathering phase, robot traverses in the network to receive packets from all sensor nodes. It moves to a location and wait for a while to receive packets from all its neighbours, and then move to the next location in its trajectory. This will continue until it receives packets from all the sensor nodes in the network. Mobile robot and sensor nodes communicate using the same protocol, IEEE 802.15.4 and sensors use unslotted Carrier-Sense Multiple Access with Collision Avoidance (CSMA-CA) channel access mechanism to transmit a packet to the robot. 

\begin{figure}
  \includegraphics[width=0.65\textwidth]{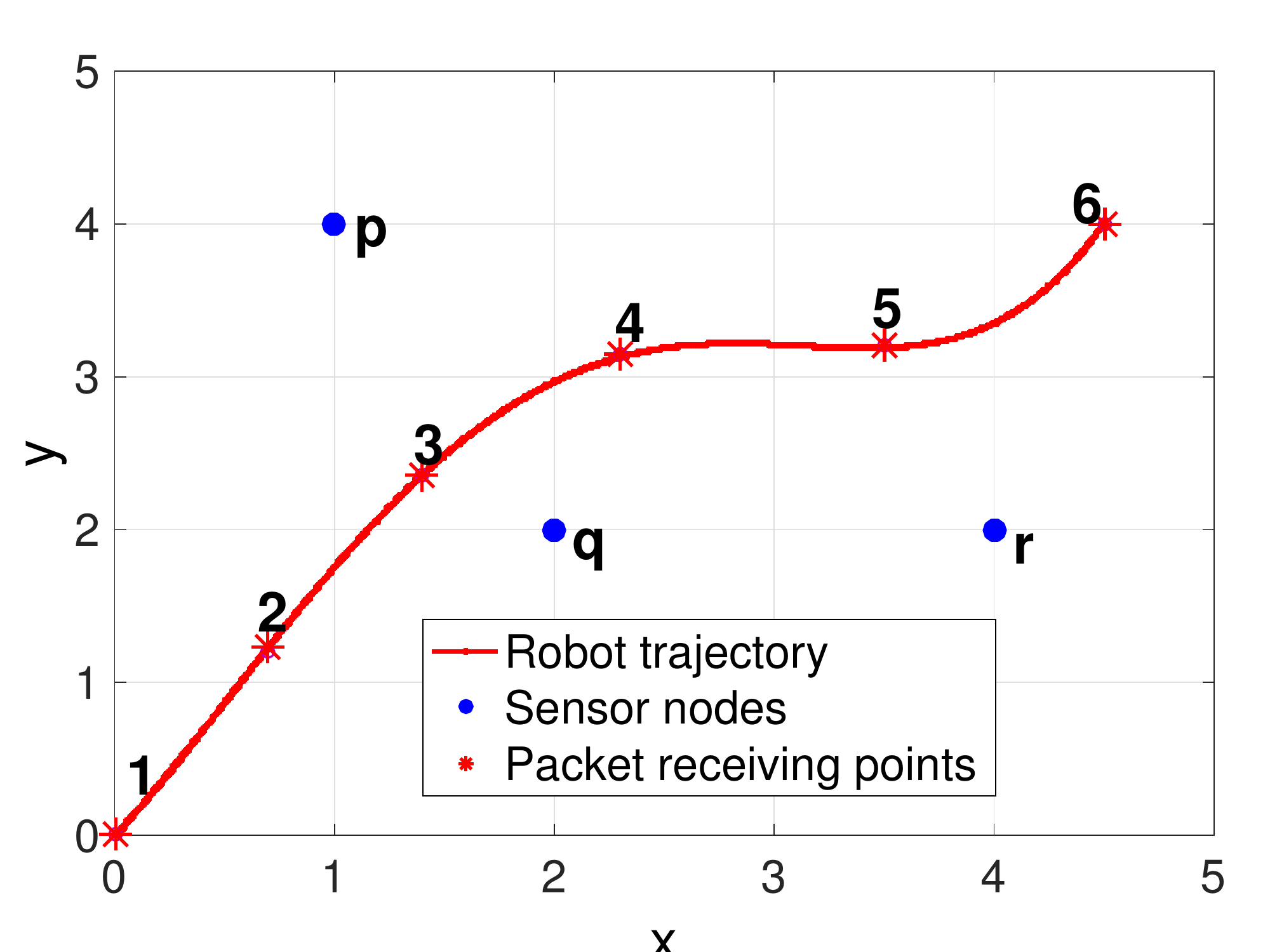}
  \caption{Sensor network with three nodes}\label{fig::mltm:examplen}
\end{figure}

As mobile robot moves on the network, let $(x_R(t_k), y_R(t_k))$ be the coordinates of the robot at time $t_k$. Consider $N$ number of steady sensors in unknown locations labelled as $s_i$, where $i = 1,2,...,N$. The robot can receive packets from the sensors at times $ t_1 < t_2 < ...< t_k<...< t_n$. A binary matrix $M$ of order $N\times n$ is introduced based on the following rule: \\
$M(i,k) = 1$, if the robot gets a packet from the sensor $s_i$ at the time $t_k$; \\
$M(i,k) = 0$, if the robot does not get a packet from the sensor $s_i$ at the time $ t_k$.\\ 
Note that the robot listens to the channel for a time slot starting at $t_k$, allowing reception from different nodes in the vicinity. The matrix entries  for $t_k$ are based on receptions during this slot. Such a binary matrix $M$ is called the packet receiving matrix. Table \ref{tab::mltm:mmatrix} refers to the M matrix of the network shown in Figure \ref{fig::mltm:examplen}. The network is consist with three nodes, $p$, $q$, $r$ and robot traverse in the trajectory shown in red color. The packet reception points are labelled as $1,2,...,6$. At $t_1$, robot doesn't receive any packet from sensor as it is located outside the communication range of all sensors. Therefore, all the elements in $t_1$ column of $M$ matrix is recorded as zero. Then robot moves one step forward in its trajectory and receives a packet from sensor node $q$. Thus, $(q,t_2)$ element of the matrix is recorded as one and remaining elements of $t_2$ column are recorded as zero. This continues until robot receives packets from all three sensor nodes. 

\begin{table}
\begin{center}
\caption{\textsc{The M matrix}}\label{tab::mltm:mmatrix}
\begin{tabular}{l| c c c c c c} 
    \hline
     &\textbf{$t_1$} &\textbf{$t_2$} &\textbf{$t_3$}&\textbf{$t_4$}&\textbf{$t_5$}&\textbf{$t_6$} \\ \hline
   $p$ & 0&0 &1&0&0&0 \\  
   $q$ & 0&1&1&1&0&0 \\
   $r$ & 0&0&0&0&1&0 \\
    \hline
\end{tabular}
\end{center}  
\end{table}

After receiving packets from all the sensor nodes and updating the M matrix, the topology coordinates of the sensors are then calculated using the packet reception probability function. Let the sensors $s_1, s_2,..., s_N$ are located at the points $l_1,l_2,...,l_N$, respectively. Then, based on the robot’s trajectory $(x_R(t_k), y_R(t_k))$, the receiving probability function $S(d)$ and the packet receiving matrix $M$, a likelihood function $P(l_1,l_2,...,l_N)$ is introduced. This likelihood function is the probability to obtain the packet receiving matrix $M$ for the robot’s trajectory $(x_R(t_k), y_R(t_k))$. The $l_1,l_2,...,l_N$ points that gives the maximum value by substituting to the likelihood function is the maximum likelihood coordinates of the sensor nodes. This is made under the assumptions that the probability or robot receiving a packet from any of the sensor nodes is described by the function $S(d)$, where $d$ is the distance between the robot and the sensor at the time of sending the packet.

\textbf{Definition}: A location set $(l^{opt}_1, l^{opt}_2, . . . , l^{opt}_N)$ is said to be an optimal location of the sensors $s_1, s_2,..., s_N$ based on the robot’s trajectory $(x_R(t_k), y_R(t_k))$, the receiving probability function $S(d)$ and the receiving matrix $M$ if,
\begin{eqnarray}
P(l^{opt}_1, l^{opt}_2, . . . , l^{opt}_N) \geq   P(l_1,l_2,...,l_N) \:\:\:\:\:\: for\:\:\: any \:\:\:(l_1,l_2,...,l_N)
\end{eqnarray}

In other words, an optimal location of the sensors are the points on the plane that maximize the probability of producing the measured packet receiving matrix $M$ for the robot's trajectory. Thus, the goal is to find an optimal location of the sensor in topology map. It can be achieve as follows.

Let the vectors $m_1, m_2, ..., m_i,..., m_N$ be the rows of the matrix $M$. Hence the vector $m_i$ is the packet receiving vector of the sensor node $s_i$ describing receiving/not receiving packets from the sensor node $s_i$ by the robot. Also, let $m_i(1), m_i(2),...,m_i(k),... , m_i(n)$ denote the elements of the vector $m_i$. Furthermore, for all $i = 1, 2, . . . , N$, introduce the function $P_i(l_i)$ which is the probability of obtaining the packet receiving vector $m_i$ for the robot’s trajectory $(x_R(t_k), y_R(t_k))$ under the assumptions that the sensor node $s_i$ is located at the point $l_i$ and the probability for the robot to receive a packet from the sensor node $s_i$ is described by the function $S(d)$ where $d$ is the distance between the robot and the sensor node $s_i$ at the time of sending the packet. Since the sensor nodes are randomly located, the event of locating a single sensor node in a particular location is mutually independent. Thus, the likelihood function $P(l_1,l_2,...,l_i,...,l_N)$ can be rewritten as in equation \ref{eq::mltm:likelihhogfn}.

\begin{eqnarray}\label{eq::mltm:likelihhogfn}
P(l_1,l_2,...,l_i,...,l_N) = P_1(l_1)P_2(l_2)...P_i(l_i)...P_N(l_N)
\end{eqnarray}

As in equation \ref{eq::mltm:likelihhogfn}, to find an optimal location of the sensors delivering the maximum value of $P (l_1,l_2,...,l_i,...,l_N)$, we just need to find independently locations $l_i$ delivering maximum values of $P_i(l_i)$. To find this value, the R-neighbourhood of all the packet received locations of node $s_i$ in the robot’s trajectory $(x_R(t_k), y_R(t_k))$ is divided into small step grids $g_1,g_2,...g_j$. Then for any grid vertex $g_j=(x_j,y_j)$, the probability of locating sensor nodes $s_i$ at that grid vertex $P_i(x_j,y_j)$ is calculated. The calculation of $P_i(x_i,y_i) $ is based on the equation (\ref{eq::mltm:probability}).

\begin{equation}\label{eq::mltm:probability}
P_i(x_j, y_j) = Z^i(d_{j1})Z^i(d_{j2})...Z^i(d_{jk})...Z^i(d_{jn})
\end{equation}

where
 \begin{equation}\label{eq::mltm:dist}
 d_{jk} := \sqrt{(x_j-x_R(t_k))^2 + (y_j -y_R(t_k))^2}
 \end{equation}
 
and the function $Z^i(d_{jk})$ is defined as:
\[
 Z^i(d_{jk}) =
  \begin{cases}
   S(d_{jk}) & \text{if } m_i(k) = 1 \\
   1-S(d_{jk})     & \text{if } m_i(k) = 0
  \end{cases}
\]
 
After calculating the $P_i(x_j,y_j)$ value for all the vertices, the grid vertex $(x_j^{opt},y_j^{opt})$ that delivers the maximum value of $P_i(x_j,y_j)$ is selected as the maximum likelihood solution. Then it is assigned as the maximum likelihood topology coordinate of sensor node $s_i$ and it is an approximation of the optimal location of the sensor node $s_i$.

\begin{table}
\begin{center}
 \caption{\textsc{$P_i(x_j,y_j)$ Calculation for the Network in Figure \ref{fig::mltm:examplen}}}\label{tab::mltm:probcal}
  \begin{adjustbox}{width=1\textwidth} \begin{tabular}{ l | c |c|c| c| c| c| c| c| c| c| c| c| c}
    \hline
   \textbf{ Grid}&  \multicolumn{2}{ |c| }{Point 1} &  \multicolumn{2}{ |c| }{Point 2} &  \multicolumn{2}{ |c| }{Point 3} &  \multicolumn{2}{ |c| }{Point 4} &  \multicolumn{2}{ |c| }{Point 5} &  \multicolumn{2}{ |c| }{Point 6}& \\ \cline{2-13}
    \textbf{ Vertex} & $d_{jk}$&$Z^i(d_{jk})$& $d_{jk}$&$Z^i(d_{jk})$& $d_{jk}$&$Z^i(d_{jk})$& $d_{jk}$&$Z^i(d_{jk})$& $d_{jk}$&$Z^i(d_{jk})$& $d_{jk}$&$Z^i(d_{jk})$&$P_i(x_j,y_j)$\\ \hline
   (0,0) &0&0&1.44&0.7064&2.64&1&3.78&1&4.68&1&6.02&1&0\\
   (1,1)&1.41&0.69&0.36&0.14&1.36&0.3368&2.42&1&3.2&1&4.6&1&0.0325\\
   (1,3)&3.16&1&1.81&0.90&0.76&0.6518&1.3&0.6316&2.50&1&3.64&1&0.3708\\
   (4,2)&4.4&1&3.2&1&2.72&0&1.97&0.985&1.21&0.5833&2.06&1&0\\
    \hline
  \end{tabular}
  \end{adjustbox}
  \end{center}
\end{table}

\begin{figure}
 \centering
 \subfigure[Node p]{
  \includegraphics[width=0.45\textwidth]{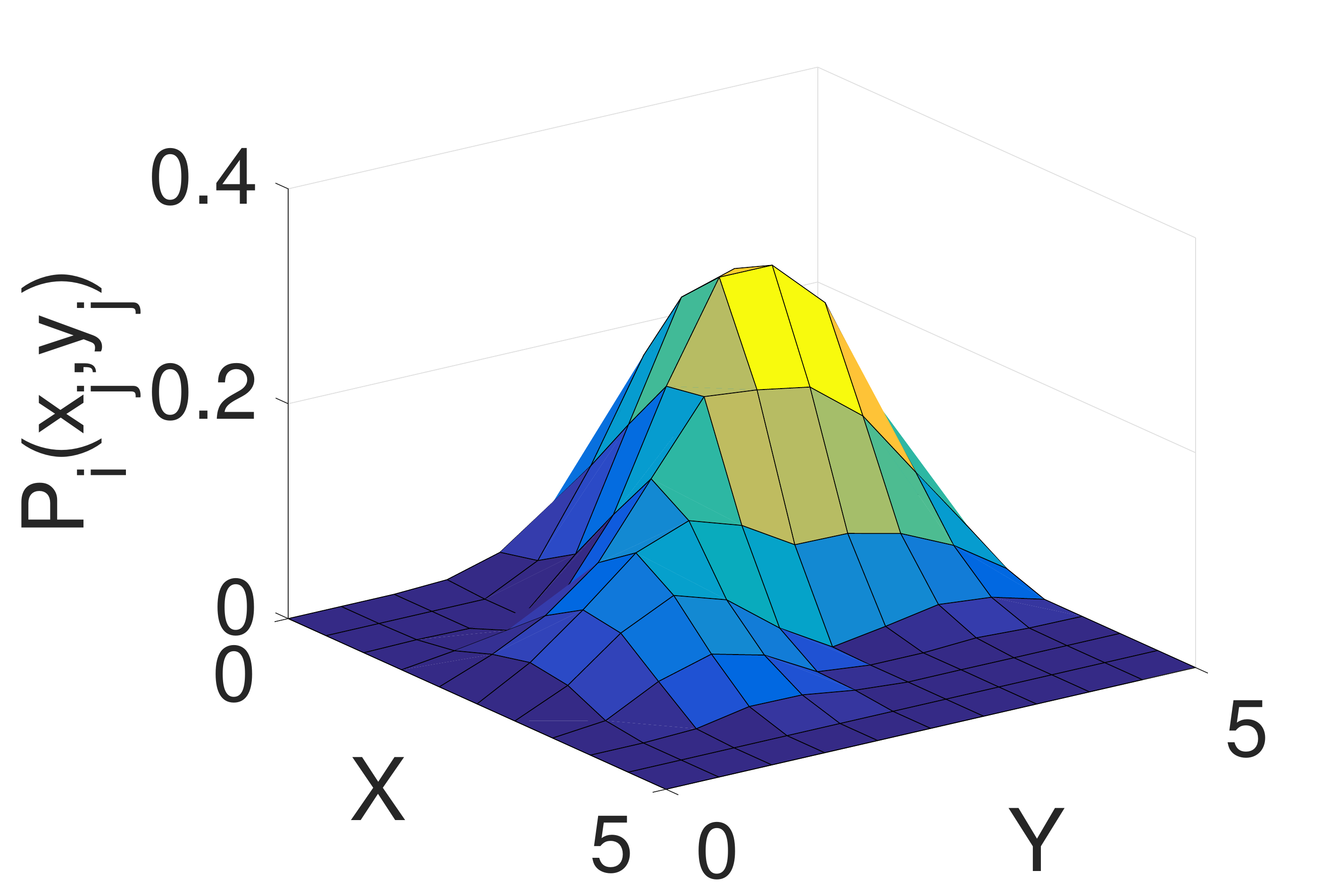}
   \label{fig::mltm:pdp}
   }
   \quad
 \subfigure[Node q]{
  \includegraphics[width=0.45\textwidth]{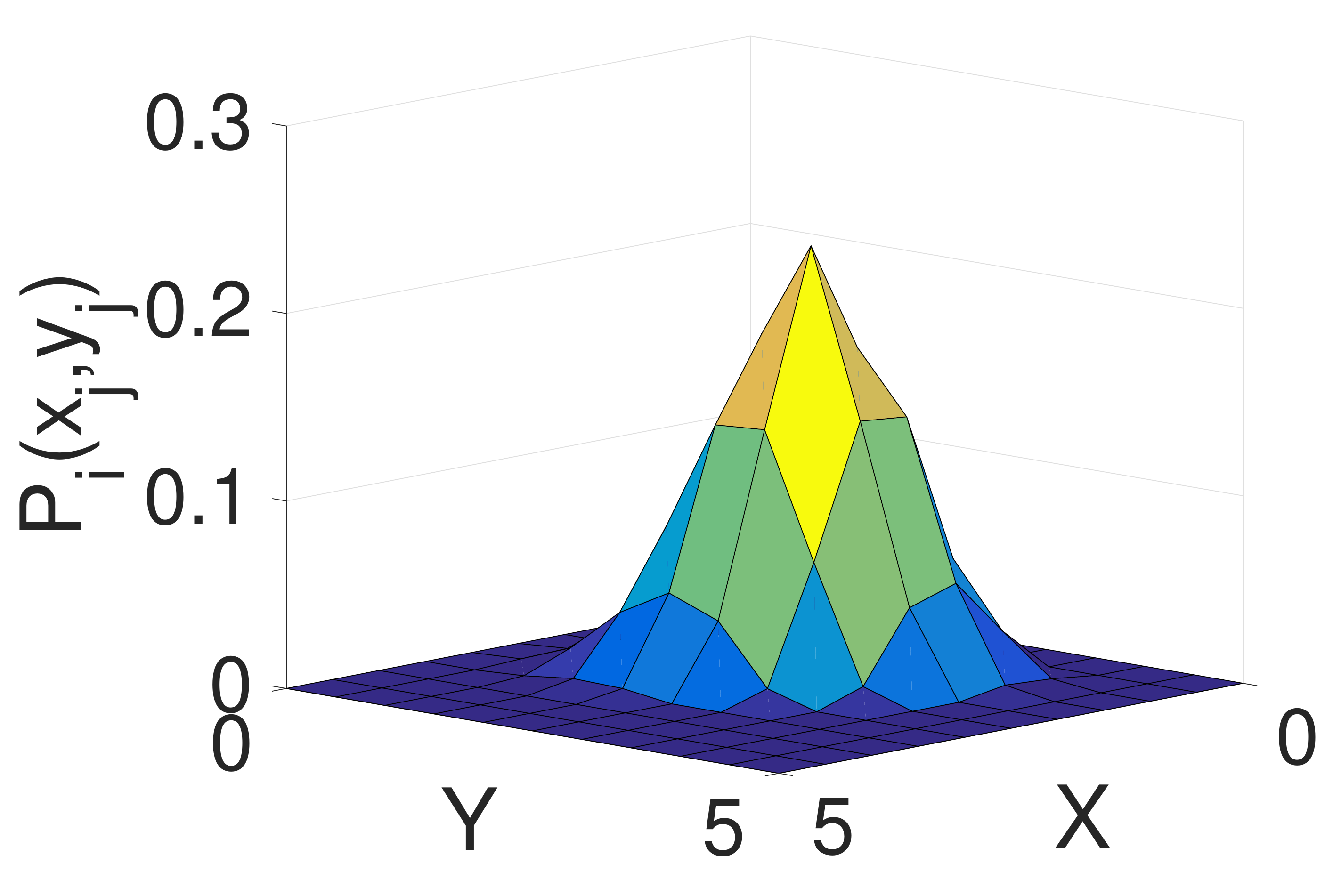}
   \label{fig::mltm:pdq}
   }

 \subfigure[Node r]{
  \includegraphics[width=0.45\textwidth]{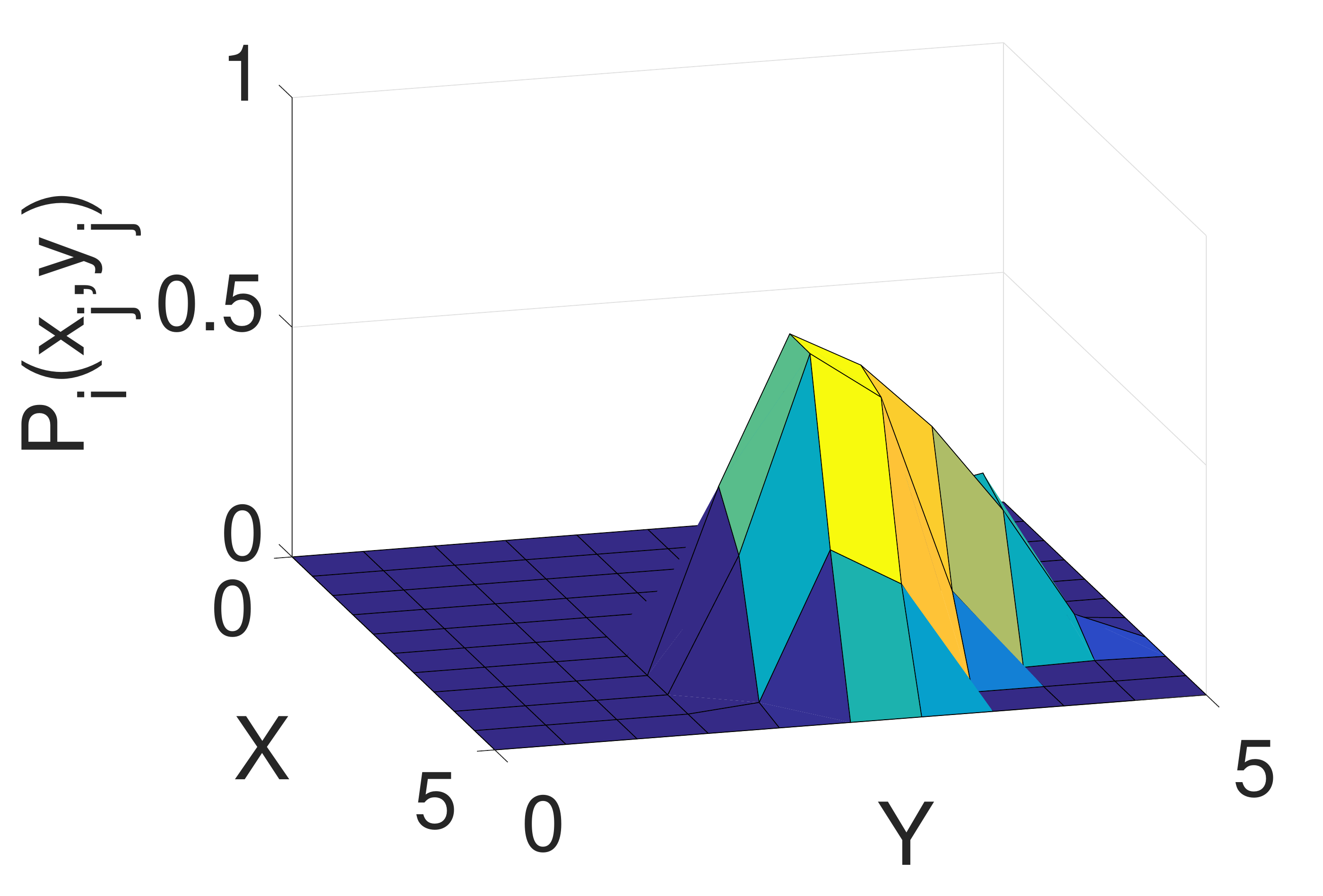}
   \label{fig::mltm:pdr}
   }
 \caption{ $P_i(x_j,y_j)$ distribution of three nodes} \label{fig::mltm:pd}
\end{figure}

Let consider the same example illustrated in Figure \ref{fig::mltm:examplen} to explain the topology coordinate calculation. As in the Table \ref{tab::mltm:mmatrix}, robot receives packets from node $p$ at $t_3$. Then the R-neighborhood of that robot location is divided into small grid. For simplicity, we consider $1\times1$ grid distribution. Select one grid vertex $g_j$ and find distance to all robot packet receiving points $1,2,...,6$. Then using the probability function (with $p_0$=1, r=0.2 and R=2) find $S(d_{jk})$ value for all distances and calculate $Z^p(D_{jk})$ as in the Table \ref{tab::mltm:probcal}. Afterwards, $P_i(x_j,y_j)$ can be calculated as in equation (\ref{eq::mltm:probability}). This can be continued to nodes $q$ and $r$ as well. Figure \ref{fig::mltm:pd} shows the $P_i(x_j,y_j)$ distribution over all the grid vertices of three sensor nodes. Finally, choose the grid vertex delivering the maximum value of  $P_i(x_j,y_j)$ and that is the optimal estimate for the  location of the sensor node. Here the optimal locations of the three sensor nodes are, $p\equiv(1,3),q\equiv(2,2),r\equiv(4,2)$.

\section{One-Hop Connectivity Error Calculation}\label{sec::mltm:error}
Evaluating the accuracy of the topology map is essential to determine the usefulness of the proposed method. Although preliminary evidence may be obtained by visual inspection, a formal mathematical approach is necessary to get the accuracy of the map to compare it with other existing algorithms. This parameter should be able to capture the failures in node connectivity with its neighbourhoods.

The purpose of the ML-TM mapping algorithms is to obtain information about node arrangement, physical layout and physical voids/obstacles. Thus, the actual physical distances are not of concern. In \cite{isomap, errorref}, the error is calculated by considering the difference of the positions in the actual physical map and the topology map. An error metric based on number of node flips is used in \cite{VC}. In this work, the arrangement of nodes is considered, but the correlation between physical map distance and topological map distance is not taken into account. However this is an issue when we use topology maps for other geographical localization applications such as target tracking, source seeking and boundary detection. 

In this section, the aim is to develop an error estimation parameter called one-hop connectivity error, which represents the percentage of nodes located incorrectly in their neighbourhood. Let, $R_c$ be the radius of communication area in the actual physical map. Then the first task is to find the communication area in the topological map. As mentioned previously, topological map layout is not same as physical map. It is a distorted version (i.e., non-linearly expanded or shrink and rotated) of physical map. Therefore, it cannot say that the communication area is a circle with radius $R_c$. However, for the simplification of the calculation, the communication area in the topological map is considered as an ellipsoid. 

\begin{figure}
  \centering
    \includegraphics[width=0.8\textwidth]{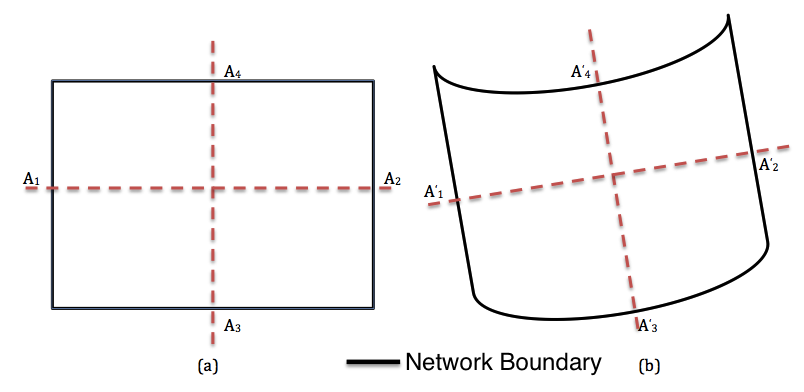}
    \caption{(a) Physical map with two perpependicular lines and (b) The two corresponding lines in topological map}\label{fig::mltm:axis}
\end{figure}

Finding the ellipsoid parameters are carried out in several steps. First, draw two perpendicular lines from the centre of the physical map and consider four points ($A_1$, $A_2$, $A_3$ and $A_4$) crossing the network boundary as shown in Figure \ref{fig::mltm:axis}(a). The topology coordinates of the four points, calculated using the ML-TM algorithm are $A'_1$, $A'_2$, $A'_3$ and $A'_4$  and shown in Figure \ref{fig::mltm:axis}(b). Then, the radius of the ellipsoid is calculated as in equation (\ref{eqn::mltm:a}) and equation (\ref{eqn::mltm:b}).

\begin{figure}
  \centering
    \includegraphics[width=0.8\textwidth]{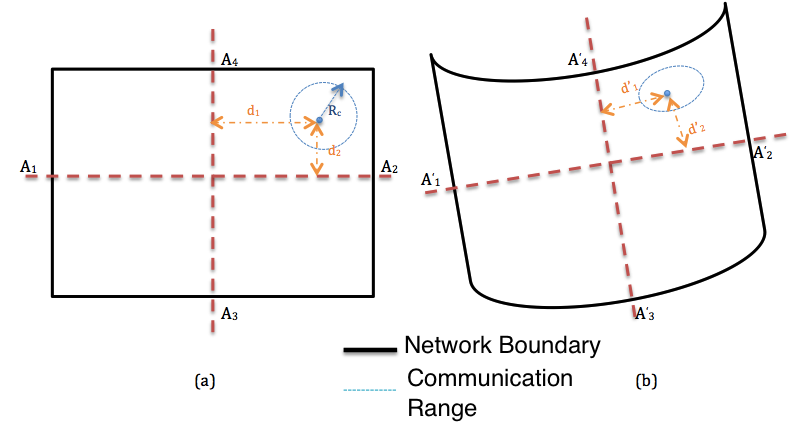}
    \caption{(a )Communication area of a sensor in physical map and (b) Communication area of the corresponding sensor in topological map}\label{fig::mltm:ca}
\end{figure}

\begin{equation}\label{eqn::mltm:a}
a_i=\frac{d'_{i,1}}{d_{i,1}}R_c
\end{equation}

\begin{equation}\label{eqn::mltm:b}
b_i=\frac{d'_{i,2}}{d_{i,2}}R_c
\end{equation}

where, $a_i$ and $b_i$ are the radius of the ellipsoid of node $s_i$ in the topological map in $A'_1A'_2$ and $A'_3A'_4$ directions respectively, $d'_{i,1}\:\: and \:\:d'_{i,2}$ are the distance from node $s_i$ to $A'_1A'_2$ and $A'_3A'_4$ axis respectively in topological map, $d_{i,1}\:\: and \:\:d_{i,2}$ are the distance from node $s_i$ to $A_1A_2$ and $A_3A_4$ axis respectively in the physical map (refer Figure \ref{fig::mltm:ca}). 

Let the node $s_i$ coordinates in the topological map be $(x'_i,y'_i)$, and then the equation of the communication area ellipsoid be as in equation (\ref{eqn::mltm:ellipeq}).

\begin{equation}\label{eqn::mltm:ellipeq}
\frac{(x'-x'_i)^2}{a_i^2}+\frac{(y'-y'_i)^2}{b_i^2}=1
\end{equation}

Therefore, one-hop neighbours in the physical map must locate within this ellipsoid. If any one-hop neighbour is not within this area, that node is located incorrectly with respect to node $s_i$. Thus, $e_i$ is defined as the set of incorrectly located nodes with reference to node $s_i$ and it can be calculated as;

\begin{equation}
e_i=N_{p,i}-\{N_{p,i} \cap N_{t,i}\}
\end{equation}

where, $N_{p,i}$ and $N_{t,i}$ are the one hop neighbour set of node $s_i$ in physical and topological map respectively.

Finally, one-hop connectivity error, $E_{total}$, is calculated as in equation (\ref{eqn::mltm:errort}). $E_{total}$ represents the node connectivity with respect to the correlation of actual map distances. Therefore, this error parameter can be used to calculate the accuracy of any type of map.

\begin{equation}\label{eqn::mltm:errort}
E_{total}=\frac{\sum _ {i=1} ^n \mid e_i \mid }{\sum _{i=1} ^n \mid N_{p,i}\mid } \times 100\%
\end{equation}
where, $\mid \cdot \mid$ is the number of elements in a set.

\section{Robot Trajectory Planning}\label{sec::mltm:robottrajectory}
\begin{figure}
 \centering
 \subfigure[Random walk]{
  \includegraphics[width=0.45\textwidth]{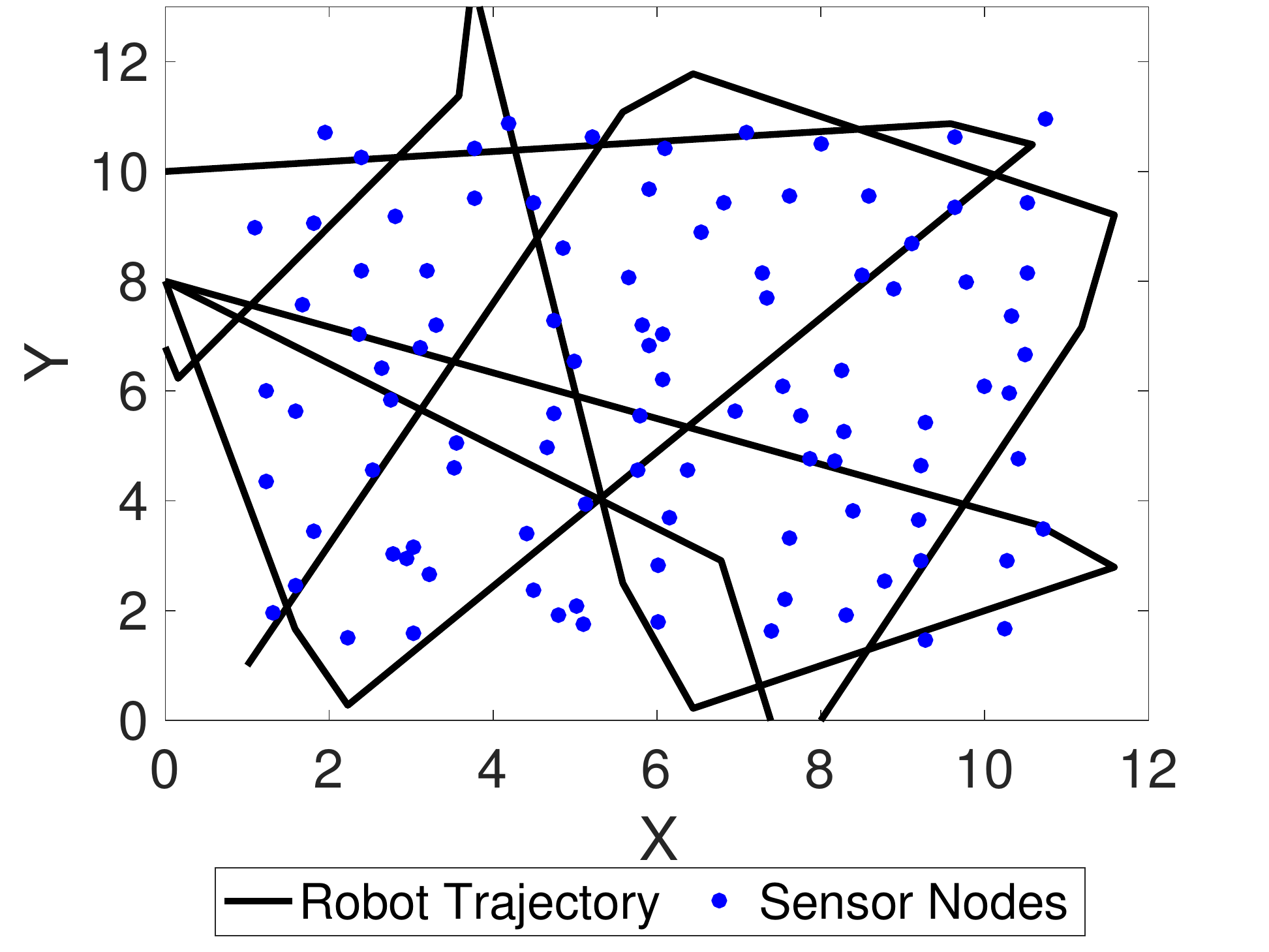}
   }
 \quad
 \subfigure[Spiral]{
  \includegraphics[width=0.45\textwidth]{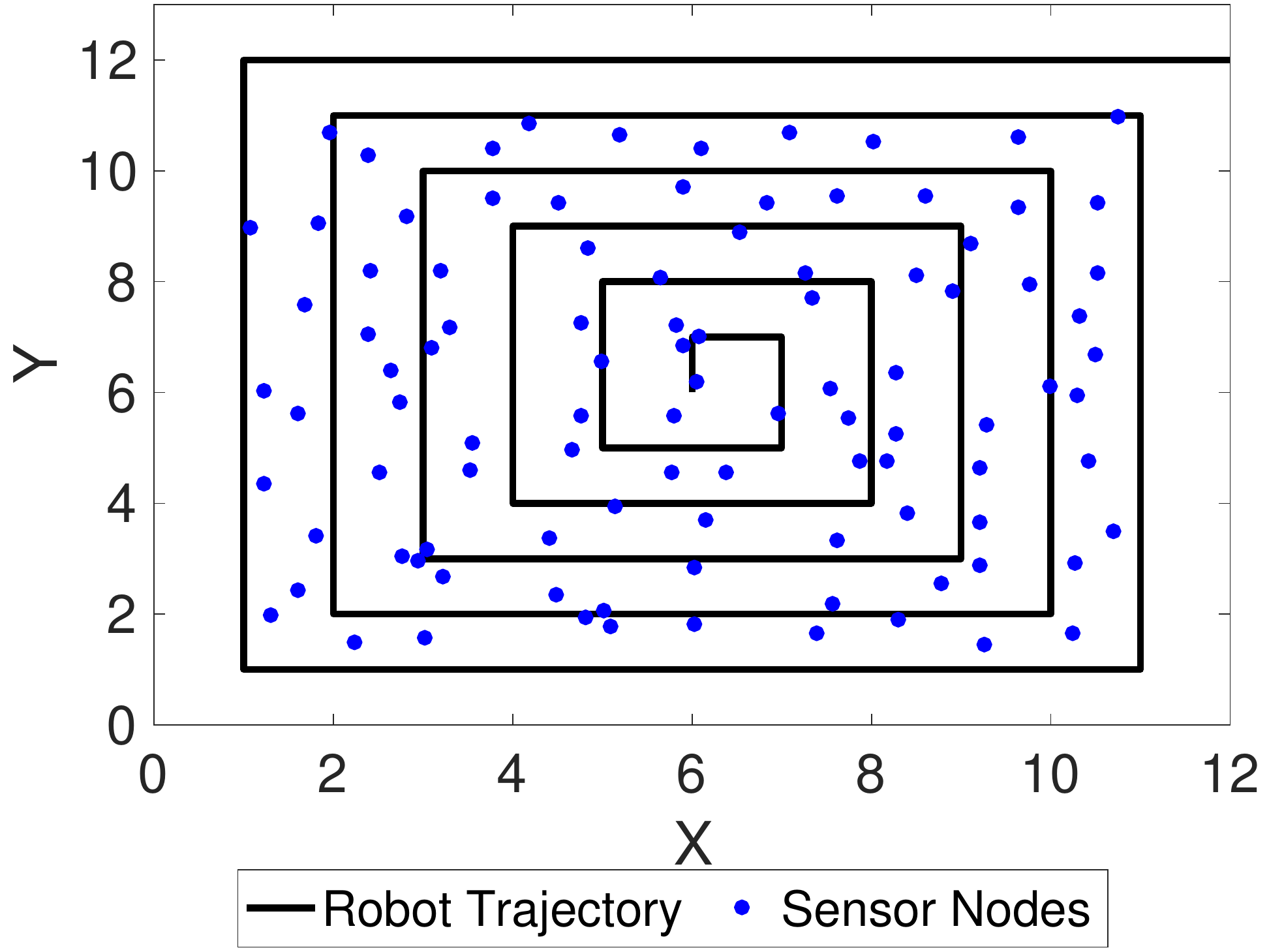}
   }
   
 \subfigure['S' shape]{
  \includegraphics[width=0.45\textwidth]{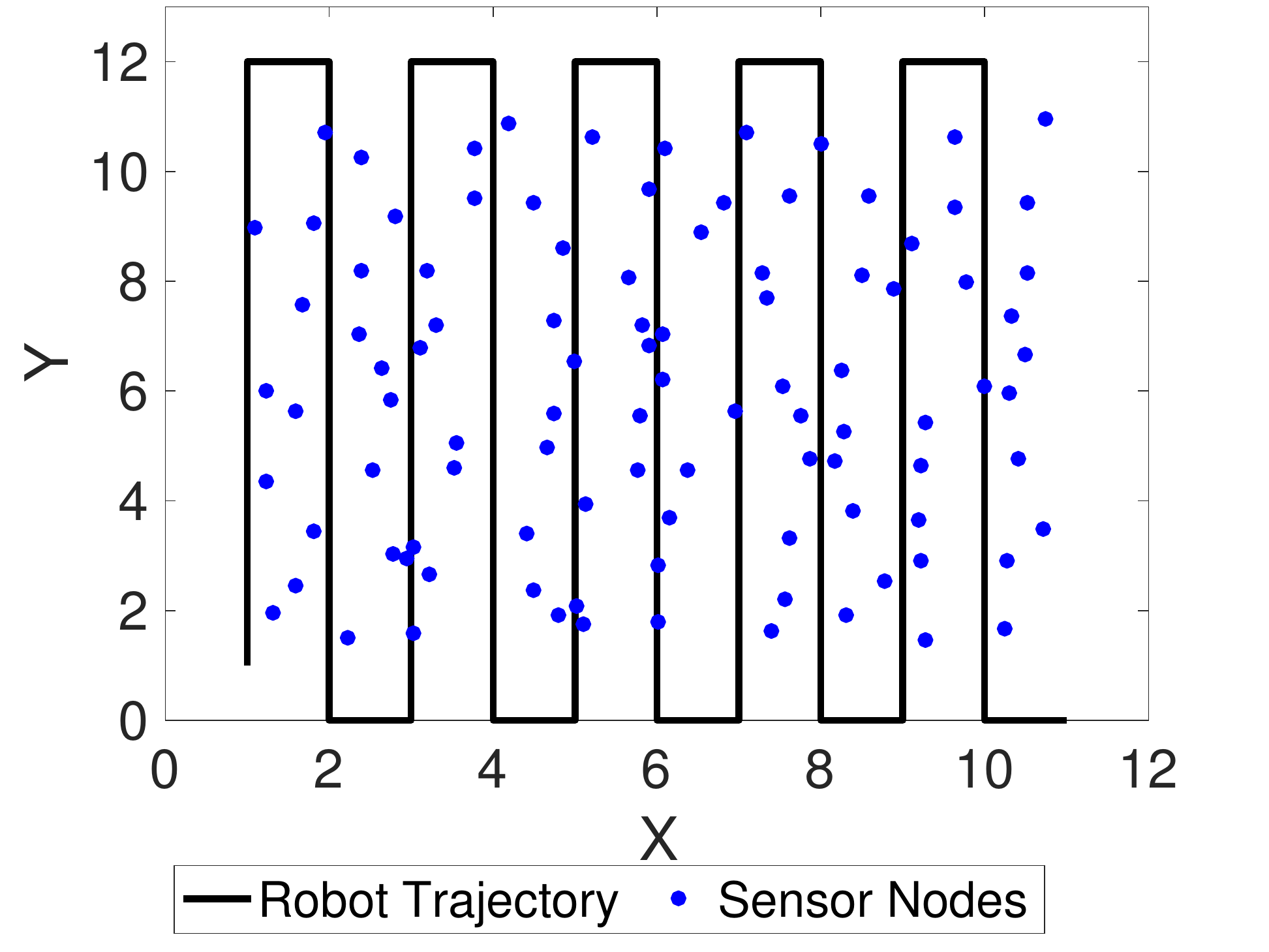}
   }
  \quad
   \subfigure[Hilbert curve]{
  \includegraphics[width=0.45\textwidth]{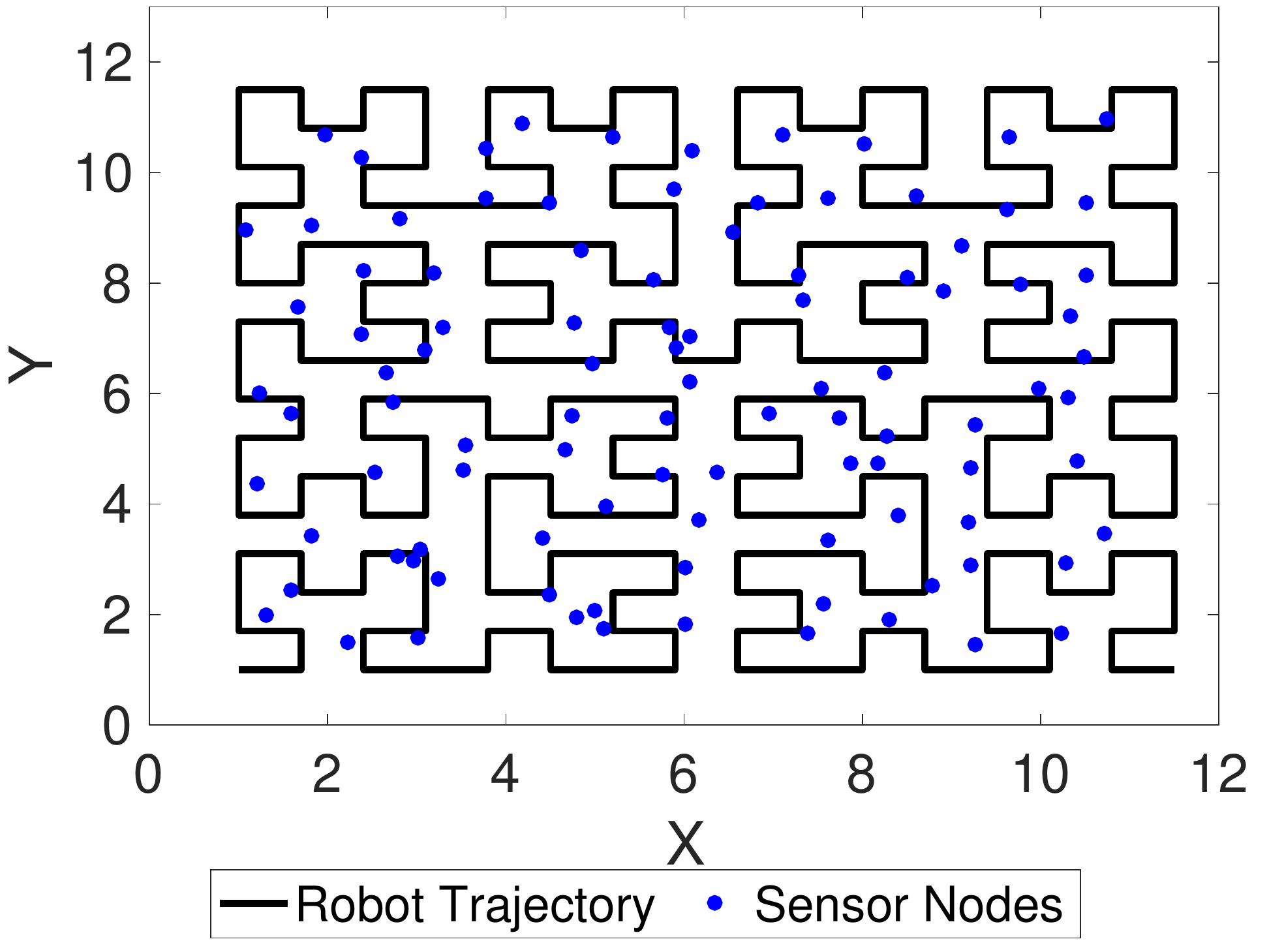}
   }
 \caption{Different robot trajectories}\label{fig::mltm:drt}
\end{figure}
In this section, the performance of the ML-TM algorithm with different robot trajectories is examined. The aim of this finding is to choose an optimal general trajectory for our proposed algorithm. The optimum trajectory is define as the one that covers the network entirely during the shortest possible time while avoiding the obstacles and provides information to calculate ML-TM accurately. Among the proposed paths for robot assisted localization in the literature \cite{robotpath, robotpath2d}, there are four common approaches namely, random walk, spiral, 'S' shape and Hilbert curve. Performance of ML-TM is examined with these four robot trajectories and the results are presented in Table \ref{tab::mltm:rtajectory}. For the evaluation, a $10m\times 10m$ network is considered with 100 nodes, each with communication range $3m$, distributed randomly in the environment. The four robot trajectories are shown in the Figure \ref{fig::mltm:drt}. 

From the Table \ref{tab::mltm:rtajectory}, it can be seen that the performance of the algorithm with different robot trajectories are nearly same. However, Hilbert curve trajectory takes more time compared to other trajectory types. With random walk, it is hard to predict a constant time value required to cover the entire network because the robot's turns are uncertain. Among 'S' shape and spiral trajectories, the performance accuracy is almost same, but 'S' shape covers the network with less time. Hence, the most effective path from the four trajectories that have examined in this chapter is 'S' shape in regards of shortest travelling time and accurate localization.
\begin{table}
\begin{center}
\caption{\textsc{Performance of ML-TM and Time Required to Cover the Network with Different Robot Trajectory Patterns}}\label{tab::mltm:rtajectory} 
  \begin{tabular}{|c| c|  c|  }
    \hline
   Trajectory Type  &\textbf{$E_{total}$} & Trajectory Length  \\ \hline
     Random walk&7\%&128$\pm$ 20s\\
    Spiral&6\%&133s \\ 
    'S' shape&6\%&130s\\ 
    Hilbert curve&5\%&256s\\ 
    \hline
  \end{tabular}
\end{center}
\end{table}
\SetKwProg{Fn}{Function}{ :}{end}

Then the next task was to automate the 'S' shape robot trajectory according to the proposed ML-TM algorithm requirements. The pseudo code of the automated robot trajectory algorithm is illustrated in the Appendix \ref{PC}. The main objective of the robot is to avoid obstacles and cover the entire network with least possible time. However, the robot is unaware about the network dimensions or the details of the obstacles. The only information that robot knows beforehand is the total number of nodes in the network. Thus, the decision about the obstacles are taken using mounted sensors such as Infra-Red(IR). Moreover, robot's moving angles and trajectory termination depends on the detected obstacles and packet reception of the nodes. In other words, robot terminates the information gathering when it receives $N_P$ number of packets from each node in the network. This  $N_p$ values is a predefined values and in the following performance evaluation section it is defined as 5. Also, robot makes a turn when obstacle detected or when it does not hear from any of the sensor nodes in the network.

\section{Performance Evaluation}\label{sec::mltm:result}
The performance of the proposed ML-TM algorithm is evaluated in this section. First the sensitivity of the algorithm is analysed and then the accuracy of the algorithm is compared with existing algorithms. MATLAB simulation software was used for the computations and the simulation setup is described below.

\subsection{Simulation Setup}\label{Preliminaries}
To emulate real communication links between sensor nodes and the robot, the following received signal model is used. The received signal strength has two components, namely path loss component and shadowing component \cite{propagation, Pathirananew1}. The commonly used propagation model of RF signals incorporating path loss and shadowing is given in equation (\ref{eqn::mltm:pm1}).   
\begin{equation}\label{eqn::mltm:pm1}
P_{rx,i} (t_k)= P_{tx,j}-10\varepsilon log d_{ij} (t_k)+ X_{i,\sigma } (t_k)                             
\end{equation}
where, the received signal strength at node $s_i$ at time $t_k$ is $P_{rx,i} (t_k)$, the transmitted signal strength of the signal at node $s_j$ is $P_{tx,j}$ , the path-loss exponent is $\varepsilon$, the distance between node $s_i$ and node $s_j$ at time $t_k$ is $d_{ij} (t_k)$, and the logarithm of shadowing component with a $\sigma $ standard deviation on node $s_i$ at time $t_k$ is $X_{i,\sigma }(t_k) $. 

However, this model is not suitable for a network with some obstacles. The MultiWall-Multifloor Model for RF communication proposed in \cite{mwm} does not consider the variation of the absorption with the thickness of the medium which signal traverses through. Therefore we updated equation (\ref{eqn::mltm:pm1}) by using the Lambert-Bouguer law.  Let $L_{ob,i}(t_k) $ be the loss due to signal absorption from obstacles that exist in the line of sight of node $s_i$ and $s_j$ at time $t_k$. Then the RF signal propagation model can be rewritten as in equation (\ref{eqn::mltm:pm2}). 
\begin{equation}\label{eqn::mltm:pm2}
P_{rx,i} (t_k)= P_{tx,j}-10\varepsilon log d_{ij} (t_k) - L_{ob,i}(t_k) +X_{i,\sigma } (t_k)                             
\end{equation}
The absorption coefficient and the thickness of the obstacle medium, which signal traverses through are $\alpha$ and $d_o (t_k)$ respectively. Then $L_{ob,i}(t_k) $ can be calculated as, 
\begin{equation}
L_{ob,i}(t_k) = \Sigma _{k=1}^{\mathcal{N}_{ij}}10\alpha d_o (t_k)log(e) 
\end{equation}
where, $\mathcal{N}_{ij} $ is the number of obstacles that exist in between node $s_i$ and node $s_j$, and e is the exponent. 

In different environmental situations, the path loss exponent ($\varepsilon$) and log-normal shadowing standard deviation ($\sigma$) are different \cite{nconstant, nconstant2}. Thus, to evaluate the performance of proposed algorithm in different environmental situation, three different scenarios have considered, namely, a suburban area, heavy tree density area and a light tree density area. The simulation parameters used in this chapter are summarized in Table \ref{tab::mltm:simpara}.

\begin{table}
\renewcommand{\arraystretch}{1.3}
\caption{\textsc{Simulation Parameters}}\label{tab::mltm:simpara} 
\center
  \begin{tabular}{| c | c |}
    \hline
     \textbf{Parameter} & \textbf{Value} \\ \hline
    Transmitted power & -50dB\\ \hline
    Sensitivity & -90dB\\ \hline
    Communication radius & 10m\\ \hline
    Suburban area  & $\varepsilon$ = 2.7, $\sigma$ =9.6\\\hline
    Heavy tree density area & $\varepsilon$ = 4.6, $\sigma$ =10.6\\
    \hline
    Light tree density area & $\varepsilon$ = 3.6, $\sigma$ =8.2\\\hline
  \end{tabular}
\end{table}

\subsection{Sensitivity of the Algorithm}
\begin{figure}[!b]
  \centering
    \includegraphics[width=0.5\textwidth]{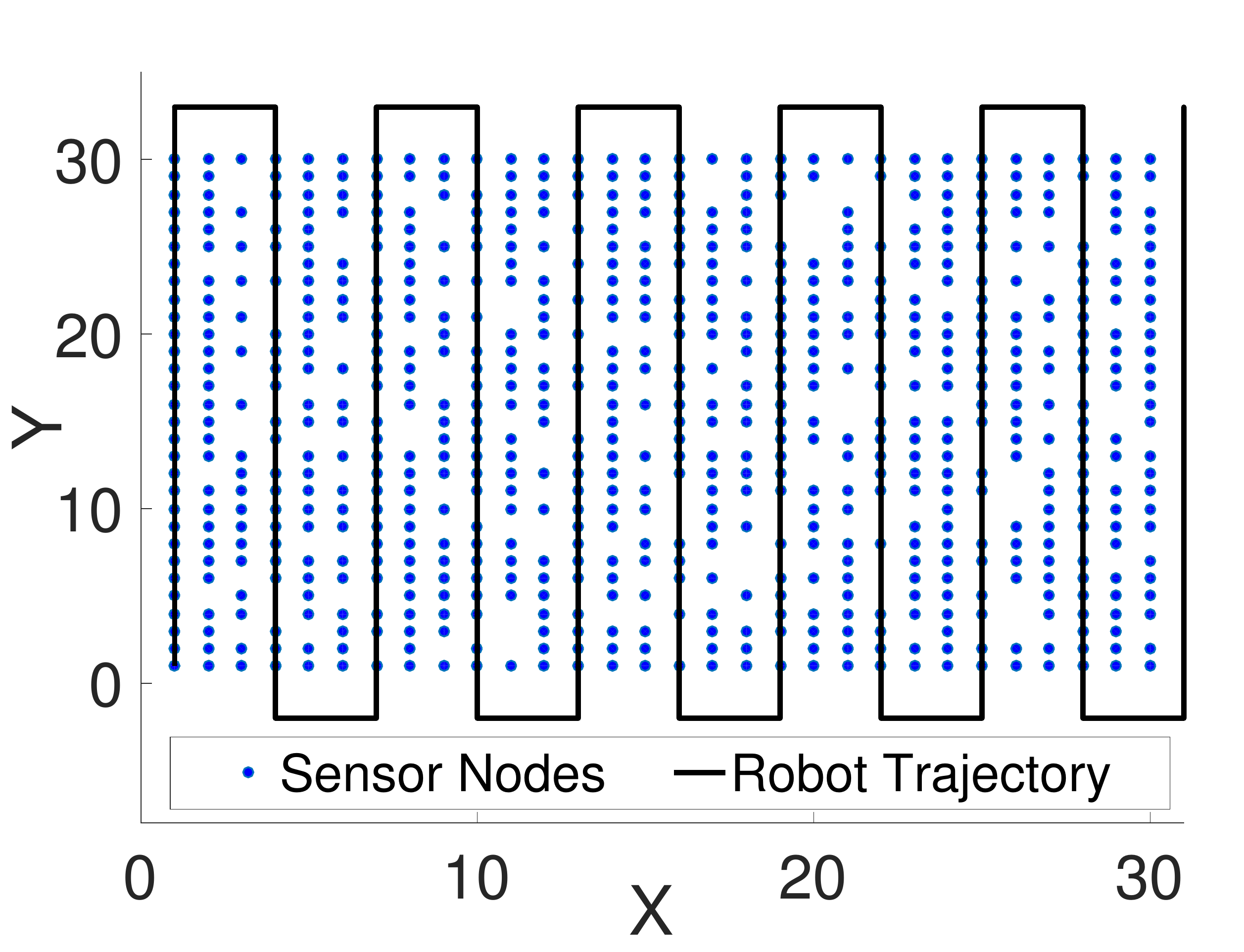}
    \caption{Sensor node distribition}\label{fig::mltm:netwrk}
\end{figure}
As the proposed algorithm depends on the packet receiving probability function described in Section \ref{subsec::mltm:pfn}, the sensitivity of the algorithm against its parameters i.e.$p_0$ and $R$, is considered in this section. To examine the sensitivity, the output of the algorithm was recorded by changing the values of $p_0$ and $R$. Also, the effect of these parameters in different network scenarios is analysed. The node distribution with the robot trajectory considered for this evaluation is shown in Figure \ref{fig::mltm:netwrk}. For the simulation, three cases with three different environment setups were considered\\
\textbf{Case 1:} Suburban environment \\
\textbf{Case 2:} Heavy tree density environment \\
\textbf{Case 3:} Light tree density environment \\
\begin{figure}
 \centering
 \subfigure[Case 1 ]{
  \includegraphics[width=0.45\textwidth]{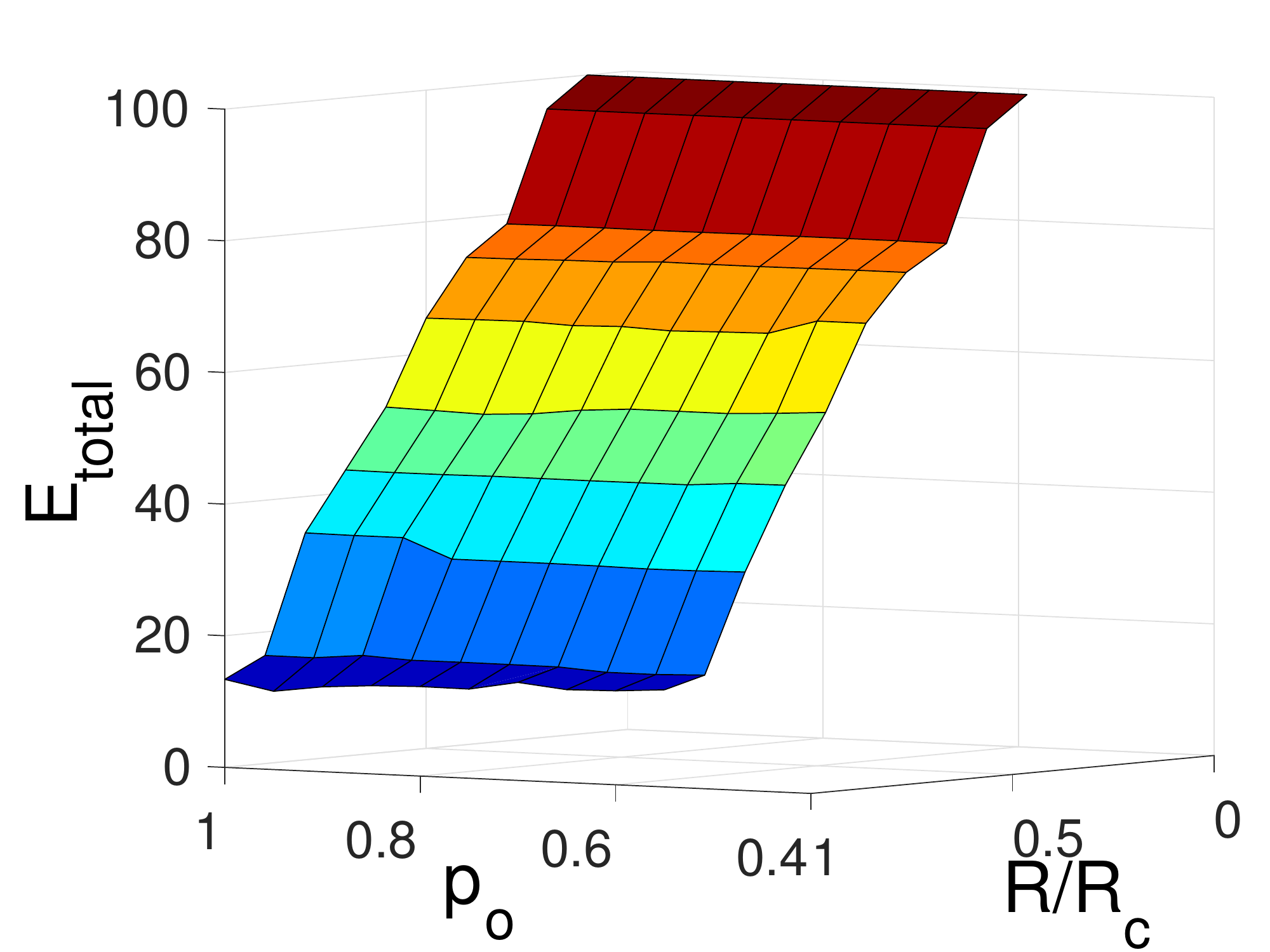}
   \label{fig::mltm:n27}
   }
   \quad
 \subfigure[Case 2]{
  \includegraphics[width=0.45\textwidth]{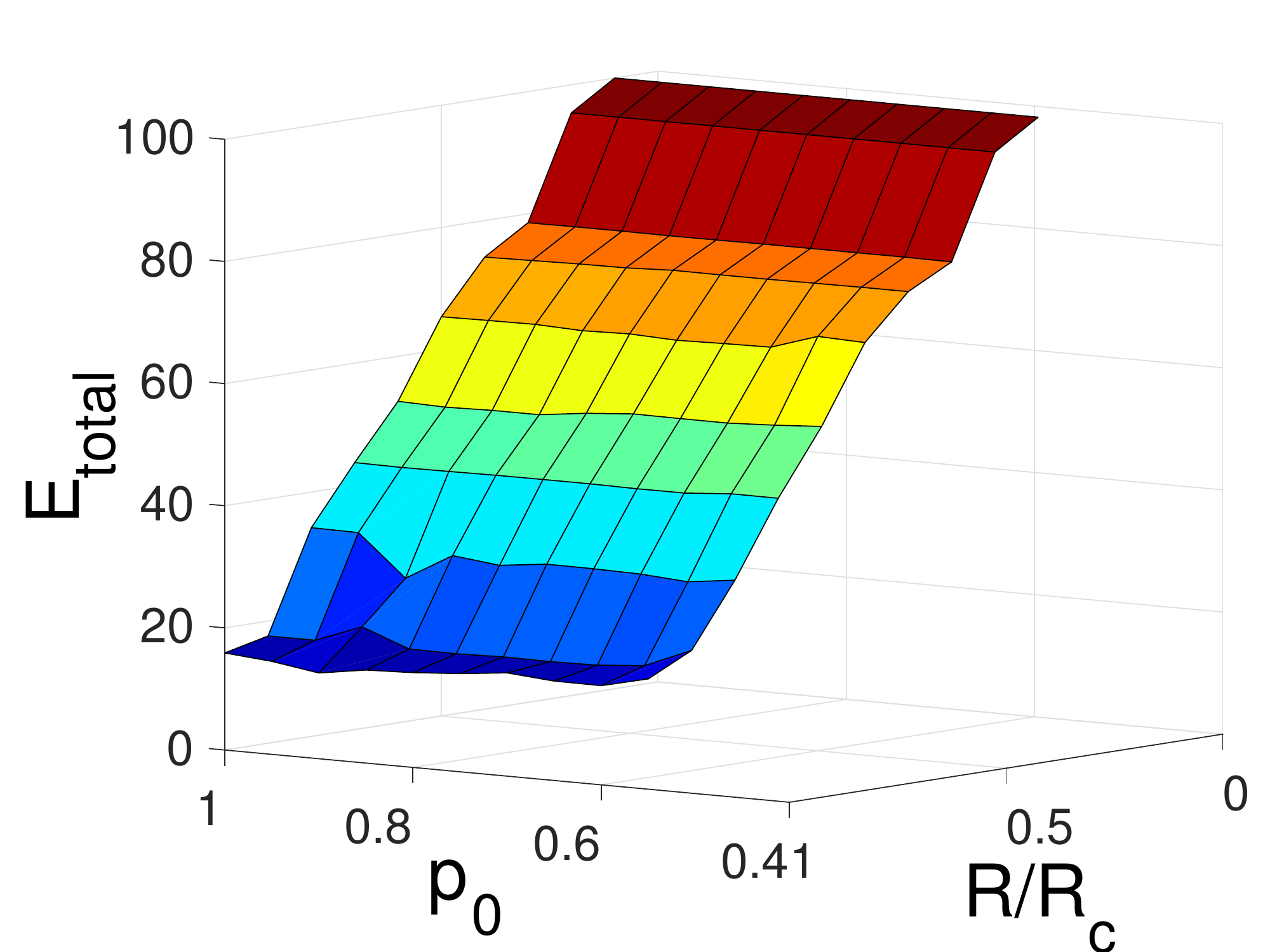}
   \label{fig::mltm:n4}
   }

 \subfigure[Case 3]{
  \includegraphics[width=0.45\textwidth]{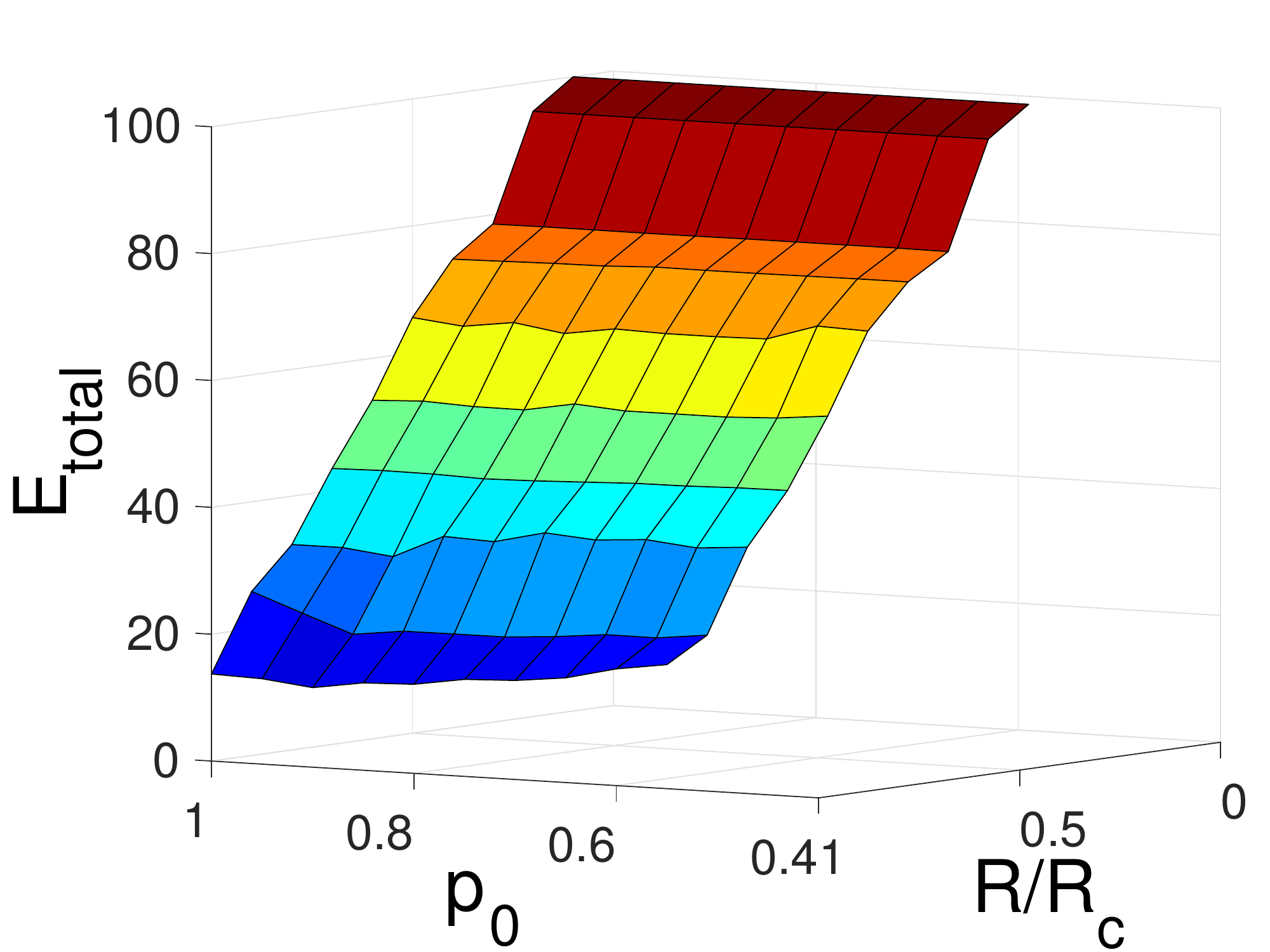}
   \label{ffig::mltm:n5}
   }
 \caption{$E_{total}$ distrubution against different $p_0$ and R values for different environments} \label{fig::mltm:nvaries}
\end{figure}
Figure \ref{fig::mltm:nvaries} shows the $E_{total}$ values obtained for the above three cases while changing the network parameters. Also, Table \ref{tab::mltm:sensitivty} presents the best $p_0$ and R values corresponding to case when $E_{total}$ is minimum, for the three cases. For the three cases the best R value is one and $p_0$ varies from 0.9 to 0.95. Furthermore, it can be seen that, for case 2 and case 3, best $p_0$ value is 0.9. The reason is, in those two cases, the disturbance introduced by the environment is high. Therefore signal receiving probability is less compared to that in Case 1. Hence, the selection of $p_0$ depends on the environmental factors. Thus, when the environment is noisy and having disturbances, $p_0$ should be closer to 0.9, otherwise it is closer to 0.95. 
\begin{table}
\begin{center}
\caption{\textsc{Best $p_0$ and R Values for the Three Cases}}\label{tab::mltm:sensitivty} 
  \begin{tabular}{|c| c | c | c|}
    \hline
     &\textbf{Case 1} & \textbf{Case 2} & \textbf{Case 3} \\ \hline
    $p_0$&0.95&0.9&0.9\\ \hline
    $R/R_c$&1&1&1\\
    \hline
  \end{tabular}
 
\end{center}
\end{table}
\subsection{Accuracy Comparison}
\begin{figure}[b!]
 \centering
 \subfigure[Physical map]{
  \includegraphics[width=0.45\textwidth]{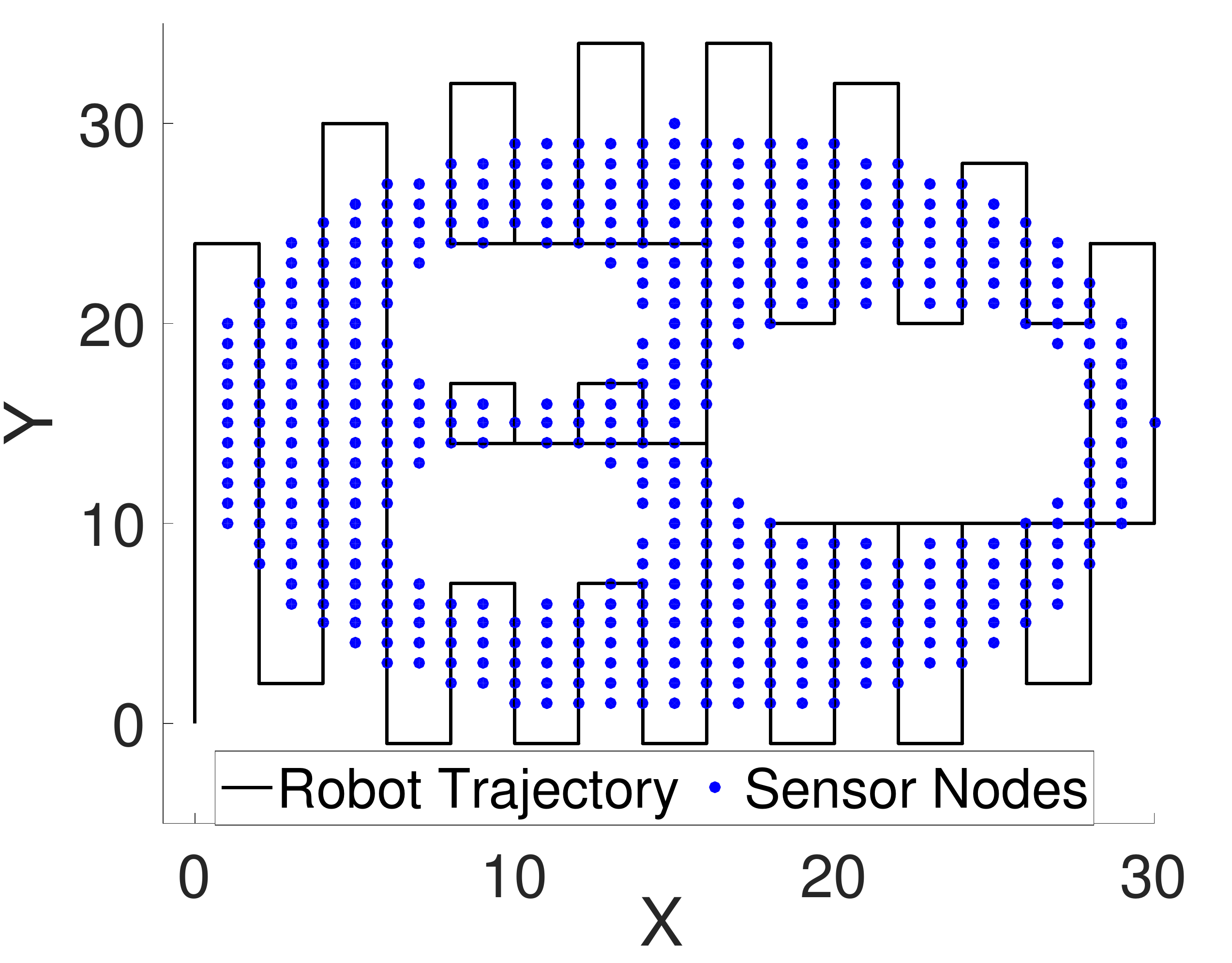}
   \label{fig::mltm:circlen}
   }
   \quad
 \subfigure[ML-TM map]{
  \includegraphics[width=0.45\textwidth]{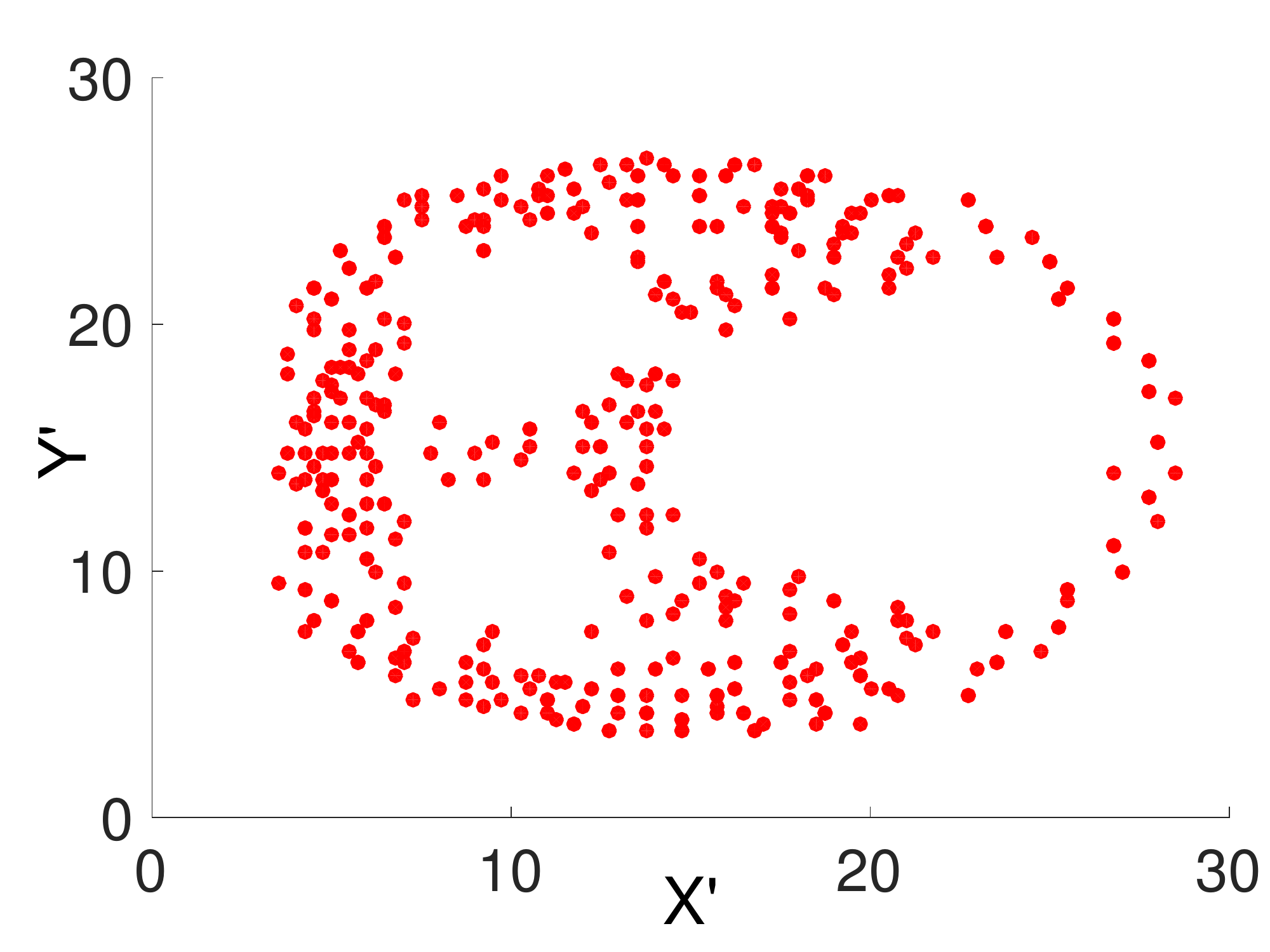}
   \label{fig::mltm:circlep}
   }
  
 \subfigure[SVD based TPM]{
  \includegraphics[width=0.45\textwidth]{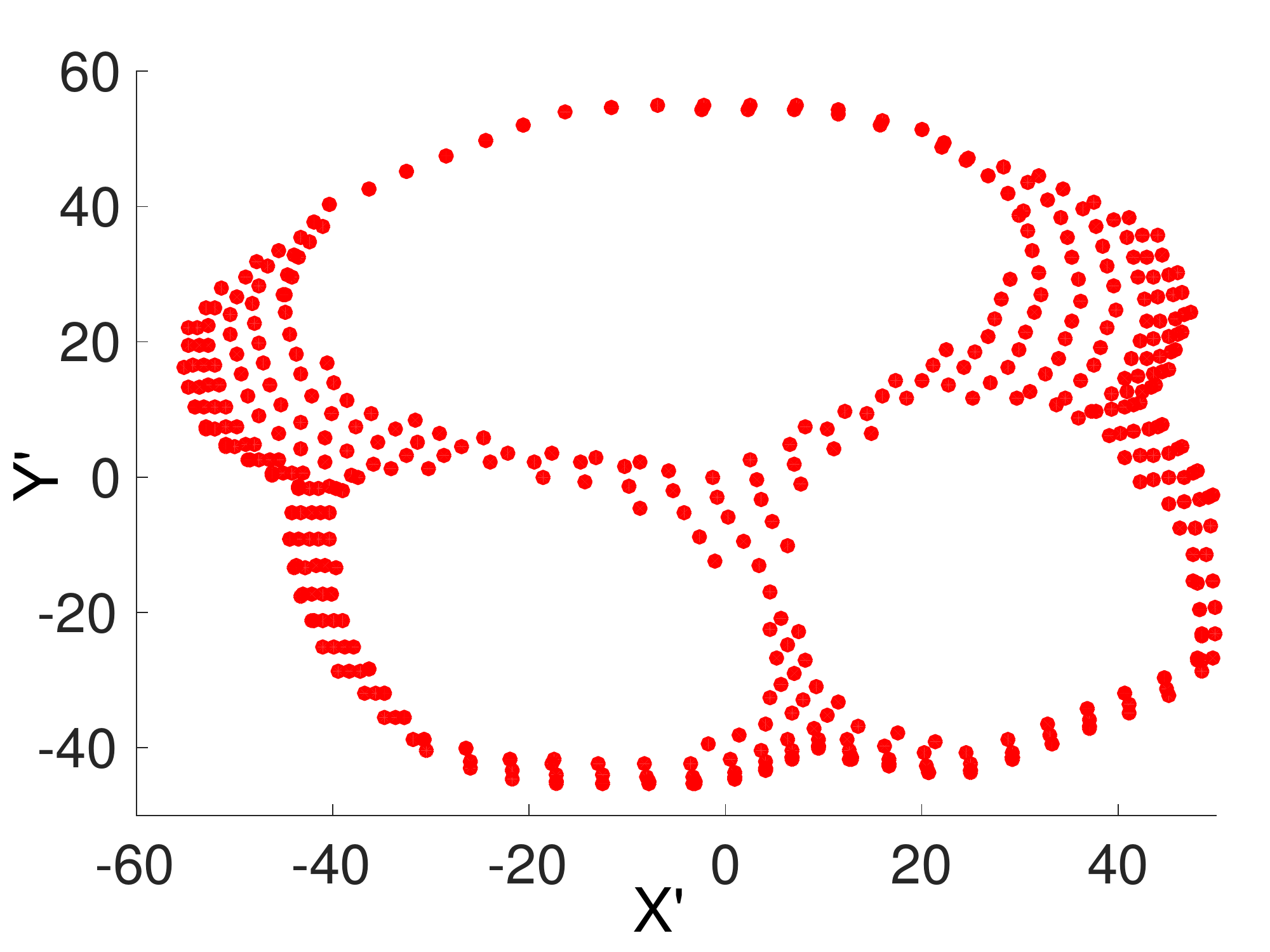}
   \label{fig::mltm:circlevc}
   }
   \quad
   \subfigure[RSSI based map]{
  \includegraphics[width=0.45\textwidth]{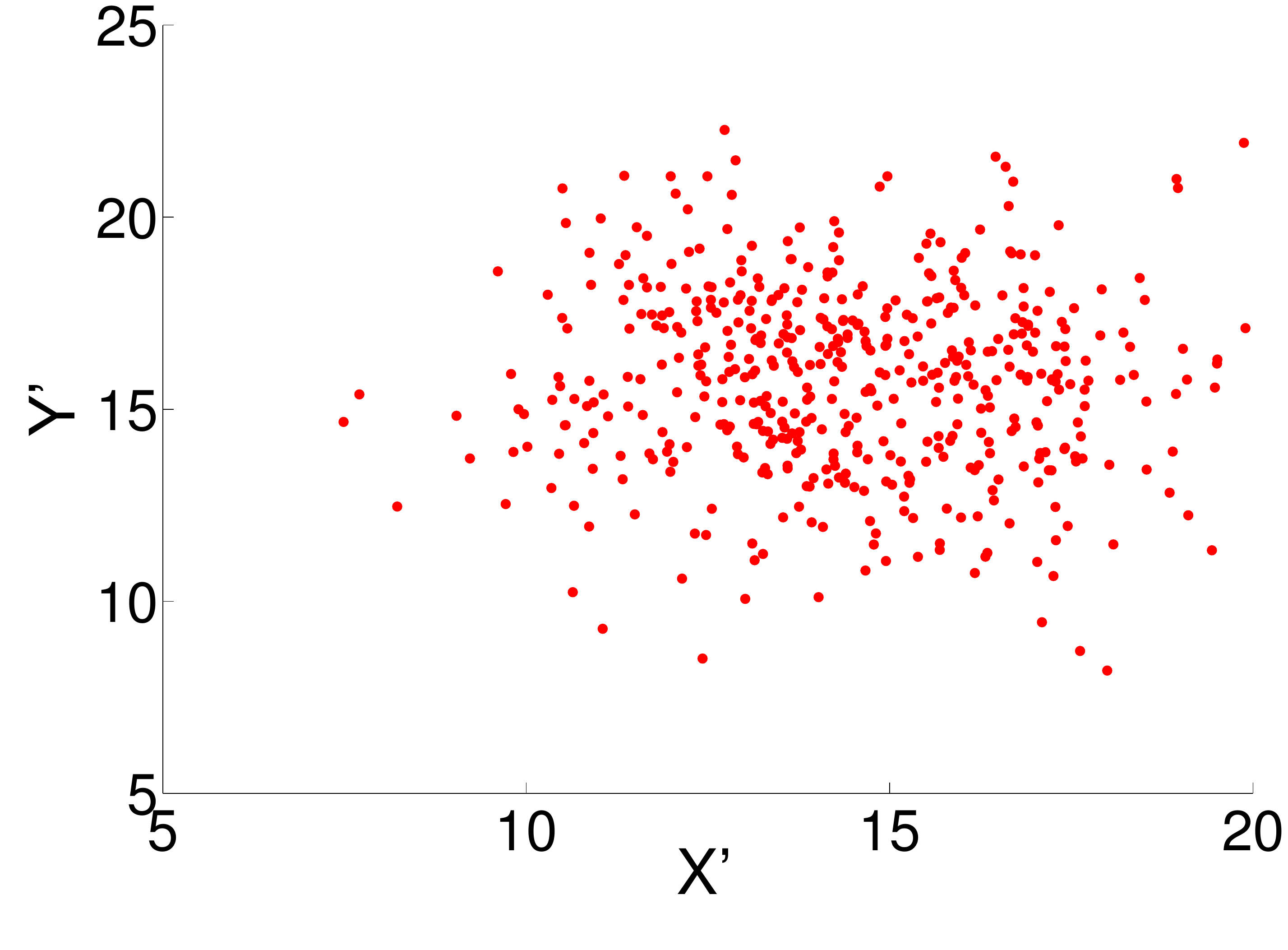}
   \label{fig::mltm:circlrrssi}
   }
 \caption{Circular-shaped network with 496 nodes} \label{fig::mltm:circle}
\end{figure}
The accuracy of the proposed ML-TM is compared with the SVD based TPM \cite{VC} and RSSI location method based on the Triangle Centroid Localization \cite{rssidirect}. SVD based TPM was selected, as it is the most recent and relevant work to the proposed algorithm. Also, to compare our results with range-based algorithm, RSSI based localization was selected. By this comparison, it can be seen that how ML-TM method eliminates the error due to RF communication effects (i.e. noise, fading etc.) as well as anchor selection. 

\begin{figure}[h!]
 \centering
 \subfigure[Physical map]{
  \includegraphics[width=0.45\textwidth]{Fig/C3/sparsenetwork.pdf}
   \label{fig::mltm:sparsen}
   }
   \quad
 \subfigure[ML-TM map]{
  \includegraphics[width=0.45\textwidth]{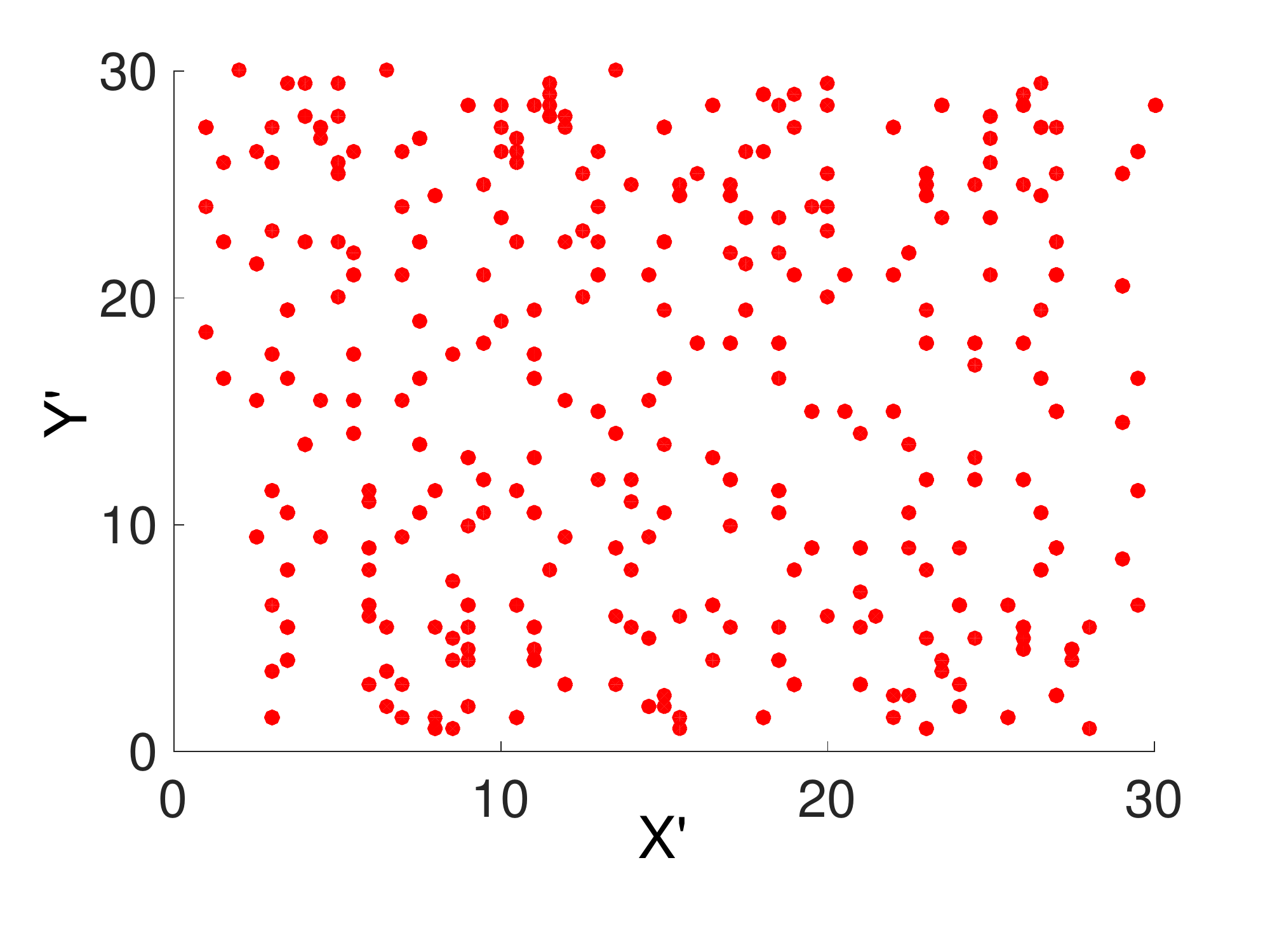}
   \label{fig::mltm:sparsep}
   }
  
 \subfigure[SVD based TPM]{
  \includegraphics[width=0.45\textwidth]{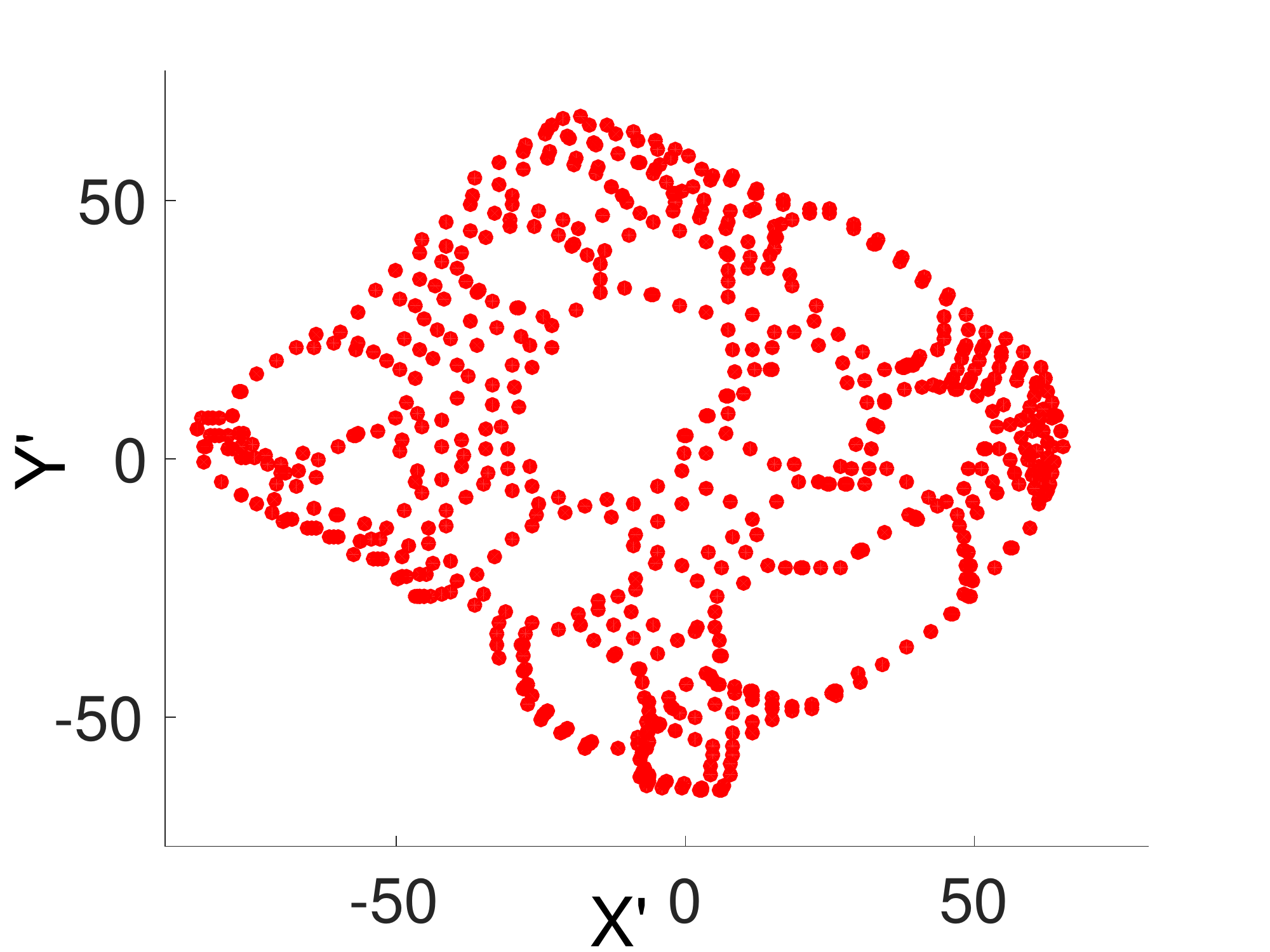}
   \label{fig::mltm:sparsevc}
   }
   \quad
   \subfigure[RSSI based map]{
  \includegraphics[width=0.45\textwidth]{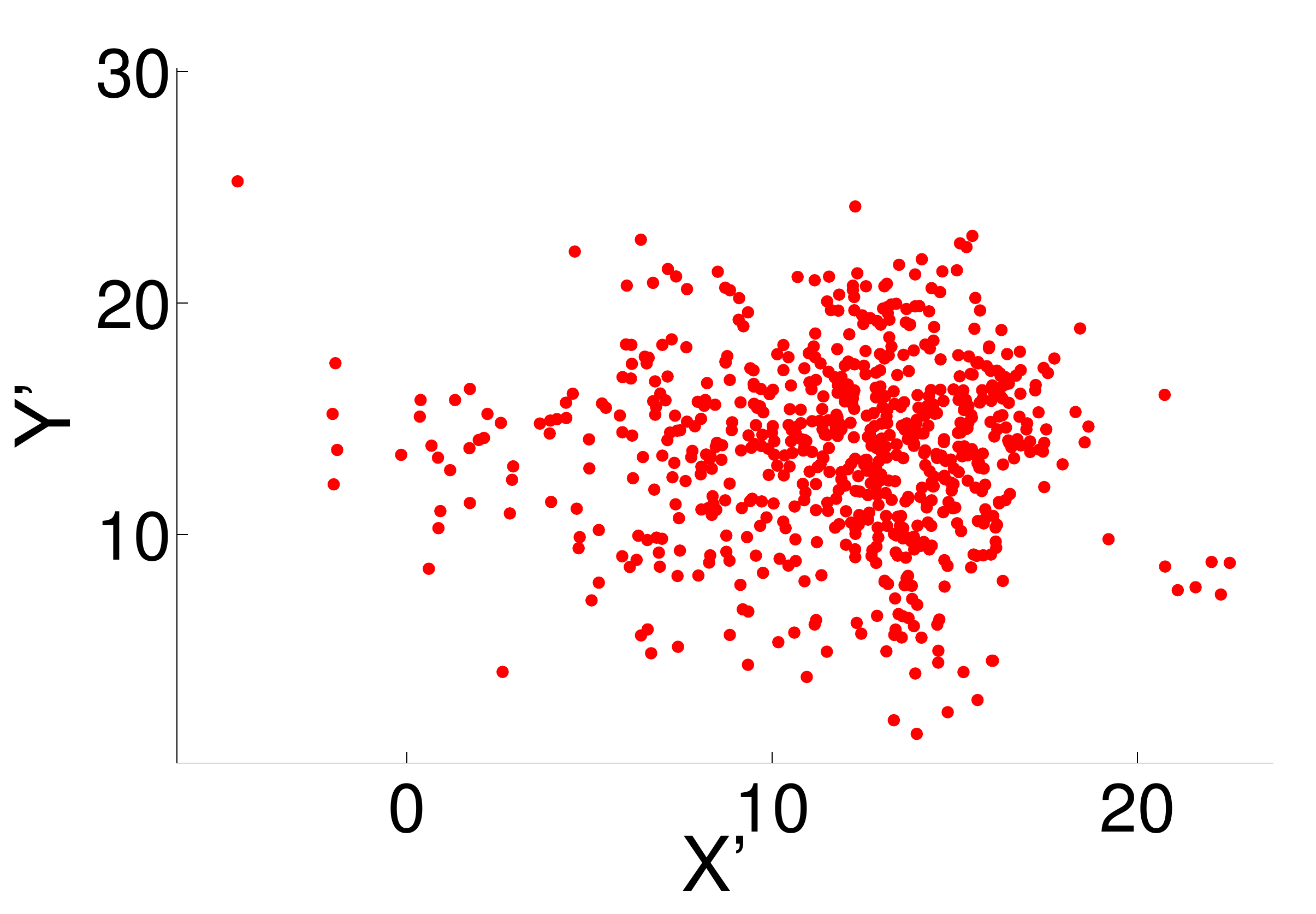}
   \label{fig::mltm:sparserssi}
   }
 \caption{Sparse grid network with 700 nodes} 
 \label{fig::mltm:sparse}
\end{figure}

\begin{figure}[h!]
 \centering
 \subfigure[Physical map]{
  \includegraphics[width=0.45\textwidth]{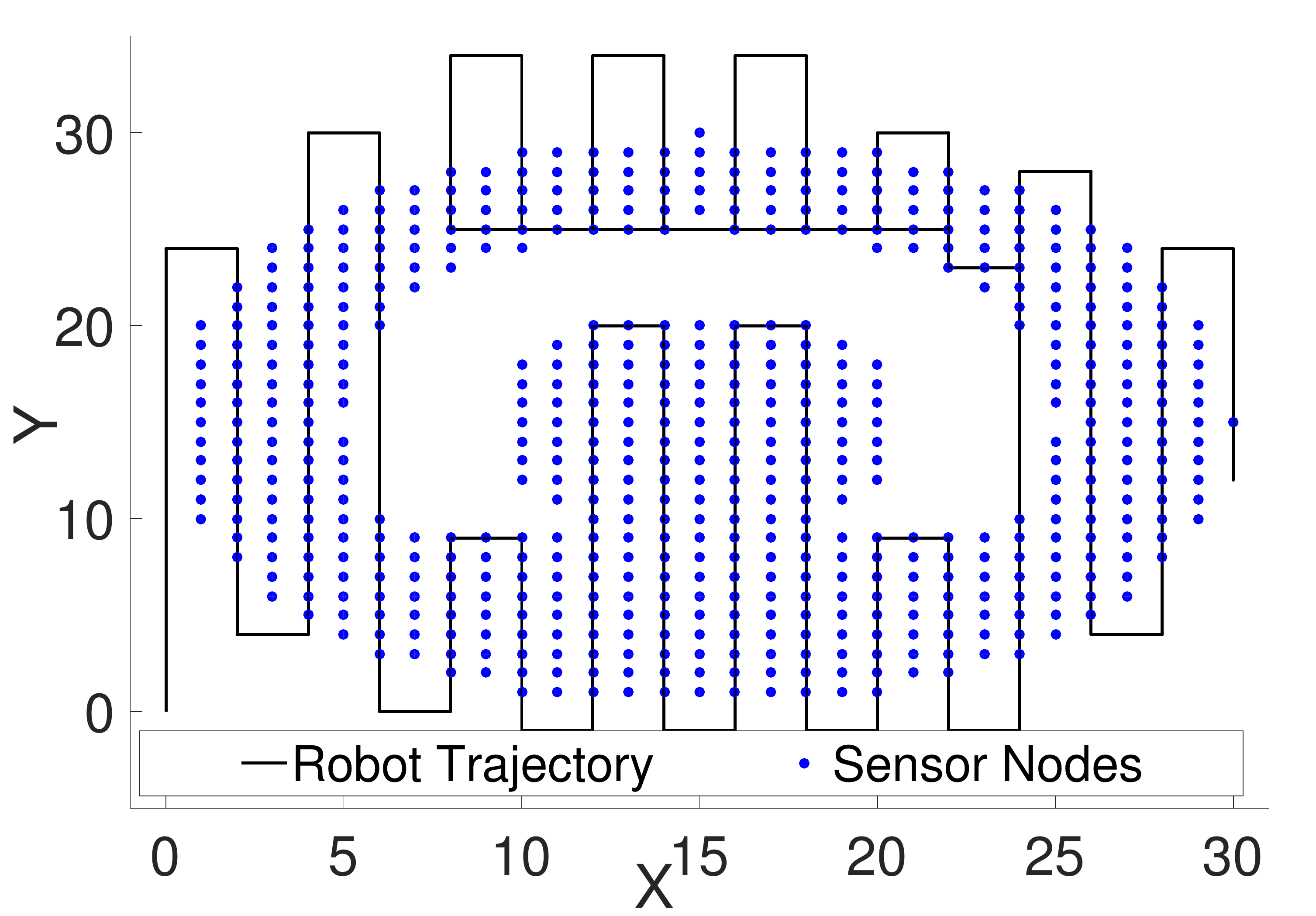}
   \label{fig::mltm:cvn}
   }
   \quad
 \subfigure[ML-TM map]{
  \includegraphics[width=0.45\textwidth]{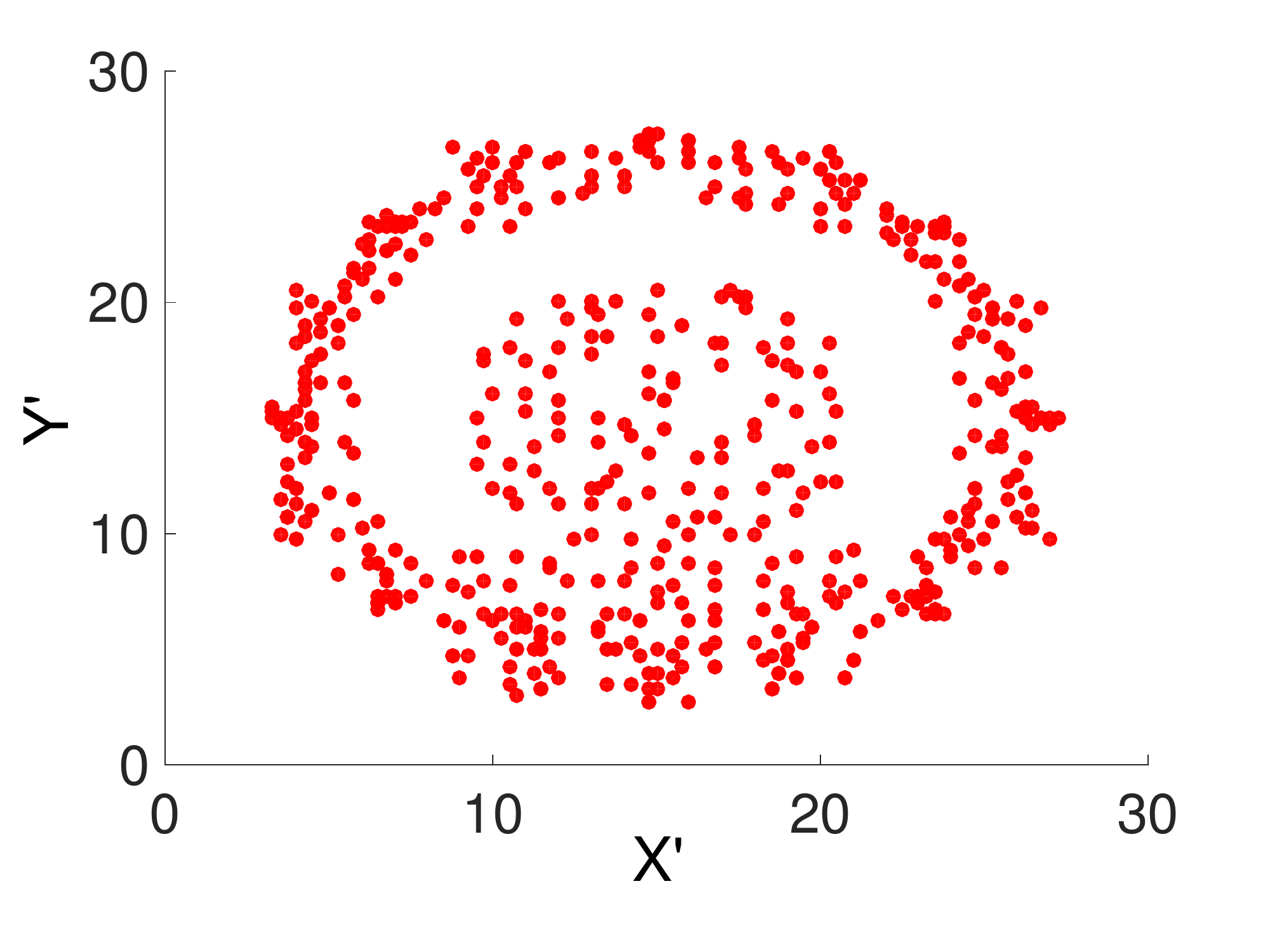}
   \label{fig::mltm:cvp}
   }
  
 \subfigure[SVD based TPM]{
  \includegraphics[width=0.45\textwidth]{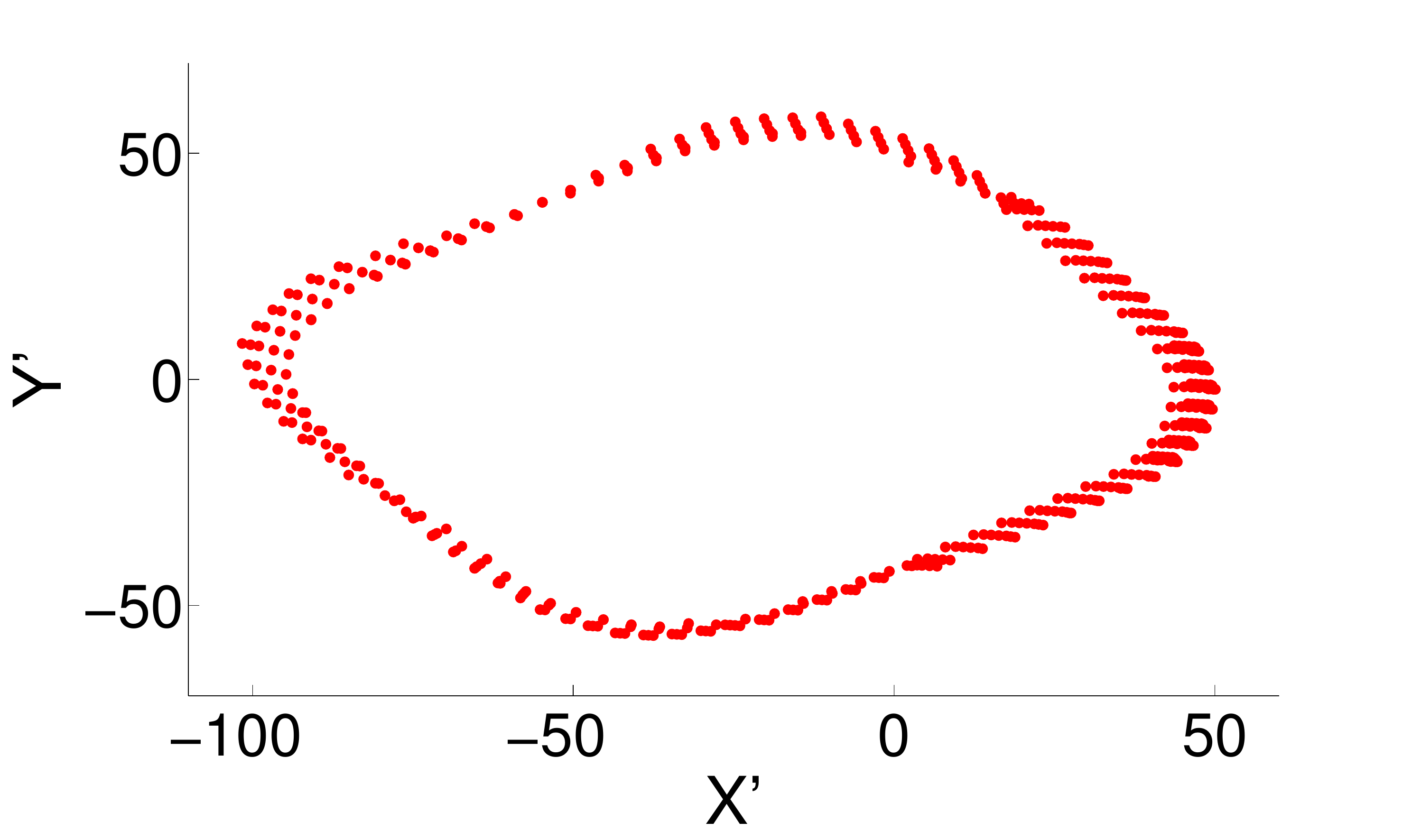}
   \label{fig::mltm:cvvc}
   }
   \quad
   \subfigure[RSSI based map]{
  \includegraphics[width=0.45\textwidth]{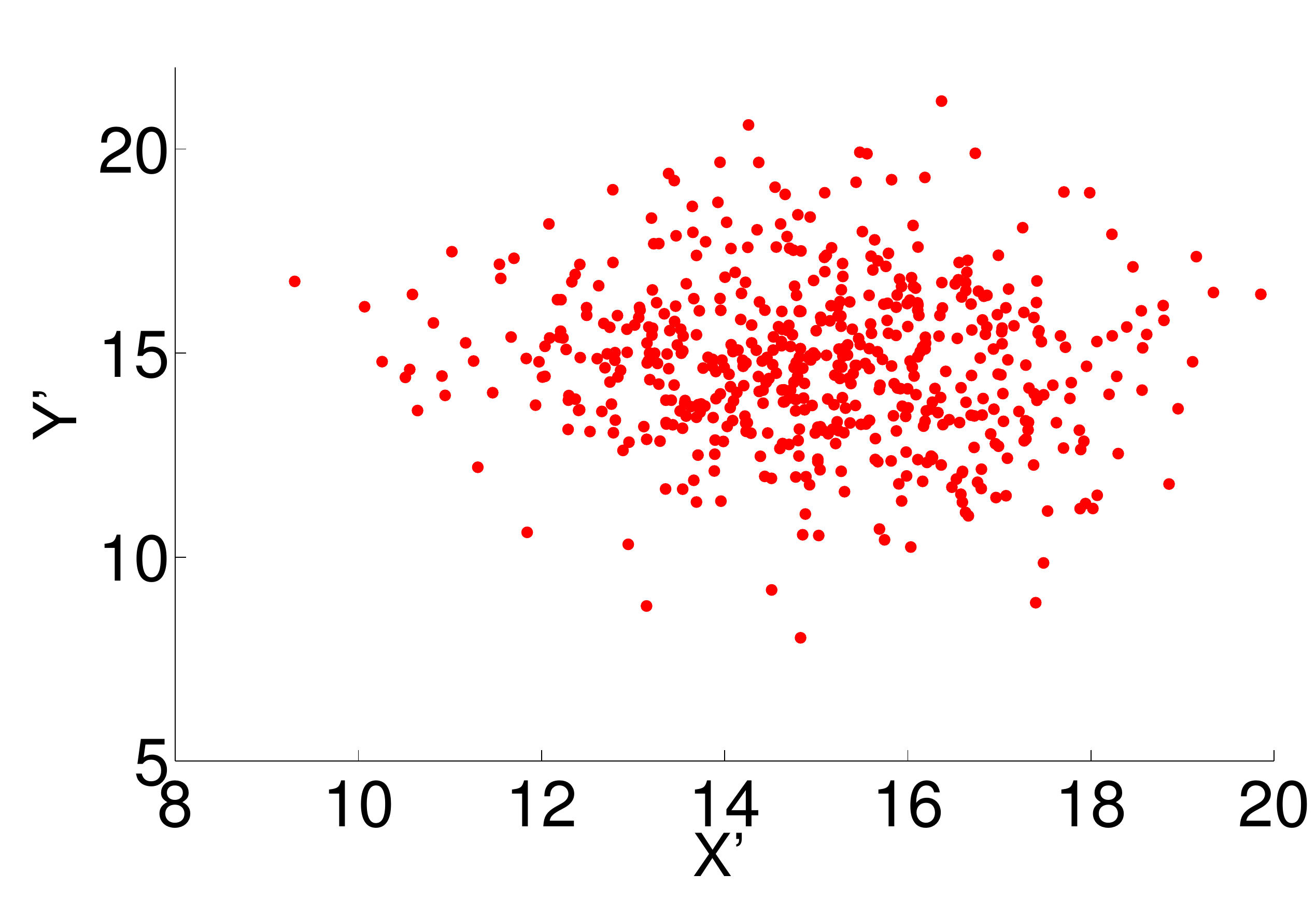}
   \label{fig::mltm:cvrssi}
   }
 \caption{ Concave void  network with 554 nodes}\label{fig::mltm:cv}
\end{figure}

The physical maps that selected for the comparison are shown in Figure \ref{fig::mltm:circlen} -\ref{fig::mltm:cvn}. Different shapes of networks deployed in different environmental conditions with and without obstacles are considered in this simulation. Figure \ref{fig::mltm:circlen} is a circular-shaped network in a suburban area with three physical obstacles (e.g., concrete barriers) and 496 sensor nodes. Figure \ref{fig::mltm:sparsen} is a sparse grid deployed in a light tree density area with 700 sensor nodes. Figure \ref{fig::mltm:cvn} is a 554 sensor nodes network with a concave void (e.g. concrete barriers) in a suburban area. 

Figure \ref{fig::mltm:circle} - \ref{fig::mltm:cv}, clearly demonstrate the effectiveness of the proposed ML-TM algorithm. The algorithm requires 444s, 423s and 438s in information gathering phase to move the robot by avoiding the obstacles in network Figures \ref{fig::mltm:circle}-\ref{fig::mltm:cv} respectively with a robot speed of 1$ms^{-1}$. The three networks cover areas of 590$m^2$, 900$m^2$ and 472$m^2$ area respectively. Since the sparse network does not have any obstacles, robot can cover the network with less amount of time. However, without any prior knowledge of geographical information, the generated topological maps have captured physical voids and boundaries of the actual physical network. Results presented above show that in all three simulation setups, the RSSI based algorithm is less accurate in capturing physical shapes and obstacles. SVD based TPM, Figure \ref{fig::mltm:circlevc}-\ref{fig::mltm:sparsevc} captures the shape of the network, but the orientation has been changed. However this can be corrected easily. In Figure \ref{fig::mltm:cvvc} SVD based TPM algorithm has generated a distorted map which does not present the network shape accurately. However, in Figure \ref{fig::mltm:circlep}-\ref{fig::mltm:cvp}, it can be seen that ML-TM is able to capture the shape of the network or obstacle without having any prior knowledge.

Moreover, to check the isomorphism between the actual physical maps and the corresponding topological maps, $E_{total}$ (described in section \ref{sec::mltm:error}) is calculated and presented in Table \ref{tab::mltm:tableerro}. From the results, it can be seen that the number of nodes located in incorrect places is less than or equal to 7\% with ML-TM. Whether obstacles exist or not, the proposed method extracts accurate maps of the networks. 

\begin{table}
\begin{center}
 \caption{\textsc{$E_{total}$ for Topology Maps in Figure \ref{fig::mltm:circle} -\ref{fig::mltm:cv}}}\label{tab::mltm:tableerro}
  \begin{tabular}{| l | c | c | c | }
    \hline
    &  \multicolumn{3}{ |c| }{$E_{total}$} \\ \cline{2-4}
    \textbf{Figure} & \textbf{Proposed } & \textbf{SVD based } & \textbf{RSSI based  } \\ 
    & \textbf{ML-TM} & \textbf{TPM } & \textbf{Map  } \\ \hline
    \ref{fig::mltm:circlen} &5\% & 10\% & 40\% \\ \hline
    \ref{fig::mltm:sparsen} & 7\% & 14\% & 33\% \\ \hline
    \ref{fig::mltm:cvn} & 6\% & 19\% & 50\%\\
    \hline
  \end{tabular}
 
  \end{center}
\end{table}

\subsection{Performance in 3D WSNs}
This section presents the performance results of ML-TM algorithm in 3D networks. While most localization algorithms proposed are applicable only to networks in a 2D plane, i.e. x and y plane, many real world WSN applications require the nodes to be distributed in 3D space requiring an additional axis, height (i.e. z-plane). Thus the simulation was carried out on networks in 3D space to evaluate the performance. 

The topology coordinates of a sensor in 3D space can be calculated without changing the approach of the algorithm. It requires only changing the coordinate system of the equation from 2D to 3D. More precisely, in 3D plane the robot trajectory at time $t_k$ is $(x_R(t_k),y_R(t_k),z_R(t_k))$ and sensor node $s_i$'s location is $(x_i,y_i,z_i)$. Then, equations (\ref{eq::mltm:probability}) and (\ref{eq::mltm:dist}) used to calculate the probability value of obtaining the packet receiving vector of node $s_i$ will change as in equation (\ref{eqn::mltm:probability3d}) and (\ref{eqn::mltm:dist3d}). The remaining calculations are same as in 2D plane and the 3D grid vertex $(x_j^{opt},y_j^{opt},z_j^{opt})$ that delivers the maximum $P_i(x_j, y_j,z_j)$ is selected as the maximum likelihood topology coordinate of sensor node $s_i$. 

\begin{equation}\label{eqn::mltm:probability3d}
P_i(x_j, y_j,z_j) = Z^i(d_{j1})Z^i(d_{j2})...Z^i(d_{jk})...Z^i(d_{jn})
\end{equation}

where
 \begin{equation}\label{eqn::mltm:dist3d}
 d_{jk} := \sqrt{(x_j-x_R(t_k))^2 + (y_j -y_R(t_k))^2+ (z_j -z_R(t_k))^2}
 \end{equation}
 
\begin{figure}
 \centering
 \subfigure[Physical map]{
  \includegraphics[width=0.45\textwidth]{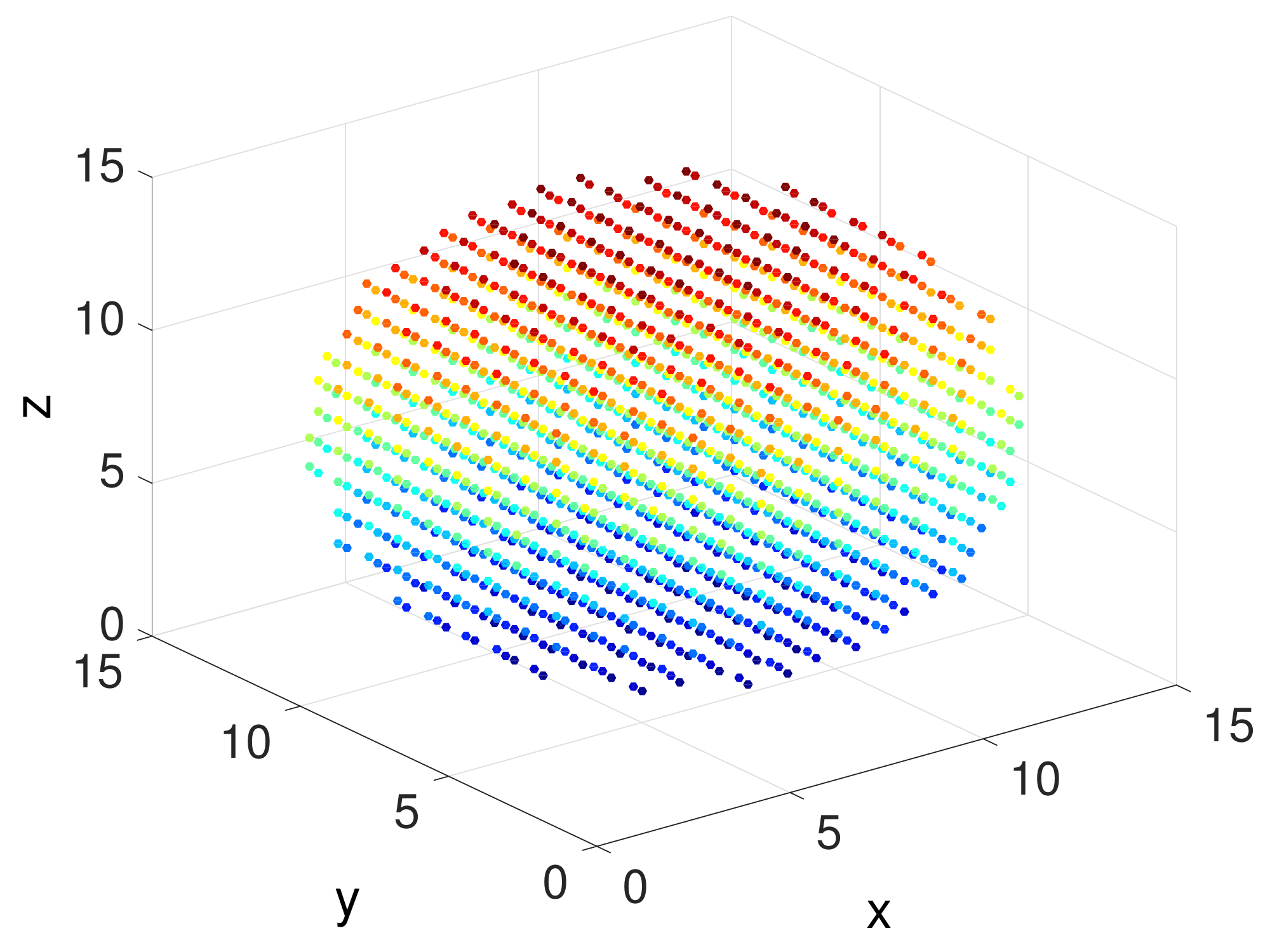}
   \label{fig::mltm:spheren}
   }
    \quad
   \subfigure[Physical map with robot trajectory]{
  \includegraphics[width=0.45\textwidth]{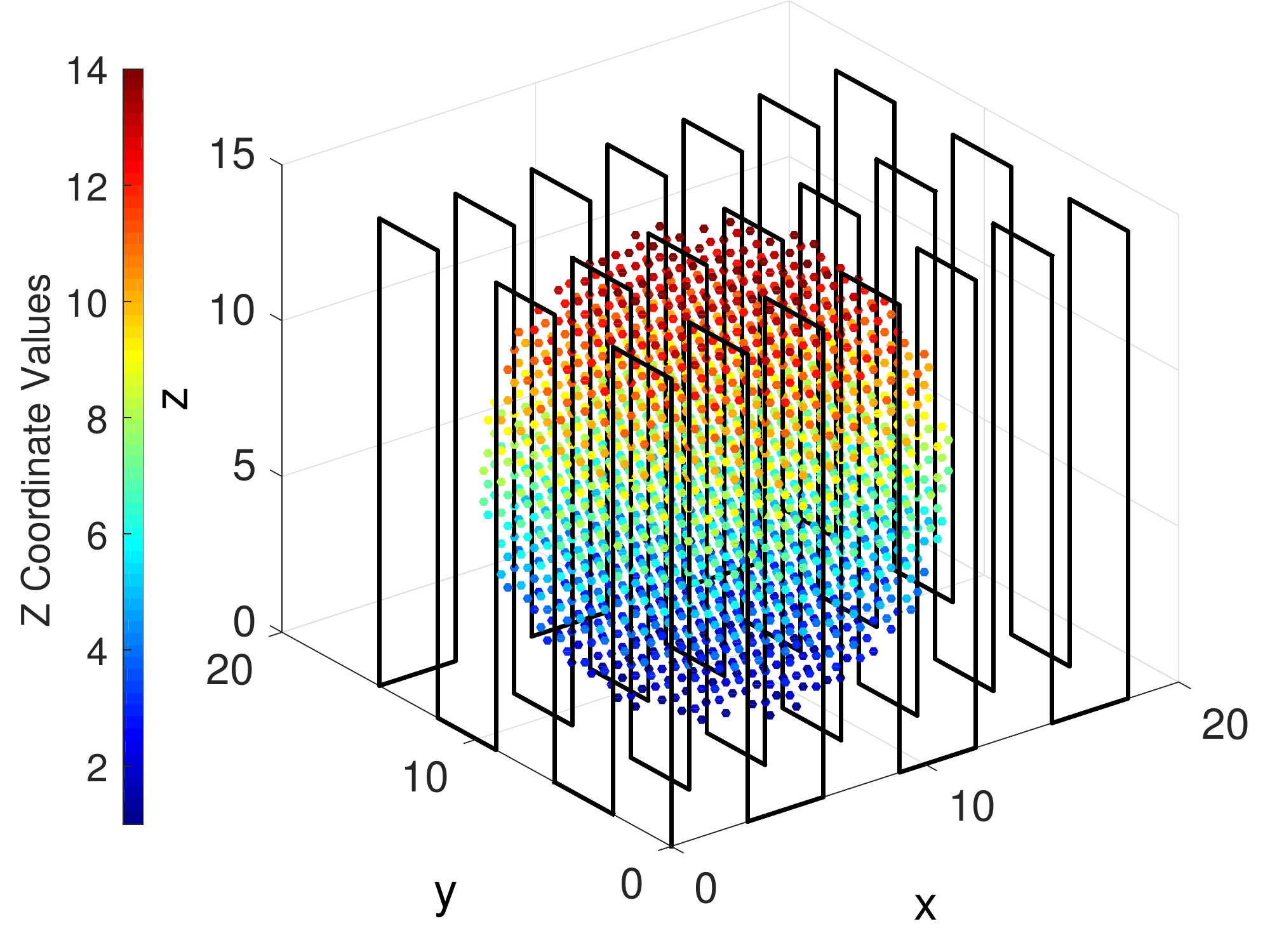}
   \label{fig::mltm:sperenrt}
   }
   \quad
 \subfigure[ML-TM map]{
  \includegraphics[width=0.45\textwidth]{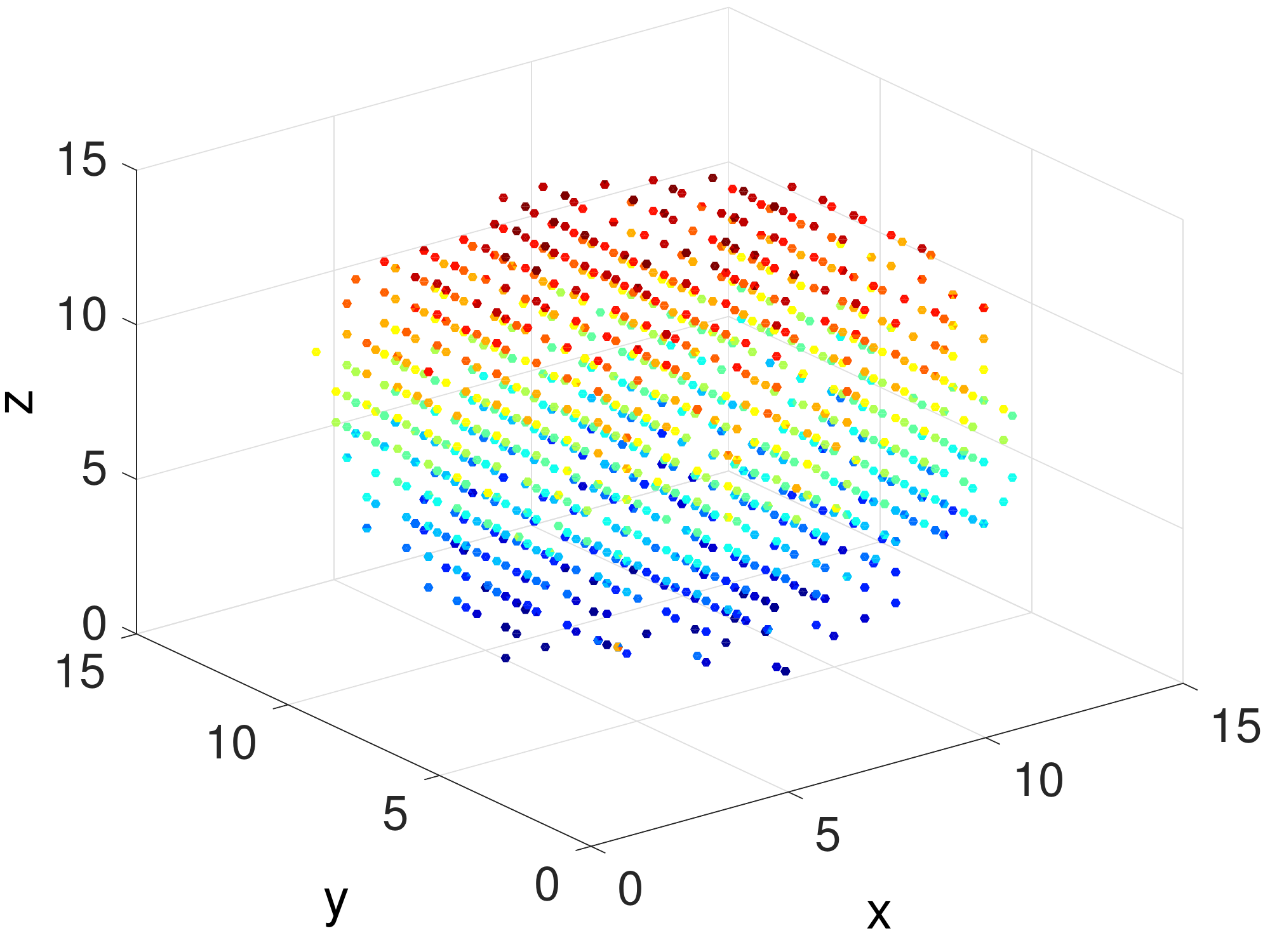}
   \label{fig::mltm:sperep}
   }
      \quad
    \subfigure[Error CDF]{
  \includegraphics[width=0.45\textwidth]{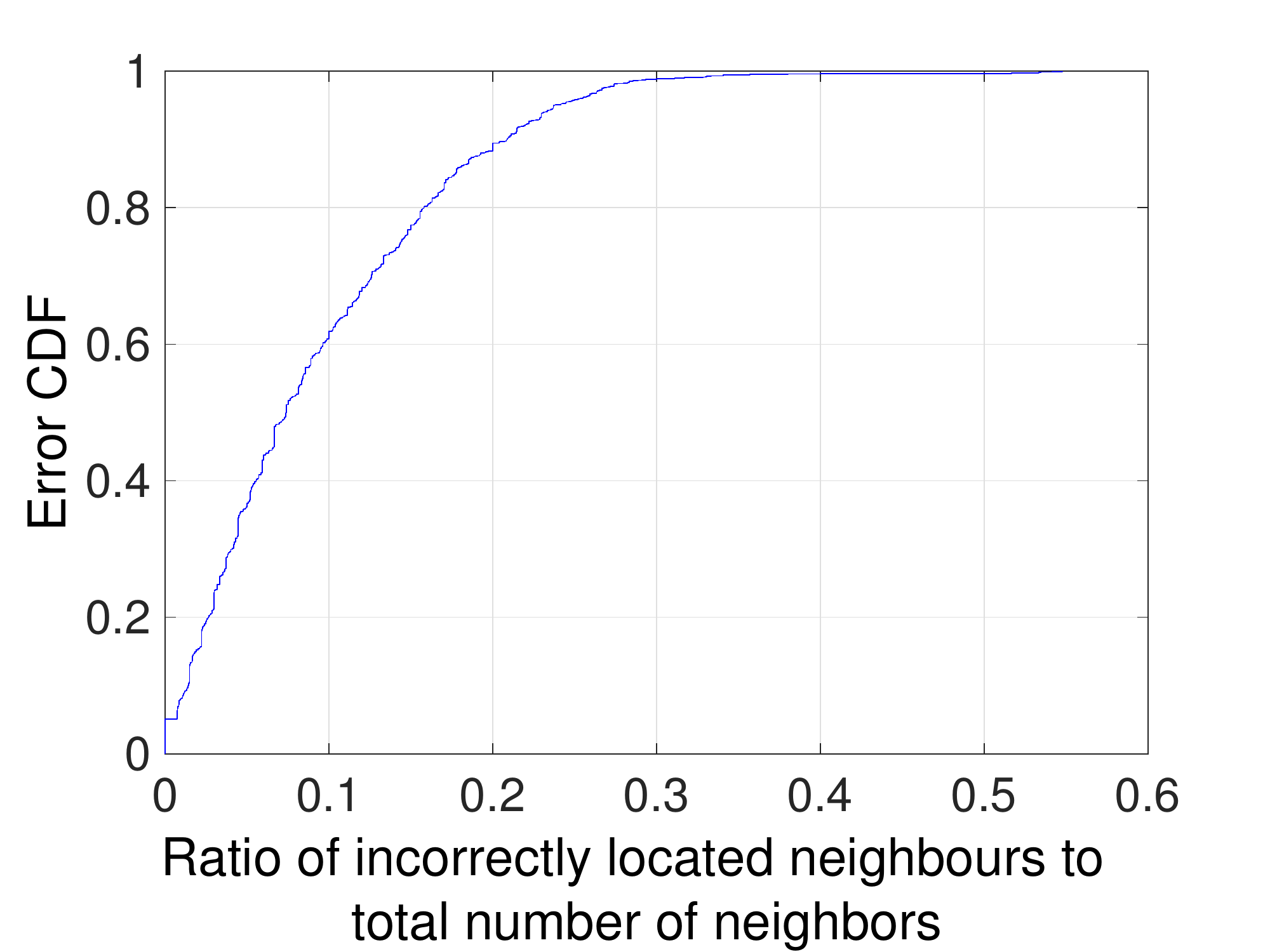}
   \label{fig::mltm:sperecdf}
   }
   \caption{Cylindrical 3D network with 688 nodes }\label{fig::mltm:spere}
\end{figure}

\begin{figure}
 \centering
\subfigure[Physical map]{
  \includegraphics[width=0.45\textwidth]{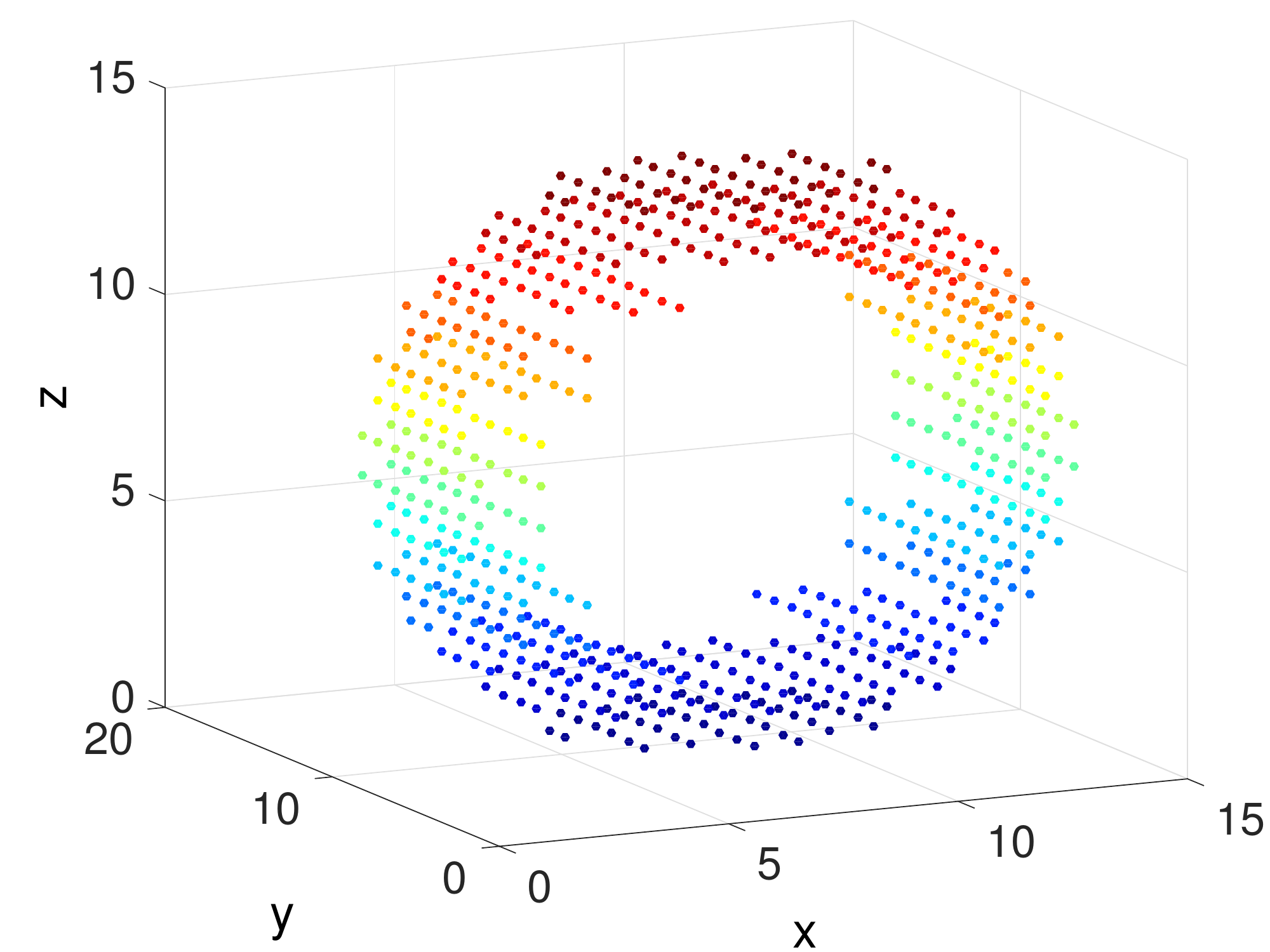}
   \label{fig::mltm:cylindrn}
   }
     \quad
    \subfigure[Physical map with robot trajectory]{
  \includegraphics[width=0.45\textwidth]{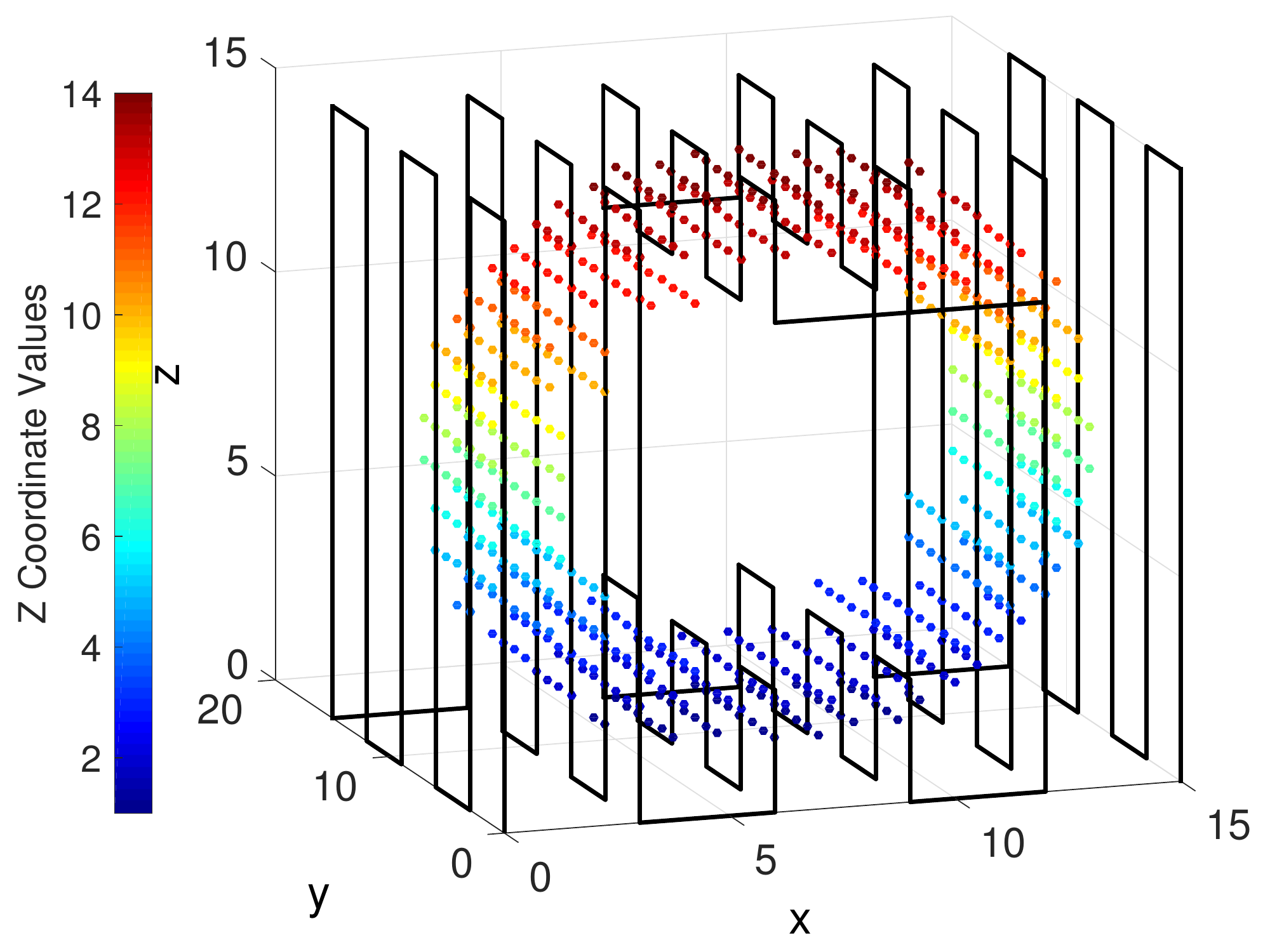}
   \label{fig::mltm:cylindernrt}
   }
   \quad
 \subfigure[ML-TM map]{
  \includegraphics[width=0.45\textwidth]{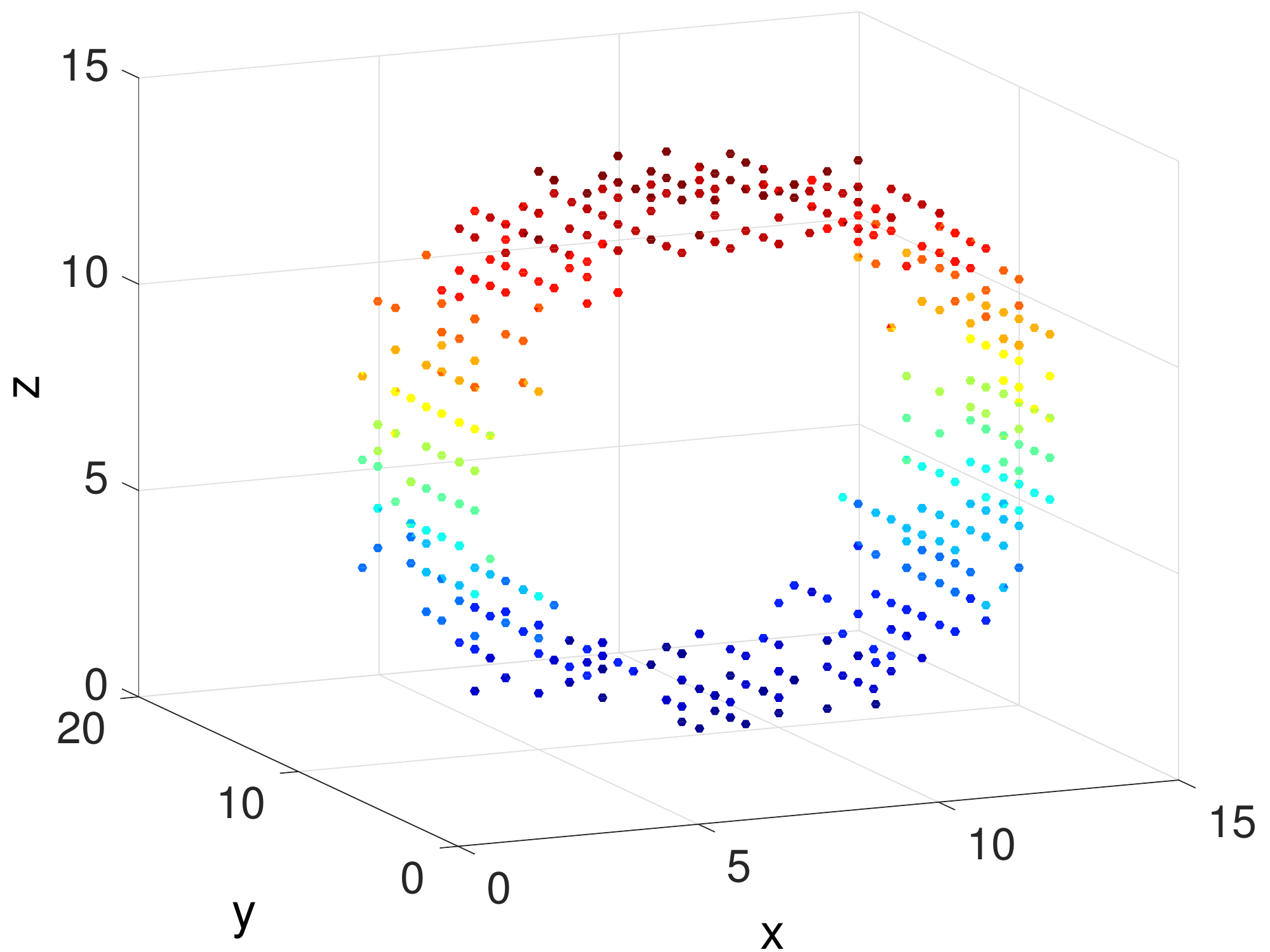}
   \label{fig::mltm:cylinderp}
   }
    \quad
 \subfigure[Error CDF]{
  \includegraphics[width=0.45\textwidth]{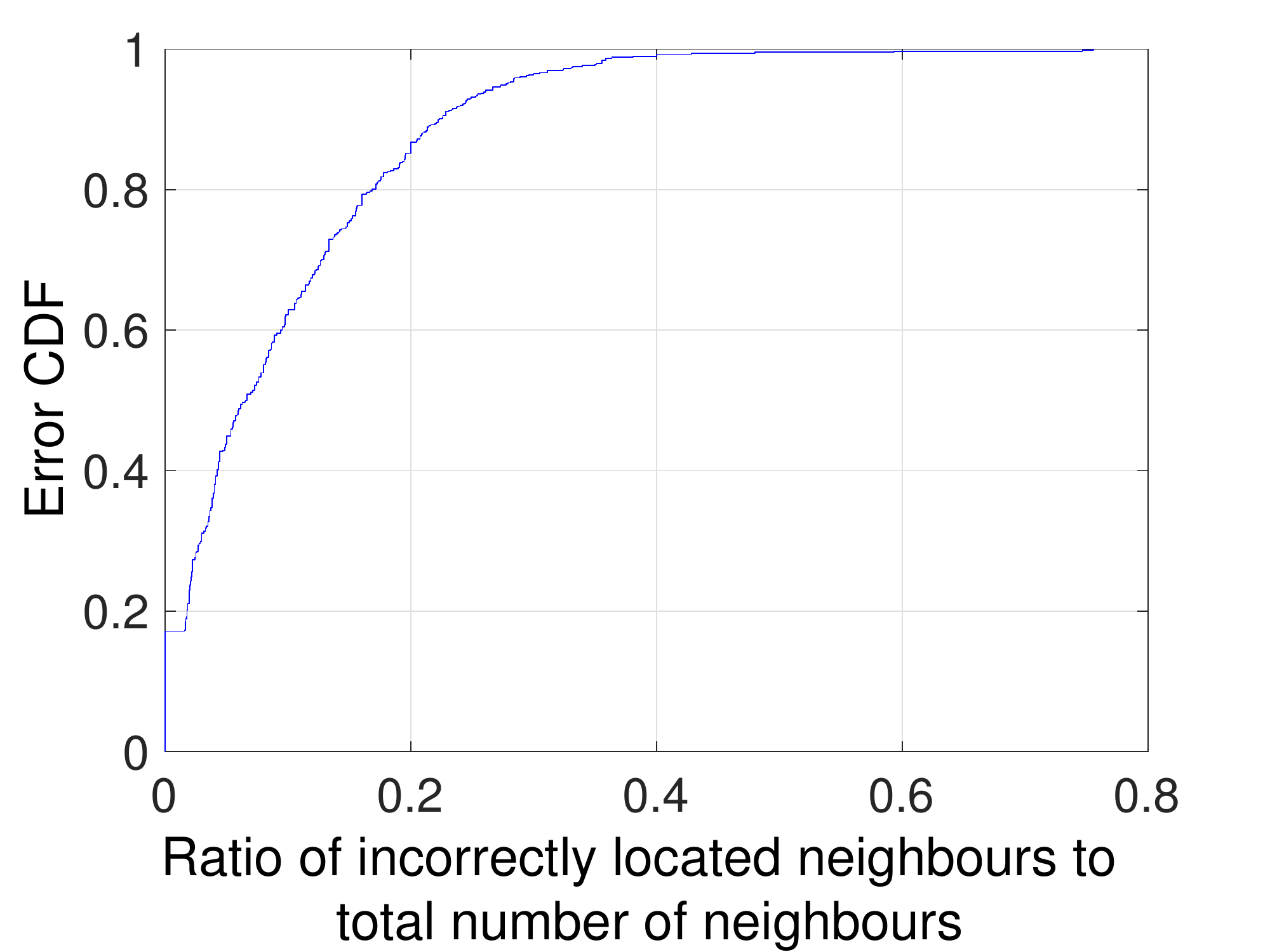}
   \label{fig::mltm:cylindercdf}
   }
   \caption{Cylindrical 3D network with 1736 nodes }\label{fig::mltm:cylinder}
\end{figure}
The 2D robot trajectory discussed in Section \ref{sec::mltm:robottrajectory} was extended to cover a 3D network with obstacles. In 3D 'S' shape robot trajectory, robot moves in z direction following the 'S' shape and makes turns when it hits an obstacle or does not receive any packet from nodes. To evaluate the performance of the algorithm in 3D space, two network setups were considered. First one is a sphere-shaped 3D network with 1736 sensor nodes distributed over a 1400 $m^3$ volume as shown in Figure \ref{fig::mltm:spheren}. The robot trajectory to cover the network is shown in Figure \ref{fig::mltm:sperenrt} and it takes 706s to cover it at a speed of 1 $ms^{-1}$. Second network setup is a cylindrical-shaped 3D network with 688 sensor nodes and a concrete obstacle in the middle as shown in Figure \ref{fig::mltm:cylindrn}. The physical maps that selected for the performance evaluation are shown in \ref{fig::mltm:cylindrn}. The robot needs 404s with the same speed to move around the network that covers a 660 $m^3$ volume. The robot trajectory is shown in Figure \ref{fig::mltm:cylindernrt}. Figure \ref{fig::mltm:sperep} and \ref{fig::mltm:cylinderp} show the generated map using ML-TM algorithm which clearly demonstrate that the proposed algorithm is able to capture the 3D network shape with/without obstacles. 

To check the isomorphism between the actual physical map and the topological maps, $E_{total}$ (described in Section \ref{sec::mltm:error}) is calculated for the two networks. The $E_{total}$ for spherical and cylindrical networks are 9\% and 10\% respectively. Moreover, the Cumulative Distribution Function (CDF) of $E_{total}$ is calculated for the two networks and presented in Figure \ref{fig::mltm:sperecdf} and \ref{fig::mltm:cylindercdf}. From the figures, it can be concluded that more than 90\% of nodes have less than 0.2 ratio of error in their neighbourhood connectivity. Thus, the proposed ML-TM method extracts accurate topology coordinates for sensor nodes deployed in 2D space and also in 3D space.

\section{Energy Awareness, Computation Overhead and the Limitations of the ML-TM Algorithm}\label{sec::mltm:copm}
Sensor nodes have scarce resources and capabilities such as energy, processing power and memory. Hence when proposing an algorithm, it is required to be compatible with the limited sensor network resources. Therefore, this section evaluates the energy awareness and computation overhead of the algorithm and compare with those of the two algorithms used in Section \ref{sec::mltm:result} for the accuracy comparison. Finally, the limitations of the proposed ML-TM algorithm are discussed.

\subsection{Energy Usage Comparison}
Since, WSNs may use batteries as their power unit, efficient energy usage is critical for many sensor network applications. A sensor node consists of three main energy consumers, namely the sensor, the processor and the RF transceiver. Energy consumption of the RF transceiver is more significant than the other two components. Therefore, this calculation considers only the RF transceiver energy  consumption for packet transmitting and receiving. 

Let the total energy consumption by the algorithm is $E_{t}$, the energy required for one packet transmission is $E_{tx}$ and energy required for one packet reception is $E_{rx}$. Then energy required by the ML-TM, RSSI based algorithm and SVD based TPM are shown in equations (\ref{eqn::mltm:proposed}),(\ref{eqn::mltm:rssi}) and (\ref{eqn::mltm:tpm}) respectively. For simplicity, it has assumed that the packet sizes used in all three algorithms are the same.

\begin{equation}\label{eqn::mltm:proposed}
E_{t} = N(nE_{tx})
\end{equation}
where, $N$ is the total number of nodes in the network and $n$ is the total number of time samples. In the proposed method, the mobile robot receives packets. Therefore, energy in the sensors reduces only due to packet transmission.

\begin{equation}\label{eqn::mltm:rssi}
E_{t} = ME_{tx}+(N-M)(mE_{rx})
\end{equation}
where, $M$ is the number of anchor nodes and $m$ is the number of anchor nodes located in the neighbourhood of non anchor nodes ($m<<M$).

\begin{equation}\label{eqn::mltm:tpm}
E_{t} = NM(pE_{tx}+qE_{rx})
\end{equation}
where, $p$ and $q$ are two constants that show the number of packets received from and transmitted to its neighbours to find the optimal hop distance to anchors. Thus those are less than or equal to the number of neighbours.

When considering the three energy equations, the RSSI based location method is the most energy-efficient algorithm. The reason is that only the anchor nodes transmit packets while others listen to them. Also, the number of anchor nodes is less than the non-anchor nodes and transmission energy is higher than the receiving energy. Due to all those reasons, RSSI method is more energy efficient than other two methods. However, the number of transmission is a controllable factor in other two algorithms. It can be decided by the application requirement.  For an example, if the application needs more accuracy, more number of samples can be used. On the other hand, if it is more important to conserve energy, the number of samples can be reduced.

\subsection{Comparison of Computation Overhead} 
Tiny sensor nodes contain limited memory and processing power. Thus, the computation overhead of the algorithm needs to be reduced to get a real-time output and to reduce the energy consumption by the processor. The method based on RSSI measurements requires $\mathcal{O}(m(n-M))$ messages to calculate the location of non-anchor nodes. The VC based TPM algorithm can perform the computation centrally or in a distributed  manner. If it is done at a central node the worst-case complexity is $\mathcal{O}(n^2)$, but there is no computational limitation at the central node. On the other hand, in the distributed case, i.e., when the coordinates are calculated at each node, the algorithm requires $\mathcal{O}(Mn)$ messages. Then this method is more complex than the RSSI based localization method. Moreover, in ML-TM, the calculation is done at the central node with $\mathcal{O}(Nn)$ messages. Since there is no computational or memory limitations at the central node, this method can be used to generate an effective and accurate topological map. 

\subsection{Limitations}
One limitation in ML-TM is that the coordinate calculation phase starts only after the information gathering from all the nodes in the network is complete. This would be a drawback in some applications in large-scale emergency environments that require sensor locations instantly. The second is that the coordinates are calculated centrally. However, ML-TM has been able to generate accurate topology maps without having any prior information about the network or special devices embedded to sensor nodes.

\section{Conclusion}\label{sec::mltm:conclusion}
This chapter presented a novel Maximum Likelihood Topology Map (ML-TM) algorithm to generate topology maps for WSNs without the need for hardware such as GPS or RSSI measurement embedded in sensor nodes. ML-TM is a more accurate map to represent 2D and 3D physical layouts with voids/obstacles compared to existing alternative topology maps. It uses a mobile robot that moves within the network to extract information from sensor nodes, and maps it to a different coordinate system by using a packet receiving probability function, which is sensitive to the distance. This function is an intermediate level between RSSI curves and VCs. Therefore ML-TM algorithm has been able to overcome negative effects related to modelling RSSI curves and node density dependency. 

Moreover, a one-hop connectivity error parameter is proposed to evaluate the accuracy of topology maps by considering the connectivity of a node in the neighbourhood. The proposed algorithm is evaluated with recently proposed RSSI localization algorithm and SVD based TPM algorithm. The results show that the error percentage is less than 7\% in ML-TM and it outperforms the other algorithms. Also, this method was demonstrated to be able to capture various network shapes with obstacles under different environmental conditions. Moreover, ML-TM scales seamlessly to 3D-WSNs thus enabling its use in networks consisting of both 3D volumes and 2D surfaces. Thus, it can be used as an alternative to geographical map in the automation of sensor network protocol.

Furthermore, the energy usage, computation overhead and limitations of the algorithm have been investigated in this chapter. The time required for the robot to explore different shape of networks is also presented. The use of ML-TM in various applications and extending it to a distributed algorithm is presented in the following chapters.

\chapter{Maximum Likelihood Topology Maps for Millimeter Wave Sensor Networks}\label{chapter:mmtm}
MmWave communication shows promise in realizing next generation WSNs for bandwidth demanding applications. However, despite its support of multi Gbps data rates, MmWave communication requires unobstructed LOS and suffers from heavy path losses. Overcoming these in complex 3D environments requires sectored antenna arrays with narrow beamwidths and adaptive beamforming. Therefore, network topology maps would be more significant than ever in MmWave sensor networks. Traditional topology mapping algorithms rely on omnidirectional transmission and reception and are therefore not tailored to such networks. A novel topology mapping algorithm, the \textbf{M}illi\textbf{m}eter Wave \textbf{T}opology \textbf{M}ap (MmTM) is proposed in this chapter for 3D deployments, which takes advantage of the directional information available from beamforming antennas as well as their beam steering capability. An autonomous robot traverses the network recording the packet reception from different nodes, along with the receiving antenna sector ID that delivers the packet with highest signal quality. The techniques used in the standard IEEE 802.11ad protocol are utilised for optimum sector selection and collision avoidance. 

The chapter is structured as follows and the main results of the chapter were originally published in \cite{mmtm}. Section \ref{sec::mmtm:Introduction} offers an introduction and motivation for the research presented in this chapter. Section \ref{sec::mmtm:algo} discusses the details of proposed ML-TM algorithm. Section \ref{sec::mmtm:rpath} explains the robot path and antenna setup to generate an accurate topology map. Section \ref{sec::mmtm:result} presents the performance evaluation and comparison of the algorithm. Section \ref{sec::mmtm:copm} examines the limitations, energy usage and complexity of the proposed algorithm. Finally, Section \ref{sec::mmtm:conclusion} provides a conclusion of the chapter.

\section{Introduction}\label{sec::mmtm:Introduction}
MmWave communication is a leading next generation communication technology that is expected to support multi Gbps data rates \cite{MMW}. MmWave frequencies also offer a possible solution for the severe spectrum shortage in RF band \cite{mmCapacity}. Therefore, increased interest is seen in utilizing MmWave for WSNs due to ever increasing high bandwidth demanding applications such as habitat monitoring, medical applications and smart cities\cite{MMWlocalization,MMWimagiingS}. 

Despite having the capability to provide high data rates, MmWave communication is practically constrained with unobstructed LOS requirement due to high path loss and Oxygen absorption  \cite{MMW, MMWlocalization}. This is overcome by using narrow beam width antenna arrays and adaptive beamforming techniques to maintain the directivity between two communicating nodes \cite{kutty16}. To the best of our knowledge, none of the existing work has addressed the use of MmWaves in WSN topology mapping. Moreover, signal attenuation is higher in MmWave compared to traditional WSN frequencies. Thus, the localization algorithms that depend on signal propagation measurements, such as RSSI, ToA and hop distance, will encounter a significant error in the coordinate calculation.  As a result, accurate sensor localization techniques have become important than ever in MmWave communication. 

Topology maps based on parameters other than direct distance measurement are a feasible alternative  when physical localization is not possible. There are some Topology Preserving Maps proposed for WSN in literature \cite{VC, mltm}. However, these existing topology mapping algorithms considered nodes with omni directional antennas, which is the case with many WSN nodes operating in frequency bands such as 2.4GHz \cite{VC, mltm}. In MmWave communication, directionality is a key feature as they use narrow beamwidth antenna arrays to communicate with each other. Although many real world applications involve WSNs deployed in 3D environments, there is a remarkable lack of work on localization related to 3D deployments. Thus, this chapter proposes a \textbf{M}illi\textbf{m}eter wave \textbf{T}opology \textbf{M}ap (MmTM), a novel topology mapping algorithm for 3D MmWave WSNs based on maximum likelihood estimation. It makes the use of narrow beam multi-sector antenna characteristics of MmWave transceivers to help achieve localization. 

This algorithm consider the problem of generating a topology map of the network by using an autonomous mobile robot, although MmTM can also be extended for other contexts, which has discussed in Chapter \ref{chapter:dmmtm}. The robot and sensor nodes are equipped with directional antennas, which has a limited effective coverage angle compared to omni directional antennas \cite{omnivsdirec}. Thus, the advantage of using directivity information available to the robot was incorporated in topology calculation. The path of the robot is automated to avoid obstacles and to gather information from all sensor nodes. The robot trajectory is modelled in three ways; i) 2D robot trajectory with Vertical Antenna Arrays (VAA), ii) 2D robot trajectory with Vertical Beam Steering (VBS) and iii) 3D robot trajectory. The robot records two matrices: first is a binary matrix that describes the packet reception from sensor nodes by the robot at each location, while second is a matrix with the antenna sector IDs that describes the best sector of the robot to communicate with each sensor node at each time instance. The maximum likelihood topology coordinates are then estimated using these two matrices and a packet receiving probability function, which is sensitive to the distance. 

The performance of the MmTM is evaluated using two simulation environments (a warehouse and a greenhouse), which have different attenuations and path losses. To emulate a real communication link between nodes and the robot, the experiment results on MmWave propagation losses obtained in  \cite{pathloss2,pathloss1} were used. Finally, the performance of MmTM is compared with two recently proposed algorithms for 3D WSN localization \cite{drmds, ntldvhop} and results show that MmTM outperformed both the techniques. 

\section{Millimeter Wave Topology Map (MmTM)}\label{sec::mmtm:algo}
This section describes MmTM algorithm, which creates topology maps for 3D WSNs based on maximum likelihood estimation. The topology mapping algorithm proposed in Chapter \ref{chapter:mltm} considers radio frequencies of sub-GHz range ($<$10GHz) for communication. The proposed MmTM algorithm deviates from ML-TM in three ways. 

\begin{enumerate}
\item Considered MmWave (30-300GHz) frequencies for communication that has higher attenuation and no penetration through obstacles compared to sub-GHz RF waves. 
\item MmWave communication requires sectored antenna arrays with narrow beam width to maintain the directivity between the transmitter and the receiver. Hence, two matrices are considered in the maximum likelihood topology coordinate calculation namely, packet reception matrix and sector ID matrix. In ML-TM, it was considered only a binary matrix based on the packets reception of nodes. 
\item 3D WSN localization is considered by modelling robot trajectory in three different ways as described in Section \ref{sec::mmtm:rpath}. 
\end{enumerate}

MmTM algorithm consists of two parts, namely, information gathering and information mapping, which are described in the remainder of this section and the psuedo code of the algorithm is shown in the Appendix \ref{PC}. The assumptions made in proposing the MmTM algorithm are, (i) the robot is equipped with a GPS and a compass to get the location and sector direction information, (ii) no power limitations in robot side (rechargeable/ have enough energy to complete its task), (iii) robot can have sectors up to 64  \cite{nit14} and (iv) sensors have fixed number of sectors and in the simulation it has considered as eight.

\subsection{Information Gathering}
\begin{figure}[ht]
\centering
\includegraphics [width=0.75\textwidth]{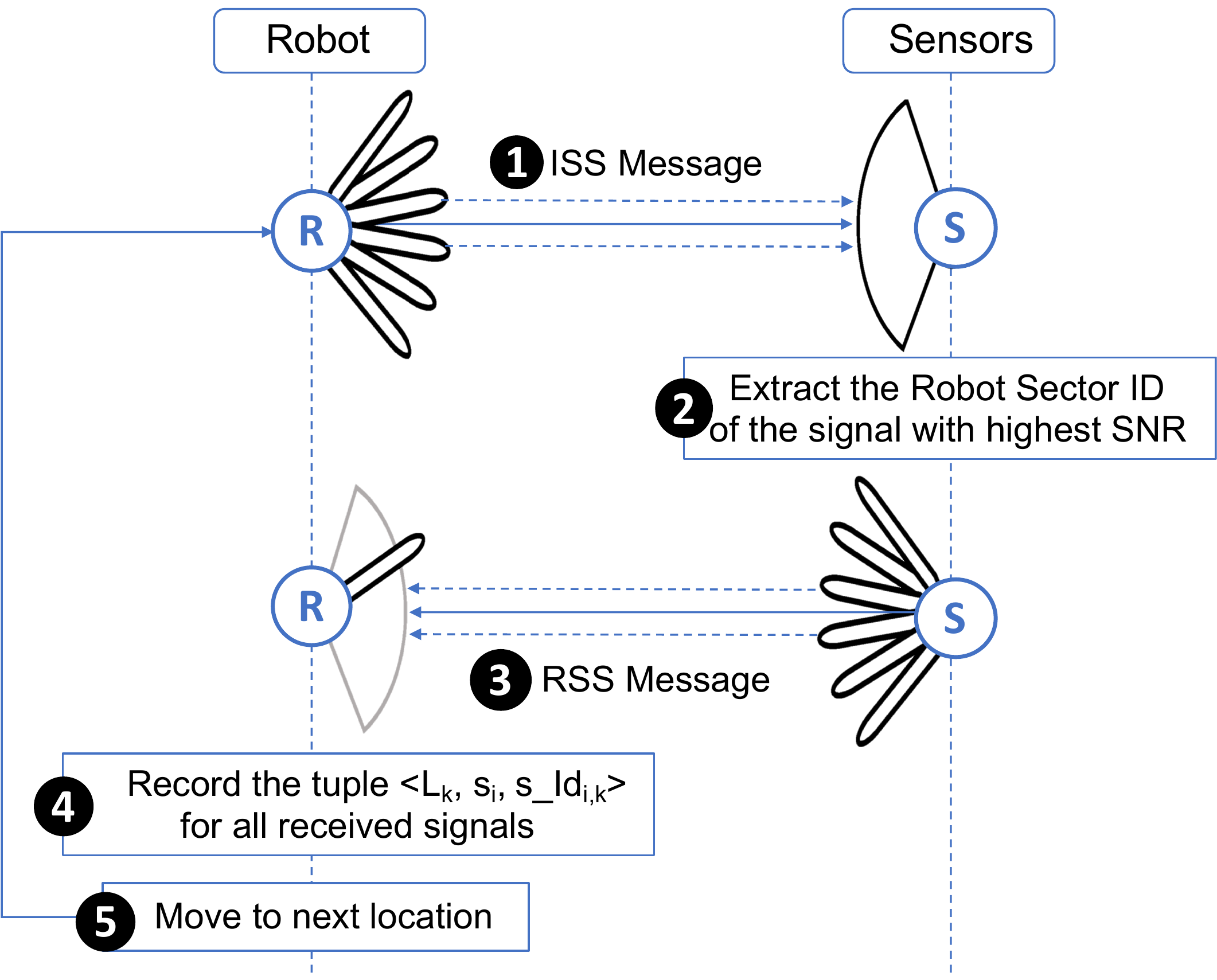} \vspace{-2mm}
\caption{MmTM information gathering protocol}
\label{fig::mmtm:protocol}
\end{figure}
This section describes the information gathering in MmTM algorithm. A protocol was proposed for the communication between robot and sensor nodes, which is based on the standard IEEE 802.11ad protocol Sector Level Sweep (SLS) phase. This proposed protocol, does not go through all the steps in beamforming concept of the IEEE 802.11ad protocol as described in the Appendix \ref{BF}. The reason is robot and sensor do not need to pair the best sectors for the communication as MmTM algorithm requires only to update packet reception binary matrix along with received sector ID. Hence, there is a energy saving with this new protocol. The proposed protocol is shown in Figure \ref{fig::mmtm:protocol} and steps are described below. 

\noindent {\it i) Step 1:} Robot broadcasts a Initiator Sector Sweep (ISS) message from all its sectors, and sensors stay on Quasi-omni pattern. In this step, a sensor node can receive packets from more than one sector of the robot. \\
\noindent {\it ii) Step 2:} Sensor node choose the best sector of the robot based on the Signal to Noise Ratio (SNR). It extracts the sector ID of the ISS message delivered with highest SNR. \\
\noindent {\it iii) Step 3:} Sensor node transmits sector ID extracted from ISS message back to the robot from all its sectors via a RSS message. At this time, robot remains on Quasi-omni pattern to hear from all the sensor nodes located in it's neighbourhood. \\
\noindent {\it iv) Step 4:} The robot updates two matrices, describes below, with the information gathered by the Responder Sector Sweep (RSS) packets sent by sensor nodes located in its neighbourhood.\\
\noindent {\it v) Step 5:}  Robot moves to the next location. 

Consider a 3D WSN consists of $N$ stationary sensors labelled $ s_1,s_2,...s_i,...,s_N$, whose locations are not known and a mobile robot traverse the network to gather information about sensor nodes. Robot is able to receive packets at $ L_1, L_2, ...L_k,..., L_n$ locations on the robot trajectory at discrete time instances, $t_1, t_2, ...,t_k,...,t_n$. Two $N\times n$ matrices are updated by the robot based on the information gathered by received RSS messages from each sensor node. First is a binary matrix $M$ that called as the packet receiving matrix which is updated by the following rule: \\\\
\indent $M(i, k) = 1$, if the robot receives a RSS packet from the sensor $s_i$ at the time $t_k$; \\
\indent $M(i,k) = 0$, if the robot does not get a RSS packet from the sensor $s_i$ at the time $ t_k$. 

The second matrix $A$ records the robot's sector ID included in the RSS packets received from sensor nodes. The matrix $A$ is updated by the following rule: \\\\
\indent $A(i, k) = s\_ID_{i,k}$, if the robot receives a RSS packet from the sensor $s_i$ at the time $t_k$; \\
\indent $A(i,k) = 0$, if the robot does not get a RSS packet from the sensor $s_i$ at the time $ t_k$. \\
where $s\_ID_{i,k}$ is the ID of the robot's sector included in the RSS packet received from sensor $s_i$ at the time $t_k$.

\subsection{Information Mapping}
This section describes the maximum likelihood topology coordinate calculation based on the gathered information in the previous step. The information gathered by the robot is mapped to the coordinates using a packet receiving probability function $S(d)$ proposed in \ref{subsec::mltm:pfn}. This function describes the probability of receiving packets from a sensor when robot is at a particular distance. Let, $S(d)$ be the probability value when robot is at distance $d$ from the sensor. Then, $S(d)$ is defined as, 
\begin{eqnarray}\nonumber
S(d) := p_0 \:\:\:\:\: \forall \: d\leq r \\ \nonumber
S(d) := 0 \:\:\:\:\: \forall \: d\geq R \\ 
S(d) := \frac{p_0(R-d)}{(R-r)} \:\:\:\: \forall \: r<d<R \label{eqn::mmtm:sd2}
\end{eqnarray}
where $0 < p_0 \leq 1$, $0 < r < R \leq R_c$ are some given constants. $R_c$ is the communication range of a sensor node. 

The goal of MmTM algorithm is to find the optimal solution for the location of the sensors in 3D space. Let consider, vectors $m_1, m_2,...,m_i,..., m_N$ and $a_1, a_2,...,a_i,..., a_N$ be the rows of the matrix $M$ and $A$ respectively. Also, let $m_i(\textit{1}), m_i(\textit{2}), . . . ,m_i(\textit{k}),..., m_i(n)$ and $a_i(\textit{1}), a_i(\textit{2}), . . . , a_i(\textit{k}),...,a_i(n)$ denote the elements of the vector $m_i$ and $a_i$ respectively. Then the R-neighbourhood of the packet received locations of each node $s_i$ is divided into small step grid $g_1,g_2,...,g_j$ with the length of $\delta k$. We introduce a function $P_i(x_j, y_j, z_j)$ which is the probability of obtaining the vectors $m_i$ and $a_i$ for the robot's trajectory under the assumptions that the sensor node $s_i$ is located at the grid point $g_j=(x_j,y_j,z_j)$. $P_i(x_j, y_j, z_j)$ can be calculated using,

\begin{eqnarray}
P_i(x_j, y_j ,z_j) &=& \lbrack Z_1^i(d_{j1})Z_2^i(j1)\rbrack \lbrack Z_1^i(d_{j2})Z_2^i(j2)\rbrack...\lbrack Z_1^i(d_{jn})Z_2^i(jn)\rbrack \label{eqn::mmtm:probability}
\end{eqnarray}

 $Z_1^i(d_{jk})$ and $Z_2^i(jk)$ are defined as,
\[
 Z_1(d_{jk}) =
  \begin{cases}
   S(d_{jk}) & \text{if } m_i(k) = 1 \\
   1-S(d_{jk})     & \text{if } m_i(k) = 0
  \end{cases}
\]
where, $L_k=(x_R(t_k),y_R(t_k),z_R(t_k))$ and
 \begin{equation}\label{eqn::mmtm:dist}
 d_{jk} = \sqrt{(x_j-x_R(t_k))^2 + (y_j -y_R(t_k))^+ (z_j -z_R(t_k))^2}
 \end{equation}

\[
 Z_2(jk) =
  \begin{cases}
   1 & \text{if } \alpha _1(ik)<=\theta (jk)<=\alpha _2(ik) and \\
   & \beta_2 (ik)<=\phi (jk)<=\beta_2 (ik)\\
   0     & \text{else } m_i(j) = 0
  \end{cases}
\]
where, $\theta (jk)$ and $\phi (jk)$ are azimuth and elevation angles respectively between robot location at time $t_k$ and grid point $j$. $\alpha$ and $\beta$ are defined as follows. 

Consider the horizontal and vertical direction angle of the sector $a_i(k)$ is $\gamma _{ik}$ and $\delta _{ik}$. The horizontal and vertical beam widths of that sector is $HBW_{ik}$ and $VBW_{ik}$. Then,
\begin{eqnarray}
 \alpha _1(ik) &=& \gamma _{ik}- HBW_{ik}/2 - \tan ^{ - 1} \{\delta k/R\} \\
 \alpha _2(ik) &=& \gamma _{ik}+ HBW_{ik}/2 + \tan ^{ - 1} \{\delta k/R\} \\
  \beta _1(ik) &=& \delta _{ik}- VBW_{ik}/2 - \tan ^{ - 1} \{\delta k/R\} \\
 \beta _2(ik) &=& \delta _{ik}+ VBW_{ik}/2 + \tan ^{ - 1} \{\delta k/R\}
\end{eqnarray}
From the calculation, it is obvious that $P_i(x_j,y_j,z_j) = 0$ for any node $s_i$ for all $(x_j,y_j,z_j)$ that are outside the R - neighbourhood of the robot's trajectory and out of the antenna lobe. After calculating $P_i(x_j, y_j,z_j)$ value for each grid vertex, the grid vertex $(x_j^{opt},y_j^{opt},z_j^{opt})$ that delivers the maximum value of $P_i(x_j,y_j,z_j)$ among all vertices of the grid is taken as the maximum likelihood topology coordinate of the sensor $s_i$. This is an approximation of the optimal location of the sensor node $s_i$.

\section{Automated Robot Path And Antenna Setups}\label{sec::mmtm:rpath}
This section describes the selection of the mobile robot path and robot's antenna setups. There are several robot path planning algorithms proposed in the literature. To formulate the path planning algorithm, robot needs to move from a given point to a goal while optimizing a defined parameter/s such as energy \cite{pathplanning1,pathplanning2} or time \cite{robotpath,robotpath2d}. In the proposed path planning algorithm, the goal is to access all the nodes in the network in least possible time. As it is assumed that robot has unlimited power, energy is not a constraint. 

In this section, 2D and 3D robot trajectories are considered to receive packets from sensor nodes. Two antenna setups were examined under the 2D robot trajectory to cover the 3D space. Those are vertical antenna arrays and vertical antenna sweeping. Since robot can cover the network vertically and horizontally in 3D robot trajectory, we deal more with trajectory planning than with antenna setup. The detailed description of robot trajectory algorithms and its antenna setups are discussed in the following subsections. 

\subsection{2D Robot Path Planning and Antenna Setups}
In the robot-assisted localization scenario presented here, one of the main issue is the path of the robot travels along. The optimum path is the one that covers the network entirely during the shortest possible time while avoiding the obstacles. Among the proposed paths for robot assisted localization in literatures such as \cite{rezazadeh14}, vacuum robot path planning is the most common algorithm for localization, with four different approaches namely, random walk, spiral, 'S' shape and wall follow \cite{robotpath}. Among them, random walk and wall follow are excluded since the former is ineffective in terms of required time and the latter is impossible for outdoor environments \cite{robotpath}. The most effective path examined in \cite{robotpath} is 'S' shape with full coverage and shortest traveling time. 2D 'S' shape robot path proposed in \cite{robotpath2d} covers the entire network by avoiding the obstacles. That algorithm was used to automate our 2D robot path. The two antenna setups that are examined with the 2D path planning are discussed follow.  

\subsubsection{Vertical Antenna Array (VAA)}
\begin{figure}
 \centering
 \subfigure[Antenna array]{
  \includegraphics[width=0.35\textwidth]{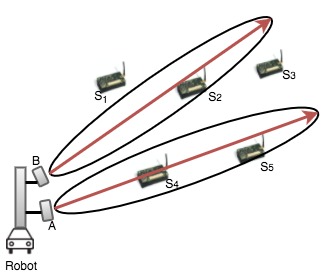}\vspace{-2mm}
   \label{fig::mmtm:VA1}
   }
 \quad
 \subfigure[Beam steering]{
  \includegraphics[width=0.35\textwidth]{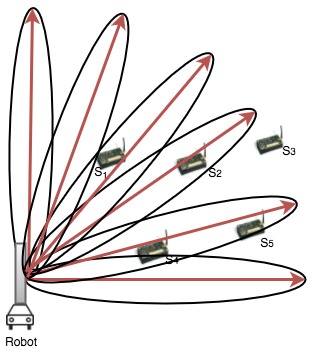}\vspace{-2mm}
   \label{fig::mmtm:AS1}
   }
 \caption{ Antenna setups for 2D robot path}
\end{figure}
This setup covers the 3D space in vertical direction while robot is moving horizontally. The elevation angle of vertical antennas are different as shown in Figure \ref{fig::mmtm:VA1}. When the robot is moving horizontally it scans the entire network and updates the two matrices.  As an example, robot receives the packets from sensor nodes $s_2$, $s_4$ and $s_5$ in its current location. But robot needs to move forward to hear from node $s_3$ and has to move backward to hear from node $s_1$. Thus, travelling along 'S' shape trajectory, receiving packets from the entire network is guaranteed.     
   
\subsubsection{Vertical Beam Steering (VBS)}
In VBS, the elevation angle of the sector antenna beam changes from $0$ to $\pi/2$ as shown in Figure \ref{fig::mmtm:AS1}. The robot stops at a point and sweeps the antenna to hear from the nodes and then moves to a new point. Since the antenna is sweeping during the sensor packet transmission time, there is a probability of packet loss compared to the VAA. 

\subsection{3D Robot Path Planning and Antenna Setups (3D)}
\begin{figure}[b!]
  \centering
    \includegraphics[width=0.65\textwidth]{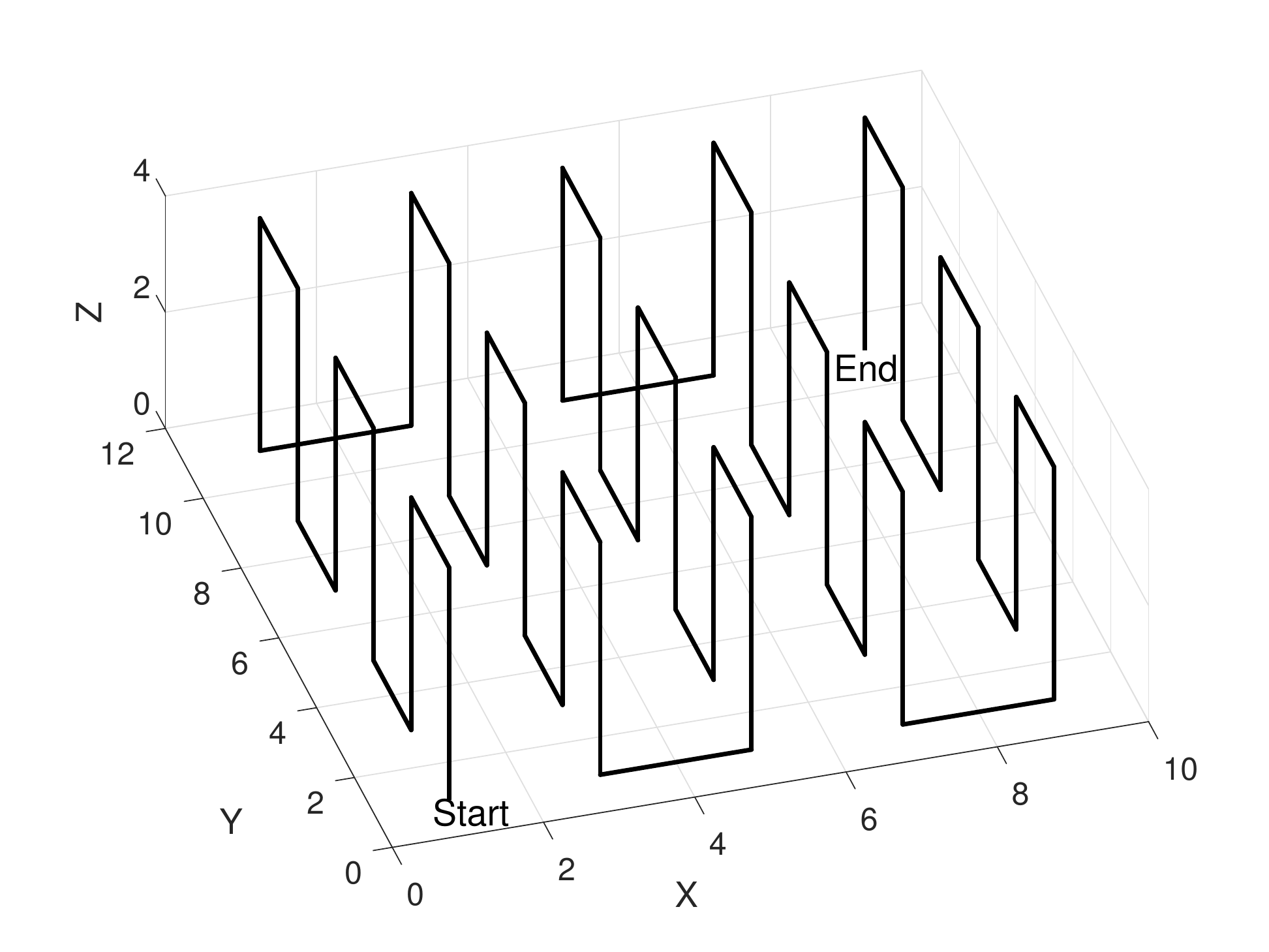}
    \caption{3D 'S' shape robot trajectory in a room}\label{fig::mmtm:3drr}
\end{figure}
In 3D robot path planning, the 2D algorithm proposed in \cite{robotpath2d} was extended to 3D 'S' shape pathway. Figure \ref{fig::mmtm:3drr} shows the robot path in a room. The robot moves in z direction following the 'S' shape and makes a turn when it hits an obstacle or ceiling/floor. In 3D robot path planning, it does not need to consider either vertical antenna arrays or vertical antenna sweeping, as robot moves through the entire network and receives packets from all the nodes. Therefore, the elevation angle of all sector antennas was set as zero.

\section{Performance Evaluation}\label{sec::mmtm:result}
\begin{figure}
 \centering
 \subfigure[Warehouse]{
  \includegraphics[width=0.45\textwidth,trim={2cm 1.5cm 2.5cm 3cm}]{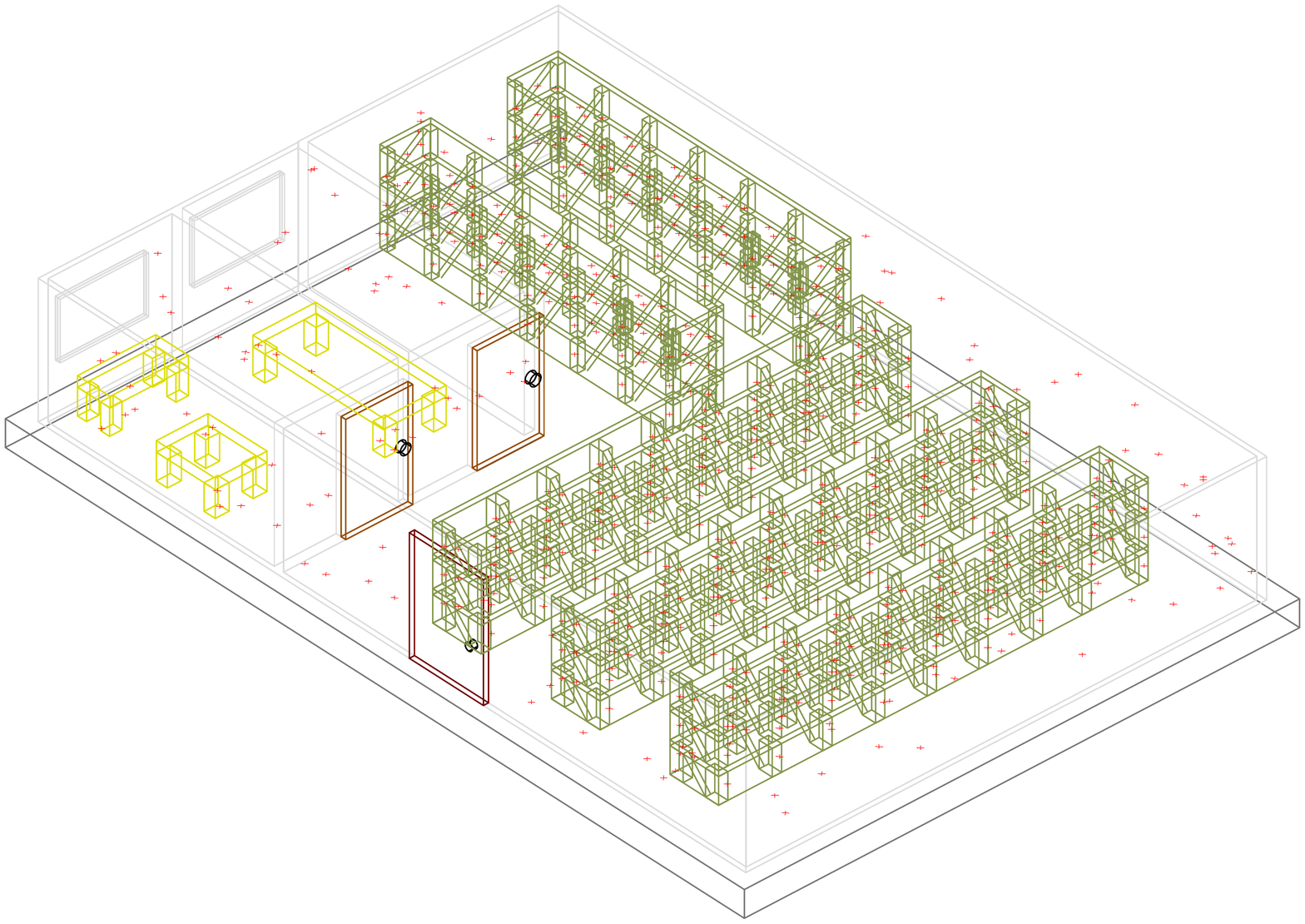}
   \label{fig::mmtm:Warehouse}
   }
 \subfigure[Greenhouse]{
  \includegraphics[width=0.45\textwidth,trim={5.5cm 2cm 5cm 5.5cm}]{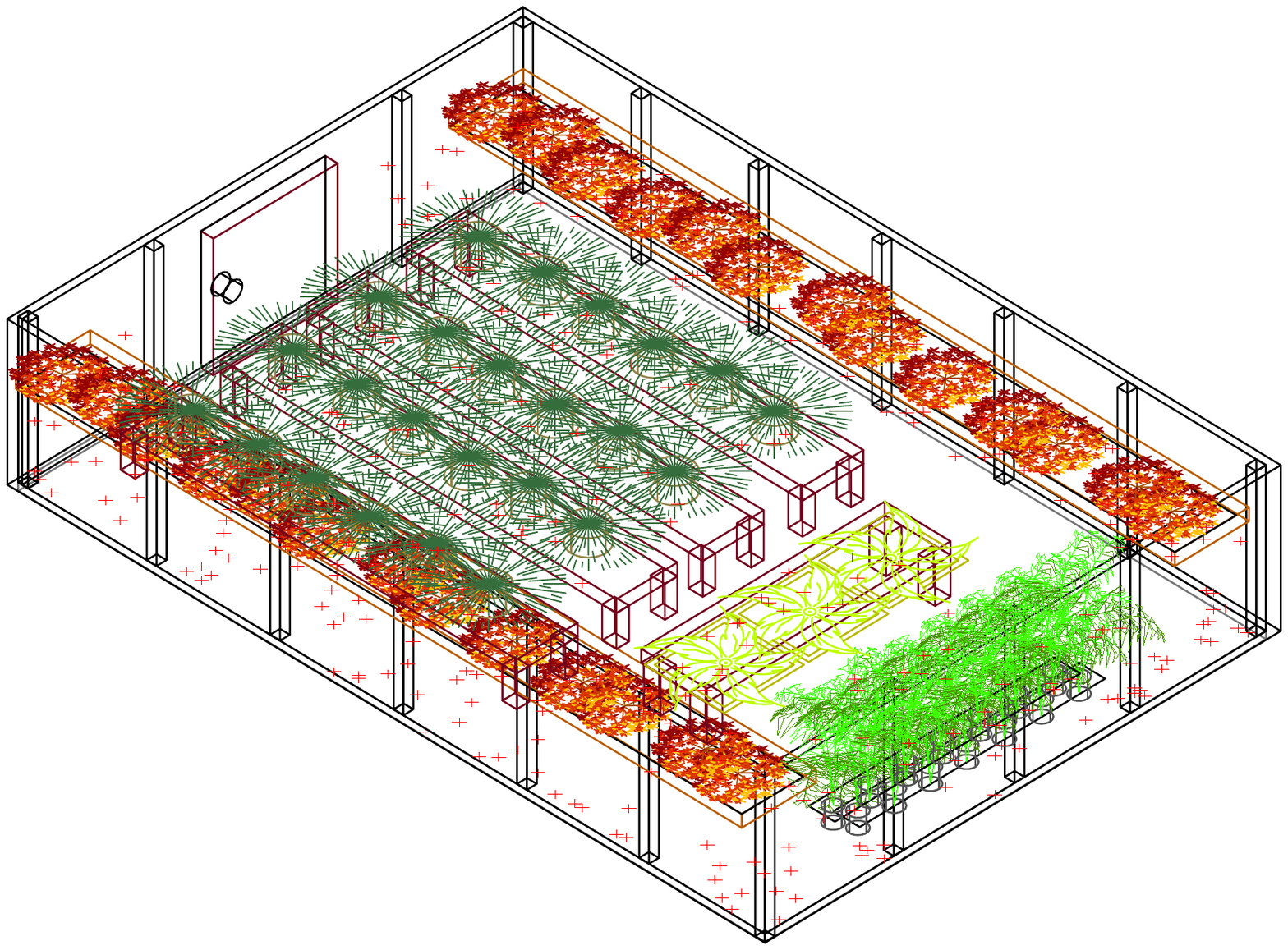}
   \label{fig::mmtm:Greenhouse}
   }
 \caption{Simulation environments with sensor locations (red dots)}
\end{figure}

The performance of the proposed MmTM algorithm is evaluated in this section using two simulation environments. One is a warehouse environment with metal racks and tables as shown in Figure \ref{fig::mmtm:Warehouse} with 591 sensor nodes distributed over the environment. Second scenario is shown in Figure \ref{fig::mmtm:Greenhouse} and it is a greenhouse full with plants. 410 sensor nodes cover the environment.

 MATLAB simulation software was used for the computations. The communication channel between robot and sensors is modelled based on a path loss model that represents propagation characteristics of MmWaves in practical environments. Several path loss models have been proposed in literature for MmWave communication and some of them are Close-in (CI) model reference distance model, Floating-intercept (FI) model and alpha-beta-gamma (ABG) model \cite{pathloss2,pathloss1}. From the experiment results proposed in \cite{pathloss2,pathloss1}, it can be seen that CI path loss model well explains the MmWave physical propagation losses in both indoor and outdoor compared to other existing models. Hence, CI path loss model was used and the model is as follows.
\begin{eqnarray}
PL(f,d)&=&FSPL(f,d_0)+10\varepsilon log_{10}(d/d_0)+X_\sigma \nonumber \\ 
&& \text{for}\:d \geq d_0,\:\:\:\text{where},\: d_0=1
\end{eqnarray}\label{eqn::mmtm:pm1}
where, $PL(f,d)$ is the path loss value in dB, $\varepsilon $ is the path loss exponent, $d$ is the distance between robot and the node, and $X_\sigma $ is a zero Gaussian random variable with standard deviation $\sigma $ in dB (this represents the large-scale channel fluctuations due to shadowing \cite{pathloss2}). FSPL is the free space path loss at distance $d_0$ and can be calculated as $10log_{10}(4\pi d_0f/ c)^2$, where $f$ is the frequency and $c$ is the speed of light.

\begin{figure*}
 \centering
 \subfigure[Actual Map]{
  \includegraphics[width=0.45\textwidth, trim={0cm 2.1cm 1cm 1cm},clip]{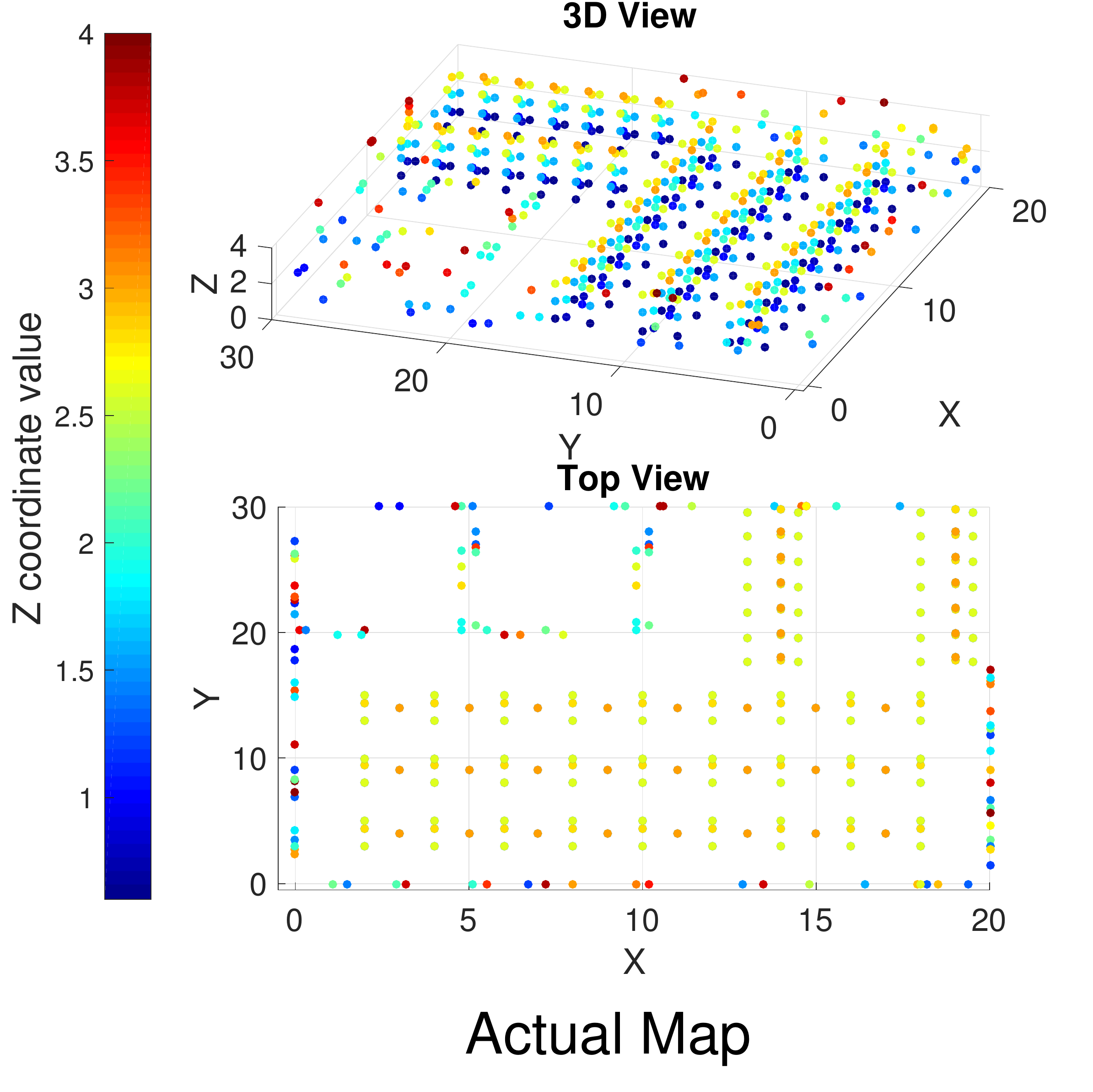}
   \label{fig::mmtm:w_network}
   }
 \quad
 \subfigure[MmTM with 2D VAA]{
  \includegraphics[width=0.45\textwidth, trim={0cm 2.3cm 1cm 1cm},clip]{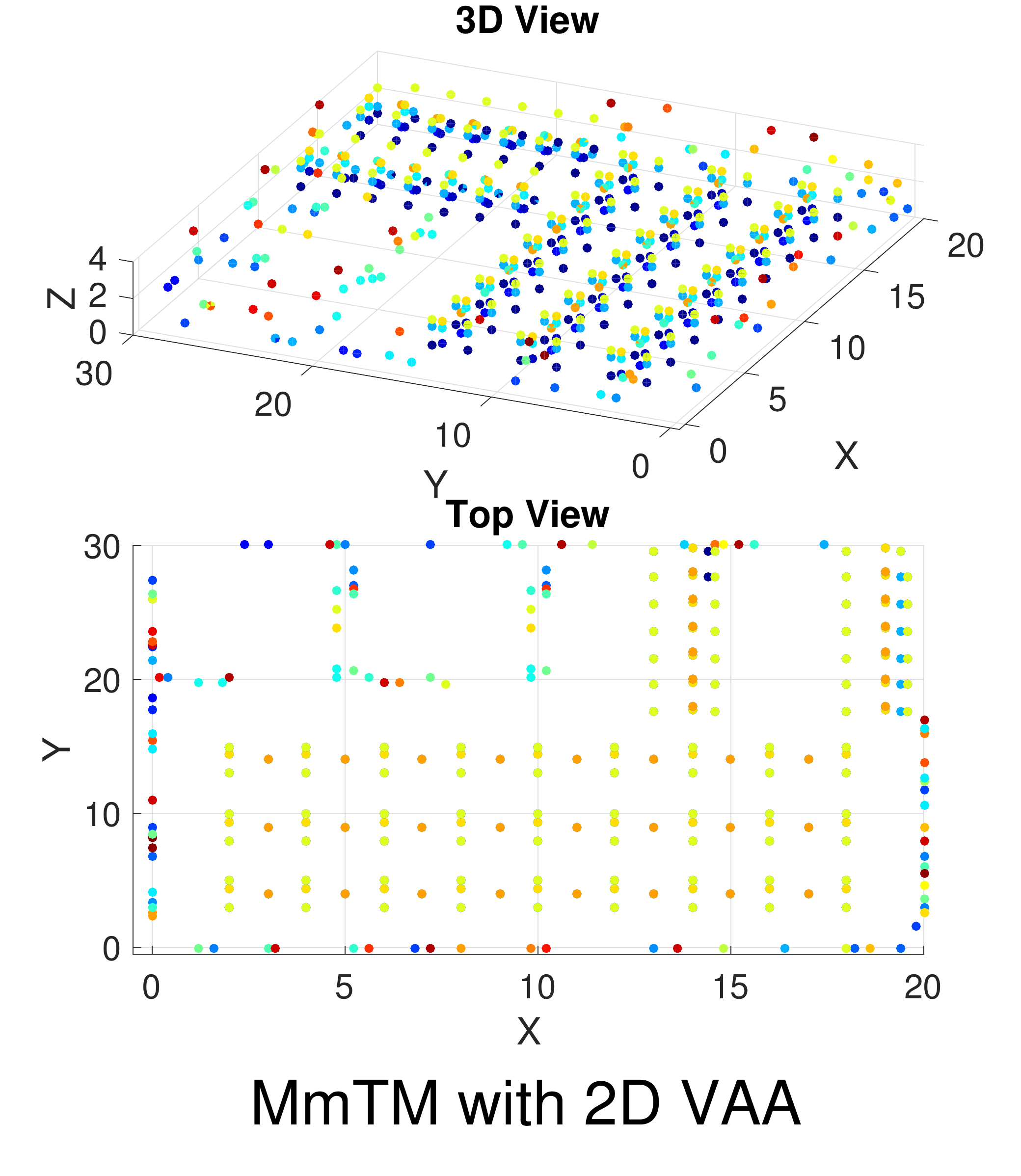}
   \label{fig::mmtm:w_tmVA}
   }
   \quad
 \subfigure[MmTM with 2D VBS]{
  \includegraphics[width=0.45\textwidth, trim={0cm 2.3cm 1cm 1cm},clip]{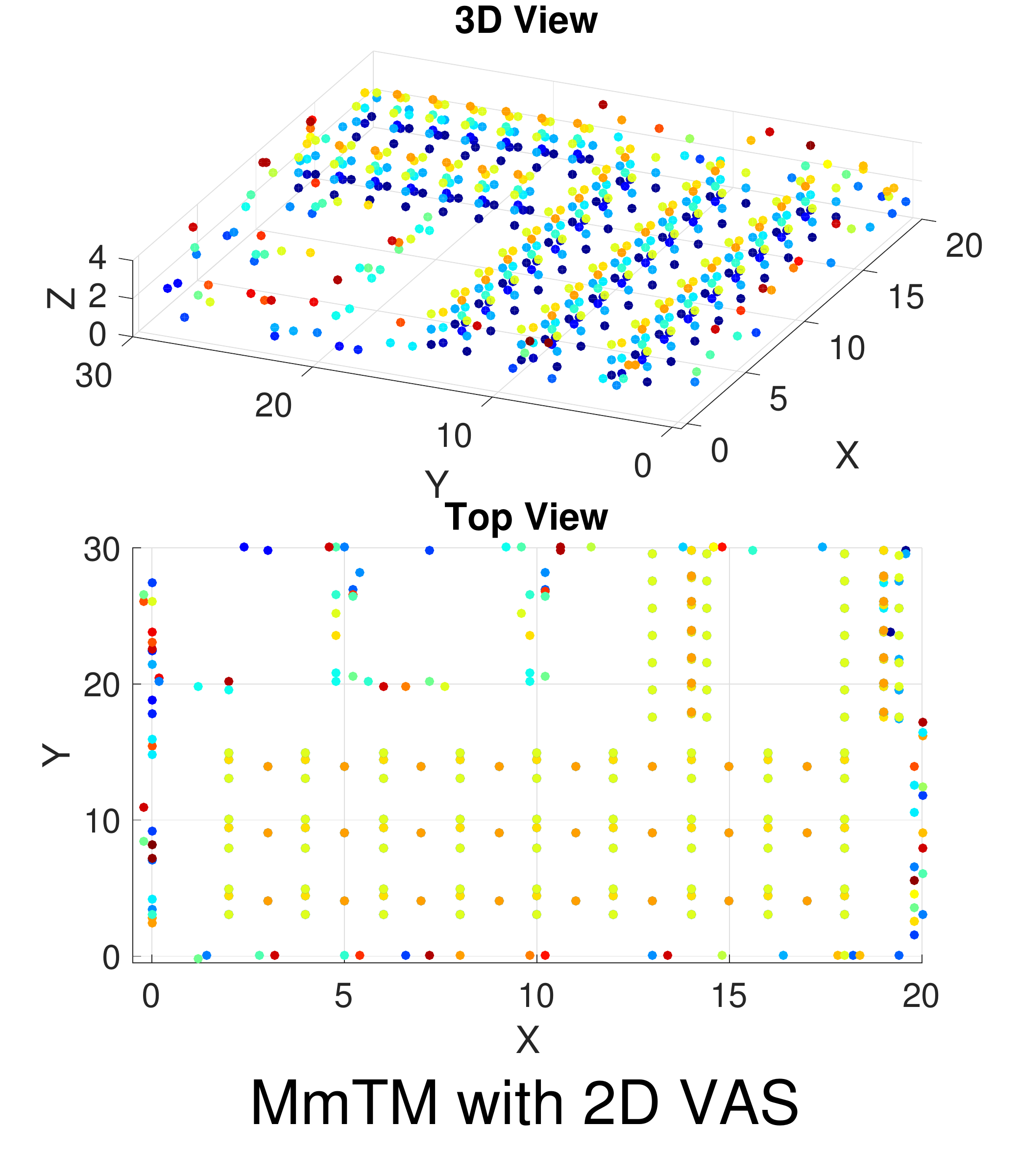}
   \label{fig::mmtm:w_tmAS}
   }
   \quad
 \subfigure[MmTM with 3D]{
  \includegraphics[width=0.45\textwidth, trim={0cm 2.3cm 1cm 1cm},clip]{Fig/C4/images/VA_warehouse_tm2.pdf}
   \label{fig::mmtm:w_tm3d}
   }
 \caption{Warehouse node distribution in actual map and calculated maps}  \label{fig::mmtm:resultmap_w}
\end{figure*}

\begin{figure*}
 \centering
 \subfigure[Actual Map]{
  \includegraphics[width=0.45\textwidth, trim={0cm 2.1cm 1cm 1cm},clip]{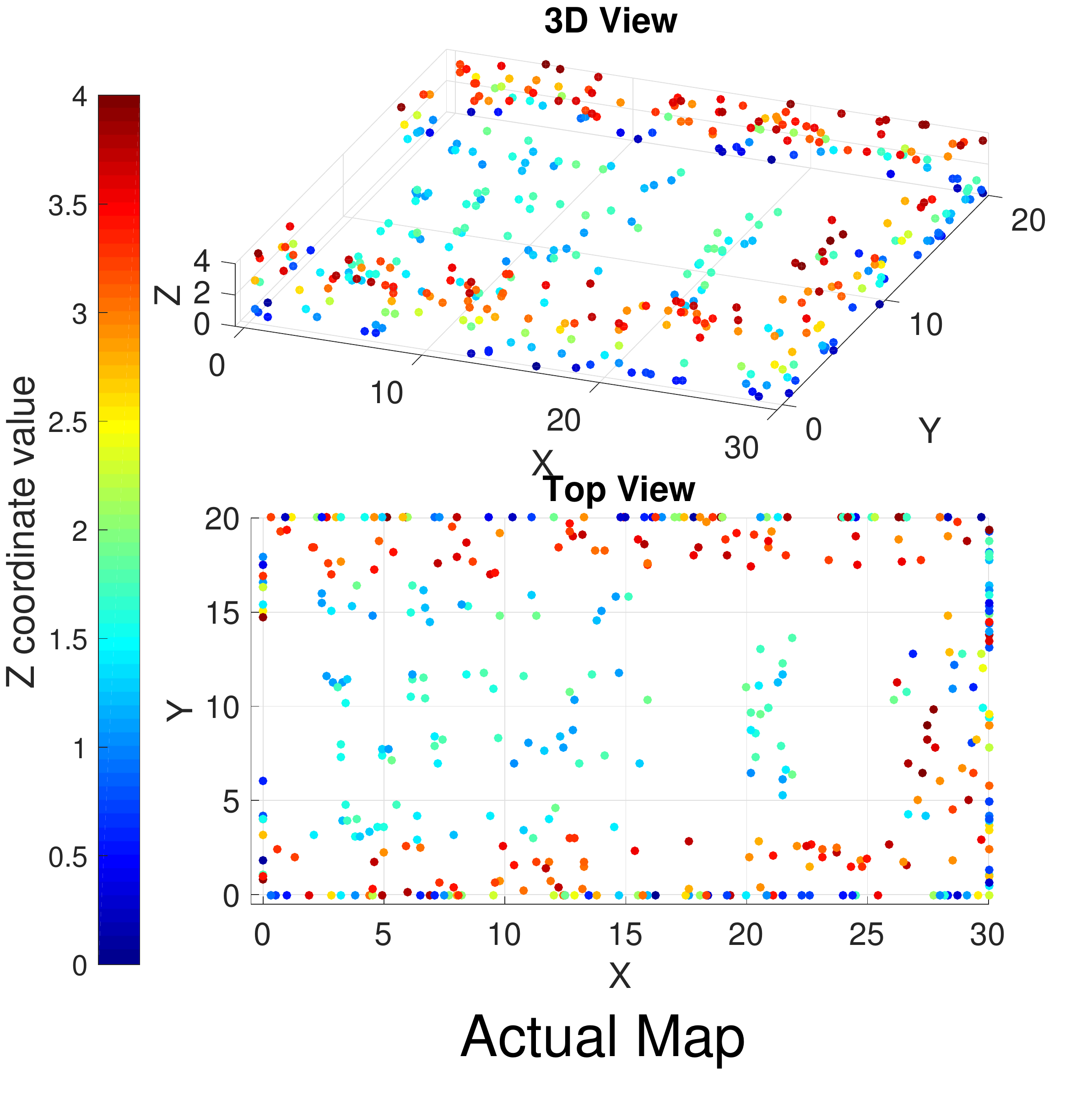}
   \label{fig::mmtm:g_network}
   }
 \quad
 \subfigure[MmTM with 2D VAA]{
  \includegraphics[width=0.45\textwidth, trim={0cm 2.2cm 1cm 1cm},clip]{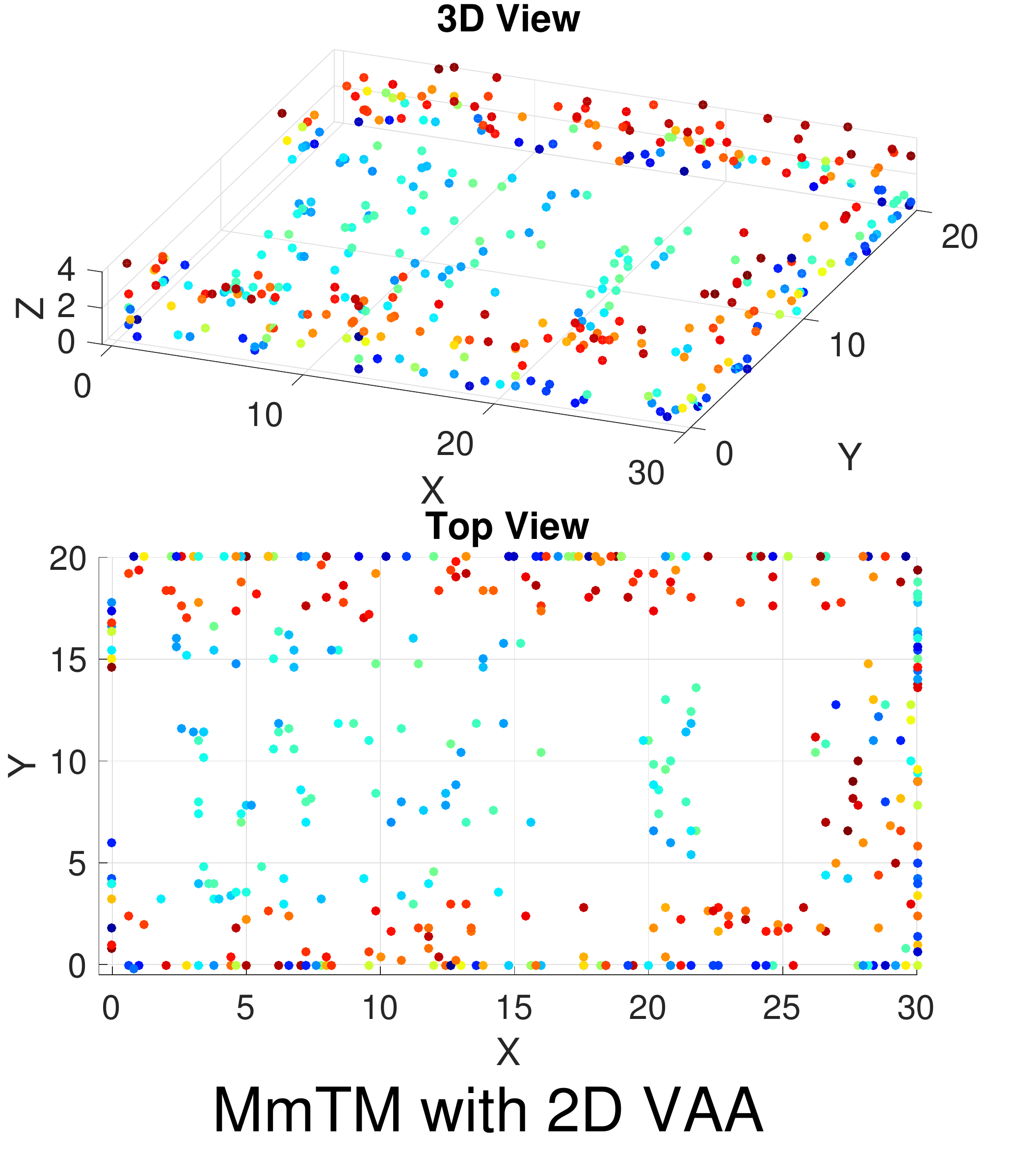}
   \label{fig::mmtm:g_tmVA}
   }
   \quad
 \subfigure[MmTM with 2D VBS]{
  \includegraphics[width=0.45\textwidth, trim={0cm 2.2cm 1cm 1cm},clip]{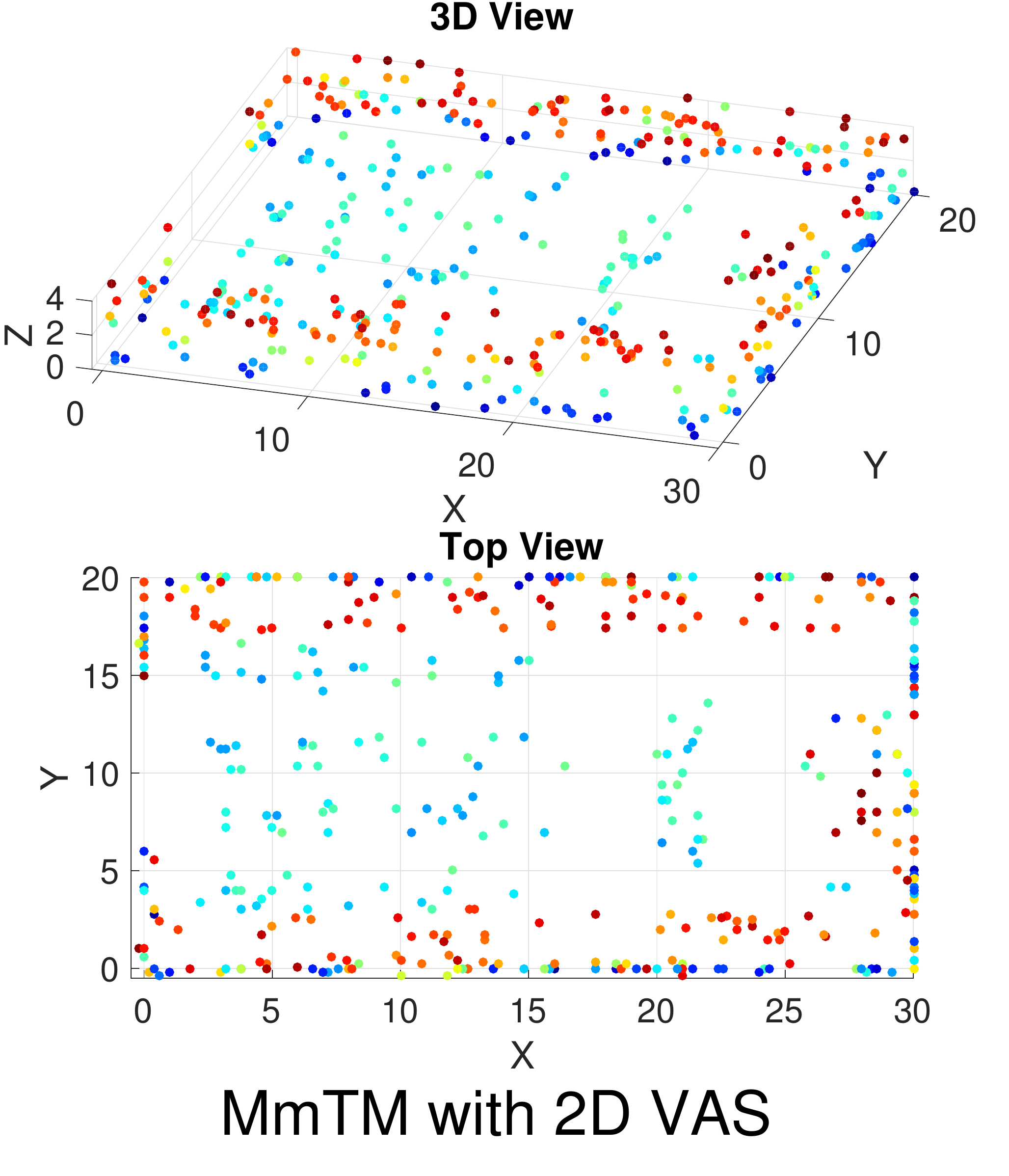}
   \label{fig::mmtm:g_tmAS}
   }
   \quad
 \subfigure[MmTM with 3D]{
  \includegraphics[width=0.45\textwidth, trim={0cm 2.2cm 1cm 1cm},clip]{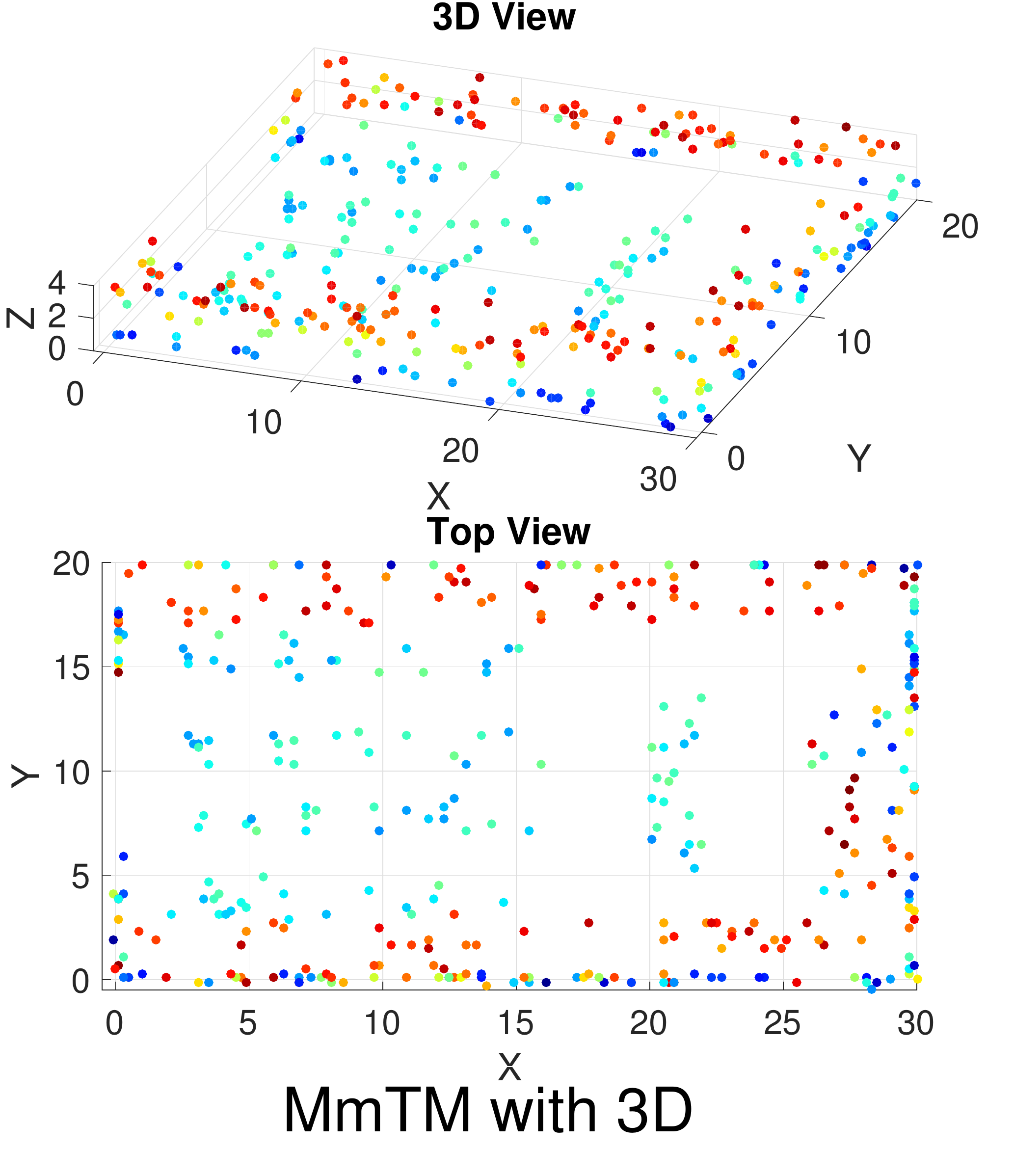}
   \label{fig::mmtm:g_tm3d}
   }
 \caption{Greenhouse node distribution in actual map and calculated maps} \label{fig::mmtm:resultmap_g}
\end{figure*}
\begin{table}
\renewcommand{\arraystretch}{1.3}
\caption{\textsc{Simulation Parameter Values}}\label{tab::mmtm:simpara} 
\center
  \begin{tabular}{ |c | c| }
    \hline
     \textbf{Parameter} & \textbf{Value} \\ \hline
    Transmitted power ($P_{tx}$)& 20dB\\ \hline
    Receiving Sensitivity & -60dB\\ \hline
    Communication Range ($R$) & 7m \\ \hline
    Frequency ($f$) & 73 Ghz \\ \hline
    Warehouse  & $\varepsilon$ = 1.7, $\sigma$ =3.2  \\\hline
    Greenhouse & $\varepsilon$ = 2.4, $\sigma$ =6.2 \\
    \hline
  \end{tabular}
\end{table}

The simulation parameter values use are listed in Table \ref{tab::mmtm:simpara}. The values for path loss model parameters are taken from the experiments described in \cite{pathloss2,pathloss1}. MmWave communication is affected by the oxygen concentration in the environment. This is one of the reasons that the attenuation is high in outdoor MmWave communication. Since there are many plants in greenhouse environment, the oxygen concentration is high compared to indoor environments (e.g. warehouse). Thus, outdoor attenuation and path loss values are considered in simulating propagation model for greenhouse. 

Figure \ref{fig::mmtm:resultmap_w} and \ref{fig::mmtm:resultmap_g} show the sensor locations in the actual map and the topology maps generated by the proposed algorithm for both environments. Most of the nodes are located in the correct position with all three robot paths, but few are deviated from the actual position. For an example, in Figure \ref{fig::mmtm:w_tmVA}-\ref{fig::mmtm:w_tm3d} some of the nodes located in the area covered by $18<x<20$ and $18<y<30$ have an error in it's $z$-coordinates. In actual map those nodes have a $z$-coordinate closer to 3, but in calculated maps it's a value between 1.5 and 3. Moreover, in Figure \ref{fig::mmtm:g_tmAS} Top view, the nodes located closer to $x=0$ and $y=5$ have a slight deviation in $x$ and $z$ coordinates compared to Figure \ref{fig::mmtm:g_network} Top view. 

To measure this deviation numerically, two metrics were used. First is a distance error metric that measures the difference between actual locations and the estimated locations. The distance error needs to be less than the distance between two adjacent nodes in actual map to maintain the connectivity and node orientation in the calculated map. Second metric is called as sector displacement metric, which calculates the number of sector deviations in topology map from the optimum sector to communicate with a node in actual network. If the number of sector displacement is zero, it ensures all the nodes are located in the correct direction and nodes can communicate with neighbours without performing a beam adjustment method. Thus, the number of nodes having a sector displacement higher than zero should be less in a well performed algorithm. Consider two neighbour nodes $i$ and $j$. Based on the actual location of the node $j$, it is located in the sector $s_{ij}$ of node $i$. On the other hand, node $j$ is located in the sector $\hat{s}_{ij}$ of node $i$ based on the estimated locations. Then the sector displacement calculation for the two nodes is, 
\[
 SD_{ij}  =
  \begin{cases}
   \vert s_{ij}- \hat{s}_{ij} \vert  \mod  (N_S/2)& \text{if }  \vert s_{ij}- \hat{s}_{ij} \vert <(N_S/2)\\
   1-\vert s_{ij}- \hat{s}_{ij} \vert  \mod  (N_S/2)    & \text{else } 
  \end{cases}
\]
where $N_S$ is the number of sectors in a sensor node..

\subsection{MmTM Performance Dependency on Number of Antenna Sectors in the Robot}
This section analyses the performance dependency of the algorithm on the number of sectors in the robot side. As mentioned earlier, the number of sectors in robot can vary from zero to 64. Theoretically, when number of sectors are increasing, the area of interest that node can be located is decreasing and as a result, the accuracy of the algorithm will increase. Thus, it was evaluated practically using the two simulation environments discussed above. The evaluation is carried out separately for the three robot paths described previously.

\subsubsection{MmTM-VAA}
\begin{figure}
 \centering
 \subfigure[Average distance error metric]{
  \includegraphics[width=0.45\textwidth, trim={0cm 0cm 1cm 1cm},clip]{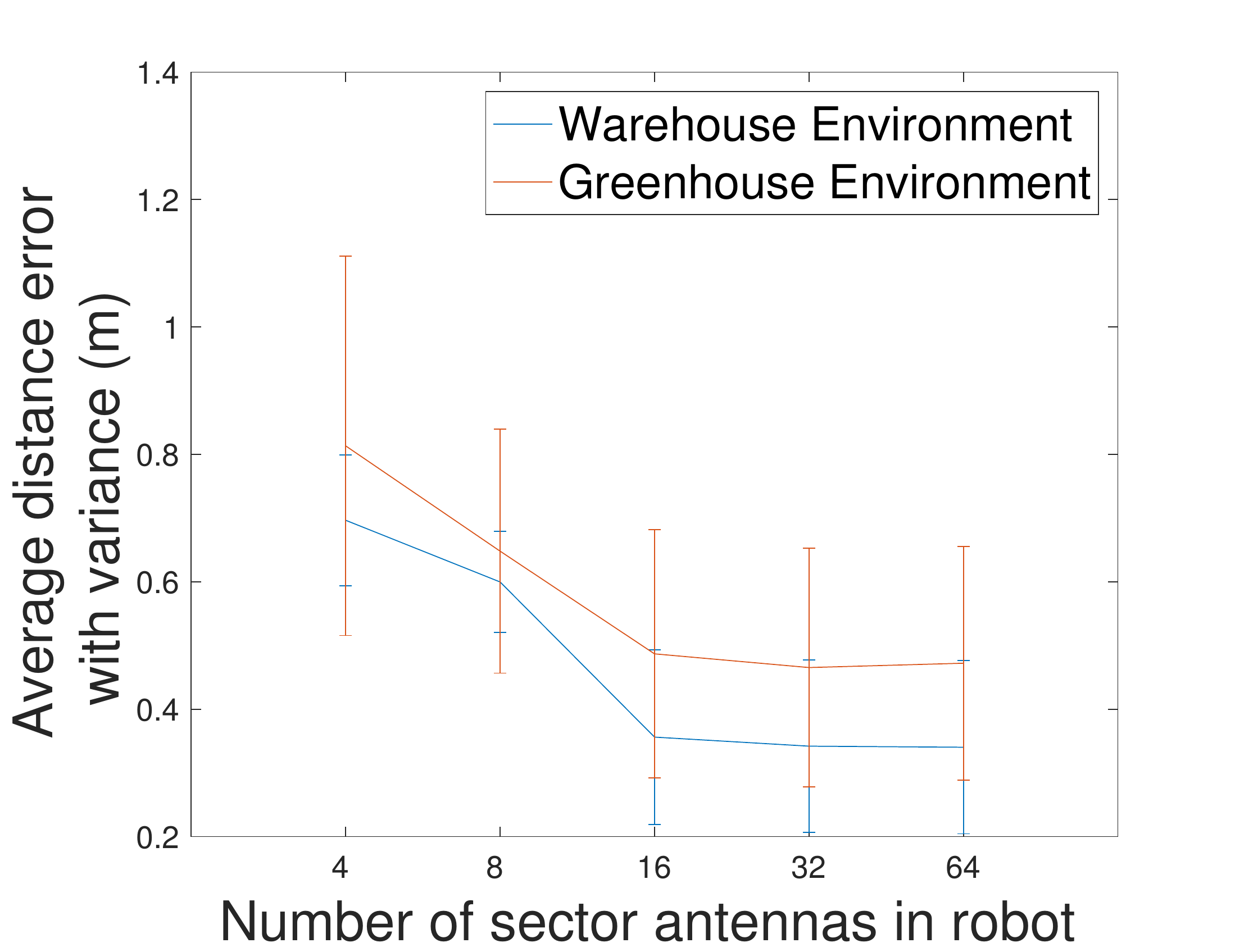}
   \label{fig::mmtm:NSdistance}
   }
 \subfigure[Sector displacement metric]{
  \includegraphics[width=0.45\textwidth, trim={0cm 0cm 1cm .5cm},clip]{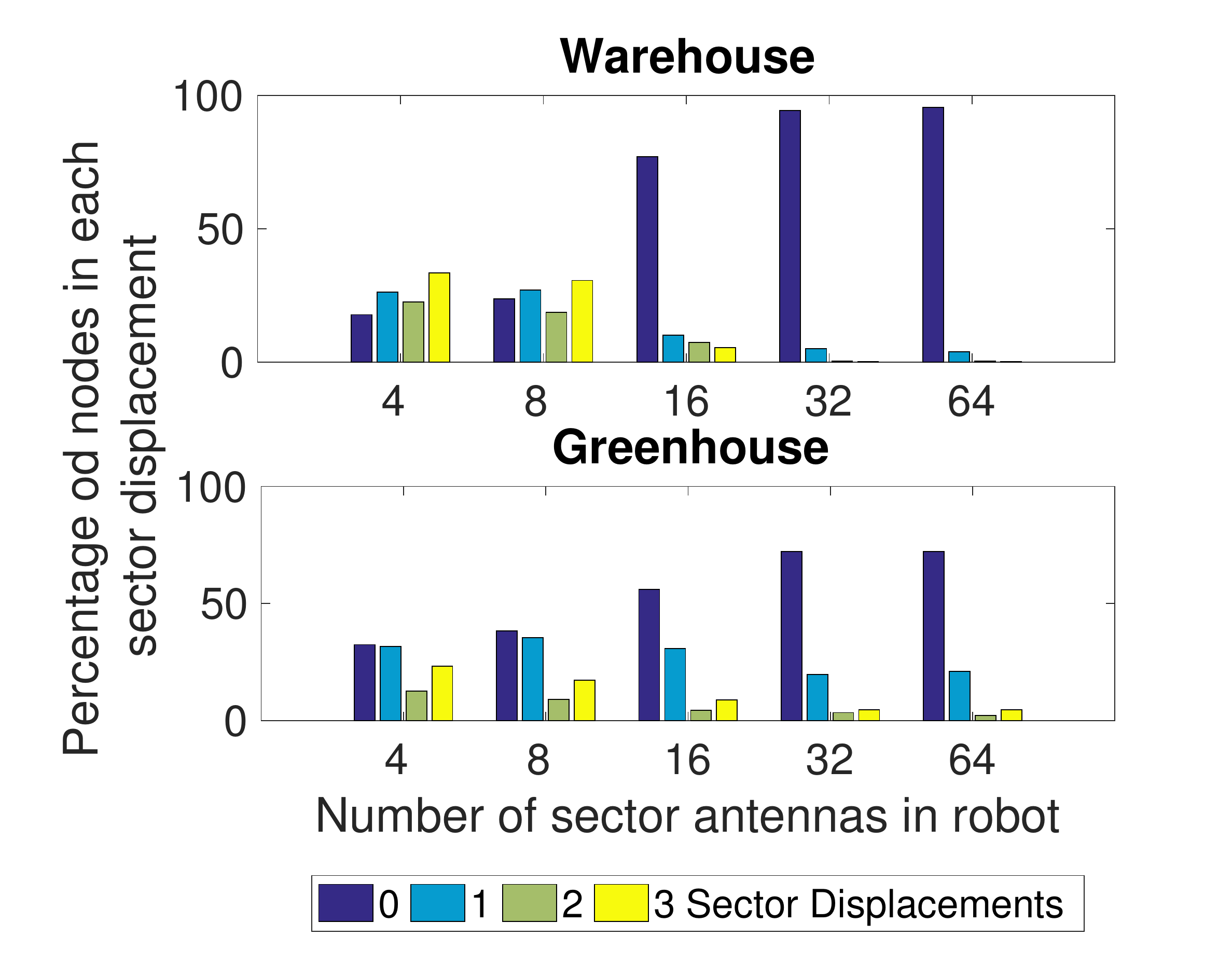}
   \label{fig::mmtm:NSsector}
   }
 \caption{2D VAA: Impact of number of sectors in robot} \label{fig::mmtm:VA}
\end{figure}

In warehouse simulation, the robot moves on the floor with two antenna elements having elevation angles of $25^0$ and $75^0$ respectively. The robot moves on the top of the building in the greenhouse and the elevation angle of vertical elements are $-25^0$ and $-75^0$. The performance of the MmTM with 2D robot path and VAA is shown in Figure \ref{fig::mmtm:VA}. The robot requires 9 minutes and 7 minutes to access all nodes in the greenhouse and the warehouse respectively. From the figures, it can be seen that when the number of antenna sectors increases, the results of both metrics improve. In Figure \ref{fig::mmtm:NSdistance}, the average distance error reduces from 0.7m to 0.35m in warehouse and from 0.8m to 0.48m in greenhouse. Figure \ref{fig::mmtm:NSsector} shows that the percentage of nodes in zero sector displacement increases from 22\% to 91\% in warehouse and from 30\% to 72\% in greenhouse. However, there is a minor impact on results when the number of sectors increases beyond 16 in both environments. 

Beside that, locating nodes in warehouse is more accurate than in the greenhouse. In Figure \ref{fig::mmtm:NSdistance} average distance error with 16 antenna sectors is 0.4m and 0.5m in warehouse and greenhouse respectively. From the Figure \ref{fig::mmtm:NSsector}, it can be seen that 76\% of nodes have a zero sector displacement in warehouse, which is 20\% greater than the greenhouse. The reason for that is robot receives less number of packets due to high attenuation in greenhouse. However, the algorithm has been able to localize the sensors with less than 0.5m average error and a 0.2m variance in this environment.

\subsubsection{MmTM-VBS}
In antenna sweeping, there is a probability of packet loss due to the misalignment of transmitter and receiver. To simulate that, a random number, $\nu _i$, is generated for each sensor node and robot is considered to receive a packet, if the conditions $ P_{rx,i}>\text{Receiving Sensitivity}$ and $\nu _i >C$ are satisfied. $P_{rx,i}$ is the receiving power of the signal send by node $i$ and $C$ is a constant between 0 and 1. In the simulation it has set to 0.6. 

\begin{figure}
 \centering
 \subfigure[Average distance error metric]{
\includegraphics[width=0.45\textwidth, trim={0cm 0cm 1cm 1cm},clip]{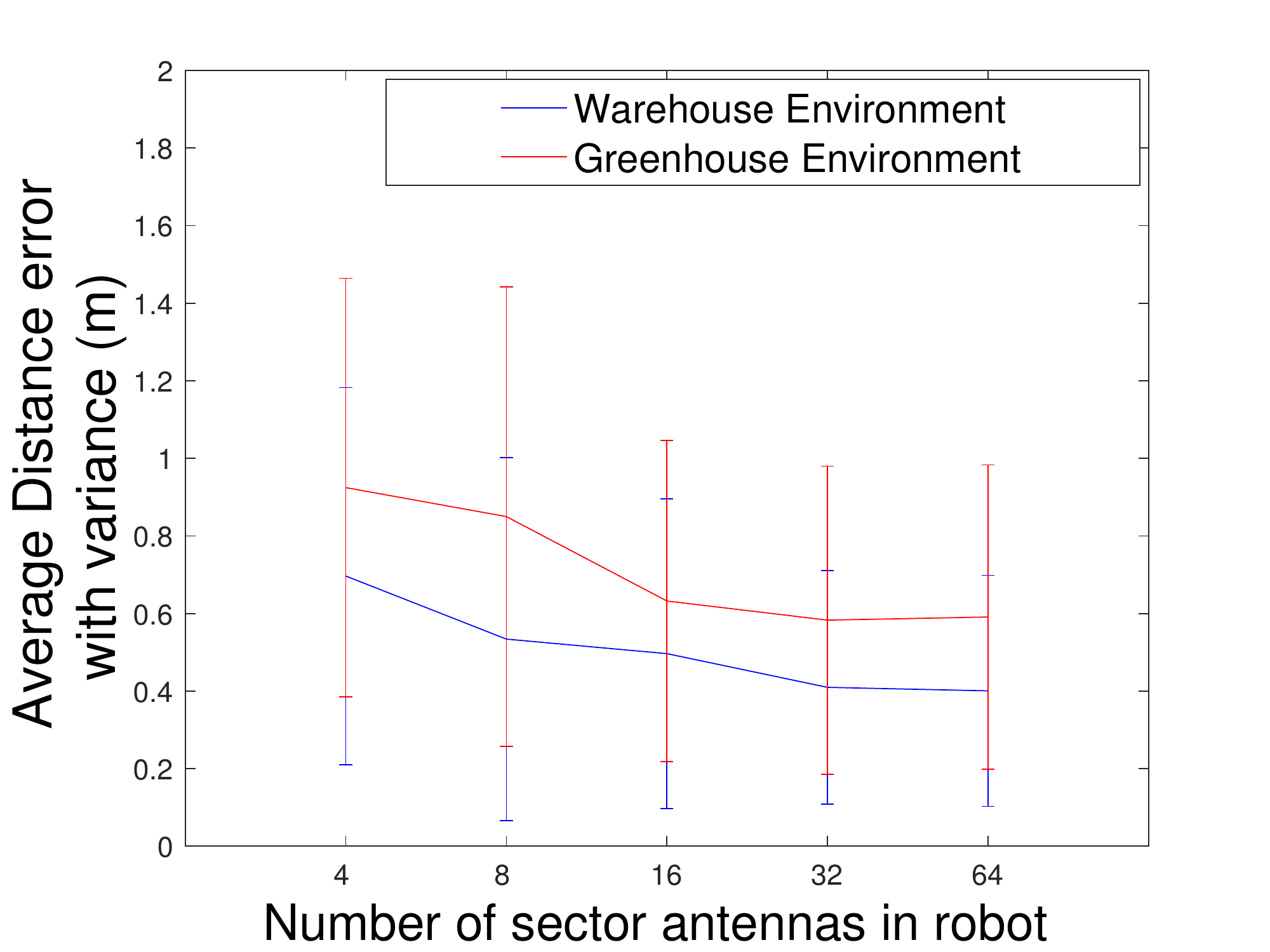}
  \label{fig::mmtm:NSdistance_AS}
   }
 \subfigure[Sector displacement metric]{
  \includegraphics[width=0.45\textwidth, trim={0cm 0cm 1cm .5cm},clip]{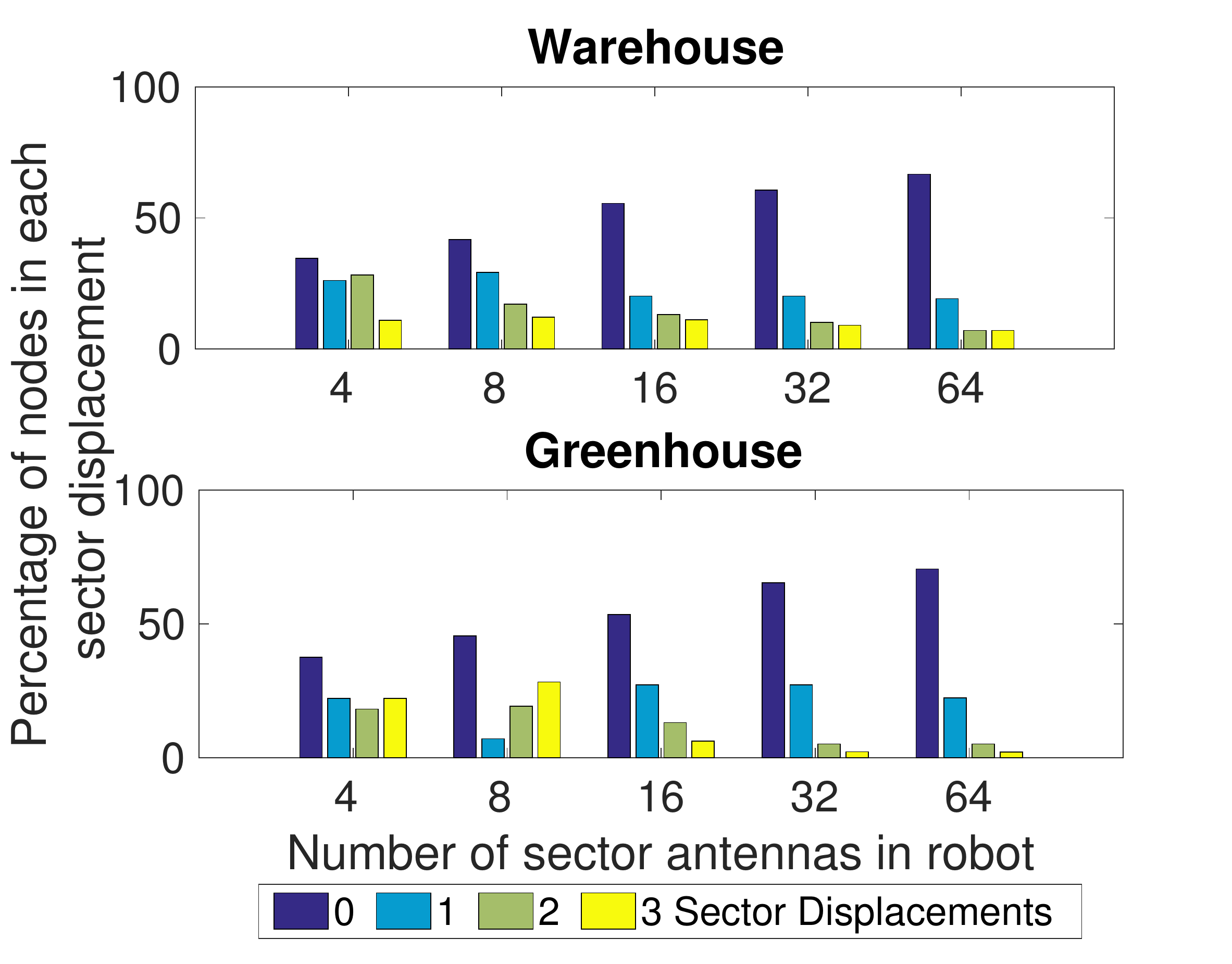}
  \label{fig::mmtm:NSsector_AS}
   }
 \caption{2D VBS: Impact of number of sectors in robot} \label{fig::mmtm:AS}
\end{figure}

Robot follows the same path as in VAA. Thus it requires same time to complete the information gathering phase. The results are shown in Figure \ref{fig::mmtm:AS}. When the number of sector antennas is increasing, the errors in both performance metrics were reduced. However, compared to VAA, VBS incurs approximately 0.1m larger distance error as well as a larger variance as shown in Figure \ref{fig::mmtm:NSdistance_AS}. Moreover, by referring Figure\ref{fig::mmtm:NSsector} and \ref{fig::mmtm:NSsector_AS}, it can be seen that  the percentage of nodes in zero sector displacement with 32 antenna sectors has reduced nearly by 25\% with VBS in both environments. The main reason is packet loss due to the antenna sweep. On the other hand, VBS does not required two antenna sets as in VAA.    

\subsubsection{MmTM 3D}
\begin{figure}
 \centering
 \subfigure[Average distance error metric]{
 \includegraphics[width=0.45\textwidth, trim={0cm 0cm 1cm 1cm},clip]{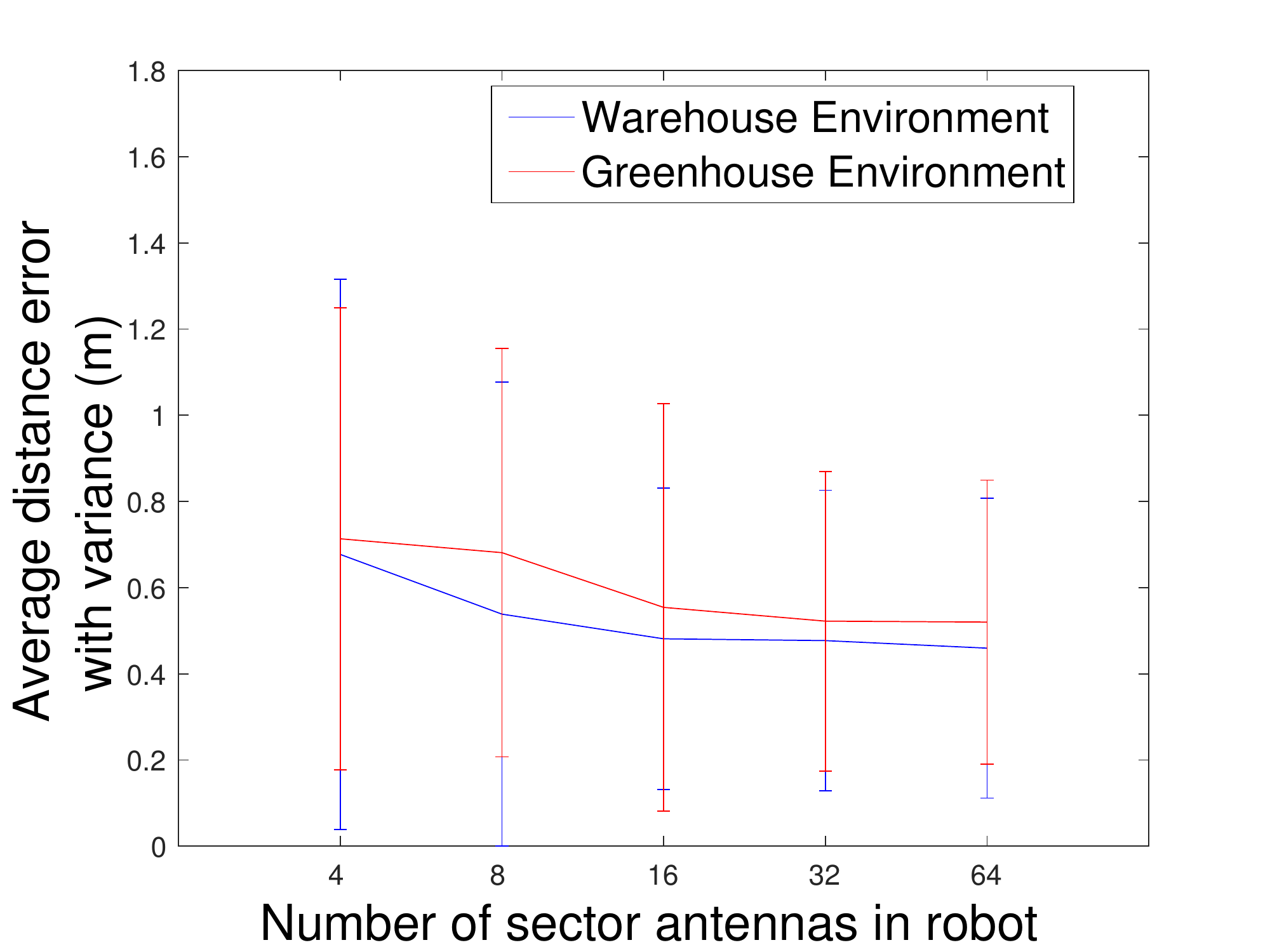}
 \label{fig::mmtm:NSdistance_3D}
   }
  \subfigure[Sector displacement metric]{
  \includegraphics[width=0.45\textwidth, trim={0cm 0cm 1cm .5cm},clip]{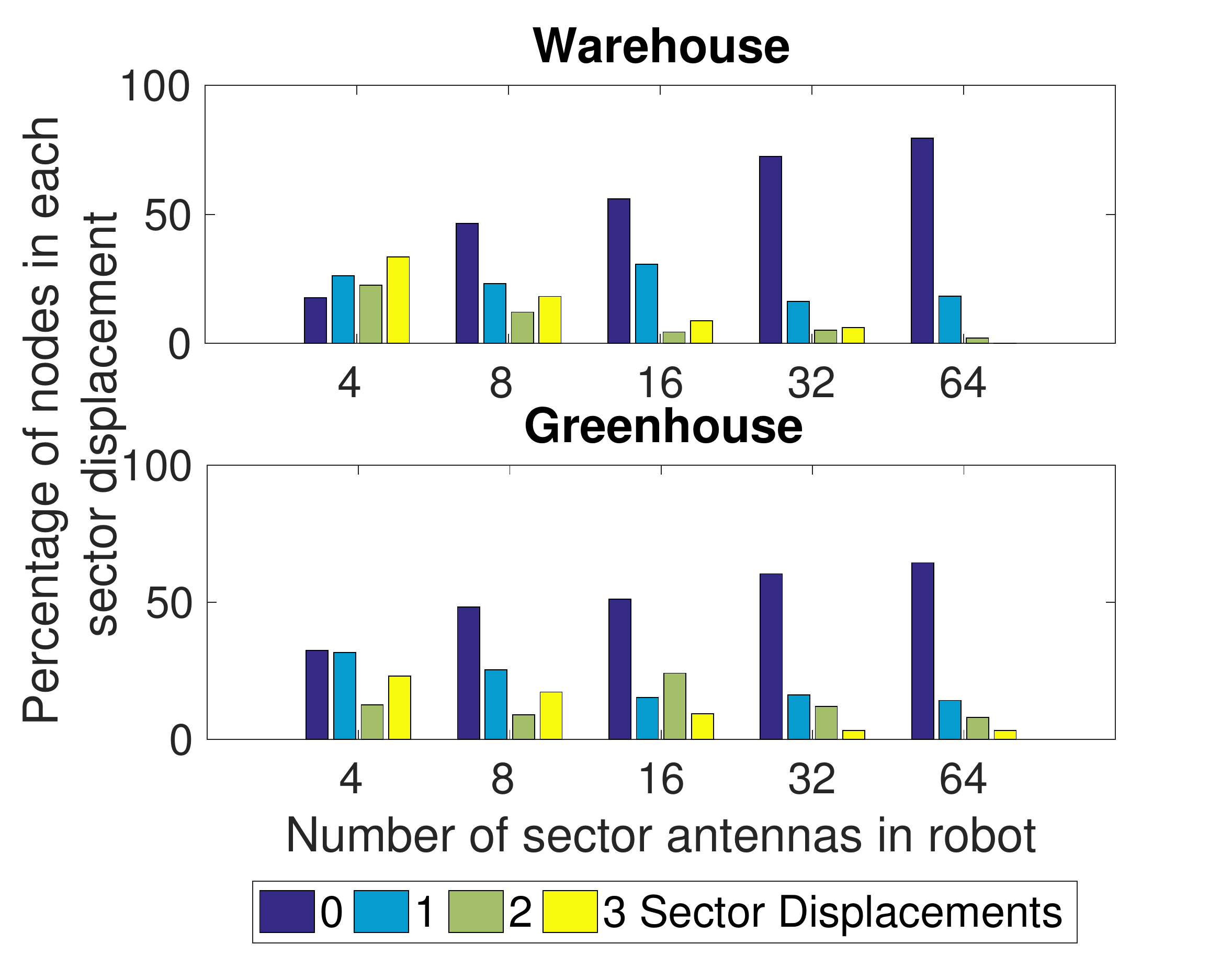}
  \label{fig::mmtm:NSsector_3D}
   }
 \caption{3D: Impact of number of sectors in robot} \label{fig::mmtm:3D}
\end{figure}

The performance of the MmTM with 3D robot path is analysed and shown in Figure \ref{fig::mmtm:3D}. Since the robot is moving in 3D space, the elevation angle of the robot antenna is set as zero. Due to the simple antenna structure in 3D robot path, it requires more time to complete the information gathering phase by accessing all the nodes in network. In greenhouse environment robot requires 21 minutes to gather information and 17 minutes in warehouse environment. Same as the 2D robot paths, the performance of the algorithm increases when the antenna sectors increase. However, in Figure \ref{fig::mmtm:NSdistance_3D} it can be seen that the difference between average distance error in two environments has reduced compared to 2D robot path. Also, in Figure \ref{fig::mmtm:NSsector}, the zero sector displacement with 16 antenna sectors is same for the two environments, but greenhouse has a higher percentage of nodes in two sector displacement compared to the warehouse. 

\subsection{Performance Comparison with Other Algorithms}
\begin{figure}
 \centering
 \subfigure[CDF of distance metric]{
  \includegraphics[width=0.43\textwidth]{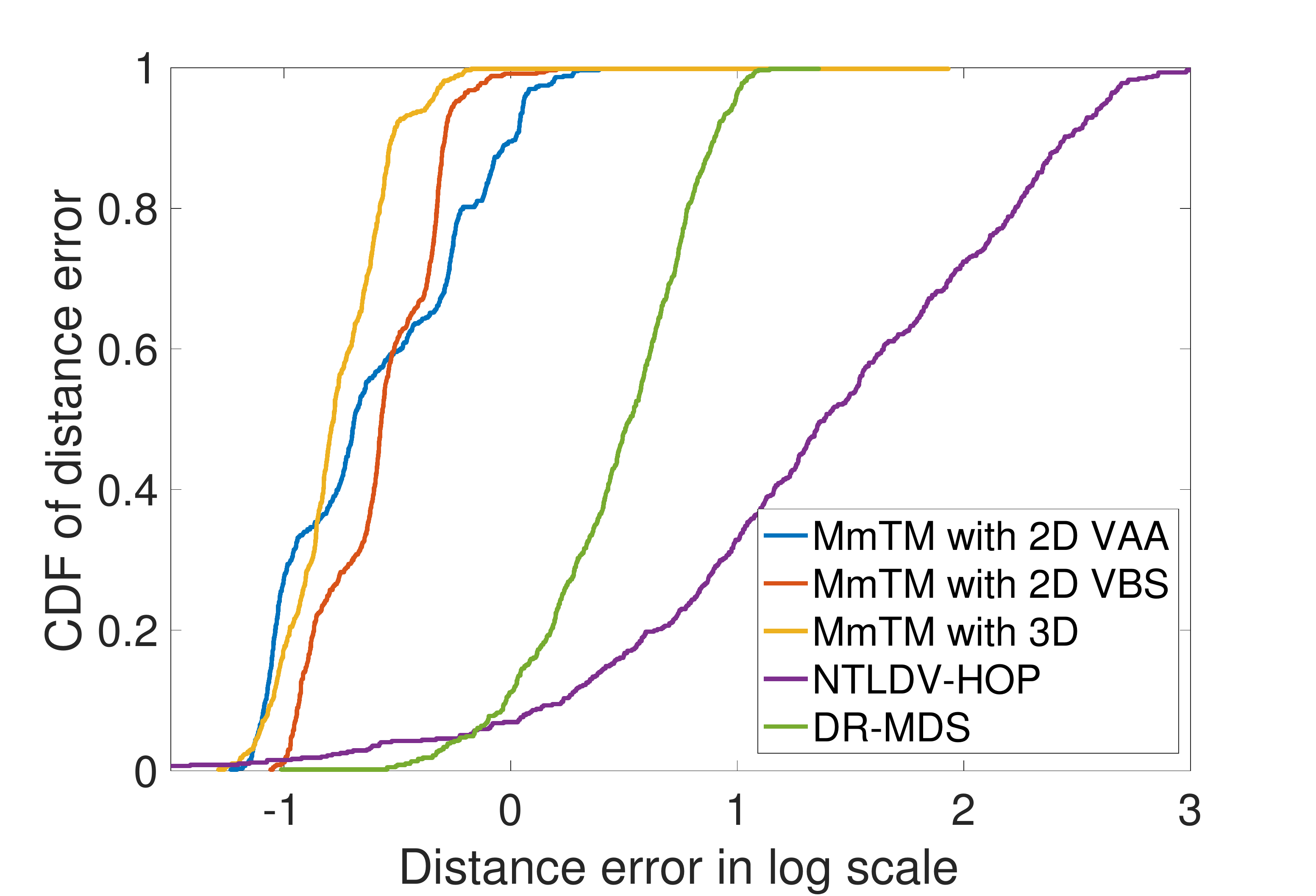}
   \label{fig::mmtm:Cdistance_w}
   }
 \quad
 \subfigure[Sector displacement metric]{
  \includegraphics[width=0.45\textwidth, trim={1.5cm 0cm 4cm 1cm},clip]{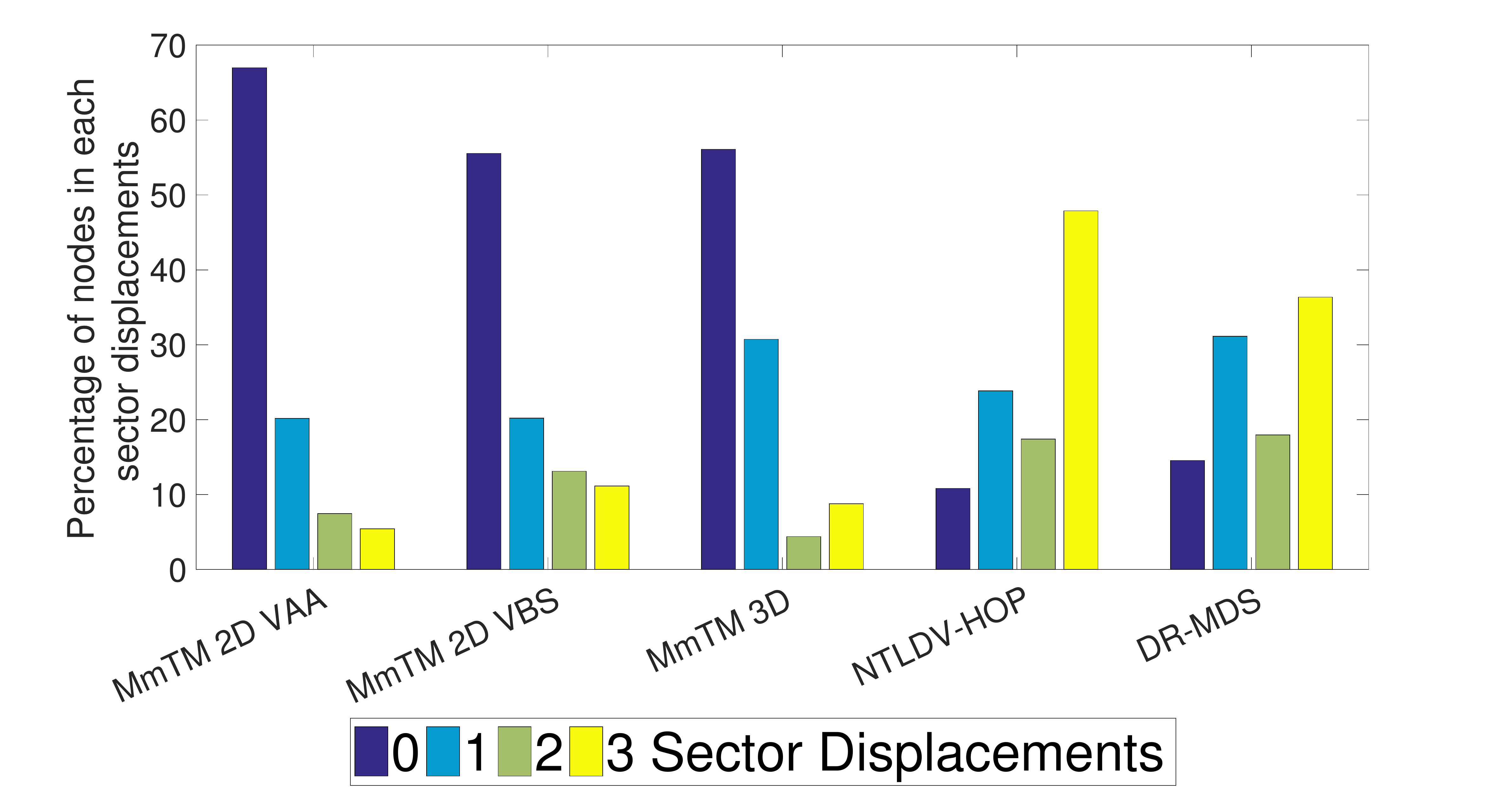}
   \label{fig::mmtm:Csector_w}
   }
 \caption{Performance comparison results for warehose  } \label{fig::mmtm:C_w}
\end{figure}

\begin{figure}
 \centering
 \subfigure[CDF of distance metric]{
  \includegraphics[width=0.43\textwidth]{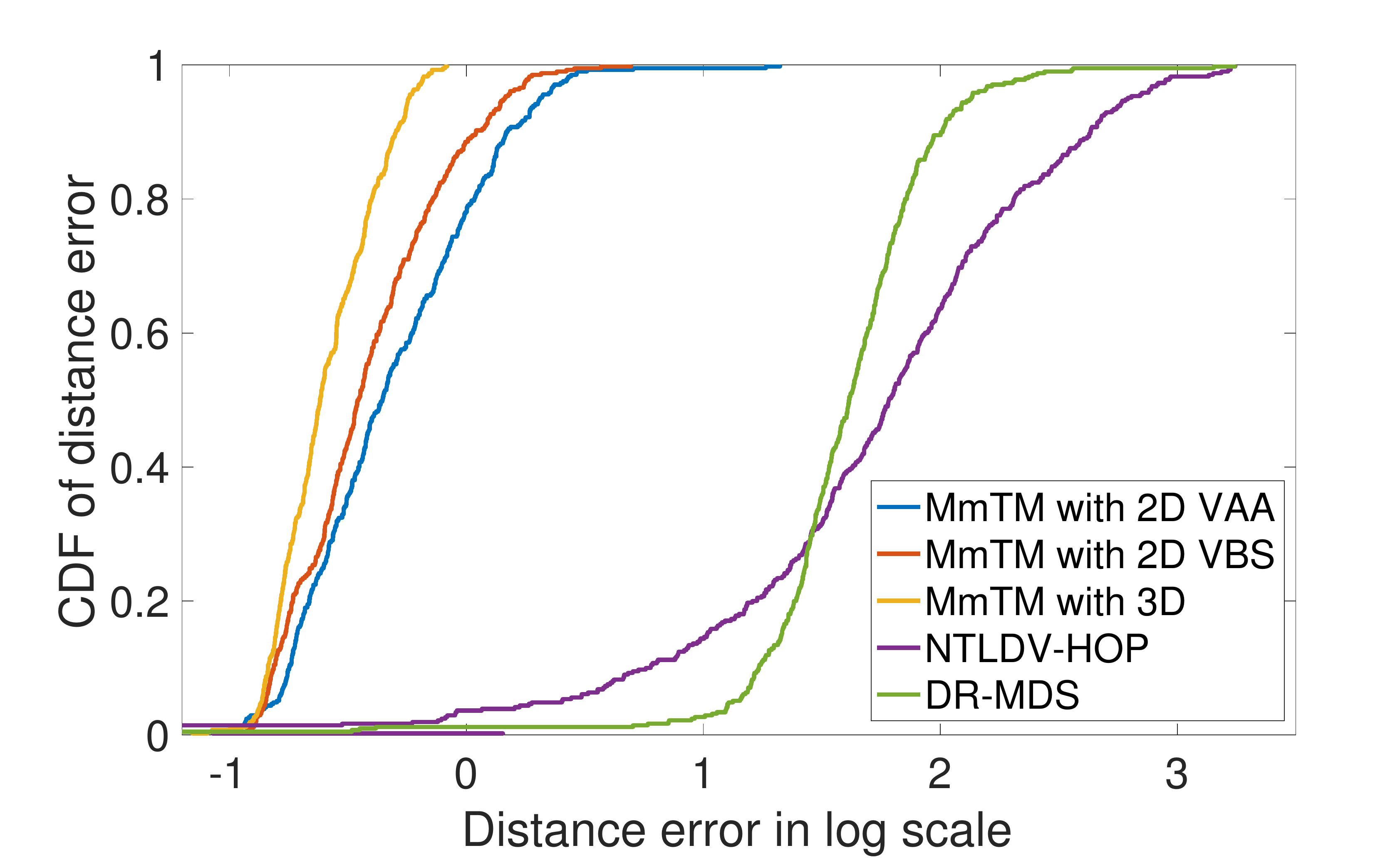}
   \label{fig::mmtm:Cdistance_g}
   }
 \quad
 \subfigure[Sector displacement metric]{
  \includegraphics[width=0.45\textwidth, trim={0cm 0cm .5cm 1cm},clip]{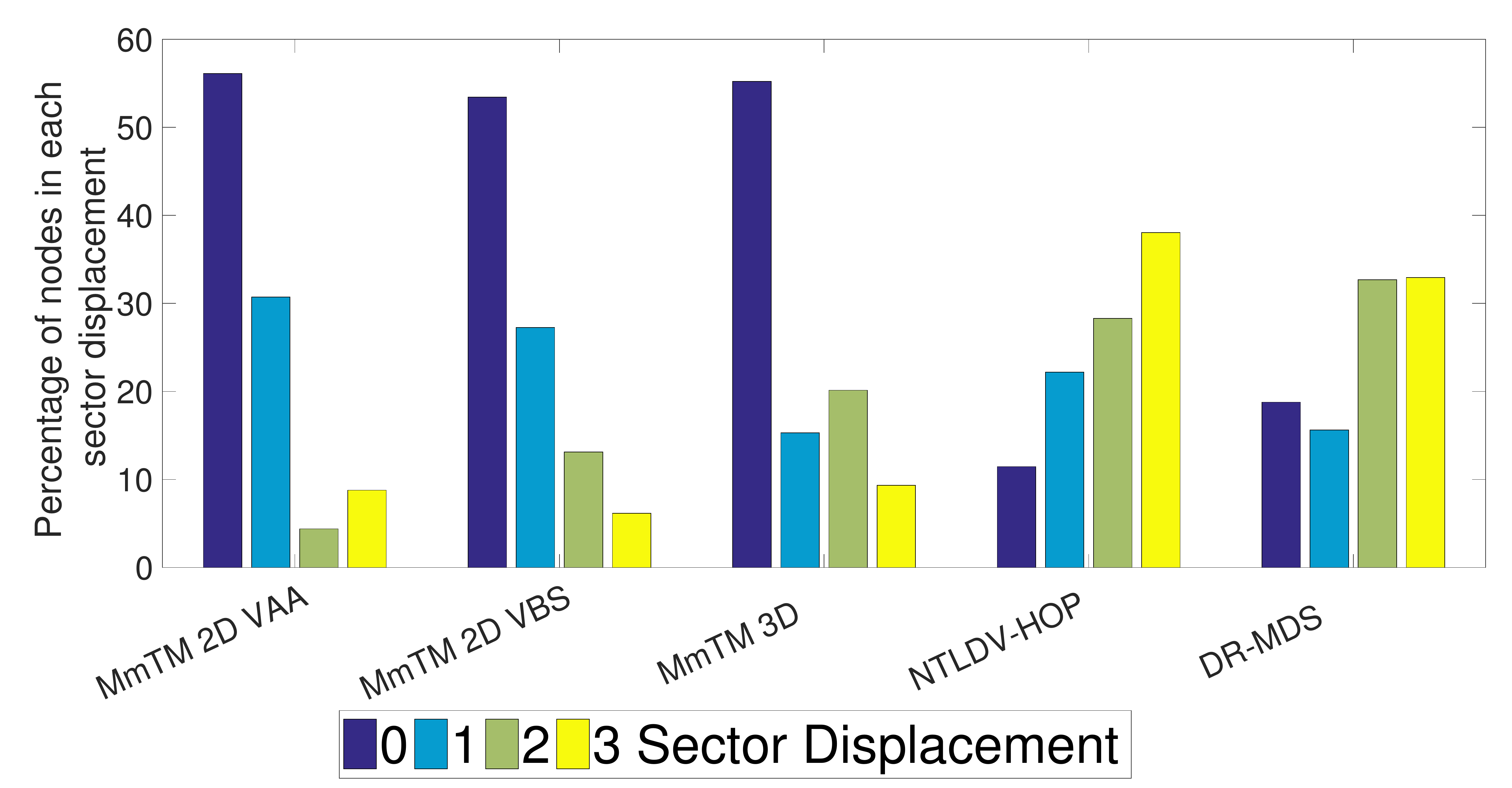}
   \label{fig::mmtm:Csector_g}
   }
 \caption{Performance comparison results for greenhouse } \label{fig::mmtm:C_g}
\end{figure}

The performance of the algorithms is compared with two existing algorithms proposed recently for 3D WSN localization. To perform a fair evaluation a rang-based algorithm named DR-MDS \cite{drmds} and a range free algorithm called NTLDV-HOP \cite{ntldvhop} were chosen. Number of sectors in robot is considered as 16 in all the cases and the results are shown in Figure \ref{fig::mmtm:C_w} and \ref{fig::mmtm:C_g}. To emphasis the differences in distance error CDF, the CDF values against the log distance error were plotted as in Figure \ref{fig::mmtm:Cdistance_w} and \ref{fig::mmtm:Cdistance_g}. According to the results, there is a great improvement in both metrics in proposed MmTM algorithm with all three robot paths and antenna setups. Referring to Figure \ref{fig::mmtm:Cdistance_w} and \ref{fig::mmtm:Cdistance_g}, it can be seen that the distance error of all nodes in MmTM is less than 5\% of the nodes distance error in NTLDV-HOP and DR-MDS. Also, 80\% of nodes in MmTM are located with less than 0.7m distance error, which is a enormous improvement compared to other two algorithms. When comparing the three antenna setups proposed for MmTM algorithm, 3D robot path antenna setup has a better performance compared to 2D robot path antenna setups. 

From Figure \ref{fig::mmtm:Csector_w} and \ref{fig::mmtm:Csector_g}, it can conclude that more than 50\% of nodes have a zero sector displacement in MmTM with different robot paths, which is a significant improvement compared to other two algorithms. For both environment set ups, NTLDV-HOP and DR-MDS algorithms have less than 20\% of nodes with zero sector displacement and more than 30\% of nodes in three sector displacement category. Thus, it required a beam training method to tune the sensor beams before communicating with its neighbours. This requires more time and energy.

\section{Energy Awareness, Computation Overhead and the Limitations of the MmTM Algorithm}\label{sec::mmtm:copm}
As sensor nodes have scarce resources and capabilities such as energy, processing power and memory, WSN algorithms required to be compatible with these limited resources. Thus, this section evaluates the energy awareness and computation overhead of the algorithm. Finally, the limitations of the proposed MmTM algorithm are discussed.

\subsection{Energy Usage Comparison}
Since, energy consumption of the transceiver is more significant than computational energy consumption and sensing energy consumption, this calculation considers only the Mmwave transceiver energy consumption for packet transmitting and receiving. 

Let the total energy consumption by the algorithm is $E_{t}$, the energy required for one packet transmission is $E_{tx}$ and energy required for one packet reception is $E_{rx}$. Then energy required by the MmTM is shown in equations (\ref{eqn::mmtm:proposed}). For simplicity, it has assumed that the packet sizes used in all the transmissions are same.

\begin{equation}\label{eqn::mmtm:proposed}
E_{t} = N(\overline{n}E_{rx}+\overline{n}sE_{tx})
\end{equation}
where, $N$ is the total number of nodes in the network, $\overline{n}$ is the average number of time that a single sensor node receives ISS message from the robot and $s$ is the number of sectors in the sensor nodes. Since there is no energy restriction in robot side, the energy consumption in robot has not considered.

When considering the above energy equations, it can be seen that $\overline{n}$ has a major effect on the algorithm energy consumption. However, it is a controllable factor that can be decided by the application requirement. For an example, if the application needs more accuracy, more number of samples can be used. On the other hand, if it is more important to conserve energy, the number of samples can be reduced.

\subsection{Comparison of Computation Overhead} 
For MmTM calculation, it requires $\mathcal{O}(Nn)$ messages. However, the calculations are done at the central and there is no computational or memory limitations at the central node. Thus, this method can be used to generate an effective and accurate topology map. 

\subsection{Limitations}
One limitation in MmTM is that the coordinate calculation phase starts only after the information gathering from all the nodes in the network is complete. This would be a drawback in some applications in large-scale emergency environments that require sensor locations instantly. The second is that the coordinates are calculated centrally. However, MmTM has been able to generate accurate topology maps without having any prior information about the network or special devices embedded to sensor nodes. 

\section{Conclusion}\label{sec::mmtm:conclusion}
In this chapter, Millimetre Wave Topology (MmTM) algorithm was presented to generate maximum likelihood topology maps for 3D MmWave WSNs. The algorithm utilizes the sector antennas used in MmWave communication for coordinate calculation. A scenario in which an automated mobile robot that extracts the information about sensor nodes by keeping track of the packet reception from sensor nodes along with the IDs of the best sectors to communicate with each node at each time instance was considered. The collected information is mapped to a coordinate system using a signal receiving probability function, which is sensitive to the distance. Three robot movements were simulated, namely a 2D robot path with VAA, a 2D robot path with VBS and a 3D robot path. 

Performance of the proposed algorithm with all three robot movement methods was evaluated. First, the dependency on the number of sector antennas in the robot is presented. It can be seen that after increasing the number of sector antennas beyond 16, the performance is almost same with a less than 0.7m error in more than 80\% of nodes and has a zero sector displacement in more than 50\% of nodes.  Then, MmTM algorithm is compared with two existing algorithms, DR-MDS and NTLDV-HOP, and results show that it outperforms those algorithms in both performance metrics.  The CDF of distance error shows that the distance error of all nodes in MmTM is less than 5\% of the nodes distance error in NTLDV-HOP and DR-MDS. Moreover, more than 35\% of nodes are having zero sector displacement compared to other two algorithms, which proves that MmTM preserves the connectivity as well as the directivity of actual physical map.

\chapter{Distributed Maximum Likelihood Topology Map}\label{chapter:dmmtm}
The two algorithms previously proposed in Chapter \ref{chapter:mltm} and Chapter \ref{chapter:mmtm} centrally calculate the topology coordinates of sensor nodes. This requires more time to gather information from sensor nodes and to calculate the coordinates for large-scale WSNs, which restricts the scalability of sensor networks. In light of this, this chapter proposes a topology mapping algorithm, the \textbf{D}istributed \textbf{M}illi\textbf{m}etre wave \textbf{T}opology \textbf{M}ap (DMmTM), which calculates the sensor coordinates at each sensor. As MmWave communication is an emerging technology that entails more restrictions and characteristics to be considered in topology map calculation, this chapter proposes a distributed topology map for MmWave WSNs. Moreover, this calculation can be easily adapted to RF WSN topology map calculation by assuming the number of sector antennas in sensor nodes as one.

DMmTM relies on a set of anchor nodes and exploits the beamforming protocol to calculate the coordinates of the surrounding nodes. The extracted information from anchors is mapped to a set of topology coordinates using a packet receiving probability function, which is sensitive to distance. As sensors get localized, a subset of such nodes is selected as new anchors to propagate the localization process throughout the network. This significantly reduces the number of initial, pre-localized anchors required. Two variants of DMmTM are proposed based on anchor placement strategies: the DMmTM-Static System (DMmTM-SS) localizes sensors using static anchors, and the DMmTM-Hybrid System (DMmTM-HS) utilizes mobile and static anchors to localize sensors. These algorithms are evaluated using two realistic sensor network environments and compared with existing localization algorithms. 

The chapter is structured as follows and the main results of the chapter were originally published in \cite{dmmtm}. Section \ref{sec::dmmtm:Introduction} offers an introduction and motivation for the research presented in this chapter. Section \ref{sec::dmmtm:algo} discusses the details of proposed DMmTM algorithm. Section \ref{sec::dmmtm:result} presents the performance evaluation and comparison of the algorithm. Section \ref{sec::dmmtm:copm} examines the energy usage and complexity of the proposed algorithm. Finally, Section \ref{sec::dmmtm:conclusion} provides a conclusion of the chapter.

\section{Introduction}\label{sec::dmmtm:Introduction}
Current radio communication frequency spectrum (sub-GHz) is already congested with numerous competing networking technologies thus limiting its availability. MmWave in 30-300 GHz frequency range with its support of multi-Gbps data rates has thus emerged as a potential solution to support WSN applications in many cases, including those requiring high-bandwidth connectivity over short distances \cite{MMW}. Even though MmWave communication supports high bandwidth demanding WSN applications and services, it requires unobstructed LoS between transmitter receiver pairs due to adverse signal propagation characteristics \cite{MMW, MMWlocalization, nit14}. Majority of signal propagation issues has been mitigated by utilising narrow beam width antenna arrays \cite{kutty16}. Therefore, adaptive beamforming has become an essential aspect of MmWave communication that determines the pair of antenna sectors with highest signal quality between the transmitter and receiver \cite{kutty16, nit14, Hosoya15}. In addition, practical feasibility of adaptive beamforming has been demonstrated through the recent indoor MmWave (60 Ghz) communication standards such as the IEEE 802.11ad \cite{11ad}. 

Thus, accurate and scalable localization algorithm for MmWave WSN becomes even more important to support beamforming techniques and to location-aware protocols. As the signal attenuation is high and propagation is sensitive to many factors in MmWaves, topology map of WSN network is a more promising and a feasible solution than distance estimation based localization techniques. Chapter \ref{chapter:mmtm} presents an MmWave Topology Map (MmTM) algorithm that exploits the antenna beamforming algorithms to derive topology maps.  However, MmTM is a centralized algorithm that evaluates topology maps using a mobile robot, which traverses the network communicating with each device in the network. MmTM requires direct accessibility to all the devices from mobile robot, and requires significant traversal time to access all devices.

This chapter proposes the \textbf{D}istributed \textbf{Mm}Wave \textbf{T}opology \textbf{M}ap (DMmTM) algorithm for locating devices in MmWave WSN networks. DMmTM algorithm relies on a set of anchor nodes, i.e., devices whose locations are known, to find the coordinates of surrounding sensor nodes at unknown locations. Both anchors and sensor nodes are equipped with antenna arrays with steerable narrow beams, which have a limited effective coverage angle compared to omni directional antennas \cite{omnivsdirec}. This chapter proposes variants of localization algorithms for two common anchor placement strategies. DMmTM for a Static System (DMmTM-SS) is proposed for static anchor deployments and DMmTM for a Hybrid System (DMmTM-HS) is proposed for a network that combines static and mobile anchor deployments.  For evaluation of algorithms, a random distribution for static anchor deployment is considered while mobile anchors follow a random path to communicate with sensors. Each sensor node records two vectors based on its communication with a subset of anchor nodes. First is a vector of neighbour anchors' location and second is a vector identifying the best antenna sectors to communicate with those anchors. Then, the maximum likelihood topology coordinates is estimated for each sensor node using these two vectors and a packet receiving probability function.

The performance of the DMmTM algorithm is evaluated using two simulation environments, a greenhouse and a warehouse, representing significantly different environmental characteristics. Even though DMmTM-SS and DMmTM-HS have similar distance errors and slightly higher direction error compared to centralized MmTM proposed in Chapter \ref{chapter:mmtm}, the distributed algorithm proposed in this chapter is more efficient in terms of communication overhead, energy consumption and computation complexity. 

\section{Distributed Millimeter Wave Topology Map (DMmTM)}\label{sec::dmmtm:algo}
This section describes the DMmTM algorithm, which creates topology maps for 3D MmWave WSNs. DMmTM calculates topology coordinates of sensors using connectivity and directional information gathered by anchors. This algorithm follows some of the steps in the IEEE 802.11ad beamforming protocol described in the Appendix \ref{CP}. DMmTM consider only the SLS phase, which identifies the optimum sector pair\footnote{two sectors with highest received signal quality}. Moreover, CSMA/CA mechanism in the IEEE 802.11ad protocol is used as the collision avoidance method \cite{nit14, Niu2015}.

This section proposes two ways of topology coordinate calculations based on the type of anchor nodes used in the network, i.e. static or mobile. In both scenarios, sensors are fixed. Following subsections describes the two algorithms.

\subsection{Distributed Millimetre Wave Topology Map for Static System (DMmTM-SS)}
\begin{figure*}
 \centering
  \includegraphics[width=0.85\textwidth]{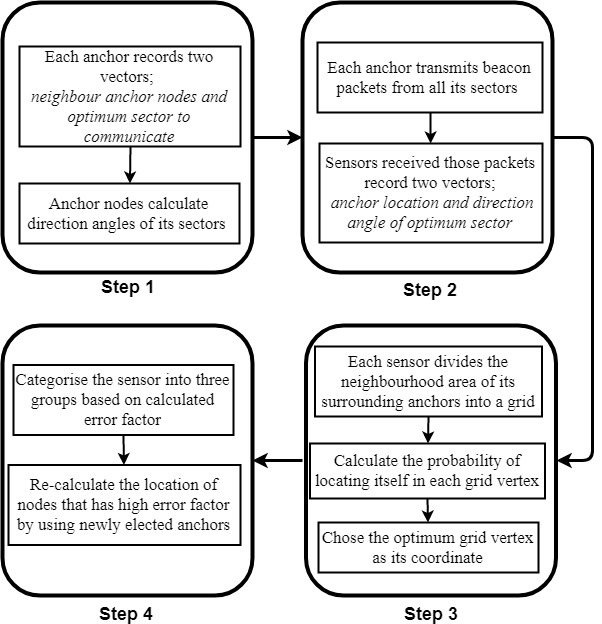}
 \caption{Work flow of DMmTM-SS}  \label{flowchart}
\end{figure*}
DMmTM-SS calculates the topology coordinates of sensors using static sensors, which are randomly distributed throughout the environment. Static anchors know their own coordinates, but are unaware of the direction of sector antennas. Also, it is assumed that elevation angles of sensor antennas are oriented in different directions to maintain the coverage of the network. Number of initially deployed anchors ($N_A$) is less compared to number of sensors ($N$) in the network. Thus, a subset of sensor nodes is selected as new anchors after calculating their topology coordinates and going through a node filtration process. These newly selected anchor nodes are then used to calculate the coordinates of non-localized sensors and sensors do not have sufficient information to localize. 

DMmTM-SS algorithm consists of four steps, namely (i) Direction estimation of anchor sectors, (ii) Information gathering, (iii) Information mapping and (iv) Node filtration. The main tasks of each step is illustrated in Figure \ref{flowchart} and described in following subsections.

\subsubsection{Direction Estimation of Anchor Sectors}
In this step, each anchor estimates its sector antenna coverage area by communicating with any other anchors located within its communication range ($R_c$) as in Step 1 of Figure \ref{flowchart}. The pseudo code of this step is illustrated in the Appendix \ref{sec::PC::dmmtm1}.

Consider $N_A$ stationary anchors are randomly distributed and labelled as $A=\{ a_1, a_2,$ $ ...a_i, ..., a_{N_A}\}$, whose locations are known but direction or orientation information are not known.  Each anchor $a_i$ has $N_S$ number of sectors and the $p^{th}$ sector (${sec_p^{a_i}}$) coverage area is defined as $\Omega _{sec_p}^{a_i}$. To estimate the direction angle of a sector antenna, both elevation angle\footnote{the direction angle of the beam in vertical plane} and azimuth angle\footnote{the direction angle of the beam in horizontal plane} need to be calculated. We assume that VBW of a sector is fixed and it is a known parameter. The HBW of a sector can be calculated as $2\pi /N_S$. 

As in the first block of the Step 1 in Figure \ref{flowchart}, each anchor records two vectors. First, the set of anchors located within the neighbourhood of anchor $a_i$, which is defined as $\kappa _{a_i}=\{k:k\in A\:\:and\:\:\Omega _{sec_p}^{a_i} \cap \Omega _{sec_q}^{k} \neq \emptyset , \:(p,q)\in \{1,...N_S\}\}$. To find $\kappa _{a_i}$, each anchor $a_i$ communicates with its neighbour anchors using the IEEE 802.11ad standard protocol \cite{11ad} and finds out its optimum sector to communicate. Then it records the second vector, which is the set of optimum sector to communicate with each neighbour anchor in $\kappa _{a_i}$ respectively. That can be defined as $\mathcal{H} _{a_i}=\{h_{k}:k\in \kappa _{a_i}\:\:and\:\:h_{k}\in {sec_p^{a_i}},\:p\in \{1,...N_S\}\}$. 
\begin{figure}
 \centering
 \subfigure[Top view of anchor $a_i$ sectors when $N_S$=4]{
  \includegraphics[width=0.7\textwidth,trim={1.3cm 14.5cm .4cm 3cm},clip]{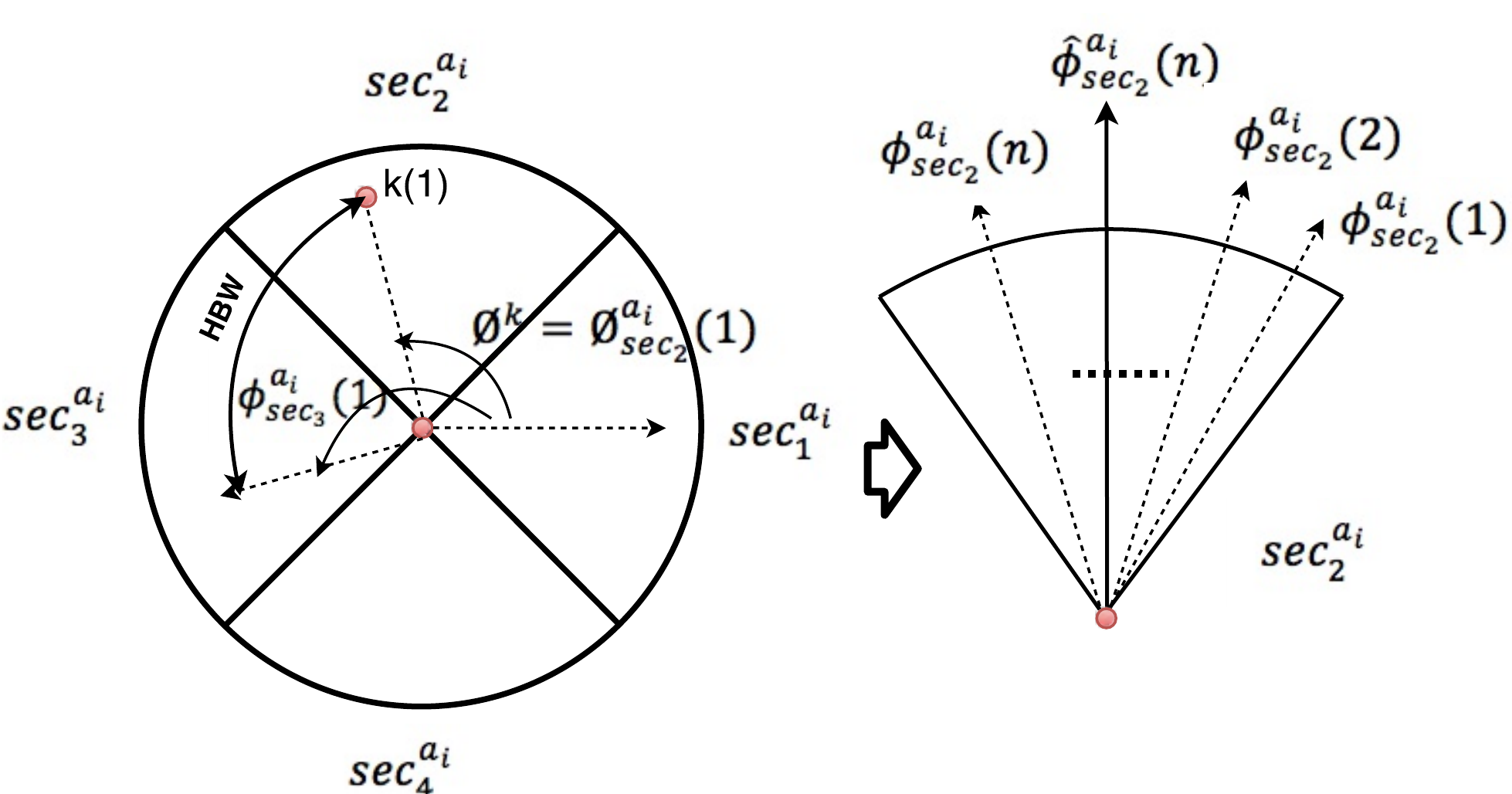}
   \label{aitopview}
   }
 \quad
 \subfigure[Side view of anchor $a_i$'s sector two]{
  \includegraphics[width=0.5\textwidth,trim={4.3cm 20cm 6.3cm 1cm},clip]{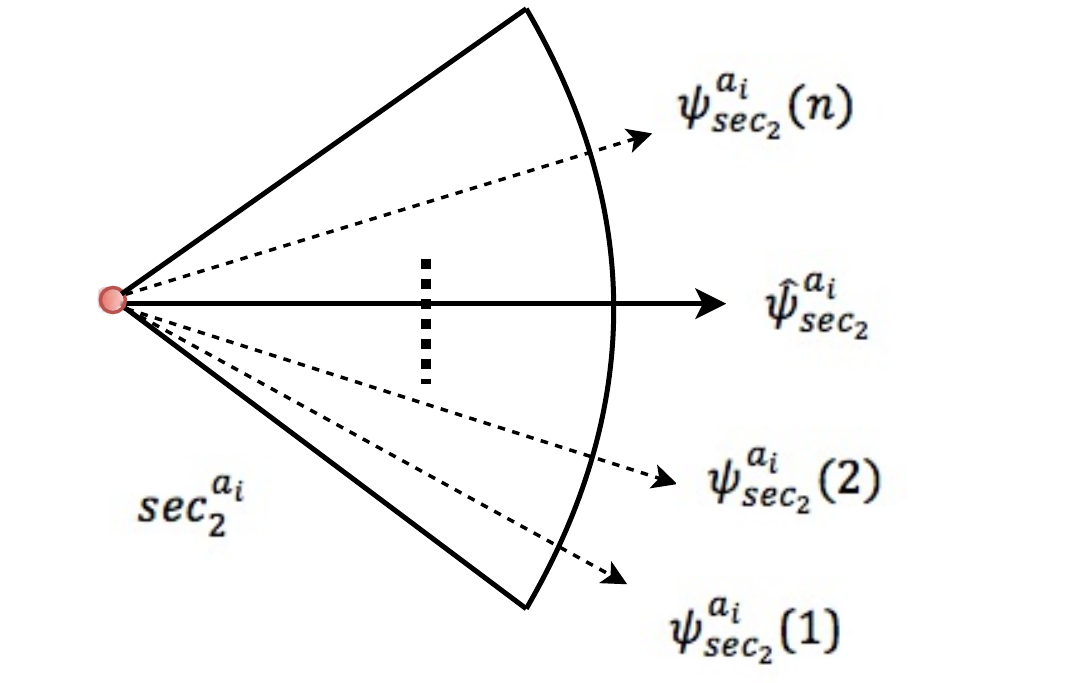}
   \label{aiavaragetopvies}
   }
 \caption{$\hat{\phi } ^{a_i}_{sec_p}$ and $\hat{\psi }^{a_i}_{sec_p}$ calculation}  \label{angles}
\end{figure}

Then the second task is to calculate the elevation and azimuth angles of all the sectors of anchor $a_i$ w.r.t. anchors in $\kappa _{a_i}$. Let $\Phi _{a_i}=\{\phi _k:k\in \kappa _{a_i}\:\:and\:\:0<\phi <2\pi \}$ and $\Psi _{a_i}=\{\psi _k:k\in \kappa _{a_i}\:\:and\:\:0<\psi <2\pi \}$ be the sets of calculated azimuth and elevation angles. Initially we assumed that calculated angles based on the anchor coordinates are the initial direction of each sector. Thus, the first element of $\mathcal{H} _{a_i}$ is considered to calculate the initial azimuth and elevation angle of sector $h_k(1)$, which is $\phi _{k}(1)$ and $\psi _{k}(1)$ that calculated using coordinates of anchor $a_i$ and $k(1)$. Then the direction angles of all sectors of $a_i$ are calculated using equation (\ref{direction}). This explains in Figure \ref{angles}.  As in the first circle of Figure \ref{aitopview}, $k(1)$ anchor located in the $sec_2^{a_i}$ and by considering the coordinate difference of two anchors, the initial direction of sector $sec_2^{a_i}$ is calculated i.e. $phi _{k}(1)$. Then using HBW, the direction angles of rest of the sectors are calculated using equation (\ref{direction}) and those are $\phi ^{a_i}_{sec_1}(1),...,\phi ^{a_i}_{sec_4}(1)$. The same procedure is followed to calculate the elevation angles and these steps are repeated for all the elements in $\mathcal{H} _{a_i}$ .
\begin{eqnarray} \label{direction}
\phi ^{a_i}_{sec_p}(1) &=& \phi _{k}(1) + HBW*\Delta _{sec_p}(1) \nonumber \\ 
\psi ^{a_i}_{sec_p}(1) &=& \psi _{k}(1)
\end{eqnarray}
where, $\phi ^{a_i}_{sec_p}(1)$ and $ \psi ^{a_i}_{sec_p}(1)$ are the initial azimuth and elevation angles of sector $sec_p^{a_i}$ that calculated based on information received from anchor k(1) as shown in Figure \ref{angles}. $\Delta _{sec_p}$ is the number of sectors between $sec_p^{a_i}$ and $h _{k}(1)$ in counter clockwise direction.

The average sector direction that satisfies all the calculated angle sets is then estimated  using equation (\ref{Fdirection}). These calculated final angles are shown in Figure \ref{angles} as $\hat{\phi } ^{a_i}_{sec_p}$ and $\hat{\psi }^{a_i}_{sec_p} $.
\begin{eqnarray} \label{Fdirection}
\hat{\phi } ^{a_i}_{sec_p} &=& \frac{\Sigma _{n=1}^{\mid \kappa _{a_i} \mid }\phi _{k}(n) + HBW*\Delta _{sec_p}(n)}{\mid \kappa _{a_i} \mid } \nonumber \\ 
\hat{\psi }^{a_i}_{sec_p} &=& \frac{\Sigma _{n=1}^{\mid \kappa _{a_i} \mid }\psi _{k}(n)}{\mid \kappa _{a_i} \mid } 
\end{eqnarray}

\subsubsection{Information Gathering}
Having estimated the direction angles each anchor transmits beacon packets, which includes the sector direction angle and the anchor coordinates. These beacon packets are transmitted from all its sectors as in the Step 2 of Figure \ref{flowchart}. The pseudo code of this step is illustrated in the Appendix \ref{sec::PC::dmmtm2}. The sensors listen for those packets in quasi-omni mode and chose the optimum antenna sector ($sec_{opt}$) to communicate. This is the same procedure as in ISS sub-phase of SLS phase in the IEEE 802.11ad \cite{11ad}.

Consider a network consisting of $N$ stationary sensors labelled $S=\{ s_1,s_2,...s_j,...,s_N\}$, whose locations are not known. $\mathcal{V}_{s_j}=\{v:v\in A\:\:and\:\:\: \Omega _{sec_p}^{s_j} \cap \Omega _{sec_q}^{v} \neq \emptyset , \:(p,q)\in \{1,...N_S\}\}$ is the set of anchors located within the communication range of sensor $s_j$. Based on the information gathered from neighbourhood anchors, node $s_j$ records two set of vectors, namely, anchor locations $L _{s_j}=\{l_v\equiv (x_v,y_v,z_v):v\in \mathcal{V} _{s_j}\:\:and\:\:x,y,z\in \mathbb{R}\}$ and optimum sector direction angle $\Theta _{s_j} =\{\theta _v \equiv (\alpha _v, \beta _v):v\in \mathcal{V}_{s_j}\:\:and\:\:\alpha _v \in \hat{\phi } ^{v}_{sec_{opt}}\:\:and\:\:\beta _v \in \hat{\psi } ^{v}_{sec_{opt}}\}$.

\subsubsection{Information Mapping}
This section describes the Step 3 of Figure \ref{flowchart}, which is the information mapping phase. The pseudo code of this step is illustrated in the Appendix \ref{sec::PC::dmmtm3}. The information gathered in previous step is mapped to topology coordinates of sensor $s_j$ using a packet receiving probability function $S(d)$ used in Chapter \ref{chapter:mltm}. This function describes the probability of receiving packets from an anchor when sensor is at a particular distance. Let, $S(d)$ be the probability value when sensor is at distance $d$ from the anchor. Then, $S(d)$ is defined as, 
\begin{eqnarray}\nonumber
S(d) := p_0 \:\:\:\:\: \forall \: d\leq r \\ \nonumber
S(d) := 0 \:\:\:\:\: \forall \: d\geq R \\ 
S(d) := \frac{p_0(R-d)}{(R-r)} \:\:\:\: \forall \: r<d<R \label{sd2}
\end{eqnarray}
where $0 < p_0 \leq 1$, $0 < r < R \leq R_c$ are some given constants. 

The sensor $s_j$ divides the R-neighbourhood of the anchors in $\mathcal{V} _{s_j}$ into a small rectangular grid $G=\{g_1,g_2,...,g_l\}$ with grid size of $\delta k$. A function $P_j(x_l, y_l, z_l)$ describes the probability of obtaining the vectors $L _{s_j}$ and $\Theta _{s_j}$ when the sensor node $s_j$ is located at the grid point $g_l=(x_l,y_l,z_l)$. $P_j(x_l, y_l, z_l)$ can be calculated using equation (\ref{probability})
\begin{eqnarray}
P_j(x_l, y_l ,z_l) = Z_j(d_{l1}) Z_j(d_{l2})...Z_j(d_{lm})...Z_j(d_{l\mid \mathcal{V} _{s_j} \mid }) \label{probability}
\end{eqnarray}

 $Z_j(d_{lm})$ is defined as,
\[
 Z_j(d_{lm}) =
  \begin{cases}
   S(d_{lm}) & \text{if   Cond(1) and Cond(2) satisfy} \\
   0     & \text{else } 
  \end{cases}
\]
where, $d_{lm}$ is the distance between grid point $g_l$ and the $m^{th}$ element of anchor in $\mathcal{V}$. The location of anchor $v(m)$ is $l_v(m)=(x_v(m),y_v(m),z_v(m))\in L_{s_j}$. Then $d_{lm}$ is calculated as,
 \begin{equation}\label{dist}
 d_{lm} = \sqrt{(x_l-x_v(m))^2 + (y_l -y_v(m))^+ (z_l -x_v(m))^2}
 \end{equation}
 The direction angles of optimum sector of anchor $v(m)$ is $\theta_v(m)=(\alpha _v(m),\beta_v(m))\in \Theta _{s_j}$.Then the condition(1) and condition(2) are define as,
\begin{eqnarray}
\text{Cond(1)} = \alpha _v(m)-HBW/2\leq\lambda (lm)\leq\alpha _v(m)+HBW/2 \nonumber\\
\text{Cond(2)} = \beta_v(m)-VBW/2\leq\gamma (lm)\leq\beta_v(m)+VBW/2 \nonumber
\end{eqnarray}
where, $\lambda (lm)$ and $\gamma (lm)$ are azimuth and elevation angles respectively between anchor of $v(m)$ and grid point $g_l$. 

After calculating $P_j(x_l, y_l,z_l)$ value for each grid vertex, we take the grid vertex $(x_l^{opt},$ $y_l^{opt},z_l^{opt})$ delivering the maximum value of $P_j(x_l, y_l,z_l)$ among all vertices of the grid. This is an approximation of the optimal location of the sensor node $s_j$.  

\subsubsection{Node Filtration}
The final step of Figure \ref{flowchart} is node filtration. In this phase, localized nodes are grouped into three categories, (i) New anchors ($C_a$), (ii) Nodes with good location estimates ($C_g$) and (iii) Nodes with bad location estimates ($C_b$). The pseudo code of this step is illustrated in the Appendix \ref{sec::PC::dmmtm4}. For the categorization an error estimation method based on the local neighbourhood nodes is used. Sensor $s_j$ transmits a beacon message to its neighbours and requests the coordinates. Then it calculates a parameter called error of neighbour scattering ($E_{N_S}^{s_j}$) as described in equation (\ref{Nsacttering}).
\begin{equation}\label{Nsacttering}
E_{N_S}^{s_j} = \frac{e_j}{\mathcal{N} _{s_i}}
\end{equation} 
where, $e_j$ is the number of nodes located outside the sensor $s_j$'s communication range based on the calculated topology coordinates i.e. the distance between two nodes is greater than $R_c$. $\mathcal{N} _{s_i}$ is the total number of nodes in sensor $s_j$ neighbourhood. This is estimated based on the packets received.

After calculating $E_{N_S}^{s_j}$, the nodes are categorized using the following rule.
\[
 \text{Node category} =
  \begin{cases}
   C_a & E_{N_S}^{s_j}\leq \tau _1 \\
   C_g     & \tau _1< E_{N_S}^{s_j}\leq \tau _2 \\
   C_b     & E_{N_S}^{s_j}>\tau _2
  \end{cases}
\]
\begin{figure}
\centering
\includegraphics [width=0.8\textwidth]{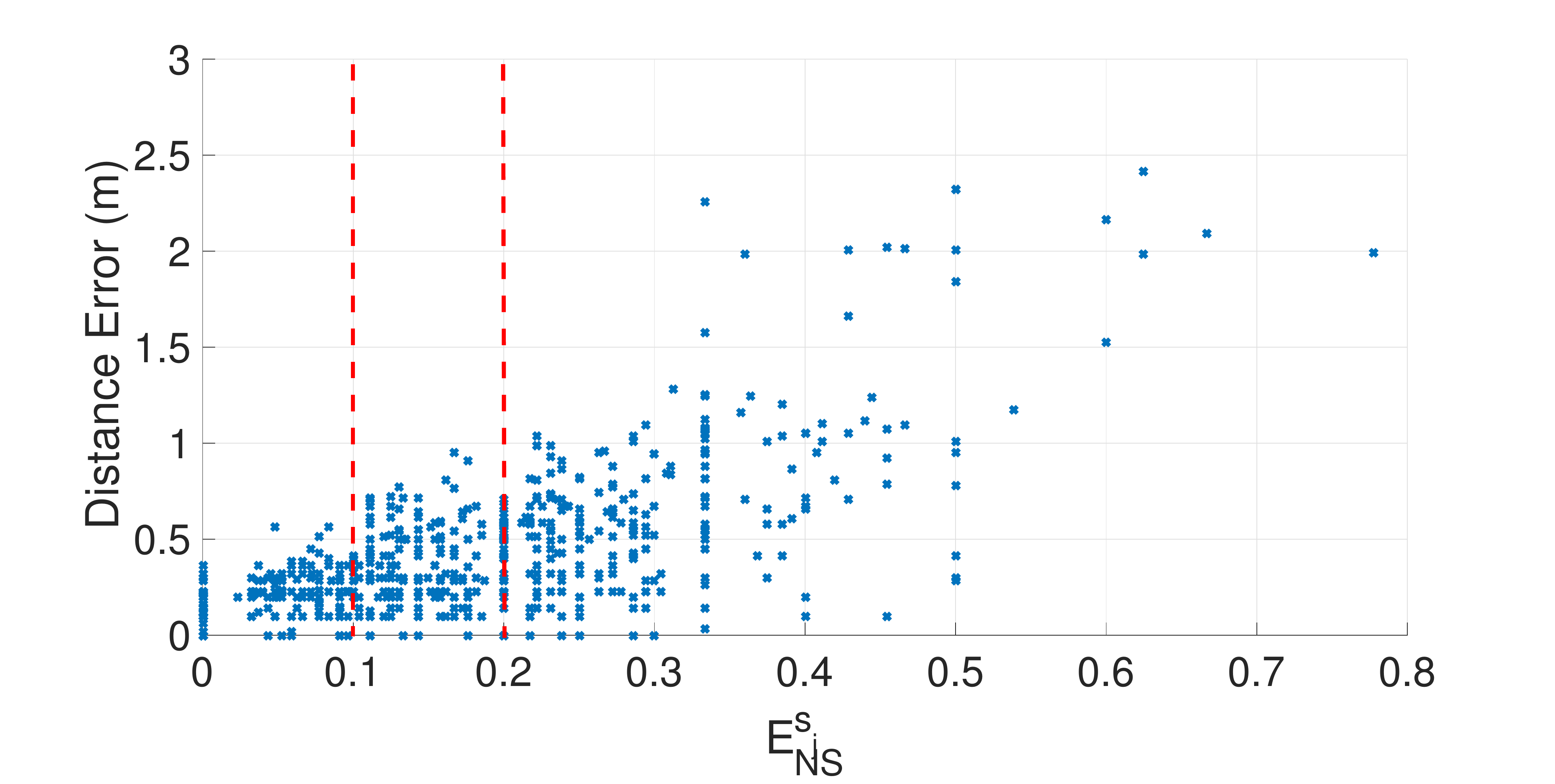} 
\caption{$E_{N_S}^{s_j}$ vs distance error}
\label{newanchorselection}
\end{figure}
where, $\tau _1$ and $\tau _2$ are predefined threshold values. To find the optimum values for $\tau _1$ and $\tau _2$, the distance error after initial topology coordinate estimation is calculated and plotted against the $E_{N_S}^{s_j}$ as shown in Figure \ref{newanchorselection}. In the Figure, it can be seen that nodes having $E_{N_S}^{s_j}$ value less than 0.1 have a distance error less than 0.5m. Also, when $E_{N_S}^{s_j}$ is less than 0.2, the distance error is less than 1m. Thus, in our simulation $\tau _1$ and $\tau _2$ are set as 0.1 and 0.2 respectively. If there are nodes in category $C_b$, the topology coordinates of those nodes are recalculated using new anchor information (i.e. nodes in group $C_a$). This process continues until zero nodes in the category $C_b$ or number of iterations exceeds the predefined limit ($iT$).

\subsection{Distributed Millimetre Wave Topology Map for Hybrid System (DMmTM-HS)}
DMmTM-HS calculates the topology coordinates of sensors using static and mobile anchors. This chapter has considered one mobile anchor following a random moving pattern to communicate with sensors in the network. However, there is no limitation on either number of mobile anchors or their movement patterns. The static anchors are randomly distributed all over the environment. The proposed algorithm assumes that mobile anchor is equipped with a compass and it know the direction angles of its sector antennas. There are two reasons for this assumption. First, as one/few mobile anchors are considered, it will not be an economical constraint to equip them with a compass. Second, the direction calculation of the sectors of mobile anchor will be time consuming as the direction changes when mobile anchor turns. However, if any application cannot equip the mobile anchor with a compass, then the same method explained in Step 1 in Figure \ref{flowchart} can be used to calculate the directional information of mobile anchor. Also, the static anchors are unaware of direction angles of its sectors and required to calculate as in DMmTM-SS. Moreover, it is assumed that elevation angle of nodes antennas are different and number of static anchors ($N_A$) is much less than the number of sensor nodes ($N$). 

DMmTM-HS algorithm introduces a new step followed by the same four steps in DMmTM-SS as described in Figure \ref{flowchart}. The pseudo code of this step is illustrated in the Appendix \ref{sec::PC::dmmtm5}. During the initial step, i.e. Information Gathering via Mobile Anchor, the mobile anchor moves around the network, transmitting ISS packets that include its current location and sector direction information from all antenna sectors. The sensors listen for these packets in quasi-omni pattern and select the optimum sector of the mobile anchor in each location as in the ISS phase of the IEEE 802.11ad standard protocol \cite{11ad}. Each sensor node $s_j$ records two sets of vectors, namely, mobile anchor locations $L _{s_j}^R=\{l_r\equiv (x_r,y_r,z_r):x_r,y_r,z_r\in \mathbb{R}\}$ and optimum sector direction angle $\Theta _{s_j}^R =\{\theta _r \equiv (\alpha _r, \beta _r):0<(\alpha _r, \beta _r)<2\pi\}$. Then, these sets are used in information mapping phase to calculate the topology coordinates of each sensor in addition to the $L _{s_j}$ and $\Theta _{s_j}$ sets gathered by static anchors.

\section{Performance Evaluation}\label{sec::dmmtm:result}
\subsection{Simulation Environments and Parameters}
\begin{figure}
 \centering
 \subfigure[Warehouse]{
  \includegraphics[width=0.44\textwidth,trim={6cm 0cm 4.5cm 5.4cm},clip]{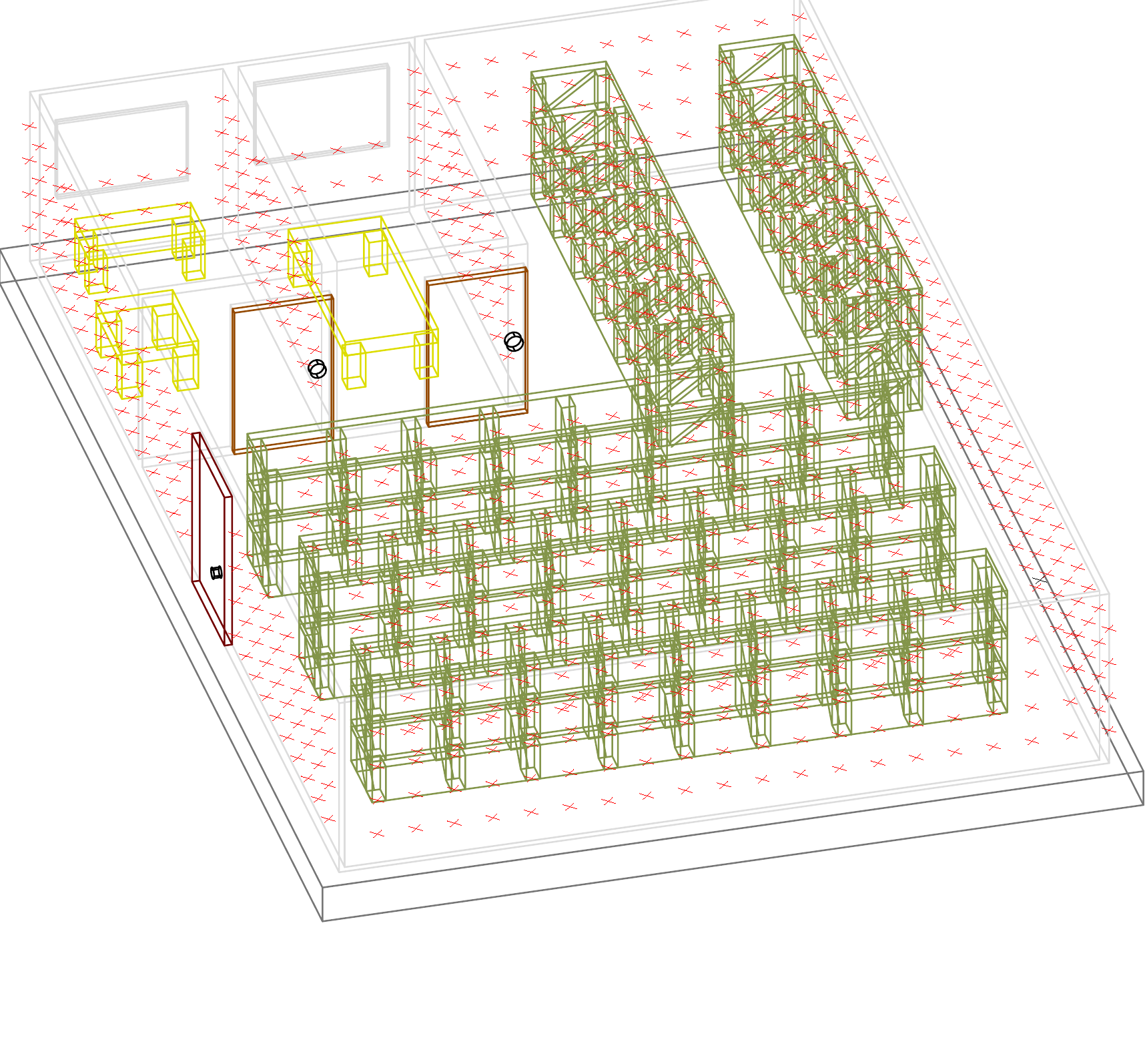}
   \label{Warehouse}
   }
 \quad
 \subfigure[Greenhouse]{
  \includegraphics[width=0.42\textwidth,trim={6.5cm 0cm 5cm 4.7cm},clip]{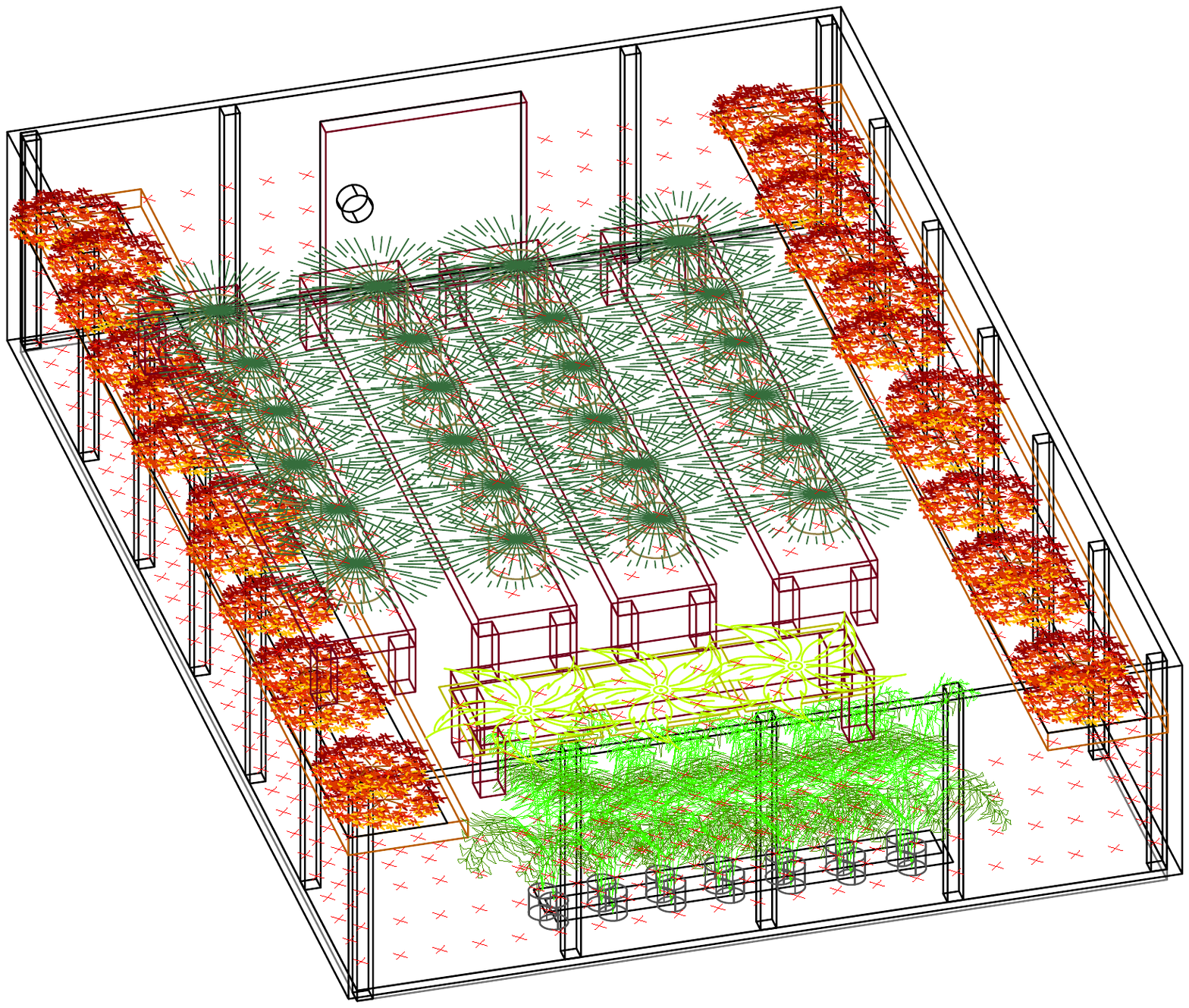}
   \label{Greenhouse}
   }
 \caption{Simulation environments}  \label{networks}
\end{figure}

The performance of the proposed algorithms are evaluated using two simulation environments; i) a warehouse environment with metal racks and tables as shown in Figure \ref{Warehouse} with 881 sensor nodes distributed over the environment and ii) a greenhouse full with plants as shown in Figure \ref{Greenhouse} which is covered by 871 sensor nodes. In both cases, the minimum distance between two adjacent nodes is 1m. MATLAB simulation software was used for the computations. The communication channel is modelled as same as in Chapter \ref{chapter:mmtm} and the equation is given below.
\begin{eqnarray}
PL(f,d)&=&FSPL(f,d_0)+10\varepsilon log_{10}(d/d_0)+X_\sigma \nonumber \\ 
&& \text{for}\:d \geq d_0,\:\:\:\text{where},\: d_0=1
\end{eqnarray}\label{pm1}
where, $PL(f,d)$ is the path loss value in dB, $\varepsilon $ is the path loss exponent, $d$ is the distance between two nodes, and $X_\sigma $ is a zero Gaussian random variable with standard deviation $\sigma $ in dB (this represents the large-scale channel fluctuations due to shadowing \cite{pathloss2}). FSPL is the free space path loss at distance $d_0$ and can be calculated as $10log_{10}(4\pi d_0f/ c)^2$, where $f$ is the frequency and $c$ is the speed of light. The simulation parameter values use are listed in Table \ref{simpara}. The values for path loss model parameters are taken from the experiments described in \cite{pathloss2,pathloss1}. Since MmWave communication is affected by the oxygen concentration in the environment, we have configured higher attenuation for the greenhouse environment compared to the warehouse.

\begin{table}
\renewcommand{\arraystretch}{1.3}
\caption{\textsc{Simulation Parameter Values}}\label{simpara} 
\center
\small
  \begin{tabular}{ |c | c| }
    \hline
     \textbf{Parameter} & \textbf{Value} \\ \hline
    Transmitted Power ($P_{tx}$)& 20dB\\ \hline
    Receiving Sensitivity & -60dB\\ \hline
    Frequency ($f$) & 73 Ghz \\ \hline
    Number of Sectors($NS$) & 8 \\ \hline
    Warehouse  & $\varepsilon$ = 1.6, $\sigma$ =3.2  \cite{pathloss1} \\\hline
    Greenhouse & $\varepsilon$ = 2.4, $\sigma$ =6.3 \cite{pathloss2}\\
    \hline
  \end{tabular}
\end{table}

\subsection{Evaluation Parameters }
To evaluate and compare the performance of DMmTM, the same performance parameters used in Chapter \ref{chapter:mmtm}, i.e. Distance error and Sector displacement are used. Distance error parameter measures the difference between the actual location and the estimated location. If the objective is to generate an identical map to the physical map, the distance error should be negligible. As the objective is to generate a isomorphic topology map of the network, the distance error should be less than the distance between two adjacent nodes. Then the topology map will make sure the connectivity of the nodes remains same as in the physical network. 

MmWave communication uses narrow beamwidth multi-sector antennas. To select the correct sector to communicate with each node, the calculated angle from the topology coordinates need to be same as in the physical map or it need to lie between the sector antenna beamwidth. Thus, sector displacement parameter measures the number of sectors deviated from its correct sector. The zero sector displacement value indicates that all the nodes are located in the correct direction and nodes can communicate without any further adjustments. Consider two neighbour nodes $i$ and $j$. Based on the actual location of the node $j$, it is located in the sector $s_{ij}$ of node $i$. On the other hand, node $j$ is located in the sector $\hat{s}_{ij}$ of node $i$ based on the estimated locations. Then the sector displacement calculation for the two nodes is: 
\[
 SD_{ij}  =
  \begin{cases}
   \vert s_{ij}- \hat{s}_{ij} \vert  \mod  (N_S/2)& \text{if }  \vert s_{ij}- \hat{s}_{ij} \vert \\
   &\indent <(N_S/2)\\
   1-\vert s_{ij}- \hat{s}_{ij} \vert  \mod  (N_S/2)    & \text{else } 
  \end{cases}
\]

\subsection{Performance of DMmTM-SS}
\begin{figure}
 \centering
 \subfigure[Actual Map]{
  \includegraphics[width=0.47\textwidth]{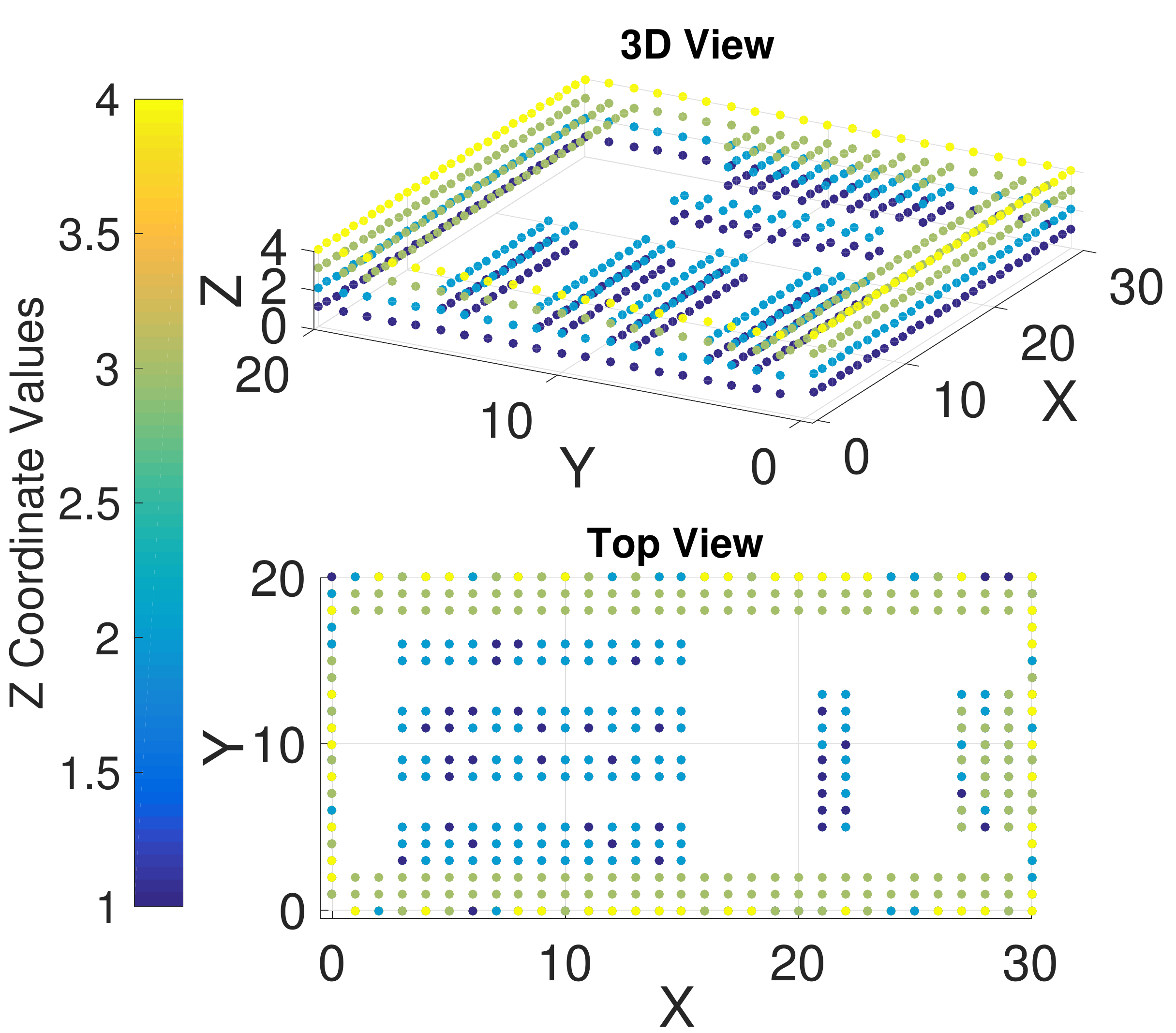}
   \label{SS_N_g}
   }
 \quad
 \subfigure[DMmTM-SS]{
  \includegraphics[width=0.4\textwidth]{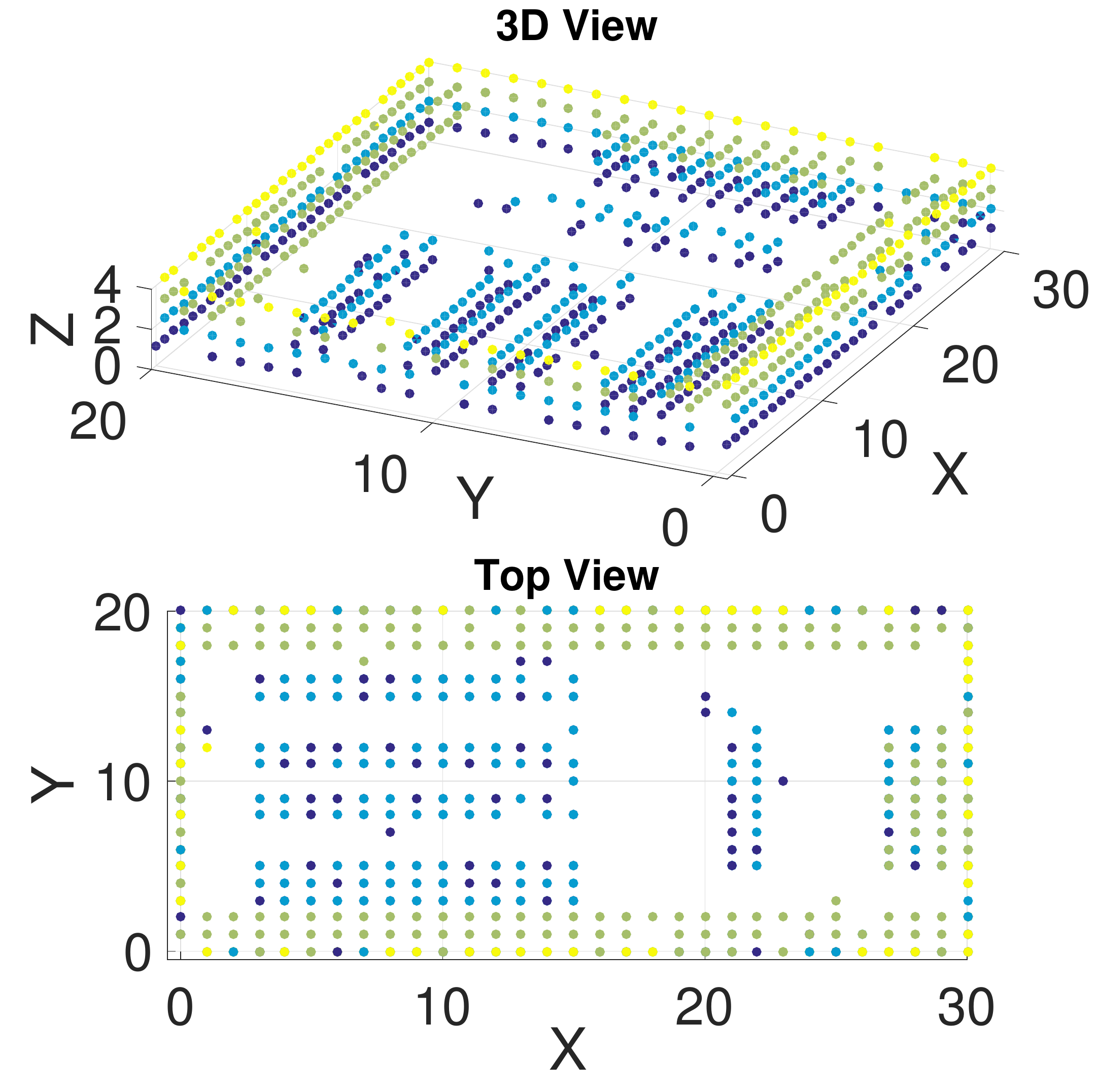}
   \label{SS_T_g}
   }
 \caption{Maps of greenhouse}  \label{SS_map_g}
\end{figure}

\begin{figure}
 \centering
 \subfigure[Actual Map]{
  \includegraphics[width=0.47\textwidth]{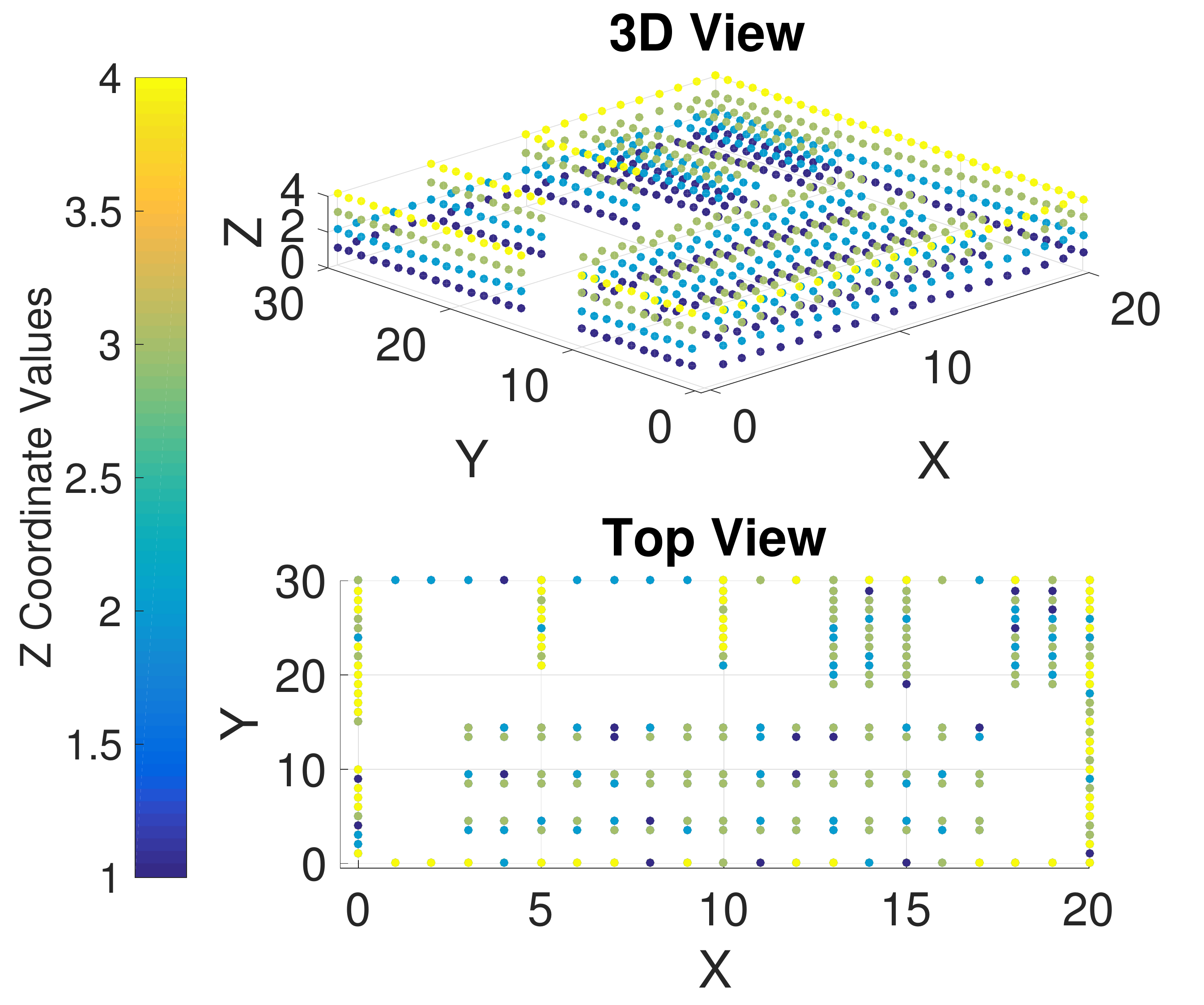}
   \label{SS_N_w}
   }
 \quad
 \subfigure[DMmTM-SS]{
  \includegraphics[width=0.4\textwidth]{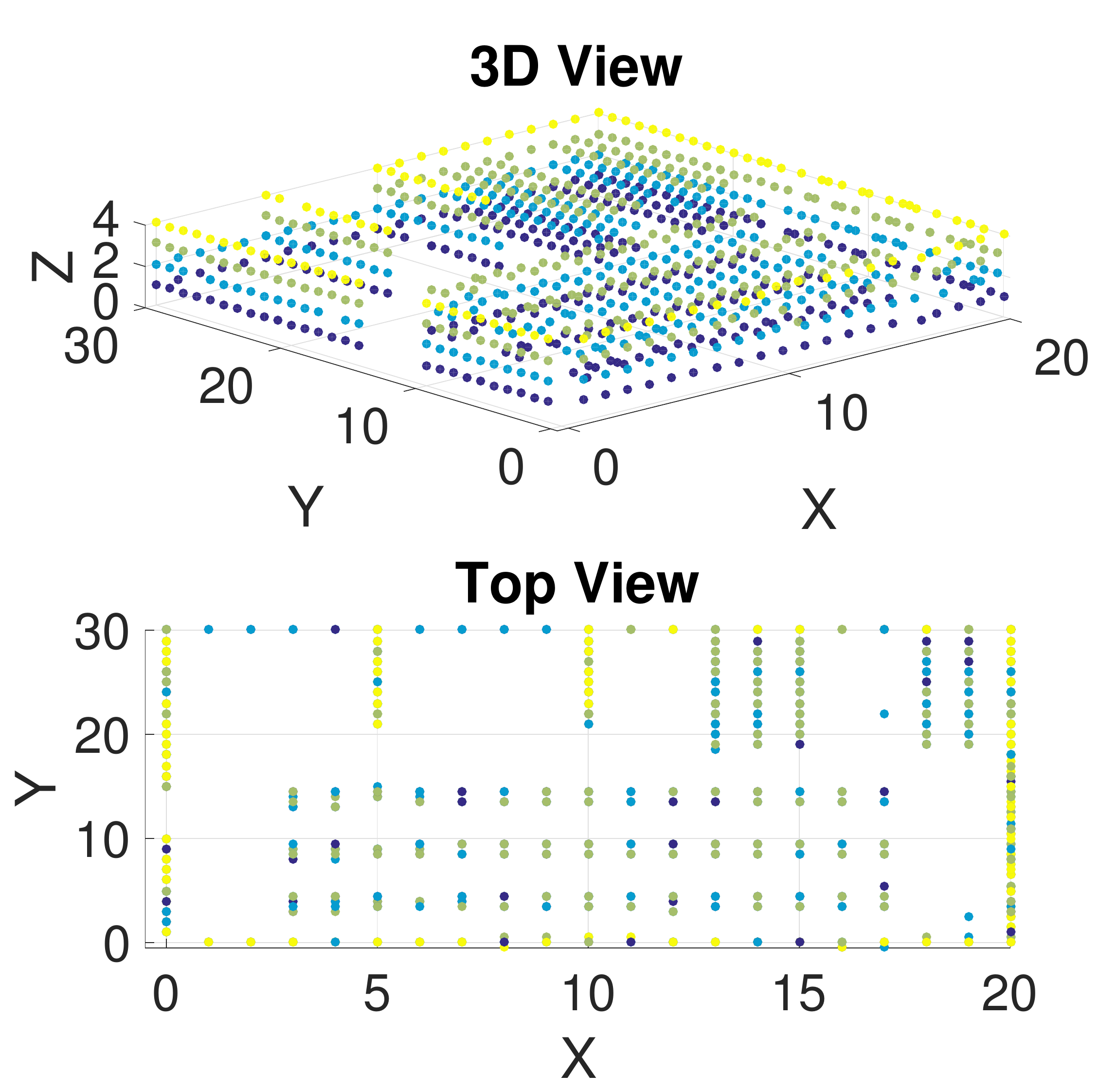}
   \label{SS_T_w}
   }
 \caption{Maps of warehouse}  \label{SS_map_w}
\end{figure}

The performance of DMmTM-SS is evaluated in this section. Figures \ref{SS_map_g} and \ref{SS_map_w} show the sensor locations in actual map and calculated topology map for greenhouse and warehouse environments respectively. In DMmTM-SS maps, most of the nodes are located in the correct position, but few are deviated from the actual position. For an example, in Figure \ref{SS_T_g} Top view, some of the nodes located in the middle area deviated from it's actual location. Moreover, there is a deviation in nodes located in x=3 and x=19 lines in Figure \ref{SS_T_w} Top view. However, most of the nodes are located in correct positions.

\begin{figure}
 \centering
 \subfigure[Greenhouse:Distance error]{
  \includegraphics[width=0.45\textwidth]{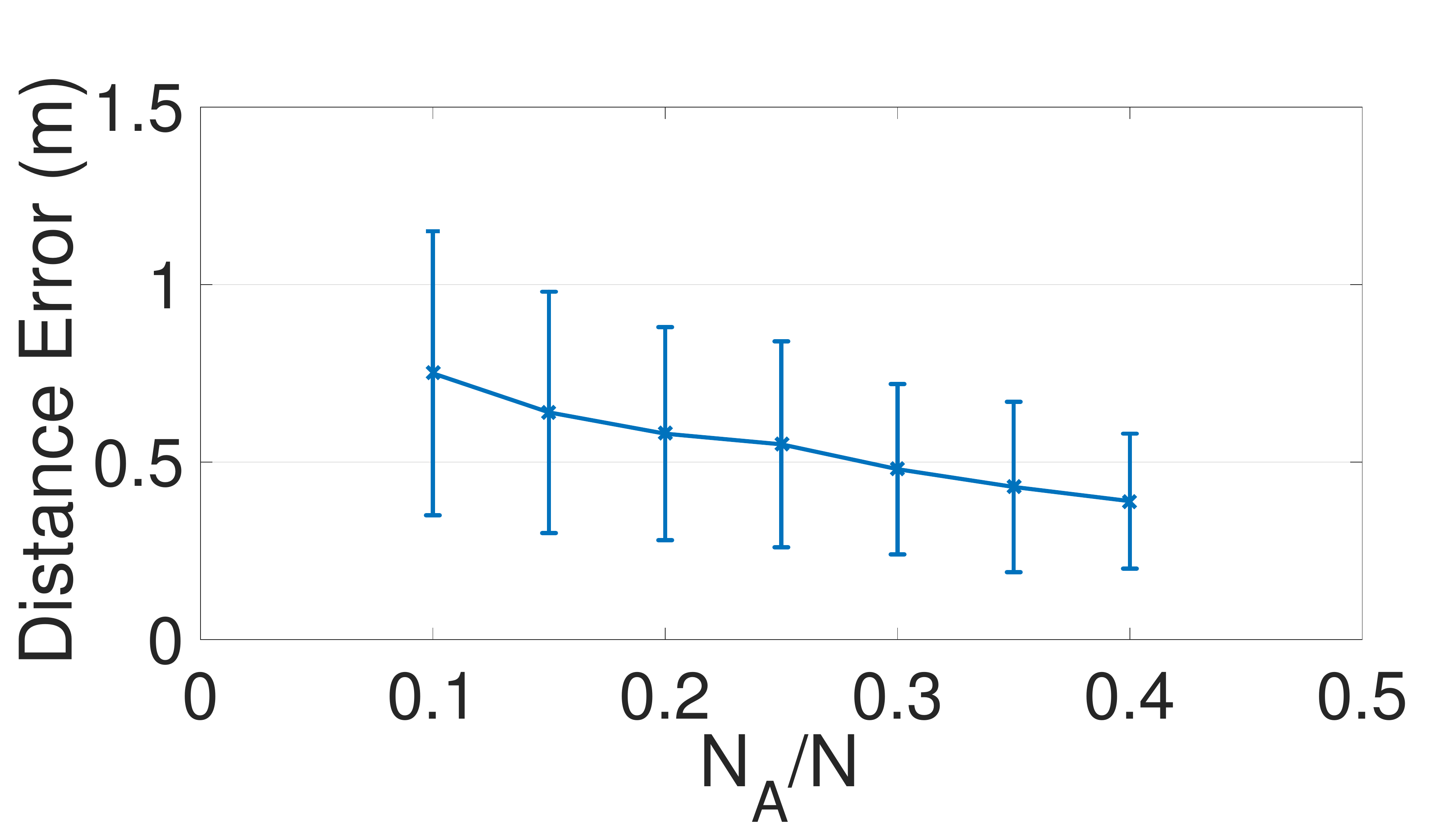}
   \label{SS_D_g}
   }
   \quad
 \subfigure[Greenhouse:Sector displacement]{
  \includegraphics[width=0.44\textwidth]{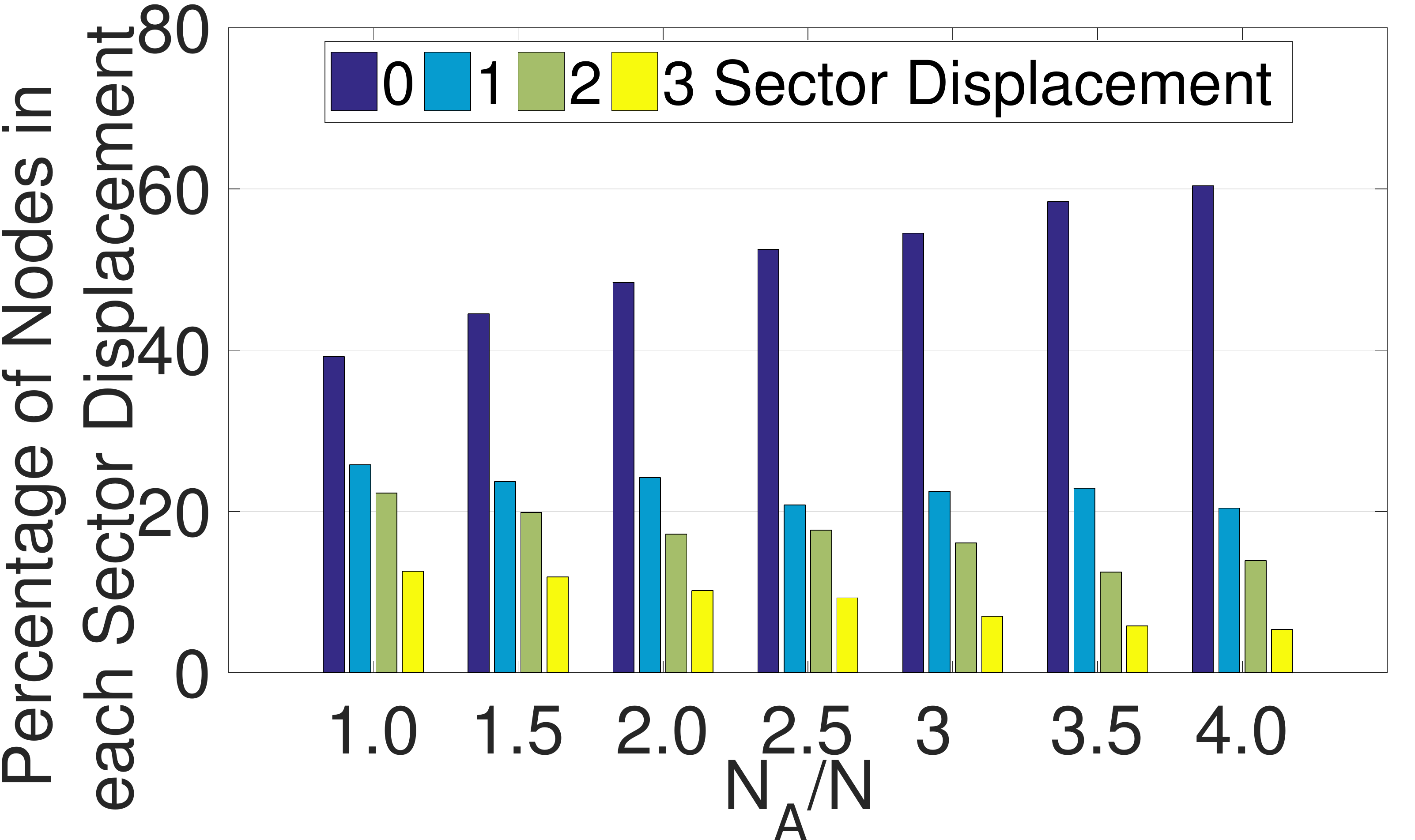}
   \label{SS_S_g}
   }
    \subfigure[Warehouse:Distance error]{
  \includegraphics[width=0.45\textwidth]{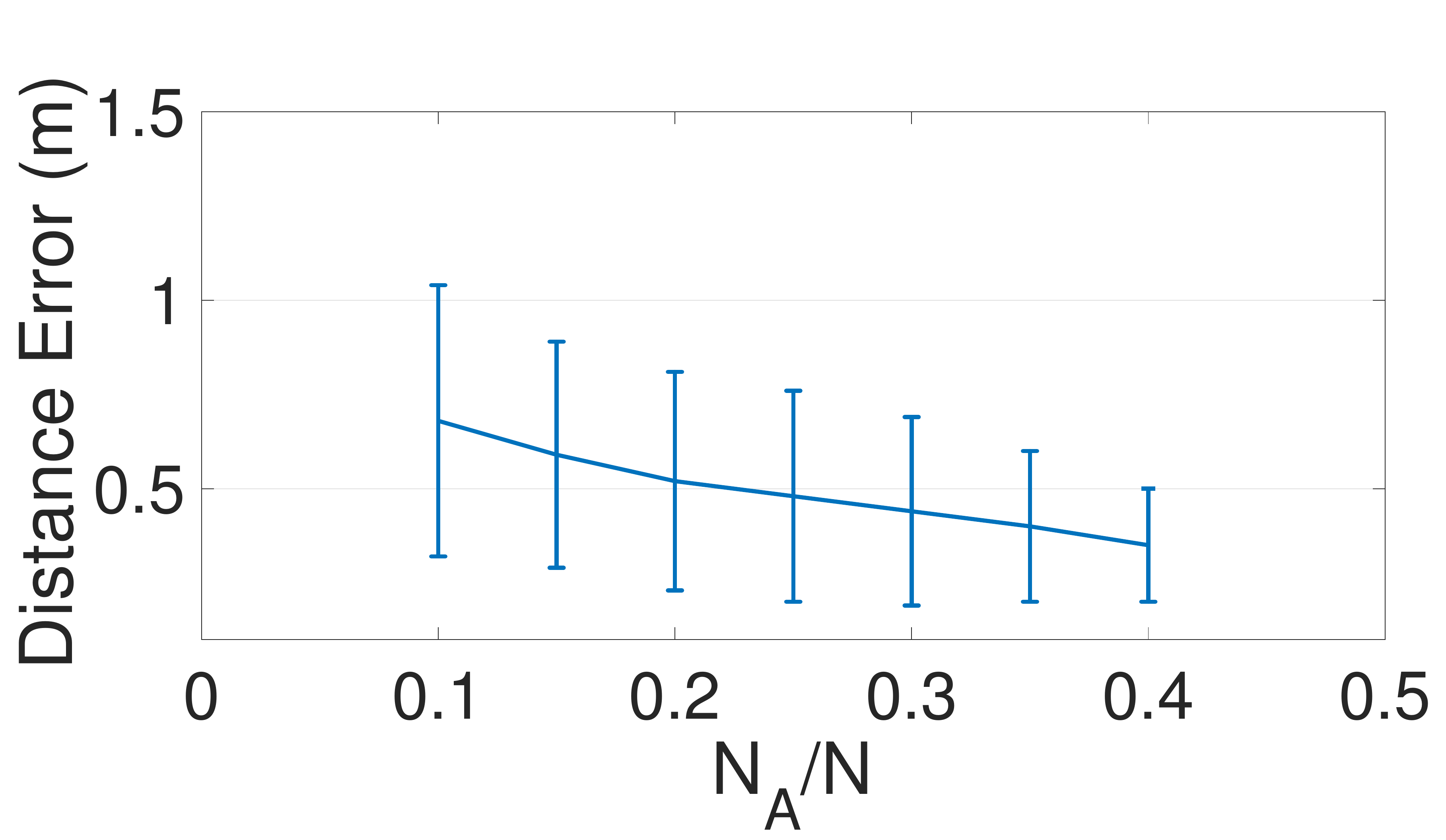}
   \label{SS_D_w}
   }
 \quad
 \subfigure[Warehouse:Sector displacement]{
  \includegraphics[width=0.44\textwidth]{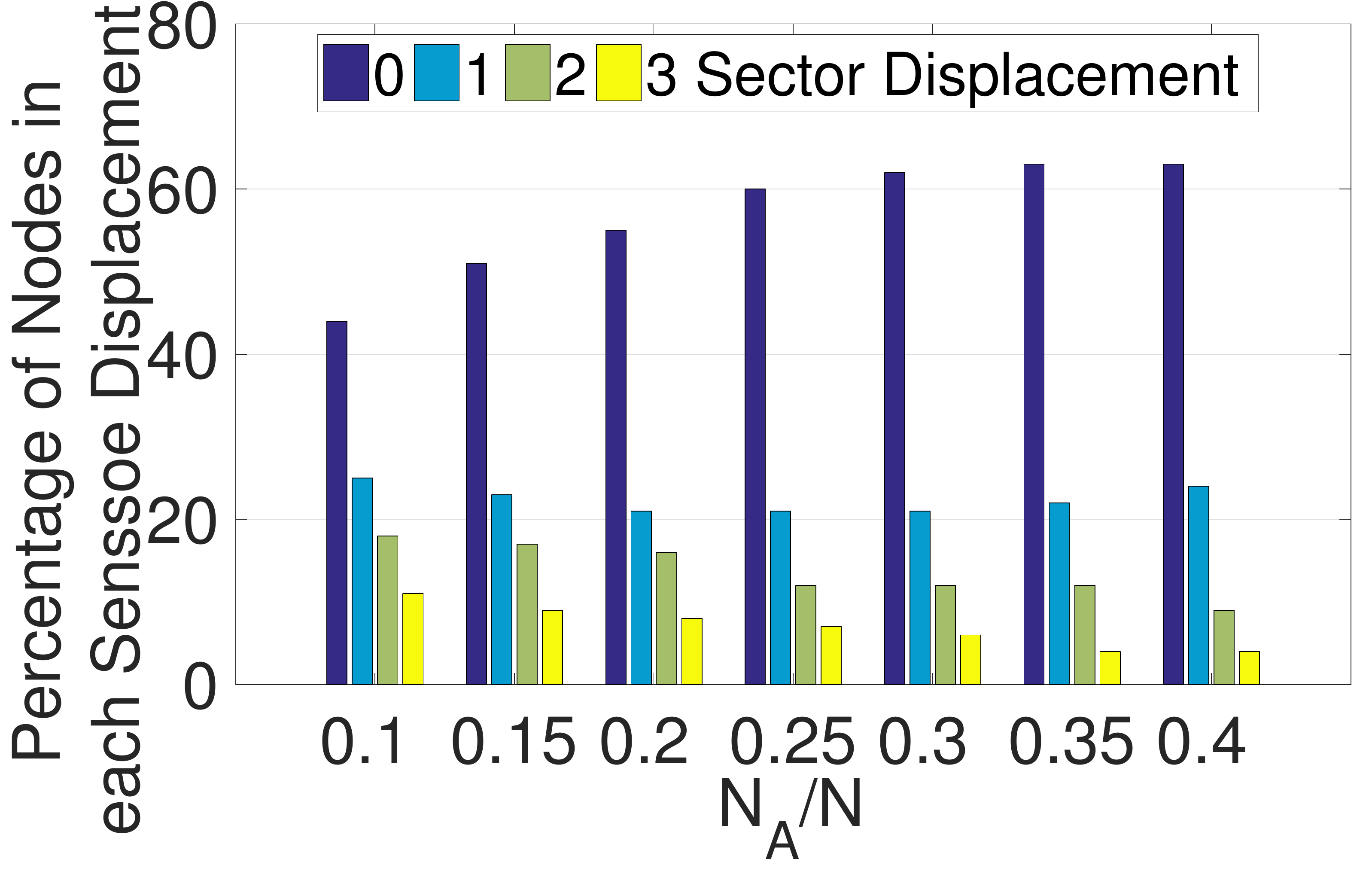}
   \label{SS_S_w}
   }
 \caption{DMmTM-SS performance against anchor ratio}  \label{SS_performance}
\end{figure}

Next the performance of DMmTM-SS is evaluated against the ratio of number of initial anchors to number of sensors in the network ($N_A/N$). Results of two evaluation metrics are shown in Figure \ref{SS_performance}. In Figure \ref{SS_D_g}, corresponding to the greenhouse, the distance error decreases when the anchor ratio increase. The algorithm locates sensors with an average distance error of 0.75m, which is less than the distance between two adjacent nodes (i.e.1m) by having 0.1 anchor ratio. When the ratio increase to 0.15, majority of sensors are located with less than 1m distance error, which ensures that the connectivity is preserved. The sector displacement for the greenhouse environment is illustrated in Figure \ref{SS_S_g}. When the anchor ratio is increasing, the percentage of nodes in zero displacement is also increasing and the percentage of nodes in three sector displacements is decreasing. Moreover, the pattern of two performance metrics are same in warehouse environment as shown in Figures \ref{SS_D_w} and \ref{SS_S_w}. However, the distance error values are comparatively less and percentage of nodes in zero sector displacement is high. This is due to the less packet losses in greenhouse because of its environmental characteristics (i.e. path loss and shadowing). 

\subsection{Performance of DMmTM-HS}
\begin{figure} 
  \centering
 \subfigure[Actual Map]{
  \includegraphics[width=0.47\textwidth]{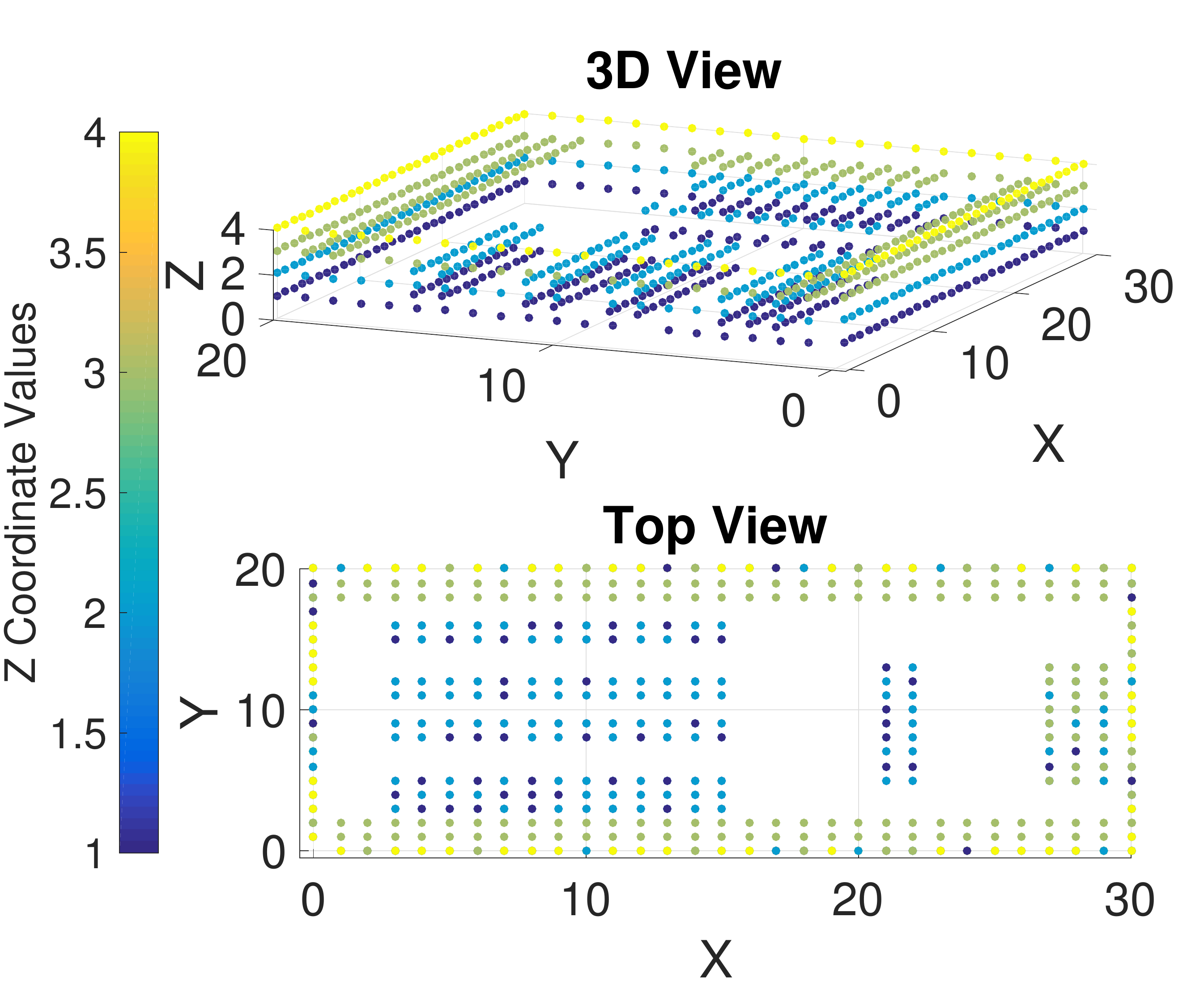}
   \label{HS_N_g}
   }
 \quad
 \subfigure[DMmTM-HS]{
  \includegraphics[width=0.4\textwidth]{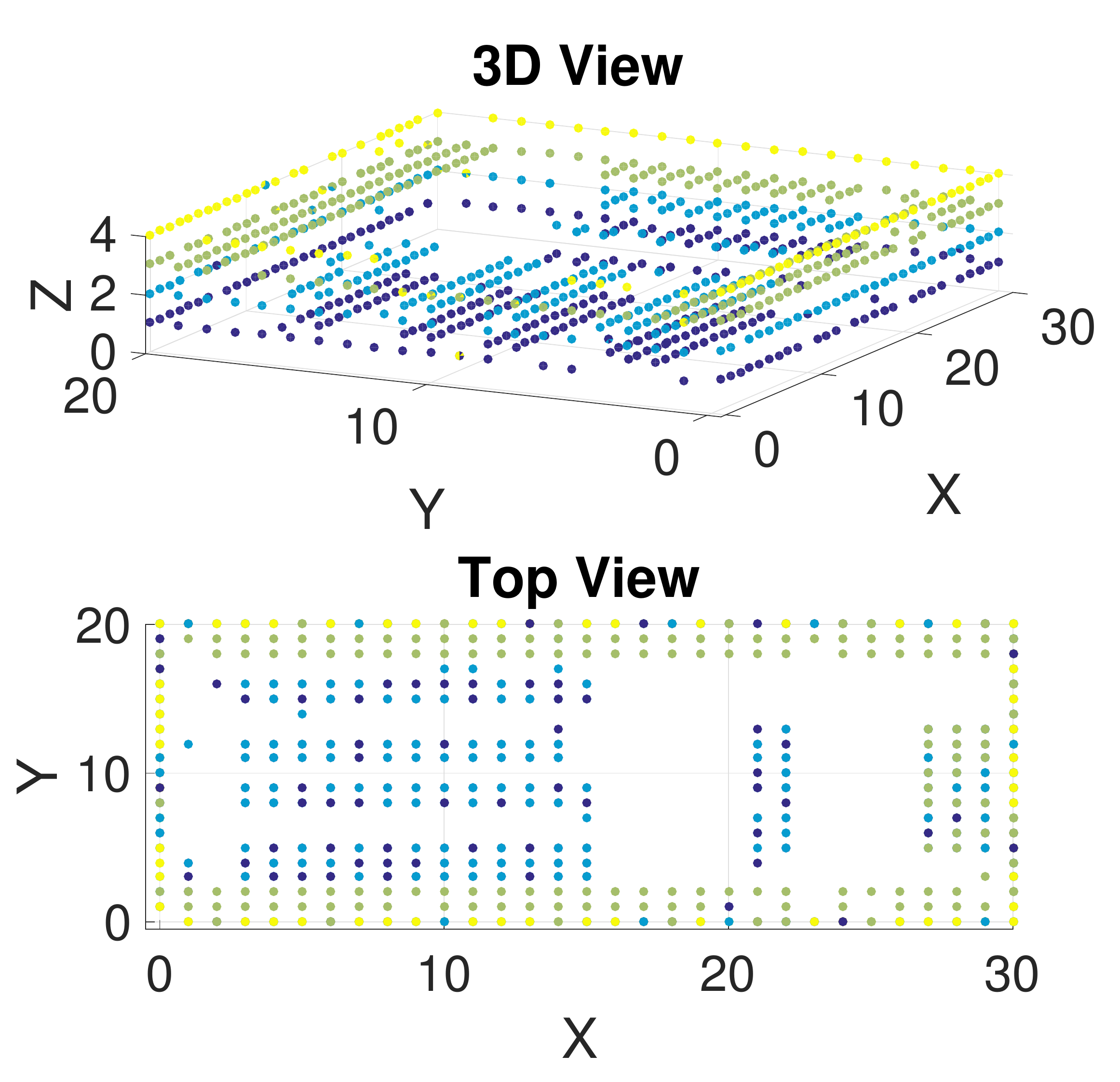}
   \label{HS_T_g}
   }
 \caption{Maps of greenhouse}  \label{HS_map_g}
\end{figure}

\begin{figure}
 \centering
 \subfigure[Actual Map]{
  \includegraphics[width=0.47\textwidth]{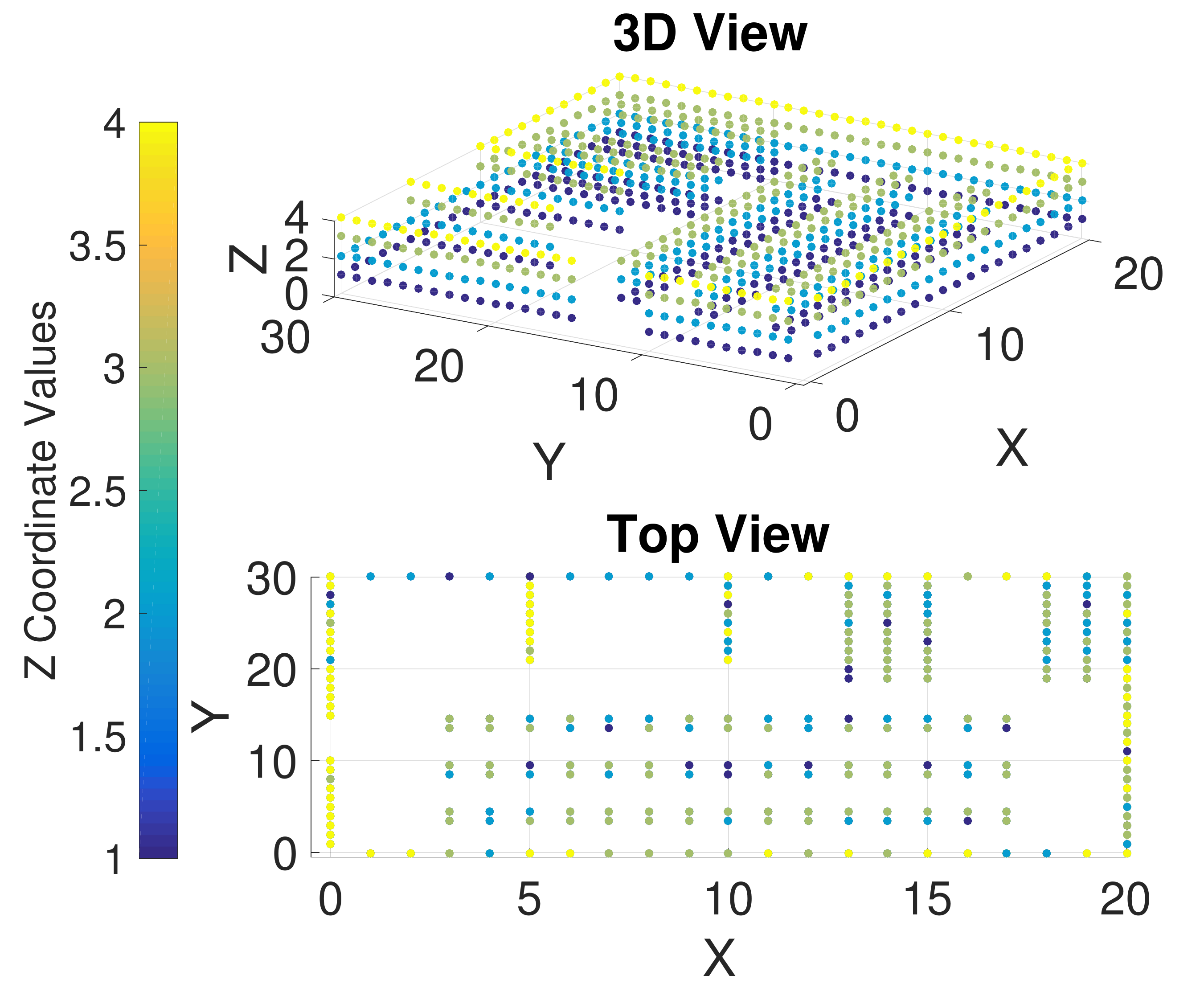}
   \label{HS_N_w}
   }
 \quad
 \subfigure[DMmTM-HS]{
  \includegraphics[width=0.4\textwidth]{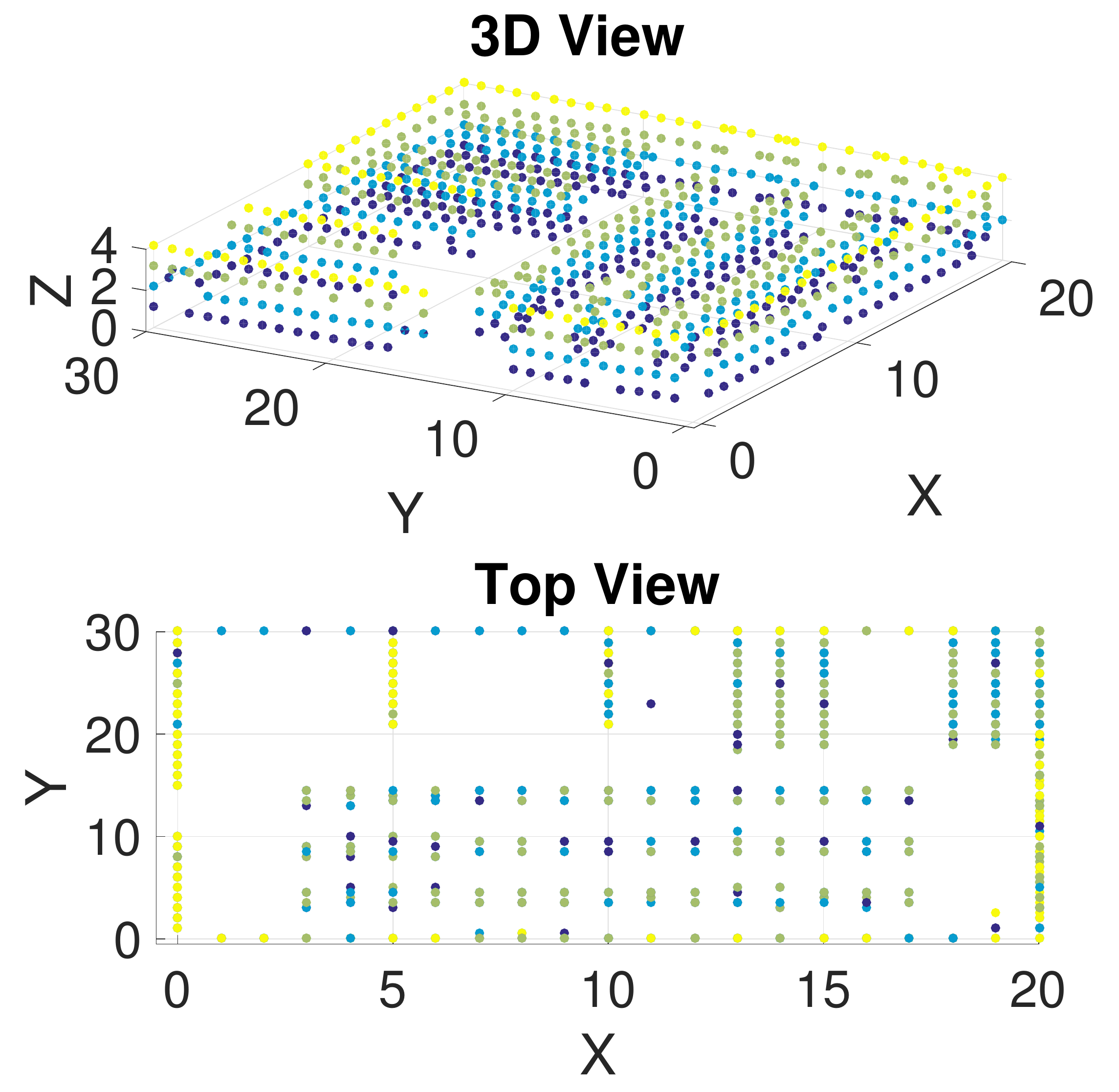}
   \label{HS_T_w}
   }
 \caption{Maps of warehouse}  \label{HS_map_w}
\end{figure}

The performance of DMmTM-SS is evaluated in this section. Figure \ref{HS_map_g} and \ref{HS_map_w} show the sensor locations in actual map and calculated topology map for greenhouse and warehouse environments respectively. It can be seen that DMmTM-HS located most of the nodes in the correct positions, but very few of them are deviated. For an example, in Figure \ref{HS_T_g} Top view, some of the nodes located in x=1 and x=2 lines deviated from its actual positions. Also, in warehouse environment, Figure \ref{HS_T_w} Top view shows that some of the nodes in x=19 line is deviated. However, the deviation between actual position and calculated position seems to be a small value.

The performance of DMmTm-HS is evaluated against two parameters. First presented with the variation of $N_R/N$, the ratio between number of sensors receiving ISS packets from mobile anchor and number of nodes in the network. Then the results of performance metrics against the number of initial anchors in the network ($N_A/N$) when $N_R/N$ is fixed. 

\begin{figure}
 \centering
 \subfigure[Greenhouse:Distance error]{
  \includegraphics[width=0.45\textwidth,trim={.2cm 0cm 1cm 0cm},clip]{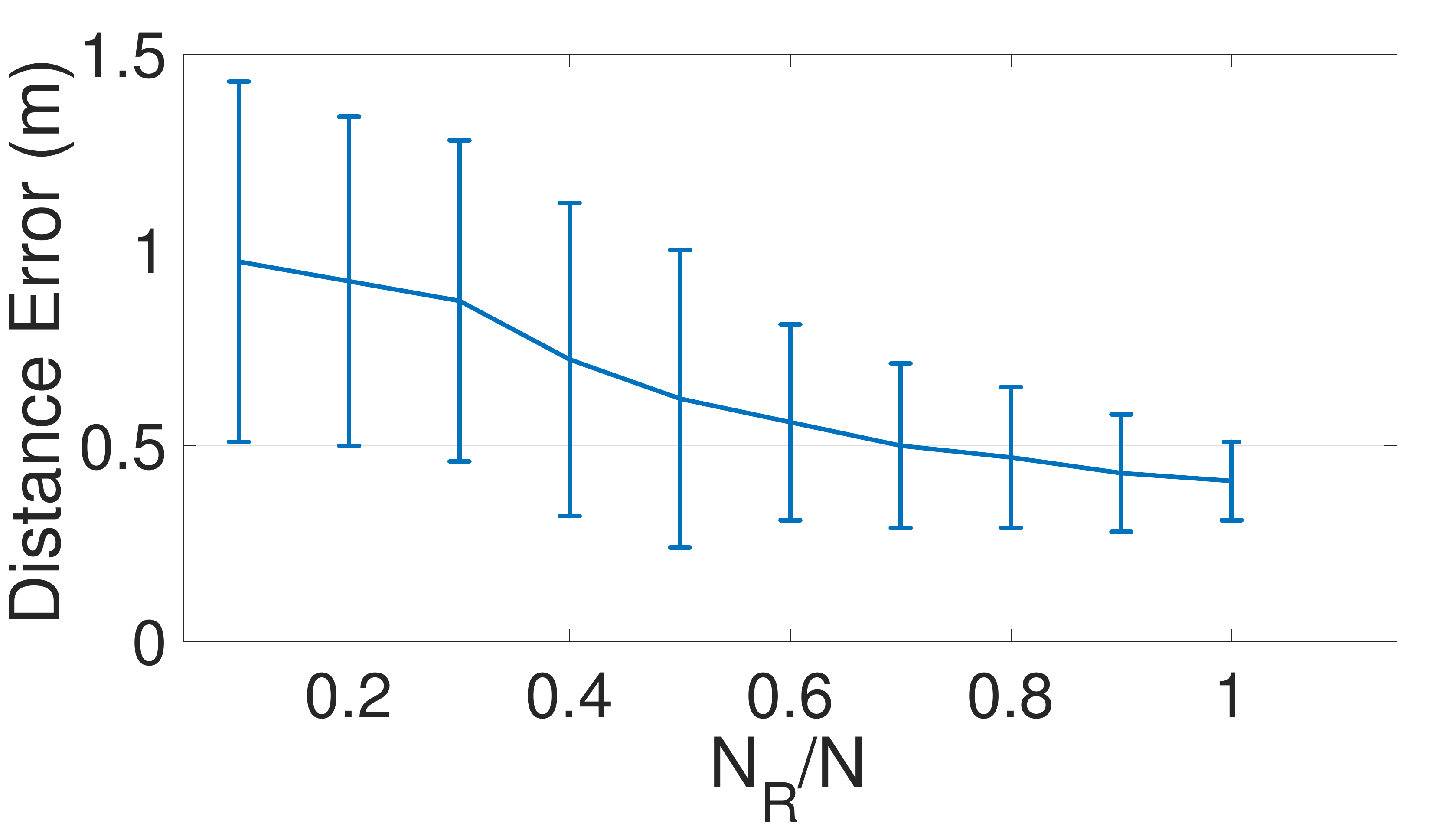}
   \label{HS_D_gR}
   }
   \quad
 \subfigure[Greenhouse:Sector displacement]{
  \includegraphics[width=0.44\textwidth,trim={.6cm 0cm 5cm 0cm},clip]{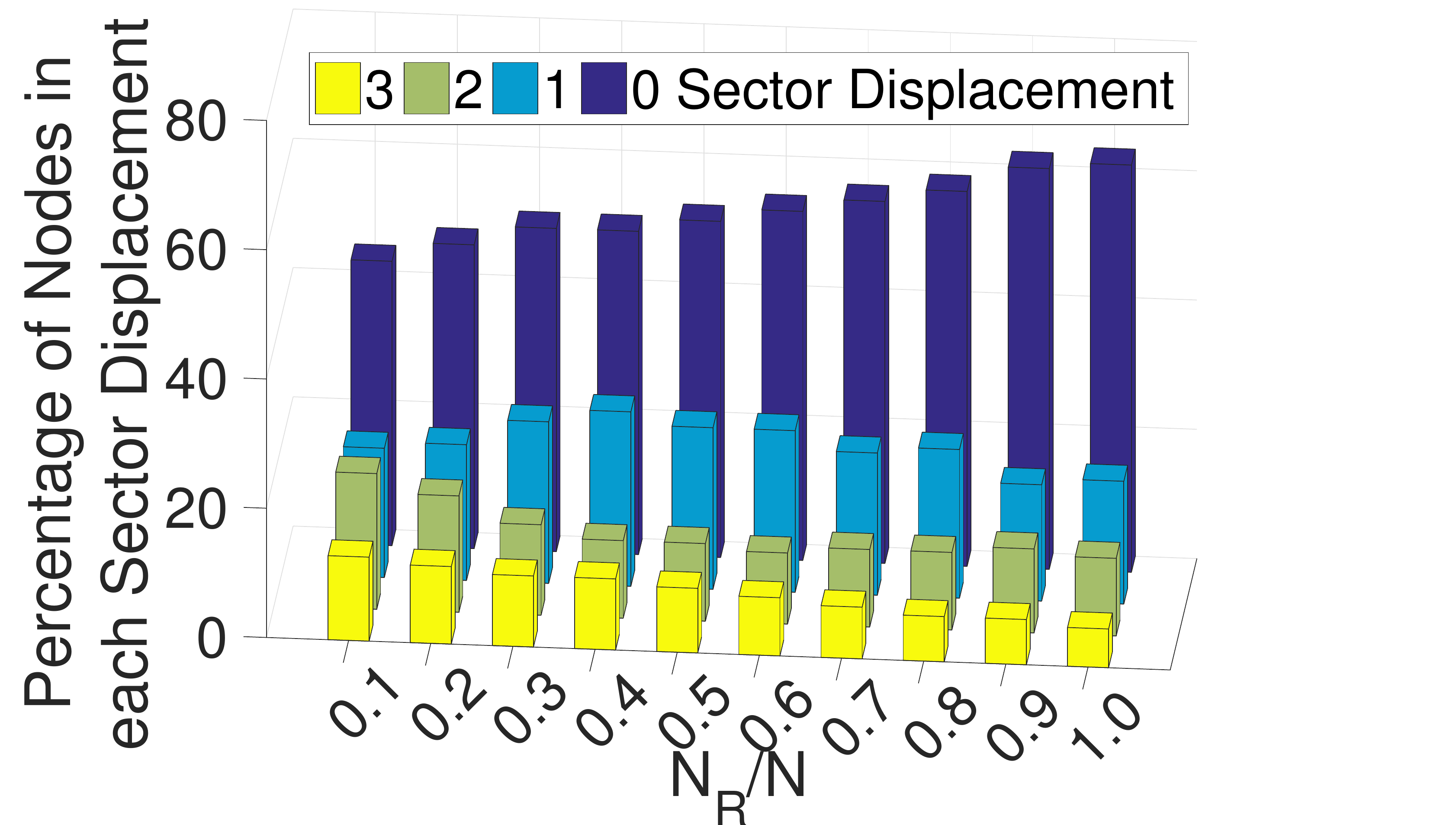}
   \label{HS_S_gR}
   }
   \subfigure[Warehouse:Distance error]{
  \includegraphics[width=0.45\textwidth]{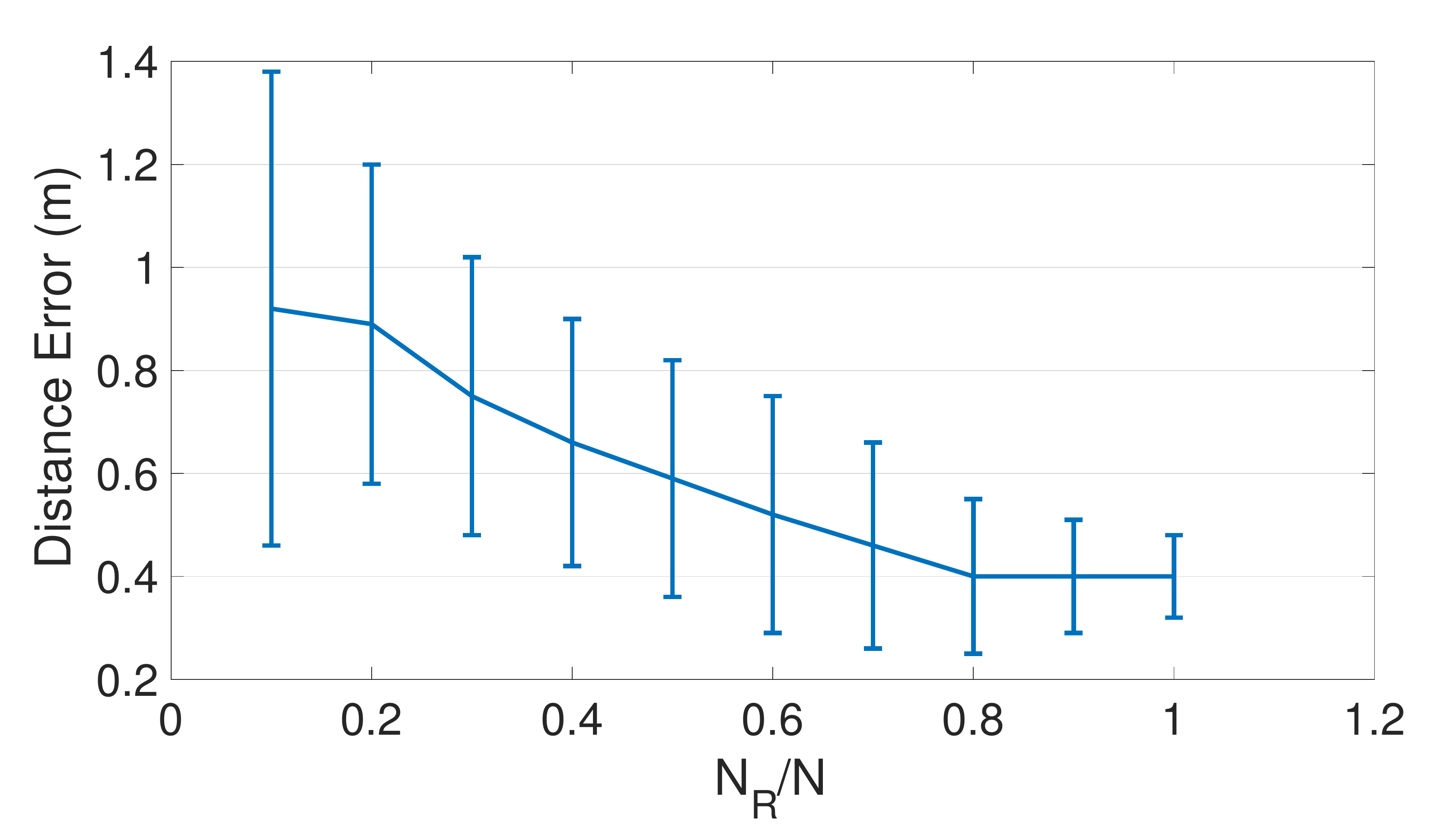}
   \label{HS_D_wR}
   }
 \quad
 \subfigure[Warehouse:Sector displacement]{
  \includegraphics[width=0.44\textwidth,trim={1cm 0cm 5cm 0cm},clip]{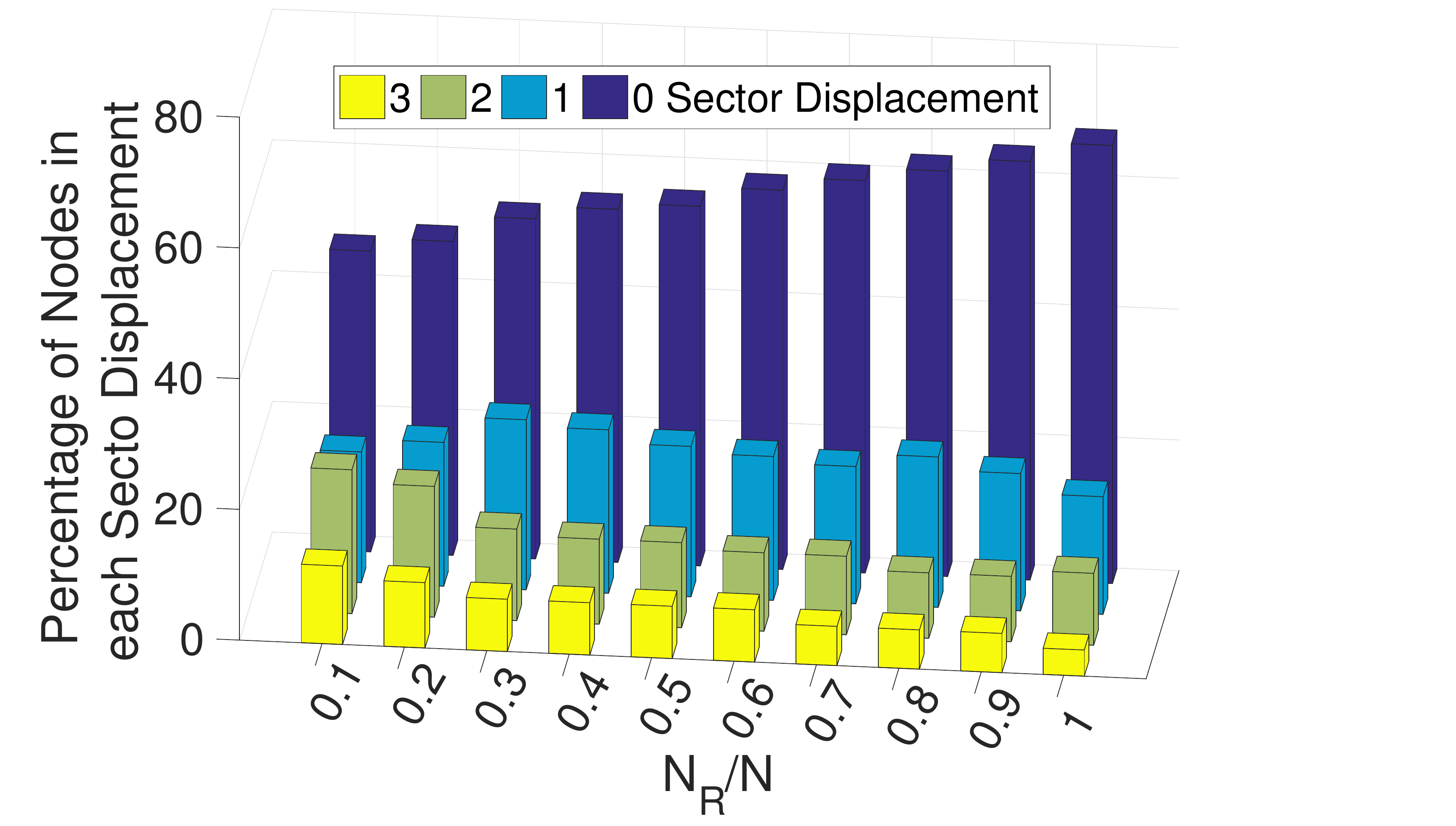}
   \label{HS_S_wR}
   }
 \caption{DMmTM-HS performance against number of nodes receive packetes from robot when $N_A/N $=0.15}  \label{HS_performance_R}
\end{figure}
Figure \ref{HS_performance_R} shows the results of performance metrics for two environments while changing the $N_R/N$ ratio. In this case $N_A/N$, the ratio between number of initial anchor nodes and number of sensor nodes, is set as 0.15. When $N_R/N$ ratio is increasing, the distance error in Figure \ref{HS_D_gR} and \ref{HS_D_wR} is decreasing, however, in Figure \ref{HS_D_wR} the distance error is almost same after $N_R/N = 0.8$. Figure \ref{HS_S_gR} and \ref{HS_S_wR} present the percentage of nodes in each sector displacement in two environments. As more sensors receive packets from mobile sensor, the percentage of nodes in zero sector displacement increases while the number of nodes with three sector displacement decreases.

\begin{figure}
 \centering
 \subfigure[Greenhouse:Distance error]{
  \includegraphics[width=0.45\textwidth,trim={2mm 0cm 1cm 0cm},clip]{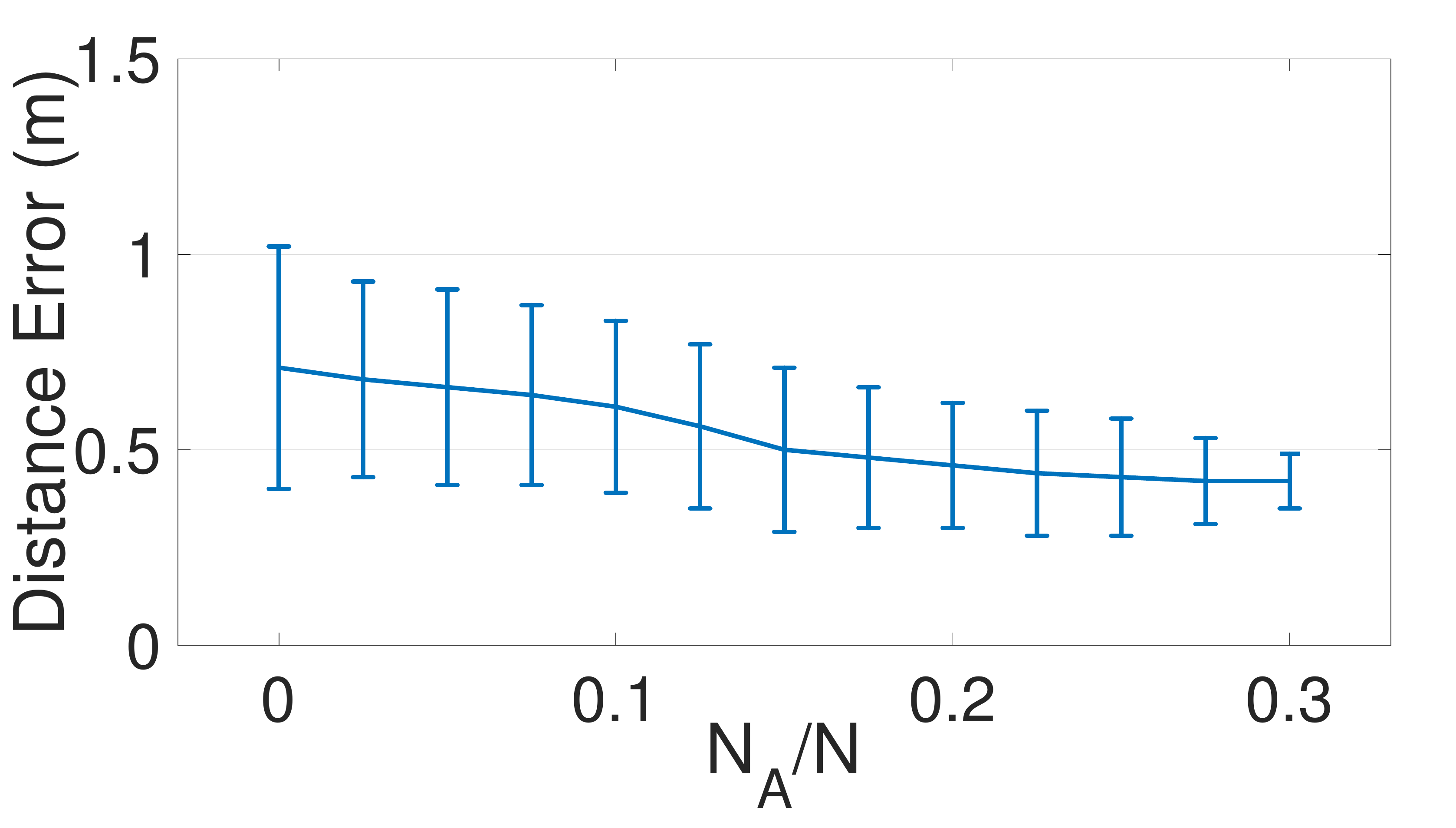}
   \label{HS_D_gA}
   }\quad
 \subfigure[Greenhouse:Sector displacement]{
  \includegraphics[width=0.45\textwidth,trim={0cm 0cm 5cm 0cm},clip]{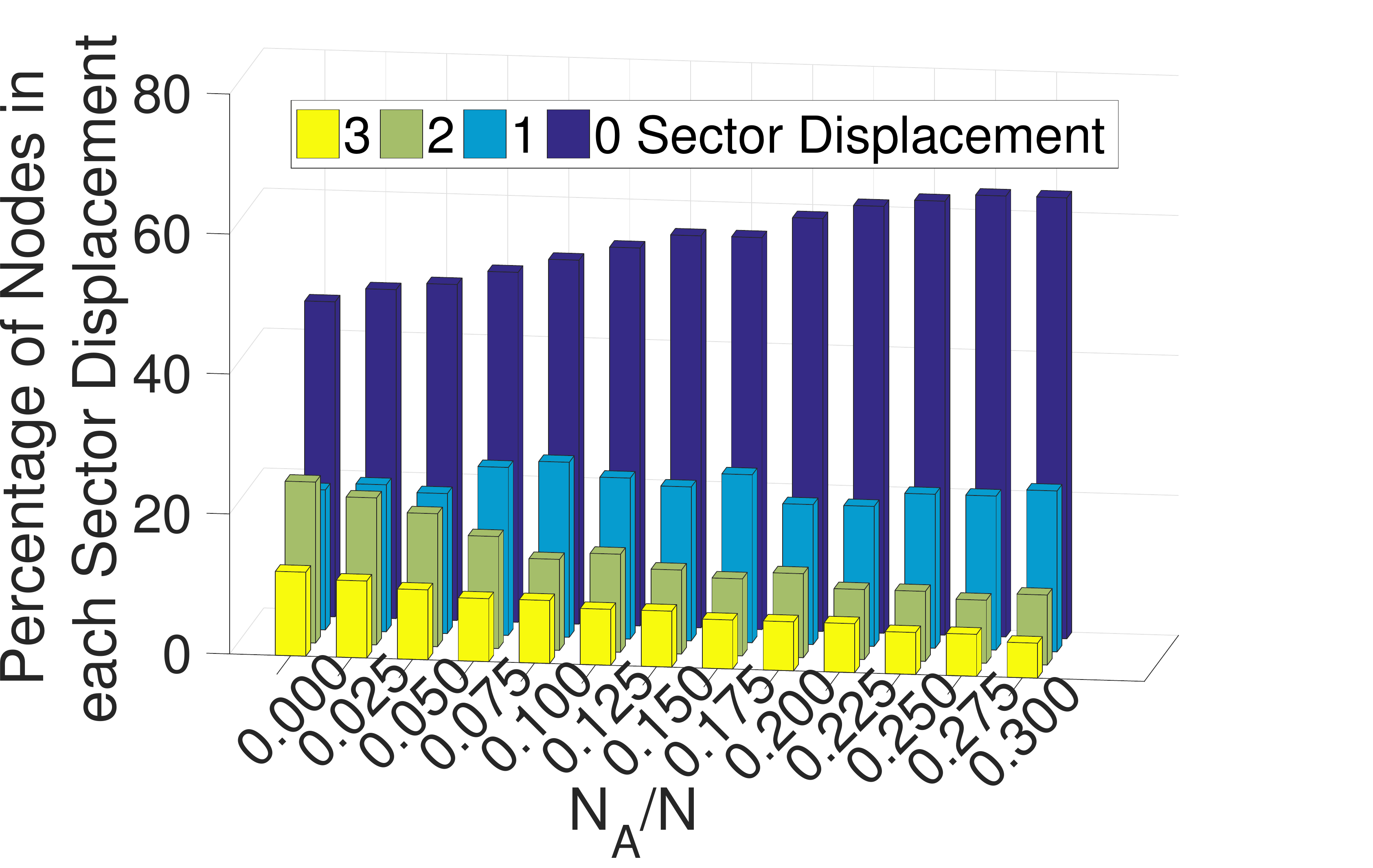}
   \label{HS_S_gA}
   }
    \subfigure[Warehouse:Distance error]{
  \includegraphics[width=0.45\textwidth]{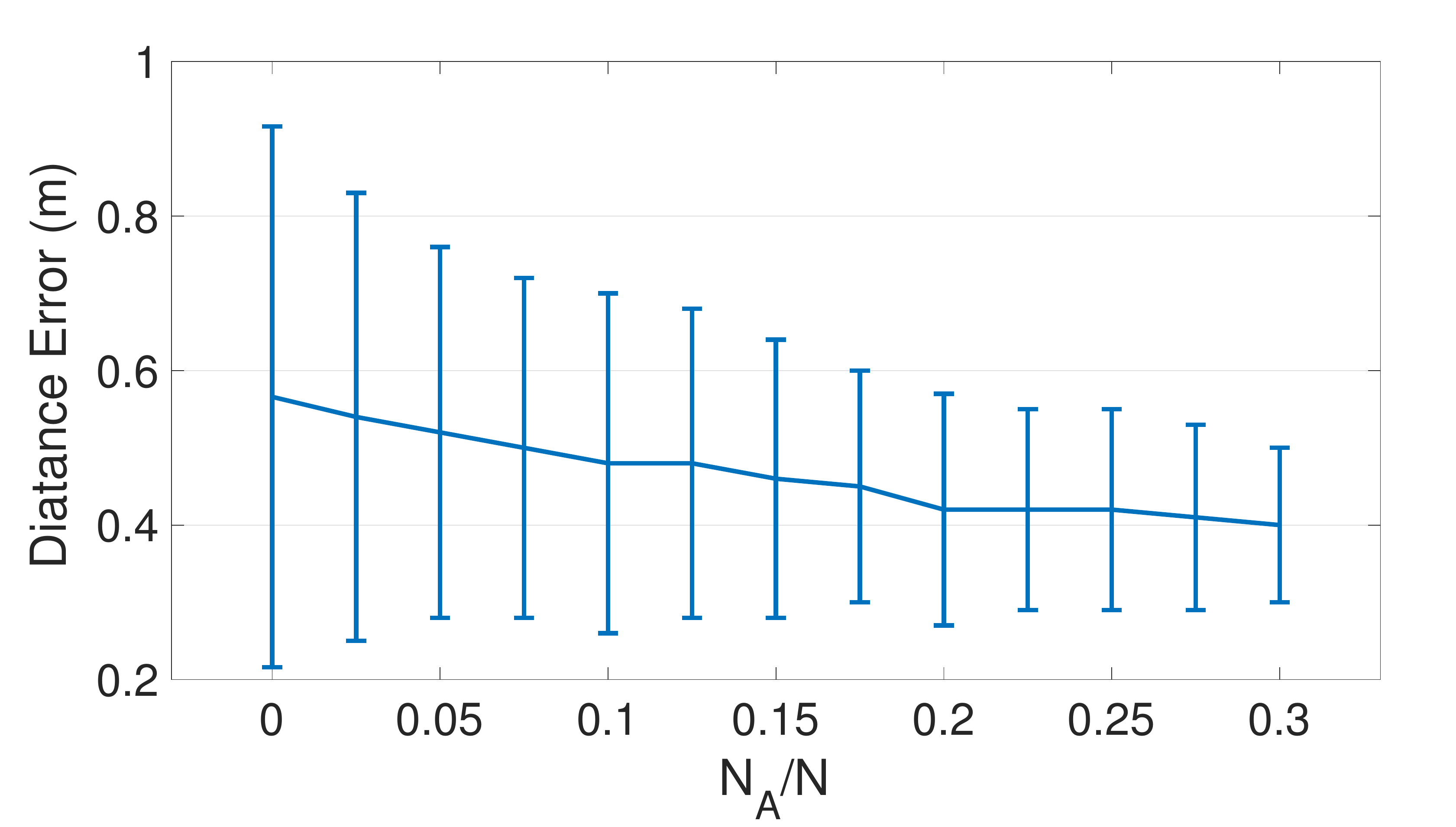}
   \label{HS_D_wA}
   }
   \quad
 \subfigure[Warehouse:Sector displacement]{
  \includegraphics[width=0.44\textwidth,trim={1cm 0cm 5cm 0cm},clip]{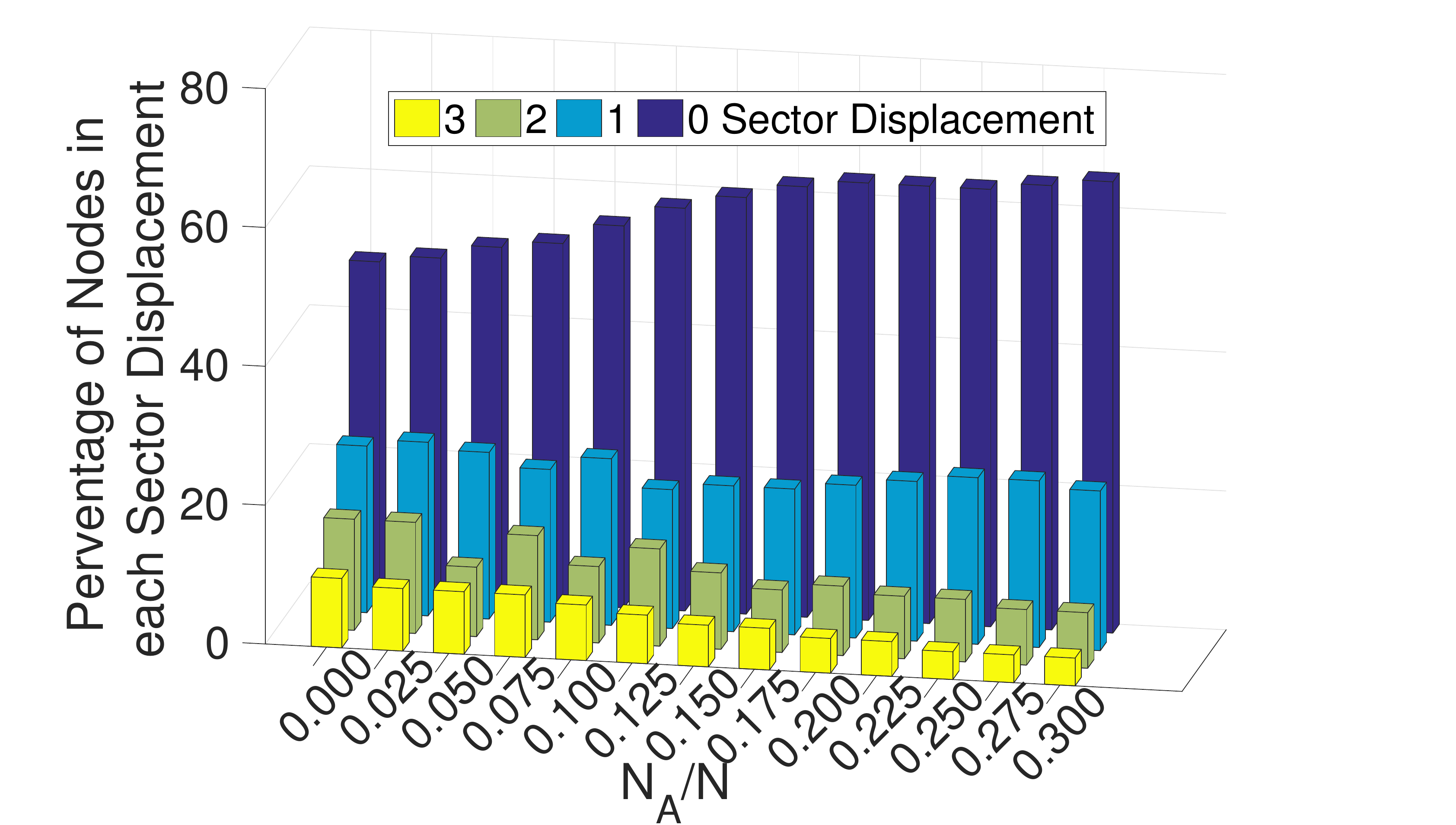}
   \label{HS_S_wA}
   }
 \caption{DMmTM-HS performance against anchor percentage when $N_R/N $=0.15}  \label{HS_performance_A}
\end{figure}
Then, Figure \ref{HS_performance_A} presents the results of performance metrics for greenhouse and warehouse against the number of initial anchors in the network when $N_R/N$ ratio equals to 0.6. Figure \ref{HS_D_gA} and \ref{HS_D_wA} show that when the number of anchors in the network increases, the distance error decreases. However, beyond 0.2 anchor ratio, the average distance error stays almost the same, while the variance of the error keeps decreasing, which ensures that the maximum distance error in the map is reducing. Moreover, Figure \ref{HS_S_gA} and \ref{HS_S_wA} illustrate that percentage of nodes located in the correct sector angle increases with the number of anchors. Similar to average distance error, the number of nodes with zero sector displacement barely changes after increasing anchors ($N_A/N$) beyond 0.2. 

\subsection{Performance Comparison}
The performance of the proposed DMmTM algorithms are compared with three existing algorithms, namely, MmTM \cite{mmtm}, DR-MDS \cite{drmds} and NTLDV-HOP \cite{ntldvhop}. For the comparison, anchor ratio is considered as 0.35 for DMmTM-SS, DR-MDS and NTLDV-HOP algorithms. In DMmTM-HS anchor ratio is considered as 0.15 and ratio of nodes receiving packets from mobile anchor as 0.6. When simulating MmTM algorithm, the 3D robot path considered in \cite{mmtm} is used.

\begin{figure*}
 \centering
 \subfigure[Distance error]{
  \includegraphics[width=0.45\textwidth,trim={.35cm 0cm .95cm 0cm},clip]{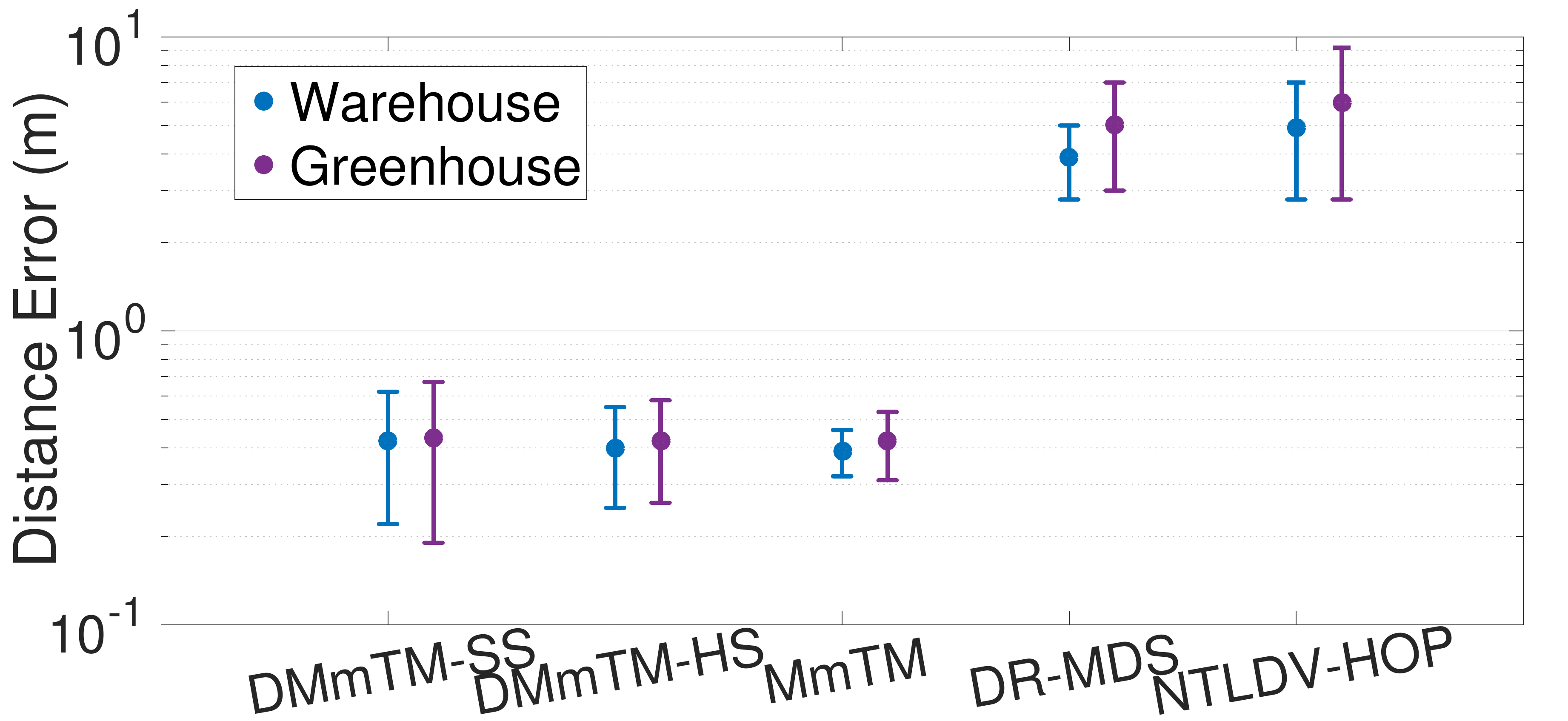}
   \label{C_D}
   }
 \quad
 \subfigure[Greenhouse:Sector Displacement]{
  \includegraphics[width=0.45\textwidth,trim={.15cm 0cm .8cm 0cm},clip]{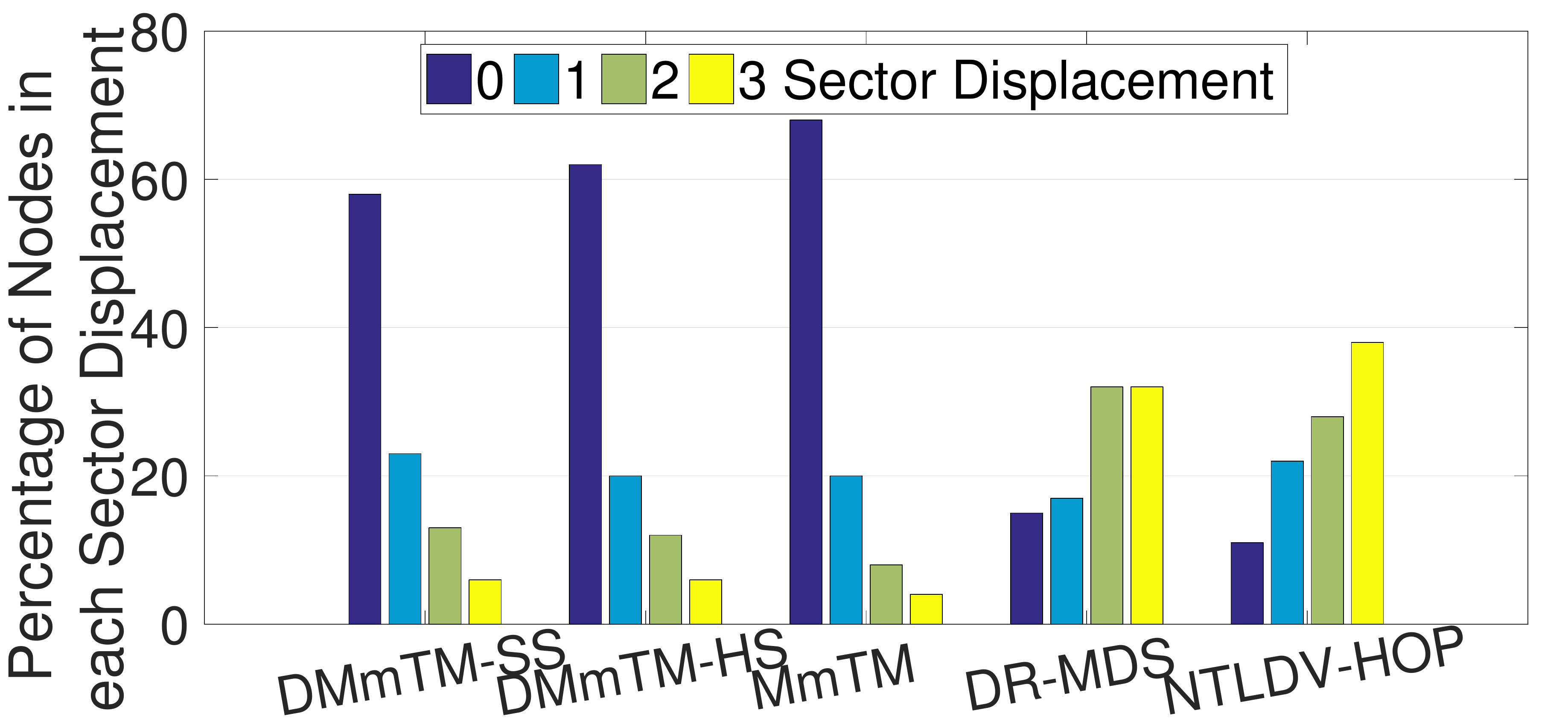}
   \label{C_SD_g}
   }
   \quad
 \subfigure[Warehouse:Sector Displacement]{
  \includegraphics[width=0.45\textwidth,trim={.15cm 0cm .8cm 0cm},clip]{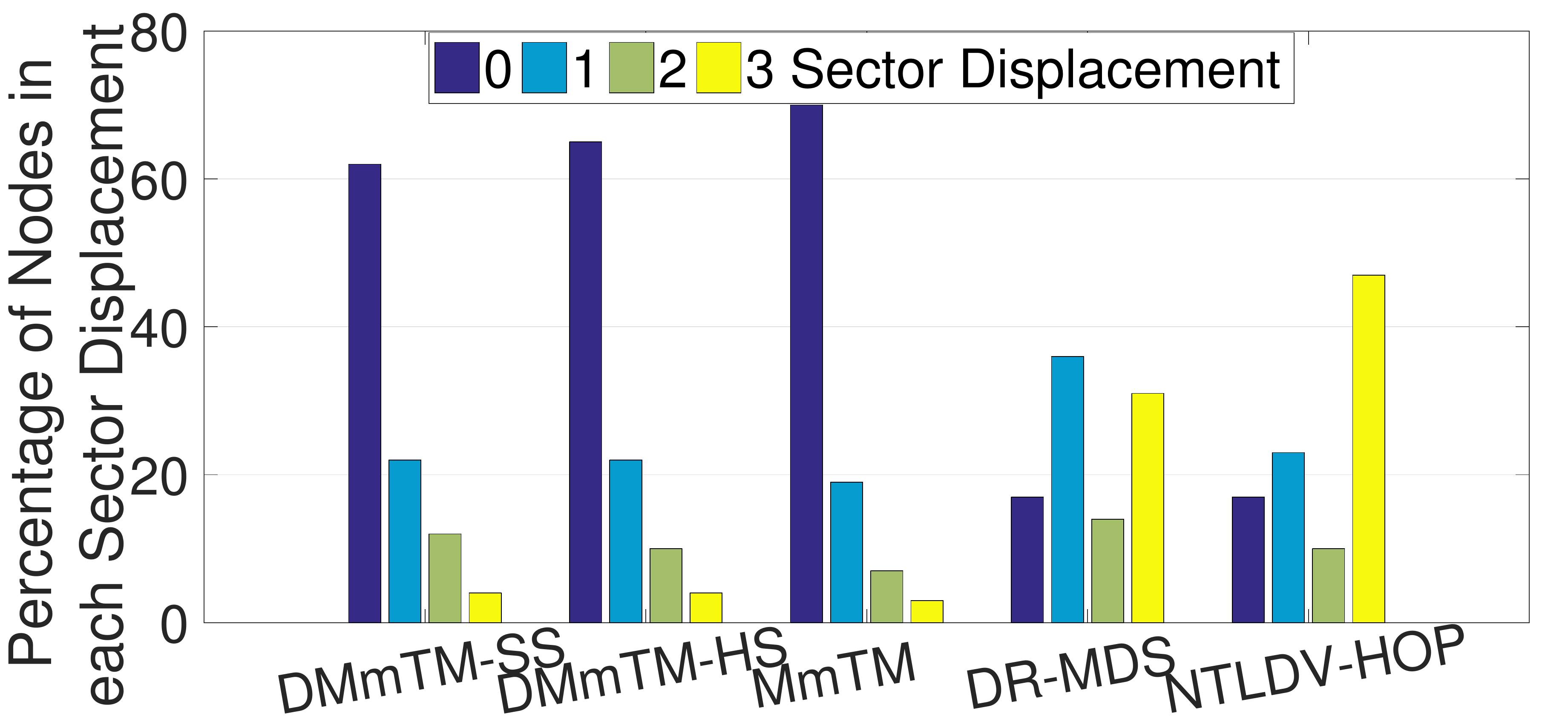}
   \label{C_SD_w}
   }
 \caption{Performance comparison}  \label{Comparison}
\end{figure*}

The performance comparison results are presented in Figure \ref{Comparison}. From the distance error comparison shown in Figure \ref{C_D}, DMmTM-SS, DMmTM-HS and MmTM algorithms have almost same average distance error, however, the variance in MmTM is relatively less. This is mainly due to the fact that MmTM algorithm uses a mobile robot traverses in the network and accesses all the nodes in the network. Therefore, for each sensor topology coordinate calculation, it has more packets received by robot. However, information gathering phase in MmTM requires more time than DMTM-SS or DMmTM-HS. Also, MmTM topology coordinates are calculated centrally, which requires more computation time. Thus, DMmTM-SS and DMmTM-HS are more efficient while resulting in almost the same average distance error as MmTM. When comparing to DR-MDS and NTLDV-HOP, both DMmTM algorithms have outperformed the two algorithms by a 5m distance error. 
The sector displacement error for the two environments are shown in Figure \ref{C_SD_g} and \ref{C_SD_w}. In both environments, DMmTM-HS has high percentage of nodes in zero sector displacement compared to DMmTM-SS. The reason is, static anchors do not know the direction information about its sectors, thus the direction calculation done in the algorithm may contain some error values. In DMmTM-HS, a mobile anchor is used for the localization, which is equipped with a compass. Therefore, the nodes receiving packets from the mobile anchor have more accurate information about its direction. Since, MmTM is using a mobile anchor to localize all the nodes, it has higher percentage of nodes in zero sector displacement. However, MmTM has some drawbacks in time required for information gathering and computation. When comparing DMmTM-SS with DR-MDS and NTLDV-HOP, which depends on static anchors, it can be seen that DMmTM-SS has outperformed the two algorithms. Compared to those two algorithms DMmTM-SS has localized more than 40\% of nodes in the correct sector. Also, DMmTM-SS has only 5\% of nodes in three sector displacement, but it is above 20\% in DR-MDS and NTLDV-HOP. Thus, it can conclude that DMmTM-SS and DMmTM-HS generates network maps more accurately and efficiently.

\section{Energy Awareness and Computation Overhead of the DMmTM Algorithm}\label{sec::dmmtm:copm}
As sensor nodes have scarce resources and capabilities such as energy, processing power and memory, WSN algorithms required being compatible with these limited resources. Thus, this section evaluates the energy awareness and computation overhead of the algorithm. 

\subsection{Energy Usage Comparison}
Since, energy consumption of the transceiver is more significant than computational energy consumption and sensing energy consumption, this calculation considers only the MmWave transceiver energy consumption for packet transmitting and receiving. 

Let the total energy consumption by the algorithm is $E_{t}$, the energy required for one packet transmission is $E_{tx}$ and energy required for one packet reception is $E_{rx}$. Then following subsection discuss the energy required by the DMmTM-SS and DMmTM-SS.

\subsubsection{DMmTM-SS}
The energy consumption of DMmTM-SS is shown in equations (\ref{eqn::mmtm:proposed}). For simplicity, it has assumed that the packet sizes used in all the transmissions are same.

\begin{equation}\label{eqn::mmtm:proposed}
E_{t} = 2N_AsE_{tx}+(N+N_A)(mE_{rx})
\end{equation}
where, $N$ is the number of sensor nodes in the network, $M$ is the number of anchor nodes, $m$ is the number of anchor nodes located in the neighbourhood of non anchor nodes ($m<<M$) and $s$ is the number of sectors in the sensor nodes. 

\subsubsection{DMmTM-HS}
The energy consumption of DMmTM-HS is shown in equations (\ref{eqn::mmtm:proposed2}).

\begin{equation}\label{eqn::mmtm:proposed2}
E_{t} = 2N_AsE_{tx}+(N+N_A)(mE_{rx})+ N(\overline{n}E_{rx}+\overline{n}sE_{tx})
\end{equation}
where, $\overline{n}$ is the average number of time that a single sensor node receives ISS message from the mobile anchor. As there is no energy restriction in mobile anchor, the energy consumption in mobile anchor is not considered.

When considering the above two energy equations i.e. equation (\ref{eqn::mmtm:proposed}) and equations (\ref{eqn::mmtm:proposed2}), it can be seen that $\overline{n}$ and $N_A$ have a major effect on the algorithm energy consumption. However, those are controllable factor that can be decided by the application requirement. For an example, if the application needs more accuracy, more number of samples and anchors can be used. On the other hand, if it is more important to conserve energy, the number of samples and anchors can be reduced.

\subsection{Comparison of Computation Overhead} 
The computation complexities of the two algorithms are discussed in this section. It is calculated based on number of messages required for each calculation as it is proportional to the time requirement.

\subsubsection{DMmTM-SS}
In the first step of DMmTM-SS, anchor nodes communicate with neighbour anchors. The worst case is all the anchor nodes are located within the neighbourhood of $a_i$ and receive packets from all the sectors. Thus the worst case complexity is $\mathcal{O}(N_AN_s)$. In the second step, anchors broadcast beacon packets from all its sectors and sensors receive those packets. The computation complexity of this step is $\mathcal{O}(N_AN_s)$, as in the worst case sensor $s_i$ may receive beacon messages from all the anchor sectors. In step three, sensor node calculates its topology coordinate by dividing it's neighbour anchors' communication area into grid points. The computation complexity of this step is $\mathcal{O}(N_G)$, where $N_G$ is the number of grid points. In the final step, which is node filtration, the nodes are grouped into categories and nodes that have higher error values re-calculate their coordinates. Thus, the computation complexity of this step is $\mathcal{O}((N_AN_s+N_G)^{p})$, where $p$ can be a value between 0 and $iT$. Also, $iT$ value is a user controllable constant value. For an example if an application requires a higher accuracy in the map, $iT$ value can be increased and if not it can be decreased. When considering the total DMmTM-SS algorithm computation complexity, it is equivalent to $\mathcal{O}((N_AN_s+N_G)^{p})$ as it is the highest complexity value. 

\subsubsection{DMmTM-HS}
When comparing DMmTM-HS to DMmTM-SS, it has only one additional step, which is information gathering via mobile anchor phase. The worst case of this step is sensor $s_i$ receives all the packets transmitted by the mobile anchor from all its sectors. Thus the worst case complexity is $\mathcal{O}(N_sT)$, where $T$ is the number of packet transmitting points. The remaining steps' complexities are same as DMmTM-SS. Thus, the total computation complexity of the algorithm is $\mathcal{O}((N_AN_s+N_G)^{p})+\mathcal{O}(N_sT)$, which equivalent to $\mathcal{O}((N_AN_s+N_G)^{p})$.

\section{Conclusion}\label{sec::dmmtm:conclusion}
Distributed Millimetre wave Topology Map (DMmTM) algorithms presented in this chapter provide localization in a coordinate space closely resembling the physical layout. DMmTM-SS is for networks with static anchors and DMmTM-HS is for networks with static and mobile anchors. Both static and mobile anchors are aware of their locations, but static anchors are unaware of their beam direction information. Thus, static anchors initially calculate their direction information, and then the sensors calculate their topology coordinates using the information extracted from the packets received by anchors. The collected information is mapped to topology coordinates using a packet receiving probability function, which is sensitive to the distance. As DMmTM algorithm calculates topology coordinates in each sensor, it is more efficient than MmTM algorithm. Furthermore, initially deployed anchors or mobile anchors in DMmTM do not require access to all the sensors in the network as they select new anchors from sensor nodes after localization. Thus the proposed algorithm does not require careful mobile anchor path planning or anchor distribution, which pose significant challenges.

Finally, the performance of DMmTM-SS and DMmTM-HS are compared with three algorithms, MmTM, DR-MDS and NTLDV-HOP. The performance results show that the proposed algorithms localize with distance errors and displacement errors similar to MmTM. However, the two algorithms are significantly more efficient than MmTM, but unlike MmTM they do not require each sensor node to be directly accessible by an anchor. Furthermore, the two algorithms outperform NTLDV-HOP and DR-MDS in both performance metrics. The results indicate the effectiveness of DMmTM-SS and DMmTM-HS generated maps to preserve the connectivity as well as the directivity of actual physical map, but without the need for distance estimation using RSSI which is extremely unreliable at MmWave wavelengths.
\chapter{Application 1: Robust Kalman Filter Based Decentralized Target Search and Prediction with Topology Maps}\label{chapter:DeTarSK}
This chapter presents a way of using topology maps in WSN based applications. Most of the accurate target search and prediction algorithms in WSNs that are accurate depend on the physical location of sensor nodes and physical distances between sensors and the target. However, there are some environments in which physical distance measurement using techniques such as signal strength is not feasible- for example in noisy environments and military environments. Under such circumstances, using a distance free mapping algorithm to locate sensor nodes and search for a target is more advantageous. Therefore, a novel distributed approach for searching and tracking of targets is presented for sensor network environments using a topology map of the network. The solution consists of a robust Kalman filter combined with a non-linear least-square method, and the ML-TM presented in Chapter \ref{chapter:mltm}. To make this algorithm distance-free, the primary input for estimating target location and direction of motion is provided by time stamps recorded by the sensor nodes when the target is detected within their sensing range. An autonomous robot is used to follow and capture the target. This robot collects the time-stamp information from sensors in its neighbourhood to determine its own path in search of the target. While the maximum likelihood topology coordinate space is a robust alternative to physical coordinates, it contains significant non-linear distortions when compared to physical distances between nodes. This also is overcome by using time stamps corresponding to target detection by nodes instead of relying on distances. 

The chapter is structured as follows and the main results of the chapter were originally published in \cite{cdetarsk, detarsk}. Section \ref{sec::DeTarSK:Introduction} offers an introduction and motivation for the research presented in this chapter. Section \ref{sec::DeTarSK:related} reviews background research conducted in target search and prediction. Section \ref{sec::DeTarSK:algo} presents the proposed DeTarSK algorithm and is followed by the results in Section \ref{sec::DeTarSK:result}. Finally, Section \ref{sec::DeTarSK:sconclusion} provides a conclusion of the chapter.

\section{Introduction}\label{sec::DeTarSK:Introduction}
Decentralized target search is an important task in many WSN applications, e.g., search and rescue, surveillance and military operations \cite{tt1, tt8}. Decentralized algorithms possess many advantages in ad-hoc WSN environments, which are susceptible to high link/node failures, uncertainty associated with multi-hop packet delivery, and inaccessibility to centralized computing resources \cite{emergency, adhoc5}. Thus, the focus of this chapter is on applications that use sensor networks to monitor details of an environment, track a target therein, and follow or rendezvous with the target with a device such as an autonomous robot or even a person carrying a mobile sensor node \cite{tt7}. Vital requirements for such applications include real time decision making related to the trajectory of a target, i.e., its current location, movement and future locations, as well as tracking the target.  

Much of the work on target tracking using WSNs rely on accurate localization information, specifically the physical coordinates of the nodes as well as the ability to measure the distance from sensor nodes to the target \cite{tt4, tt3, adhoc8}. However, localization of sensor nodes and distance estimation using strategies such as RSSI measurement or time delay is not feasible in many complex and harsh environments \cite{emergency}. Thus novel coordinate systems have emerged that do not rely on distance estimations in place of physical (geographical) coordinates \cite{VC, mltm}. Topology maps presented in Chapter \ref{chapter:mltm}, which we rely on in this Chapter, are an alternative to physical maps but with the distances among nodes significantly distorted. While such coordinate systems have been used extensively for sensor networking protocols such as routing and placement, they have found only limited traction for target tracking due to the fact that such systems do not provide accurate physical information such as position and velocity. Although they can be easily and accurately generated, use of such topology coordinates require overcoming significant non-linear distortions between the physical coordinate space and topology coordinate space. This approach overcomes the non-linear distance distortions in topology coordinate space compared to physical coordinates using robust Kalman filtering. However, Robust Kalman Filter (RKF) based target tracking algorithms \cite{propagation2, propagation} require centralized operation, i.e., an environment where all the information is  processed in a centralized location. As indicated above, distributed solutions possess many advantages over centralized solutions in WSN based environments, thus we develop a distributed approach. 

This chapter presents a novel algorithm, \textbf{De}centralized \textbf{Tar}get \textbf{S}earch based on Robust \textbf{K}alman Filter and Behavior Formula Extraction (DeTarSK), to track and predict target locations in an environment where accurate physical distance estimation is not feasible. Instead of using physical coordinates and distances, DeTarSK relies on maximum likelihood topology coordinates and corresponding distances.  The algorithm uses RKF uncertain mobility model in a distributed manner to filter the error calculated in target’s previous locations. Then, it predicts the movement, direction and the future locations of the target, using an approach in which RKF is combined with nonlinear least square method \cite{NLS}. Here, it considers the scenario of an autonomous robot searching and tracking the target using DeTarSK by communicating with sensors within its communication range. 

The search considered in this chapter is discrete. The robot search is performed on an  $n$ by $n$ grid, and its next available location is a function of its current location. Also, the target’s movement is independent of the robot's movement and can take any form. The autonomous robot calculates the target detection points from the time stamp sets received from sensors in its locality. These time stamp sets describe the times at which certain node(s) detected the target in its vicinity.  A major challenge is that the gathered information is not current, i.e., the information gathered from a region may be the target’s past locations and moving directions, but not their current values. Also, the received information about the target’s past behaviour may not be accurate. However, this is the information that is available for predicting target’s location. Therefore, the objective of this algorithm is to predict the target’s current location and calculate the shortest path to catch the target using the available information in real time.  The main contributions of this chapter are:  
\begin{itemize}
\item A decentralized target-search algorithm, 
\item Use of RKF in a decentralized manner to predict future target locations, and 
\item Overcoming the distortion between topology coordinates and physical coordinates (which are unknown) for effective target detection.
\end{itemize}

\section{Background on Target Search and Prediction}\label{sec::DeTarSK:related}
The problem of searching for targets can be characterized based on different aspects, such as a one-sided or two-sided search, tracking a stationary vs. a moving target, discrete or continuous time search, and real-time or off-line search \cite{tt3, tt2}. Tracking can also be categorized by the underlying the query routing structure, e.g., tree-based \cite{treebased}, hierarchical cluster-based \cite{hierar}, geometrical \cite{geometrical} and hash-based \cite{bashbased}. 

Tracking a target using a large scale network is challenging due to limitations of WSNs such as limited energy, processing and memory resources  \cite{adhoc6}. In addition, message losses are common and nodes are prone to failures \cite{adhoc5}. Therefore, target searching algorithms have to meet constraints, including energy-efficiency, distance-sensitivity, scalability and fault-tolerance \cite{adhoc5}. Chong et al. \cite{adhoc6} proposed a target localization algorithm in visual sensor networks based on a certainty map that described an area in which the target was occupied or non-occupied. This algorithm aims at an optimum solution considering complexity, energy efficiency and robustness. Cluster based target tracking algorithm is proposed in \cite{ietR3} to minimize the energy consumption in the network by the use of novel communication protocol they have proposed. The proposed communication protocol reduces the packet transmission in the target tracking process to optimize the energy usage. However, the decisions are made in a central node, which requires packet transmission from sensors to a sink node in timely manner. 

An algorithm for sensor selection for target tracking is presented in \cite{adhoc4} with low estimation error and low computation power. Njoya et al. \cite{ietR1} proposed a sensor placement algorithm to cover a target with minimum number of sensors in a reasonable time. A sensor selection algorithm is proposed in \cite{ietR5} to improve the accuracy of target localization while optimizing the energy. Sensor allocation for target tracking in heterogeneous networks is discussed in \cite{adhoc1}.

A generalized search for the best path selection based on NP-complete optimal search path problem is presented in \cite{tt4}. It considers a target moving within a known indoor environment partitioned into interconnected regions. The search path problem is modelled as a discrete search that concerns both cell connectivity and transmit time. The extended tracking approach for a ultra-wideband indoor sensor network proposed in \cite{sensorJ1} is based on a Bernouli filter and considers the uncertainties of data association, target measurement rate, detection and noise. A real-time algorithm based on a probabilistic version of a local search with estimated global distance is presented in \cite{tt3}. The initial target location probabilities and the transition matrix for possible moves of target are the algorithm inputs in the proposed method. The decision making at each step of this search applies probabilistic distance estimation to find the path trajectory of the agent that minimizes the average number of search steps.

Locating a single target using synchronous measurements from multiple sensors is discussed in \cite{aoa10}. This approach is based on forming a geometric relationships between the measured parameters and their corresponding errors. This relationship is then used to formulate the localization task as a constrained optimization problem. Moreover, a geometrically constrained optimization approach to localize a stationary target with AoA and TDoA of sensors are proposed in \cite{newA31}. A Learning Real-Time A* algorithm (LRTA*) for a static target was proposed in \cite{LRTA} and for a Moving Target in \cite{MTS}. Information Moving Target Search (IMTS) algorithm \cite{tt6}  has enhanced the MTS algorithm with informational distance measures based Rokhlin metric\footnote{found in \cite{Rokhlin} to represent both the conditional entropy and the orthogonality measures in an effective way } and Ornstein metric\footnote{lower bound of Rokhlin metric}, which gives the necessary distance measures. 

Locating stationary target in a ultra wideband network using time of arival parameter is proposed in \cite{ietR4}. Also, Wang et al. \cite{adhoc8}  proposed an algorithm to locate a single static target using Kalman filter and a least squares algorithm. This algorithm is based on a distance model and an angel model. Dong et al. \cite{A1} proposed a mobility tracking algorithm for cellular network that uses RSSI measurements and a velocity matrix. Moreover, Mahfouz et al. \cite{sensorJ2} proposed a moving target tracking method based on RSSI models and a Kalman filter. The clustering based fusion estimation target tracking approach in \cite{sensorJ3} calculates the  filter estimators at each cluster head, and it is suitable for sensor networks with multiple sampling intervals. Mobile object tracking in a randomly deployed binary sensor network is addressed in \cite{adhoc2} with a location aware algorithm in which  the output dynamically changes according to the object’s movements. Location prediction of mobile objects in WSNs is proposed in \cite{A2}. This method is based on Gauss-Markov mobility model and maximum likelihood technique, which assumes that sensors are capable to measure the velocity of the target. Mobility estimation in WCDMA network using TDoA and AoA is proposed in \cite{A3}. However, the distance uncertainty calculation based with range measurements in complex and harsh environments limits it applicability. 

\begin{figure}
 \centering
 \subfigure[Centralized target search algorithm]{
  \includegraphics[width=0.75\textwidth]{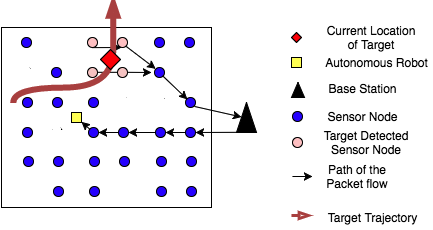}
   \label{fig::DeTarSK:c}
   }
   \quad
 \subfigure[Decentralized target search algorithm]{
  \includegraphics[width=0.75\textwidth]{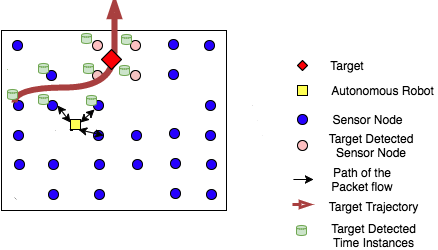}
   \label{fig::DeTarSK:d}
   }
   \caption{The packet flow of target search algorithms } \label{fig::DeTarSK:cd}
\end{figure}

The proposed approach deviates from prior work in two ways. First, a WSN topology map is used instead of a physical map thus overcoming disadvantages associated with physical localization and distance measurements. Second, a decentralized target search algorithm is proposed. As shown in Figure \ref{fig::DeTarSK:c}, in centralized algorithms, the sensed information has to be  routed to base station, processed, and then to the robot \cite{ietR3}. Thus, data is routed through multi-hops, which increase the network traffic and it is subjected to uncertainties in the network reducing the reliability of the algorithm. Hence, this Chapter proposes a decentralized algorithm in which robot directly communicate with its neighbours and move towards the target with available information as indicated in Figure \ref{fig::DeTarSK:d}. Thus decentralized algorithms overcome the disadvantages associated with multi-hop communication in centralized algorithms. Moreover, in a situation that links with the base station are lost, centralized algorithms are unable to search the target any further. But in decentralized algorithms, as robot is mobile and using single-hop communication, it can move around and re-store the communication links easily. Also, if the robot is unable to function, it can be easily replaced by another robot as navigation paths are completely independent from each other and depend only on time stamps gathered from robot's surrounding nodes. But in centralized algorithms, replacing a base station is not a feasible solution in an emergency situation. However, there is a disadvantage relates with decentralized algorithms, which is robot may not get latest information about the target as in centralized algorithms. But it has overcome in this algorithm by using a prediction method based on Robust Kalman filter combined with a non-linear least-square method.  

\section{Robust Kalman Filter based Decentralized Target Tracking and Target Behavior Formula Extraction (DeTarSK )}\label{sec::DeTarSK:algo}
\begin{figure*}
  \centering
    \includegraphics[width=1\textwidth]{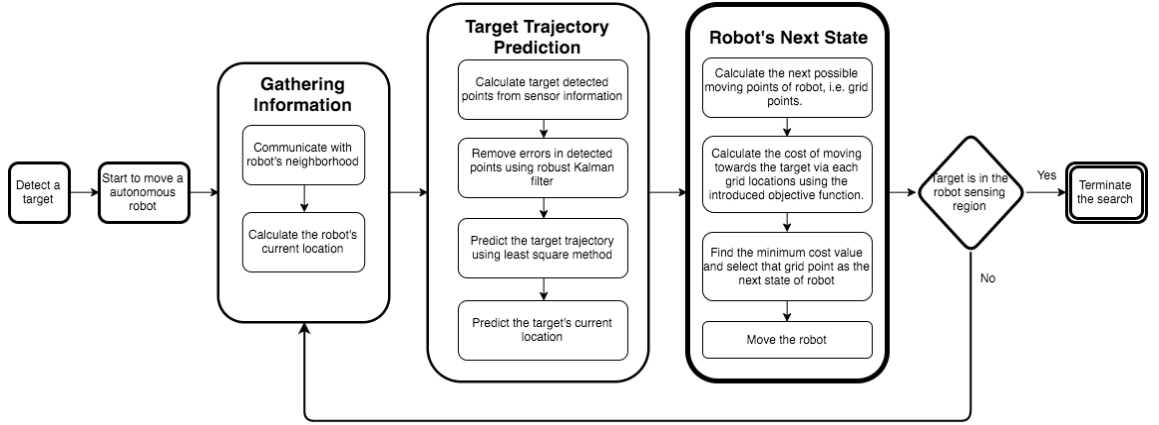}
    \caption{DeTarSK  algorithm work flow}\label{fig::DeTarSK:flow}
\end{figure*}

DeTarSK algorithm based on ML-TM of a WSN is described in this section. The objective of this algorithm is to construct a real-time path to find a target with least possible time using the information gathered by robot's local neighbourhood. The basic idea of the algorithm is described in Figure \ref{fig::DeTarSK:flow}. This algorithm uses standard the IEEE 802.15.4/Zigbee point to point protocol to communicate with sensors and robot \cite{ietR2}. Keywords used in this algorithm are described below,

\indent\textbf{Target :} a robot/person enters to the network and moves around in an independent pattern.\\
\indent\textbf{Autonomous robot/agent :} a robot/person with a wireless mote that is searching for the target by communicating with sensor nodes. \\
\indent\textbf{Target initial detection point :}  location the  target enters the area covered by the WSN and is  detected by sensor nodes.\\
\indent\textbf{Target tracking : } sensor nodes sense for the target and keep track of time instances that they have detected the target.\\
\indent\textbf{Target prediction:} autonomous robot predicting target’s future locations based on available tracking information.\\
\indent\textbf{Target search:} autonomous robot looking for the target based on tracking and prediction information.

Let consider, $N$ number of nodes are randomly deployed over an environment and $N_t (<N)$ number of nodes are detected a target entering to the network at time $t$. Then the detected nodes, $S_t=\lbrace s_1,...,s_k\rbrace$, transmit an initial message to the sink node informing the detection. Sink node calculates the initial target location $L_{t_0}=(X_{t_0},Y_{t_0})$ using the equation (\ref{eqn::DeTarSK:tlocation}) and send a message to autonomous robot including the calculated target detection point. Since, it takes time for these message transmission, it was assumed that the robot starts the search after a $t_a$ time from target detection. Thus, robot does not have any data about target's current location or moving direction. This information is gathered by its neighbourhood nodes in a decentralized way as described in following subsections.
\begin{equation}\label{eqn::DeTarSK:tlocation}
X_{t_0}=\frac{\displaystyle\sum_{s_i\in S_t}x_i}{N_t}\:\:\:\:\:and\:\:\;\;\:Y_{t_0}=\frac{\displaystyle\sum_{s_i\in S_t}y_i}{N_t}
\end{equation}

\subsection{Information Gathering}
Information about target is gathered by robot's local neighbourhood. A message is transmitted to its $V_a+\varepsilon$-neighbourhood \footnote{nodes located within a circle centred at the robot's location with a radius of $V_a+\varepsilon$} to request their topology coordinates, transmitting power and time sets $Z_{s_i}^{t_j}$ that each node has captured the target. The  $\varepsilon$ is a value within 0 and $V_a/2$. Then, robot calculates its topology coordinates $L_{a_i}$ at time $t_i$ using equation (\ref{eqn::DeTarSK:alocation}). 
\begin{eqnarray}\label{eqn::DeTarSK:alocation}
X_{a_i}={\displaystyle\sum_{j=1}^{\mid N_{a_i}\mid }w_j x_j}\Bigg /{\displaystyle\sum_{j=1}^{\mid N_{a_i}\mid }w_j }\:\:\:\:\:and\:\:\;\;\: \nonumber \\
Y_{a_i}={\displaystyle\sum_{j=1}^{\mid N_{a_i}\mid }w_j y_j}\Bigg  /{\displaystyle\sum_{j=1}^{\mid N_{a_i}\mid }w_j }
\end{eqnarray}
where, $s_j\in N_{a_i}$ and $w_j=\frac{P_{rx_j}}{P_{tx_j}}$. $P_{tx_j}$ is the transmitting power of node $s_j$ and $P_{rx_j}$ is the receiving power of the message sent by node $s_j$.

\subsection{Target Trajectory Prediction}
In this stage, the target trajectory is predicted using the information gathered in previous stage. From $Z_{s_i}^{t_j}$ time sets, the subset of nodes that detected the target at each time instance can be obtained. Then, $L_{t}^{t_j}(k)=(X_{t}^{t_j}(k),Y_{t}^{t_j})(k)$, the target locations at time $t_k$ with available information at $t_j\: (>t_k)$ is calculated using the equation (\ref{eqn::DeTarSK:ptloc}). This location information is stored in robot memory for future trajectory prediction as well. 
\begin{eqnarray}\label{eqn::DeTarSK:ptloc}
X_{t}^{t_j}(k)&=&\left [\displaystyle\sum_{i=1}^{N }c_i^k x_i+X_{t}^{t_{j-1}}(k)n_{t_{j-1}}\right ]\Bigg /\left [\displaystyle\sum_{i=1}^{N }c_i^k +n_{t_{j-1}}\right ],\nonumber \\
Y_{t}^{t_j}(k)&=&\left [\displaystyle\sum_{i=1}^{N }c_i ^kx_i+Y_{t}^{t_{j-1}}(k)n_{t_{j-1}}\right ]\Bigg /\left [\displaystyle\sum_{i=1}^{N }c_i^k +n_{t_{j-1}}\right ]
\end{eqnarray}
where, $c_i^k =  1$, if $s_i$ node is in the robot neighbourhood and detected the target at time $t_k$ and $c_i^k =  0$, if it's not in the neighbourhood or does not detected the target at time $t_k$. $n_{t_{j-1}}$ is the number of nodes used to calculate $L_{t_k}^{t_{j-1}}$ at time $t_{j-1}$.

Therefore, the RKF is considered with an uncertainty mobility model to remove the errors in target detection points. Kalman filter provides a method for constructing an optimal estimate of the state that consists of a linear dynamical system driven by stochastic white noise processes \cite{kalmann}. However, it does not address the issue of robustness against large parameter uncertainty in the linear process model. Therefore, a RKF-based state estimation algorithm is used for target detected point calculation \cite{propagation2}\cite{propagation}.

Since the target detected points are calculated from its current and previous neighbourhood information, those might have encountered some errors. Kalman filter has attracted attention on target tracking in sensor networks in recent past because it provides an optimal way of extracting a signal from noise by exploiting a state space signal model \cite{propagation}. It provides a method for constructing an optimal estimate of the state that consists of a linear dynamical system driven by stochastic white noise processes \cite{kalmann}. However, it does not address the issue of robustness against large parameter uncertainty in the linear process model. In this problem, model dynamics are unknown but bounded. Therefore, Kalman filter may lead to poor performance. RKF that focuses on the uncertainties of the system has been proposed in \cite{kalmann}, however, originally this framework was presented in \cite{newA14}. RKF was applied to derive an estimate of the mobile target's location in mobile sensor network \cite{propagation2} and in Delay-Tolerant Sensor Networks \cite{propagation}, nevertheless, these algorithms are calculated target locations in a centralized approach. In this algorithm, the RKF is used in a decentralized manner to remove errors in calculated target positions and to predict the target trajectory up to $t_K$ time instances, i.e., the largest time instance value of all $Z_{s_i}^{t_j}$ time sets that robot received. 

The uncertainty mobility model and measurement model used in this paper is as follows.
\begin{eqnarray}
x_t(k+1)&=&(A+B\Delta (k) K)x_t(k)+w(k)\label{eqn::DeTarSK:kalmanstate}\\
y(k)&=&Cx_t(k)+v(k)\label{eqn::DeTarSK:kalmanmes}
\end{eqnarray}
%
where $x_t=[X_t^{t_j},Y_t^{t_j},\dot{X}_t^{t_j},\dot{Y}_t^{t_j}]^T$. $(X_t^{t_j},Y_t^{t_j})$ and $(\dot{X}_t^{t_j},\dot{Y}_t^{t_j})$ are the position and velocity of the target in topology map. $\Delta (k)=[\gamma (k)-1 0;0 \beta (k)-1]$ is the uncertain matrix. $\gamma (k)$ and $\beta (k)$ satisfy the constraint of $1-\zeta \leq \gamma (k), \beta (k)\leq 1+\zeta $, where $0<\zeta <1$. Thus $\Delta (k)$ satisfy the bound $\Delta (k)^T\Delta (k)\leq I$. $w(k)$  is the process noise with $Q$ covariance that denotes the  driving/acceleration command of the target. $v(k)$ is the measurement noise with $R$ covariance that denotes the nodes' topology coordinates error component. $A$, $B$, $C$ and $K$ are as follows.
\[
A=
\begin{bmatrix}
    1&0&t&0 \\
    0&1&0&t \\
    0&0&1&0 \\
    0&0&0&1 
\end{bmatrix}
,\:\:\:\:\:\:B=
\begin{bmatrix}
    t&0\\
    0&t\\
    0&0\\
    0&0
\end{bmatrix}
\]
\[
C=
\begin{bmatrix}
    1&0&0&0 \\
    0&1&0&0 
\end{bmatrix}
,\:\:\:and \:\:\: K=
\begin{bmatrix}
   0&0&1&0\\
   0&0&0&1
\end{bmatrix}
\]
where $t$ is the sampling time.

The non-linear least square method is used to calculate the target trajectory from the output of the RKF. The trajectory equation varies with time is a 3D Space-time equation when a 2D network is considered. Hence the space-time equation is divided into two equations that describe how x and y coordinates of the target varies with time and those are named as $t-x$ curve and $t-y$ curve respectively. If the network is 3D, three equations must be calculated namely $t-x$, $t-y$ and $t-z$.  


The next challenging thing is to find the degree of the $t-x$ and $t-y$ curves. Let consider p and q are the degree of $t-x$ and $t-y$ curves respectively. Then the sign changes of $X_t^{t_j}(k)$ and $Y_t^{t_j}(k)$ coordinates are considered to calculate p and q as shown below. 
\begin{equation}
p=\sum _{k=3}^j sign_x^{t_k}
\end{equation}
where,
$ sign_x^{t_k}= $
\[
  . 
\begin{cases}
    1,& \text{if } \frac{(X_t^{t_j}(k)-X_t^{t_j}(k-1))}{\vert X_t^{t_j}(k)-X_t^{t_j}(k-1)\vert }\times \frac{(X_t^{t_j}(k-1)-X_t^{t_j}(k-2))}{\vert X_t^{t_j}(k-1)-X_t^{t_j}(k-2)\vert }=-1 \\ &and {\vert X_t^{t_j}(k)-X_t^{t_j}(k-1)\vert } >1\\
    0,              & \text{otherwise}
\end{cases}
\]

\begin{equation}
q=\sum _{k=3}^j sign_y^{t_k}
\end{equation}
where,
$sign_y^{t_k}= $
\[
 .   
\begin{cases}
    1,& \text{if } \frac{(Y_t^{t_j}(k)-Y_t^{t_j}(k-1))}{\vert Y_t^{t_j}(k)-Y_t^{t_j}(k-1)\vert }\times  \frac{(Y_t^{t_j}(k-1)-Y_t^{t_j}(k-2))}{\vert Y_t^{t_j}(k-1)-Y_t^{t_j}(k-2)\vert }=-1 \\ &and {\vert X_t^{t_j}(k)-X_t^{t_j}(k-1)\vert } >1\\
    0,              & \text{otherwise}
\end{cases}
\]
The $t-x$ and $t-y$ curve equations are given by equations (\ref{eqn::DeTarSK:tx}) and (\ref{eqn::DeTarSK:ty}) respectively.

\begin{eqnarray}
x&=&a_0+a_1t+...+a_pt^p \label{eqn::DeTarSK:tx}\\
y&=&b_0+b_1t+...+b_qt^q \label{eqn::DeTarSK:ty}
\end{eqnarray} 

Equation (\ref{eqn::DeTarSK:a}) and equation (\ref{eqn::DeTarSK:b})  can be used to calculate ${a_0,a_1,...a_p}$ and ${b_0,b_1,...b_p}$, which is obtained by using the least square method.
\begin{eqnarray}
[a_0\:\: a_1 \cdot\cdot\cdot a_p]^T &=& inv(T_x)X \label{eqn::DeTarSK:a} \\ \relax
[b_0\:\: b_1 \cdot\cdot\cdot b_q]^T &=& inv(T_y)Y \label{eqn::DeTarSK:b}
\end{eqnarray}
where,
\[
T_x=
\begin{bmatrix}
    \sum 1 & \sum t_i &\dots& \sum t_i^p \\
    \sum t_i & \sum t_i^2 &\dots &\sum t_i^{p+1} \\
    \vdots & \vdots & \ddots & \vdots \\
    \sum t_i^q & \sum t_i^{p+1}& \dots &\sum t_i^{2p} 
\end{bmatrix}
\]
\[
T_y=
\begin{bmatrix}
    \sum 1 & \sum t_i& \dots &\sum t_i^q \\
    \sum t_i & \sum t_i^2 &\dots &\sum t_i^{q+1} \\
    \vdots & \vdots & \ddots & \vdots \\
    \sum t_i^q & \sum t_i^{q+1} &\dots& \sum t_i^{2q} 
\end{bmatrix}
\]

\[
X^T=
\begin{bmatrix}
    \sum x_i & \sum t_ix_i \dots \sum t_i^px_i 
\end{bmatrix}
\]
\[
Y^T=
\begin{bmatrix}
    \sum y_i & \sum t_iy_i \dots \sum t_i^qy_i 
\end{bmatrix}
\]
The $t-x$ and $t-y$ equations above describe the behaviour of the target (e.g.: changing directions) up to $t_k\: (<t_j)$ time. However, the requirement of the algorithm is to predict the target location at time $t_j$, which cannot be obtained by substituting the current time in the equation. Therefore, to incorporate future behaviours, the time at which the target changes its direction is considered. Let consider the $t-x$ equation. First, differentiate the equation with respect to time and equate it to zero to find the set of time values $\{t_{d1},t_{d2},...\}$ that changes the moving direction. Then, the average time of target moving in same direction $ \overline{\Delta t}$ is calculated and the total set of estimated time instances  that target changes it direction is $TD=\{ t_{d1},t_{d2},...,t_k+\overline{\Delta t},t_k+2\overline{\Delta t},...\}$..

After that obtain the final estimated $t-x$ equation that incorporates the all behaviours as shown below.
\begin{equation}
Final\_t-x = a\times int\_t-x+C
\end{equation}

where, a and C are constants and
\begin{equation}
int\_t-x = \prod _{i=1}^{\mid TD\mid }(t-TD(i))
\end{equation}

To calculate $a$ and $C$ two points at which the target was detected can be substituted to $Final\_t-x$ equation and solve them. The $Final\_t-y$ equation also can be obtained by following came steps mention above. Then, the target’s current location $L_{t\_current}=(X_{t\_current},Y_{t\_current})$ can be obtained by substituting $t_j$ value to the $Final\_t-x$ and $Final\_t-y$ equation.

\subsection{Next State of the Autonomous Robot}
In this stage, algorithm calculates the next point for the robot in its move towards the target as in A* algorithm \cite{tt9}. The next point calculation is based on grid search as in \cite{mytarget} and the number of grids in it's neighbourhood is considered as eight. The grid locations $G_i$ is based on the robot's current location $L_{a_i}$ and it can be calculated as in equation (\ref{eqn::DeTarSK:gridlocation}).

\begin{eqnarray}\label{eqn::DeTarSK:gridlocation}
X_{G_k} &=& X_{a_i} + V_a cos\{(k-1)\pi /4\} \nonumber \\
Y_{G_k} &=& Y_{a_i} + V_a sin\{(k-1)\pi /4\}
\end{eqnarray}
where, $k=1,...,8$.

Once the grid layout is calculated, robot finds the grids that are free with obstacles. Then it selects one grid as it's next point to move based on following equation used in \cite{mytarget}.
\begin{equation}
F(n)=G(n)+H(n)
\end{equation}

where, G(n) is the cost of moving from the initial step to the next step on the grid and H(n) is the cost of moving from a grid point to the target's predicted current $L_{t\_current}$ location.

For the H(n) calculation, the Euclidean distance in the ML-TM is used. However, the actual length of the path cannot be calculated because obstacles can be in the way of the target. Thus, H(n) is a approximate value. The robot calculates F(n) value for all the grid locations and chooses the grid point that has minimum F(n) value as it's next state. 

\section{Performance Evaluation}\label{sec::DeTarSK:result}
This section discusses the performance of the DeTarSK algorithm. MATLAB based simulation was carried out and two network setups were considered in the evaluation. One is a sparse network with 1400 sensor nodes and the other one is a circular shape network with three obstacles and 496 sensor nodes. The circular shape network has three obstacles in the middle of the network. 
To model a real communication link between nodes and robot, a propagation model proposed in Chapter \ref{chapter:mltm} is used and the equation is shown below. 
\begin{equation}\label{eqn::DeTarSK:pm2}
P_{rx_i} = P_{tx_j}-10\varepsilon log d_{ij} - L_{ob_i} + X_{i,\sigma }
\end{equation}
where, the received signal strength at node $i$ is $P_{rx_i} $, the transmitted signal strength of the signal at node $j$ is $P_{tx_j}$ , the path-loss exponent is $\varepsilon$, the distance between node $i$ and node $j$ is $d_{ij} $, $L_{ob_i}$ is the loss due to signal absorption from obstacles exist in the line of sight of node $i$ and $j$, and the logarithm of shadowing component with a $\sigma $ standard deviation on node $i$ at is $X_{i,\sigma }$. The shadowing values are selected from a normal distribution with zero mean parameter and standard deviation parameter. The absorption coefficient and the thickness of the obstacle medium, which signal traverses are $\alpha$ and $d_o $ respectively. Then $L_{ob_i} $ can be calculated as, 
\begin{equation}
L_{ob_i} = \Sigma _{k=1}^{n}10\alpha d_o log(e) 
\end{equation}
where, e is the exponent and n is the number of obstacles exist in between node $i$ and node $j$. 

To evaluate the performance of proposed target search algorithm, three cases with different target motion patterns are considered. In all three cases the target and robot are moving with a speed of 1.5$ms^{-1}$, 2$ms^{-1}$ respectively.

\indent \textbf{Case 1:} Target enters to the sparse network at $(1, 20)$, randomly moves around and leaves the network after 82s at $(1, 10)$. The autonomous robot starts to search the target from $(30,30)$ after 5s of target detection \\
\indent \textbf{Case 2:} Target enters to the sparse network at $(1, 25)$, randomly moves around by changing the direction more frequently and leaves the network after 84s at $(60, 25 )$. The autonomous robot starts to search the target from $(30,30)$ after 10s of target detection \\
\indent \textbf{Case 3:} Target enters to the circular shape obstacle network at $(2,15)$, randomly moves around and leaves the network after 44s at $(28, 22)$. The autonomous robot starts to search the target from $(3,20)$ after 5s of target detection \\
\begin{figure*}
 \centering
 \subfigure[Physical map]{
  \includegraphics[width=0.48\textwidth]{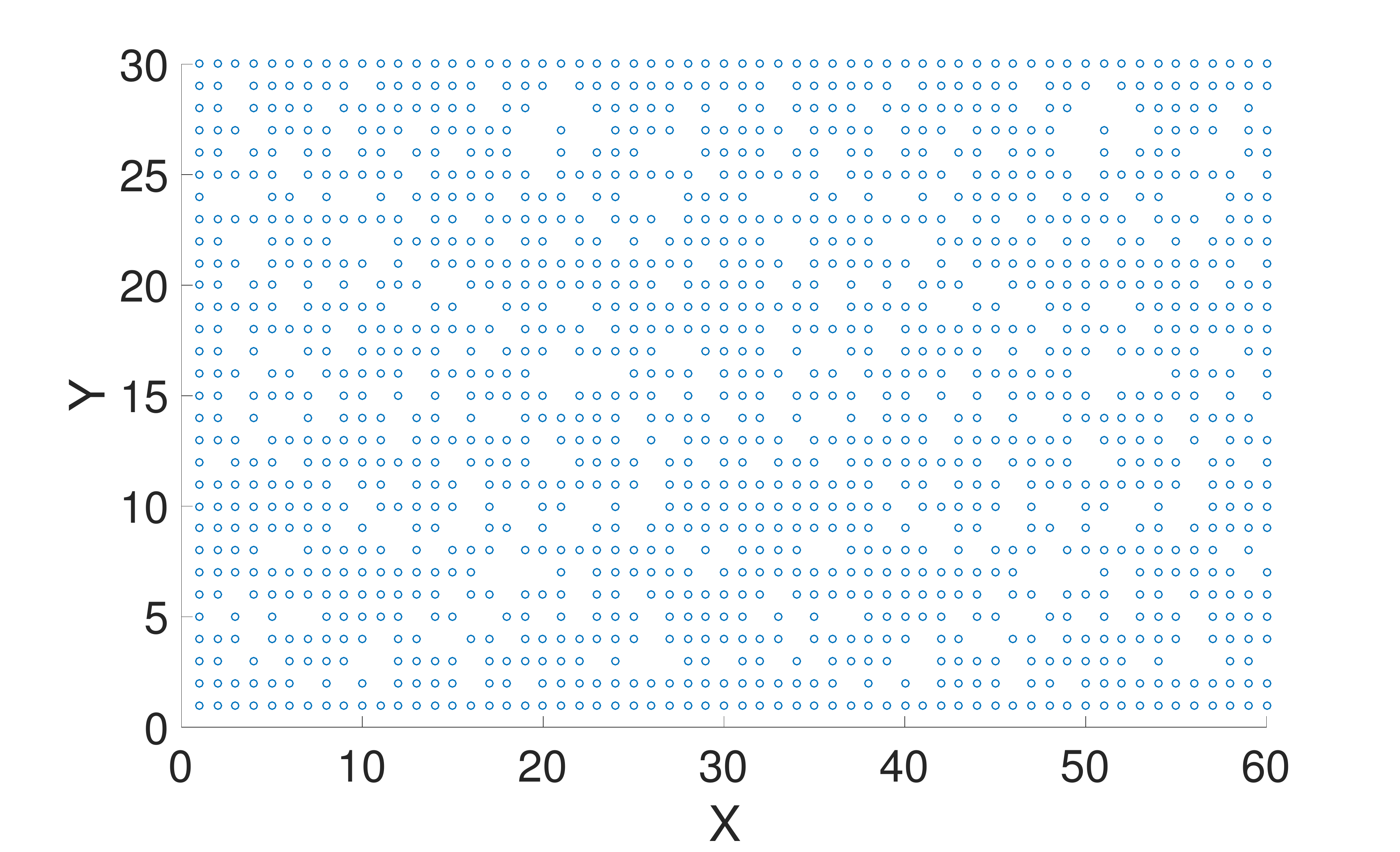}
   \label{fig::DeTarSK:spm}
   }   
 \subfigure[Calculated ML-TM map]{
  \includegraphics[width=0.48\textwidth]{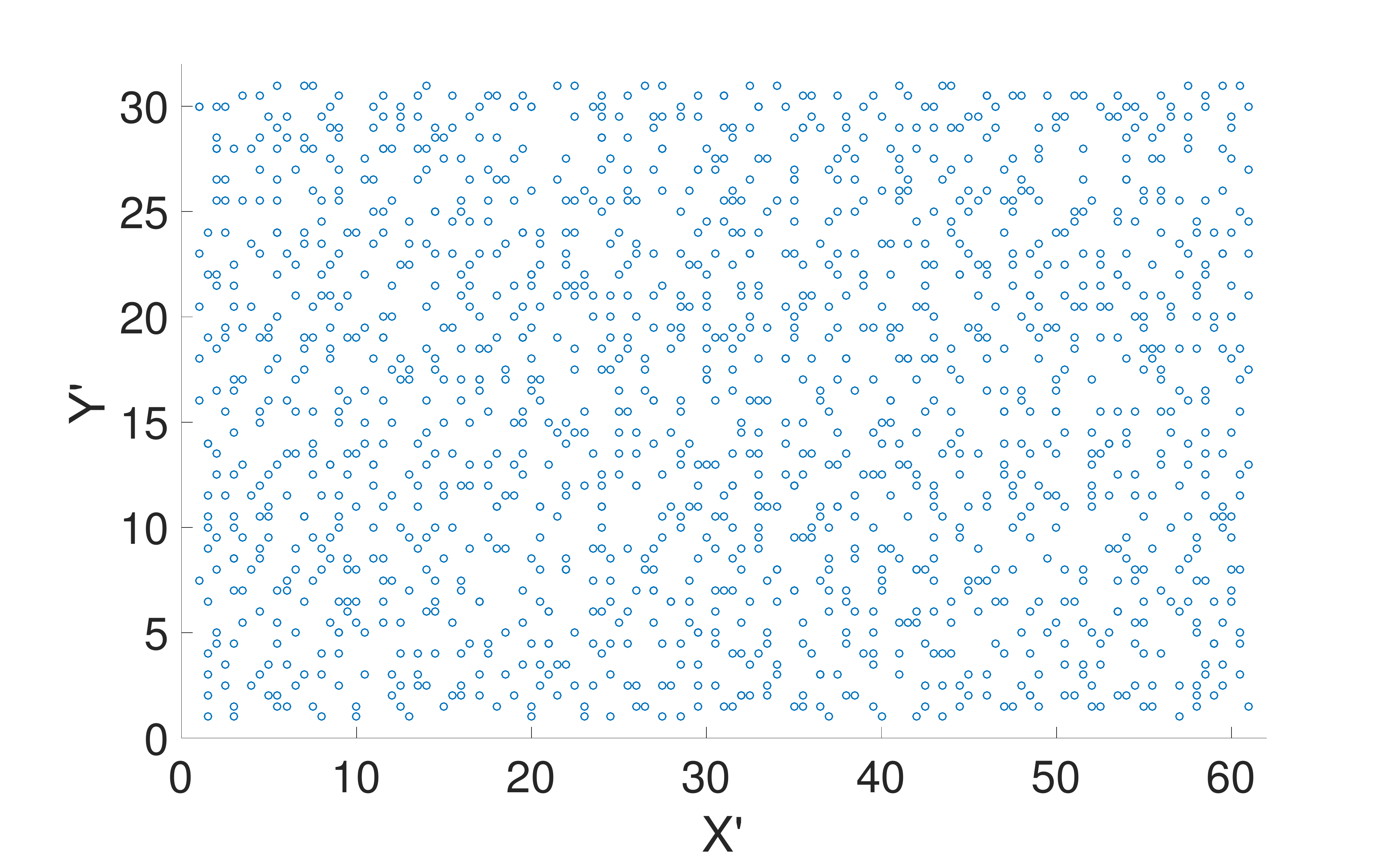}
   \label{fig::DeTarSK:stm}
   }
 \subfigure[Search using DeTarSK ]{
  \includegraphics[width=0.48\textwidth]{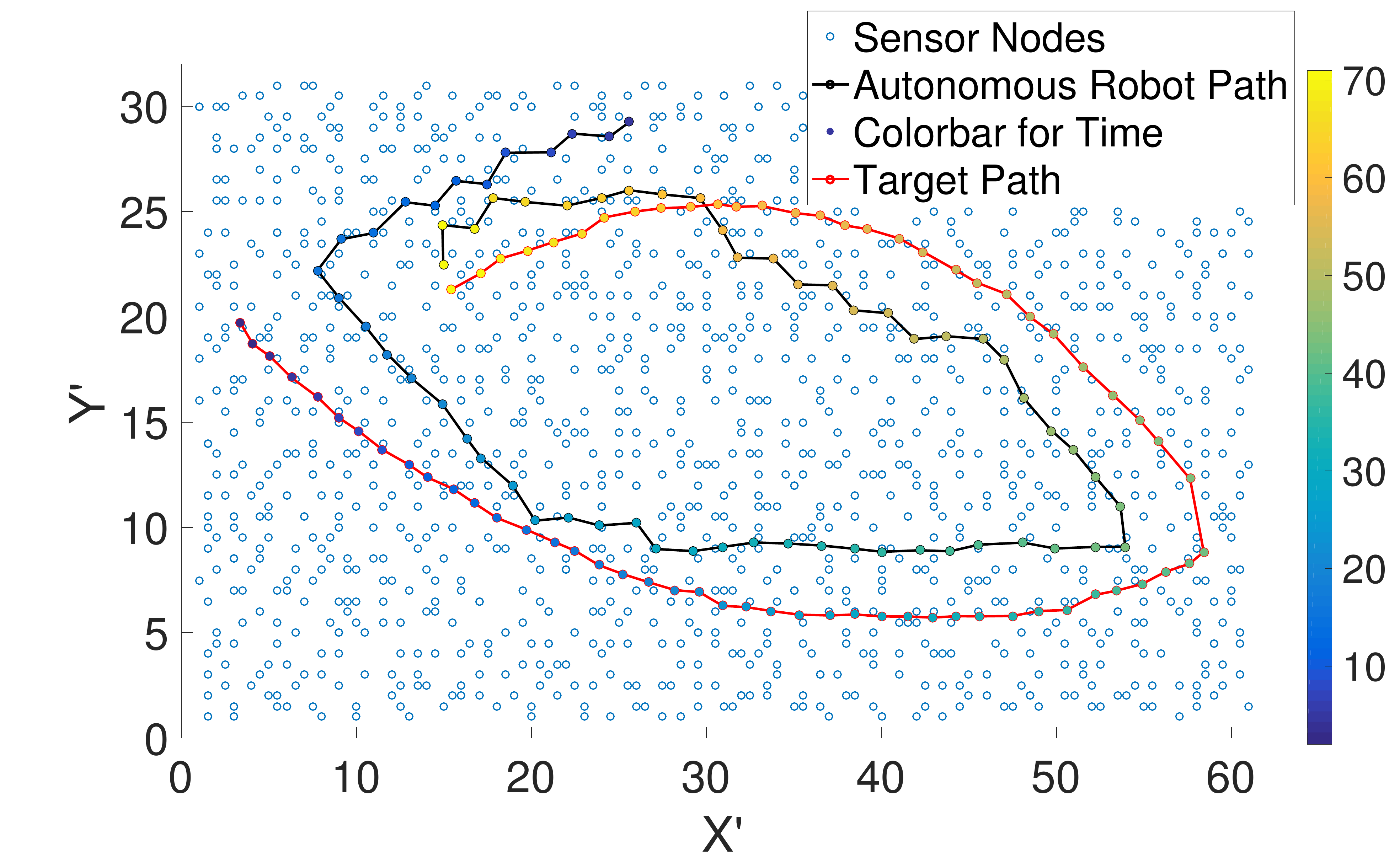}
   \label{fig::DeTarSK:et}
   }   
 \subfigure[Target prediction error]{
  \includegraphics[width=0.48\textwidth]{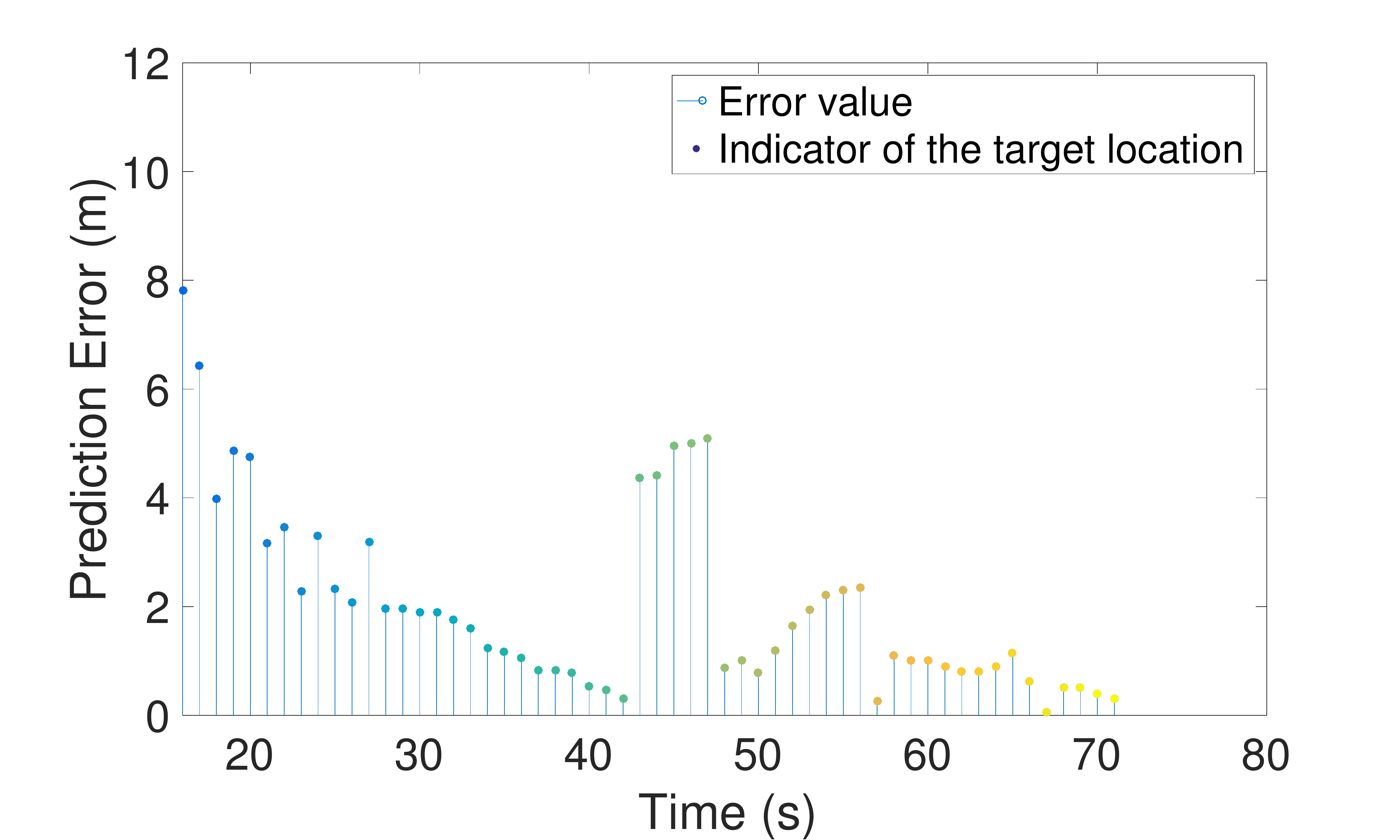}
   \label{fig::DeTarSK:epre}
   }
   \caption{Results of DeTarSK for case 1} \label{fig::DeTarSK:se}
\end{figure*}

\begin{figure}
 \centering
 \subfigure[Search using DeTarSK]{
  \includegraphics[width=0.48\textwidth]{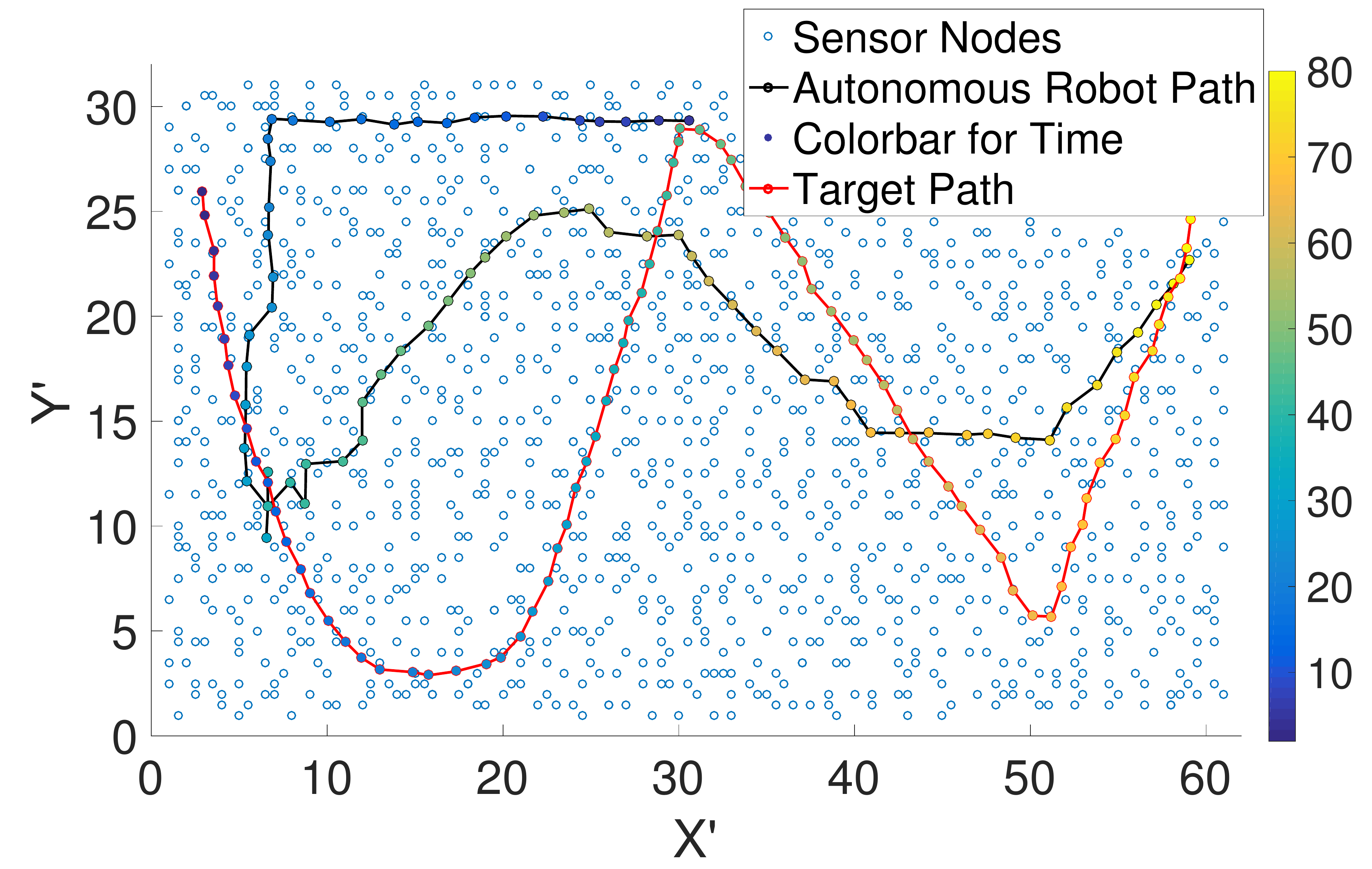}
   \label{fig::DeTarSK:rt}
   }
 \subfigure[Target prediction error]{
  \includegraphics[width=0.48\textwidth]{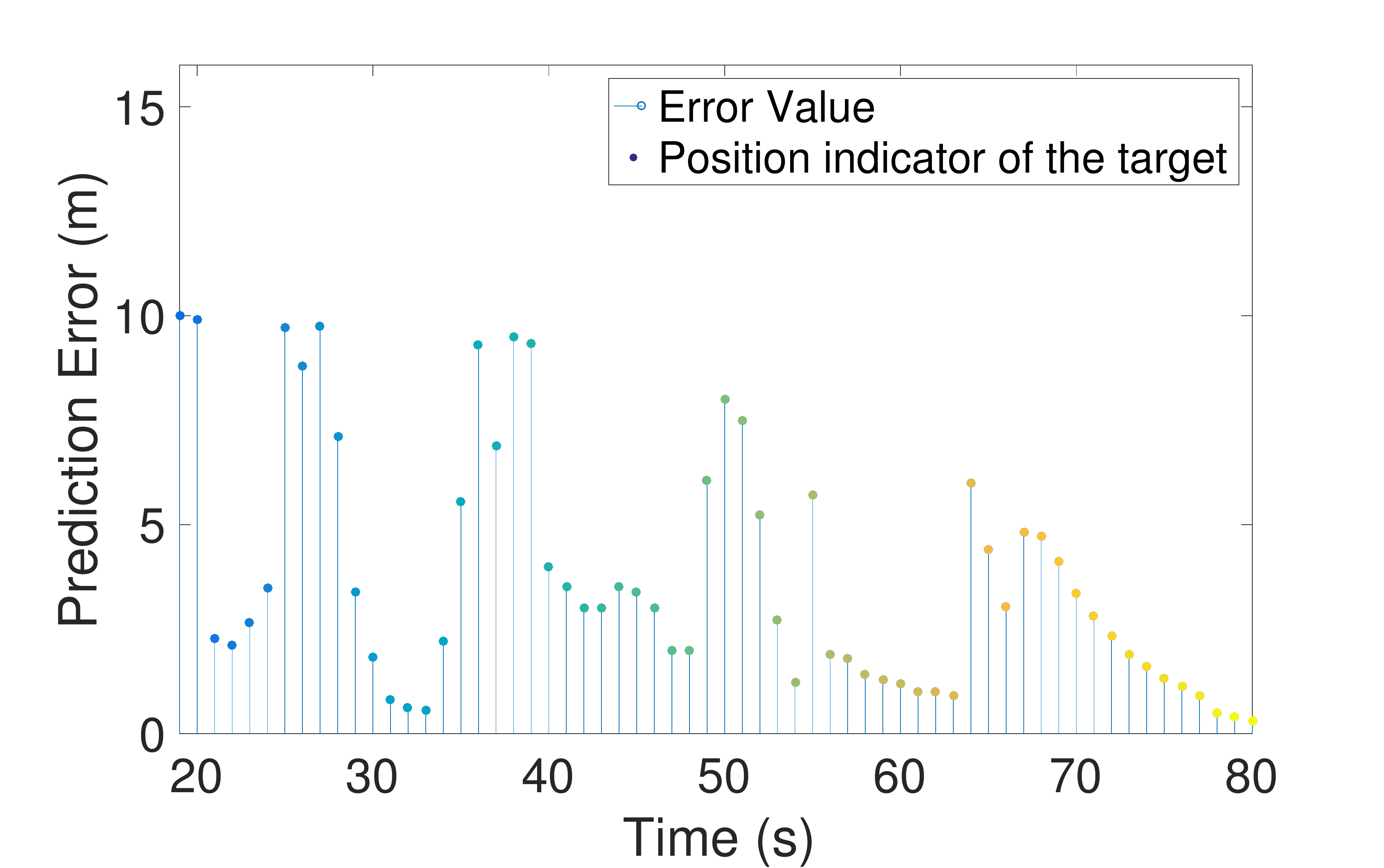}
   \label{fig::DeTarSK:rpre}
   }  
  \caption{Result of DeTarSK for case 2} \label{fig::DeTarSK:sr}
\end{figure}

\begin{figure*}
 \centering
 \subfigure[Physical map]{
  \includegraphics[width=0.48\textwidth]{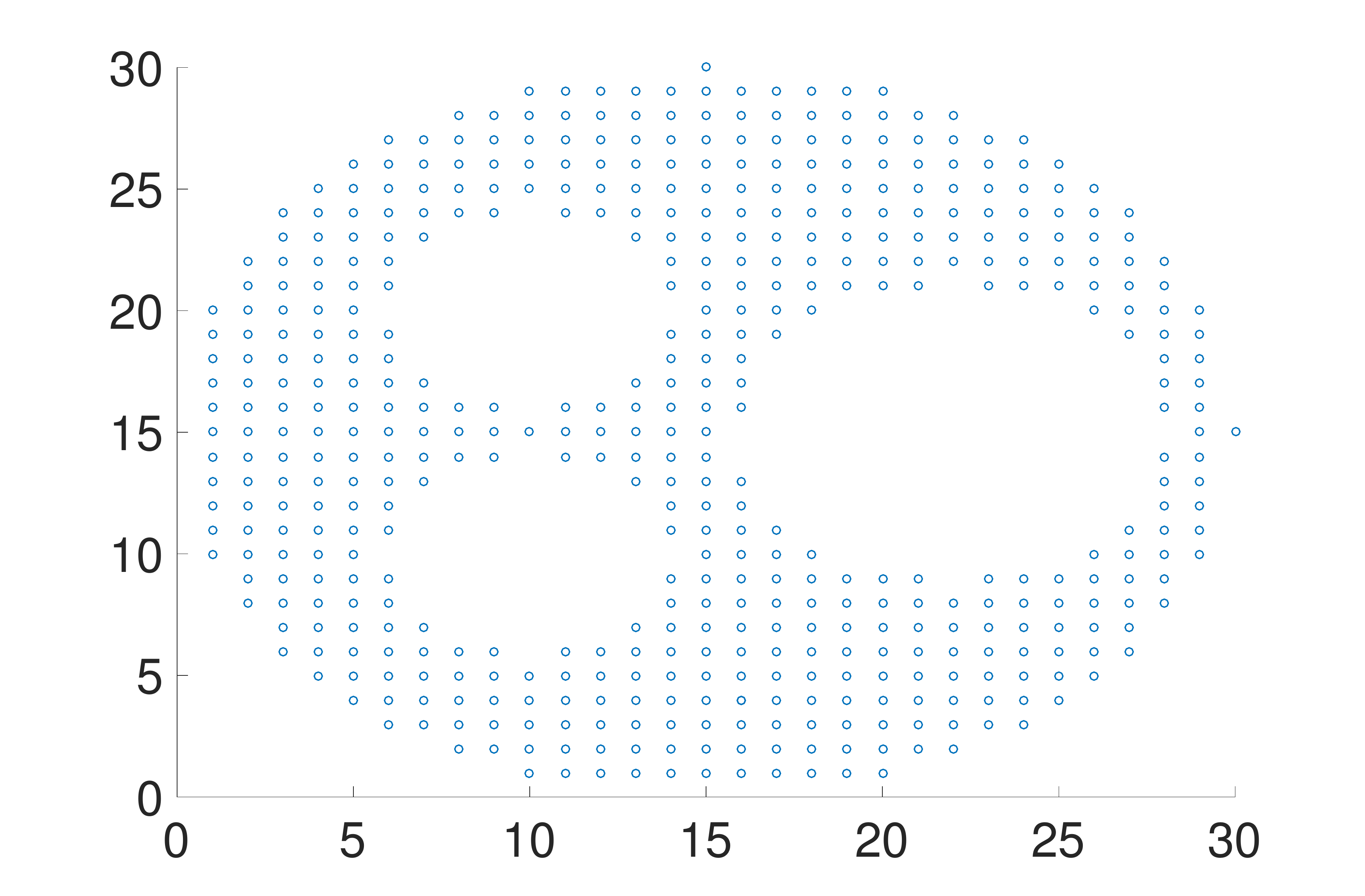}
   \label{fig::DeTarSK:cpm}
   }   
 \subfigure[Calculated ML-TM map]{
  \includegraphics[width=0.48\textwidth]{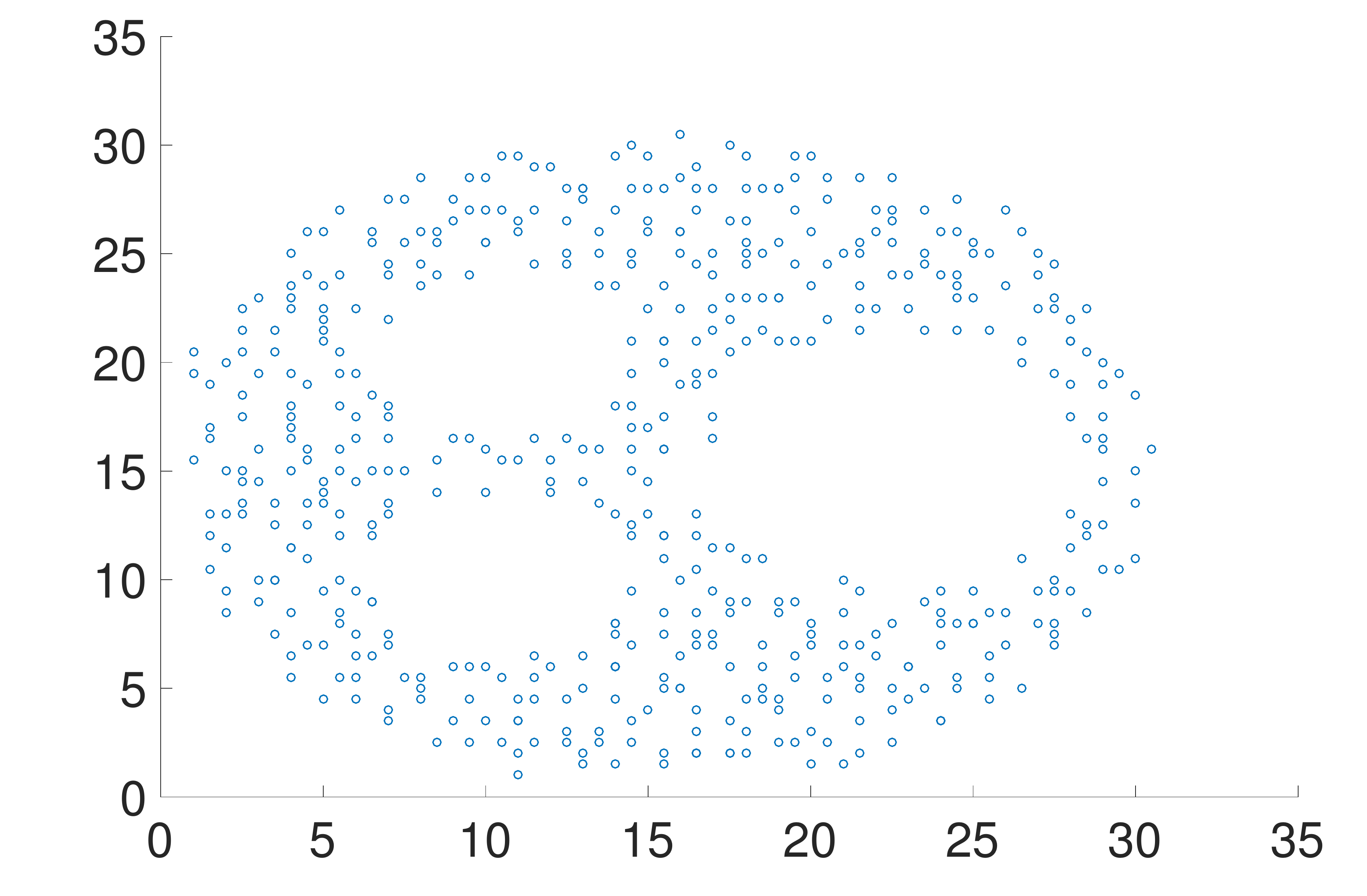}
   \label{fig::DeTarSK:ctm}
   }
 \subfigure[Search using DeTarSK]{
  \includegraphics[width=0.48\textwidth]{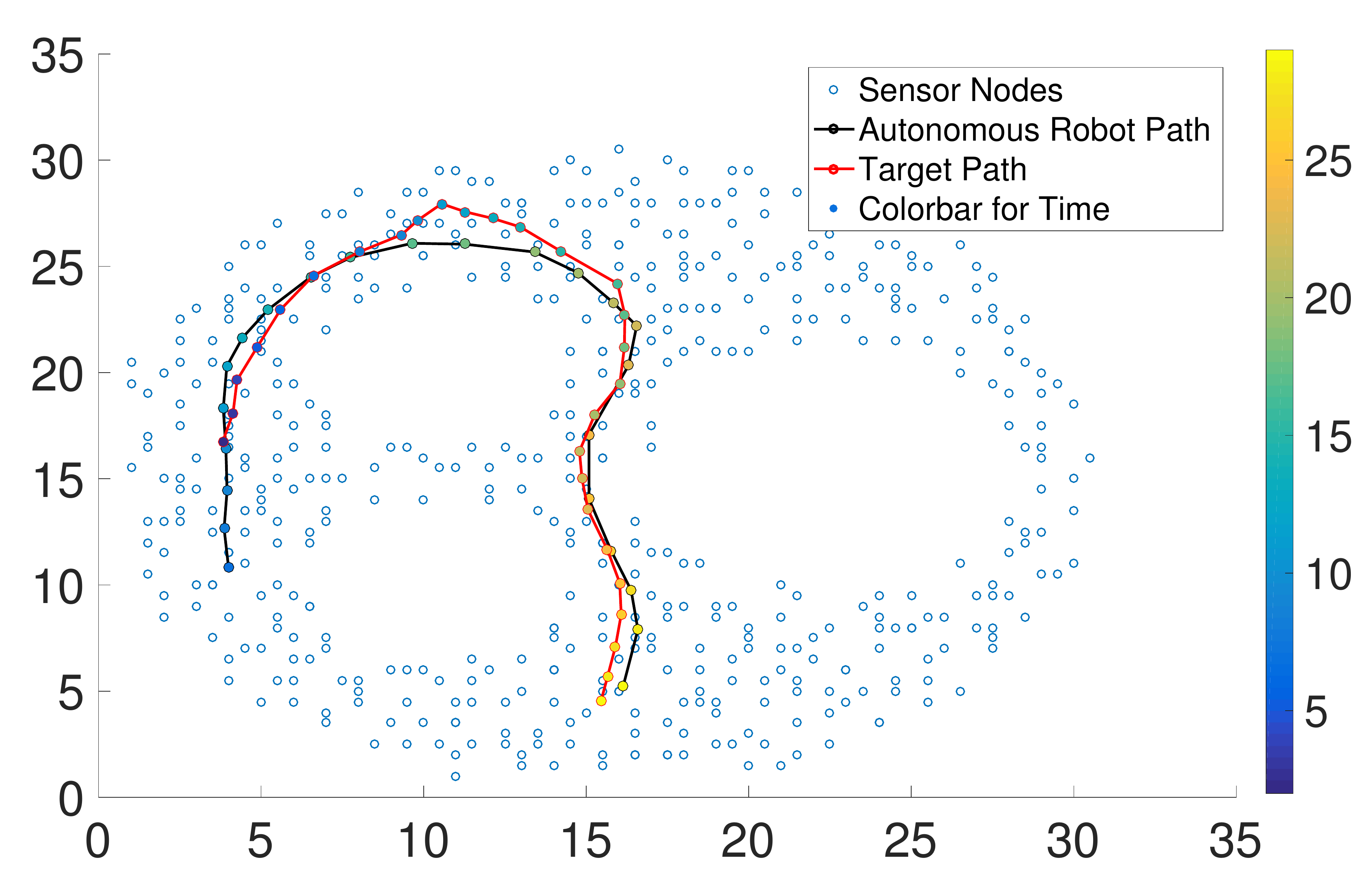}
   \label{fig::DeTarSK:ct}
   }   
 \subfigure[Target prediction error]{
  \includegraphics[width=0.48\textwidth]{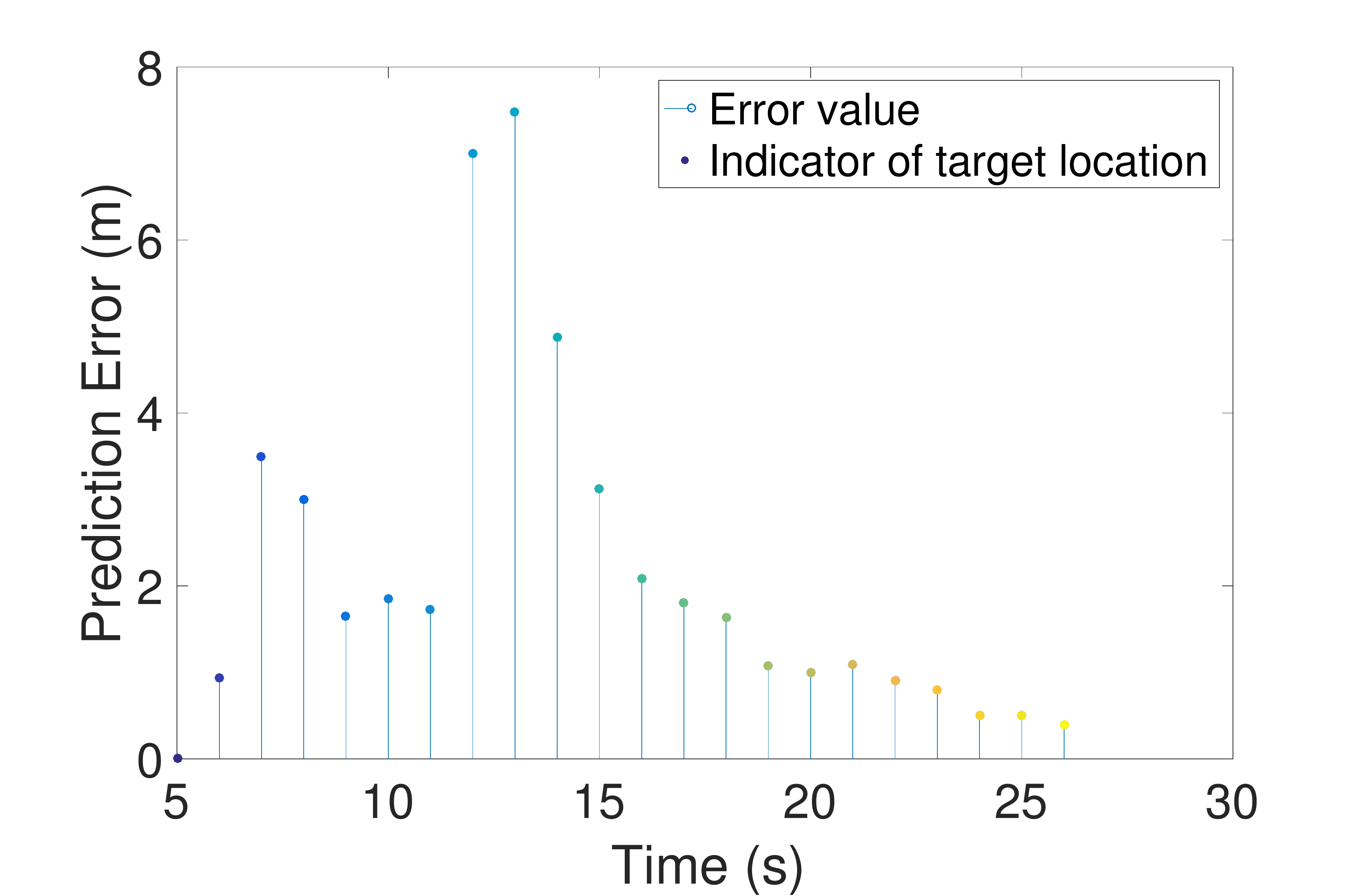}
   \label{fig::DeTarSK:cpre}
   }
  \caption{Result of DeTarSK for case 3} \label{fig::DeTarSK:cr}
\end{figure*}

For each case, target prediction error that describes the difference between predicted value and actual location of the target was calculated. Let's consider $\hat{L}_{t_i}=[\hat{X}_{t_i},\hat{Y}_{t_i}]$ as the predicted target location at time $t_i$ using the proposed algorithm. Then the target prediction error can be calculated as,
\begin{eqnarray}
&&E_{pr} =\sqrt{(\hat{X}_{t_i}-X_{t_i})^2+(\hat{Y}_{t_i}-Y_{t_i})^2}
\end{eqnarray}
Figure \ref{fig::DeTarSK:se}-\ref{fig::DeTarSK:cr} show the results for the  three cases. The prediction error plots describe how the proposed method adjusts for the sudden movement changes of the target. For an example, in Figure \ref{fig::DeTarSK:se}, it can be seen that the prediction error reduces with the time. However, when the target makes a sudden direction change (at 43s), the prediction error increases, but it fine tunes using the proposed prediction method and brings down the error. It can be seen in Figure \ref{fig::DeTarSK:sr} and Figure \ref{fig::DeTarSK:cr} also. In Figure \ref{fig::DeTarSK:sr}, the target follows a random path, which has many directional variations. However, the proposed method captures those variations in a few seconds and predicts the target location with an error less than 5m.   
\subsection{Performance Comparison}
The comparison of DeTarSK with other existing algorithm is presented in this section. For that, a recently proposed P-G algorithm \cite{tt8} was chosen. It is based on pseudo-gradient approach that requires a communication hop count and RSSI in the node neighbourhood, which haven't considered in the proposed algorithm. Additionally, a central node does not control this algorithm, but it is needed to calculate pseudo-gradient in each target movement. For the comparison, the same three cases described above are used.
\begin{figure*}
 \centering
 \subfigure[Target search using DeTarSK]{
  \includegraphics[width=0.7\textwidth]{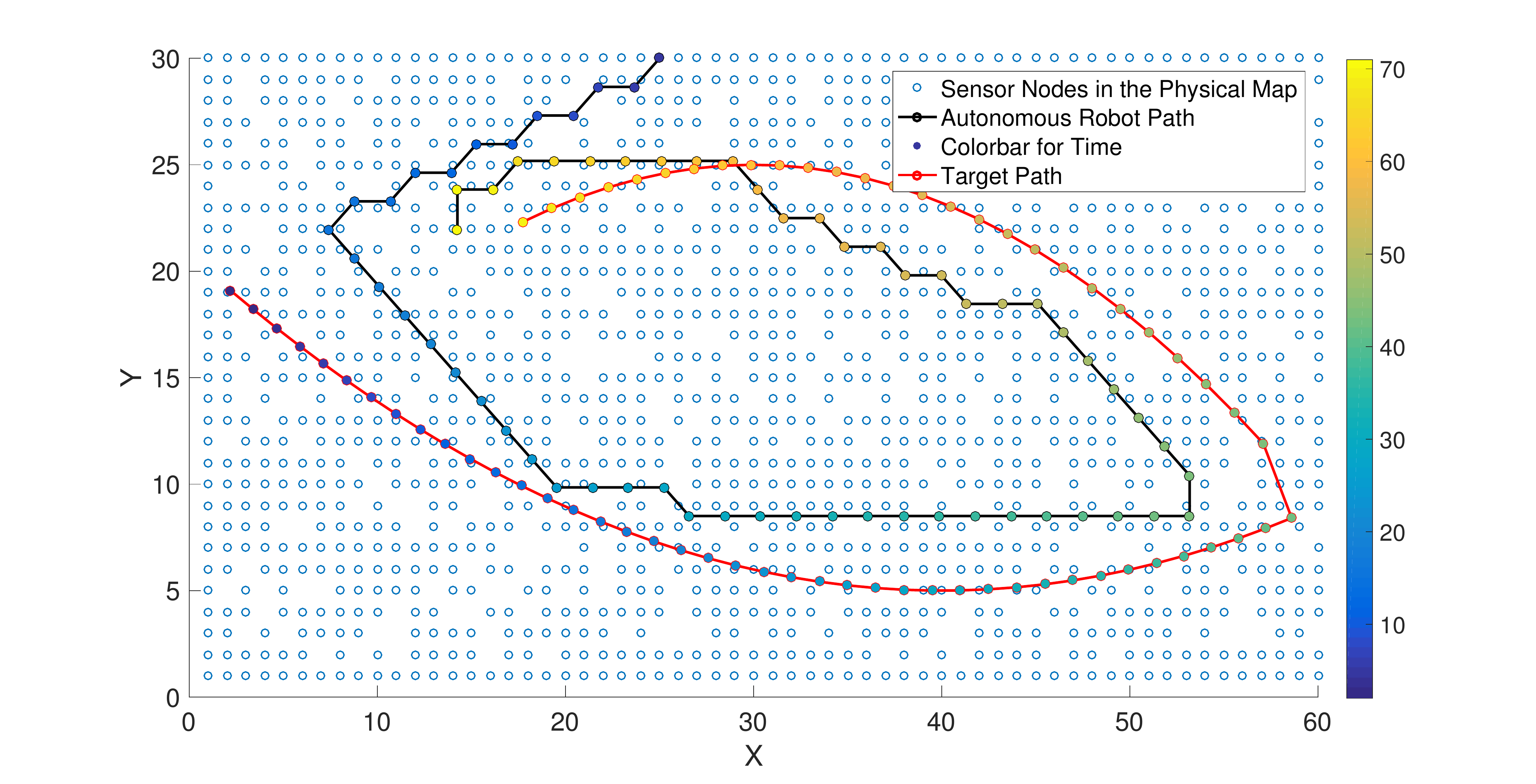}
   \label{fig::DeTarSK:ep}
   }
    
 \subfigure[Target search using P-G algorithm]{
  \includegraphics[width=0.7\textwidth]{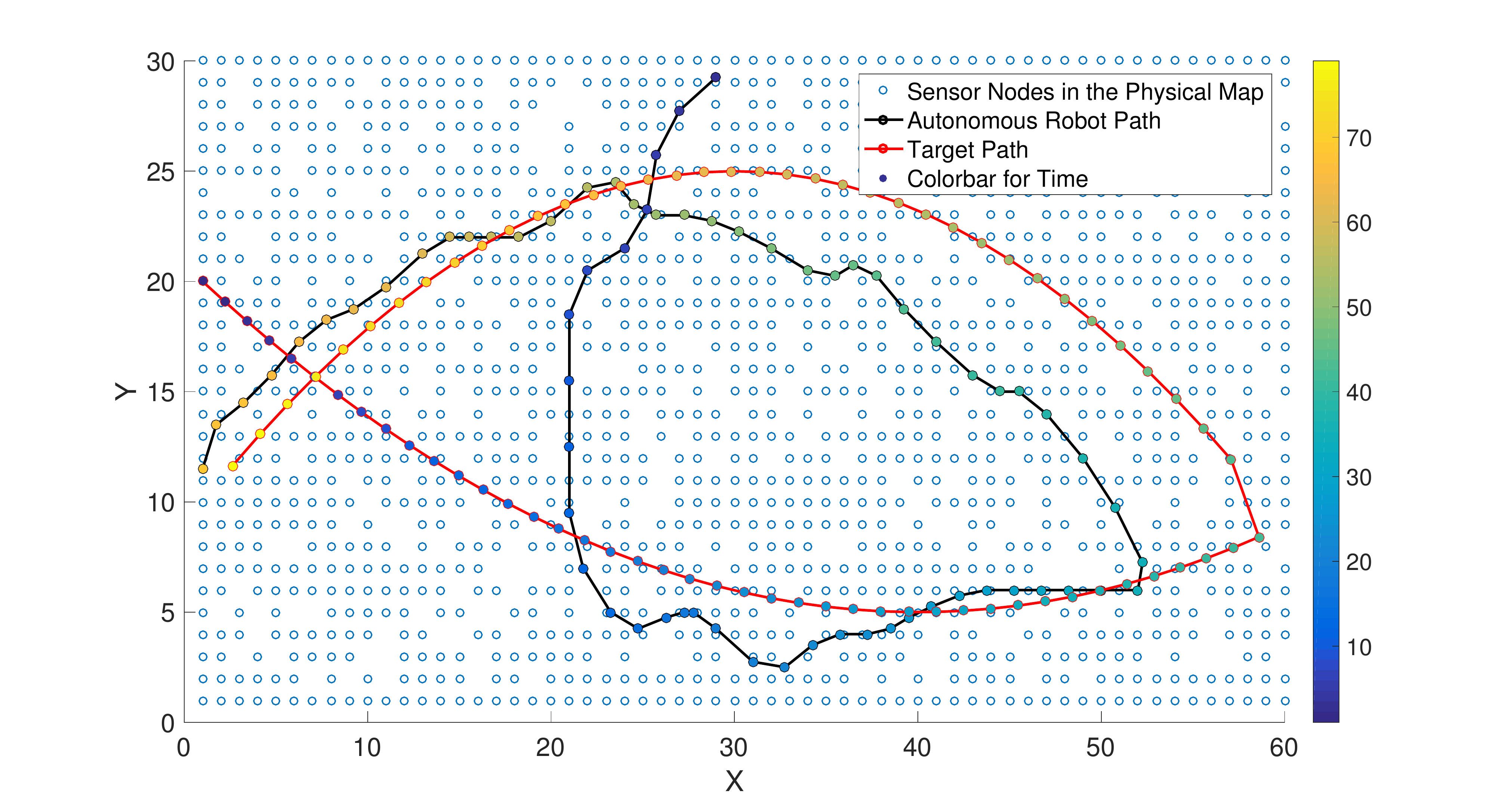}
   \label{fig::DeTarSK:ec}
   }
   \quad 
   \subfigure[Distance between target and robot ]{
  \includegraphics[width=0.7\textwidth]{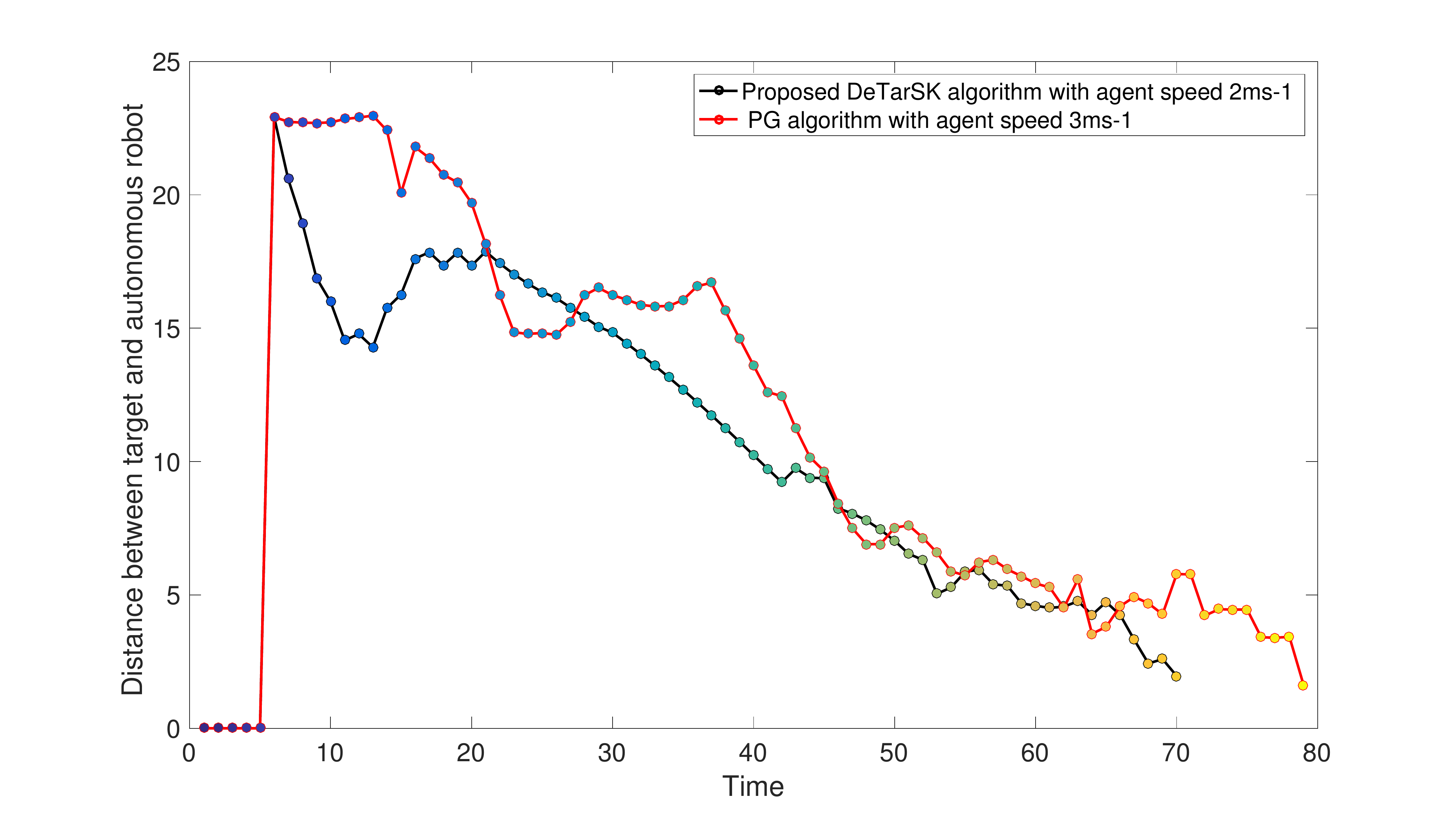}
   \label{fig::DeTarSK:edis}
   }
   \caption{ Performance comparison for Case 1} \label{fig::DeTarSK:se1}
\end{figure*}
\begin{figure*}
 \centering
 \subfigure[Target search using DeTarSK]{
  \includegraphics[width=0.7\textwidth]{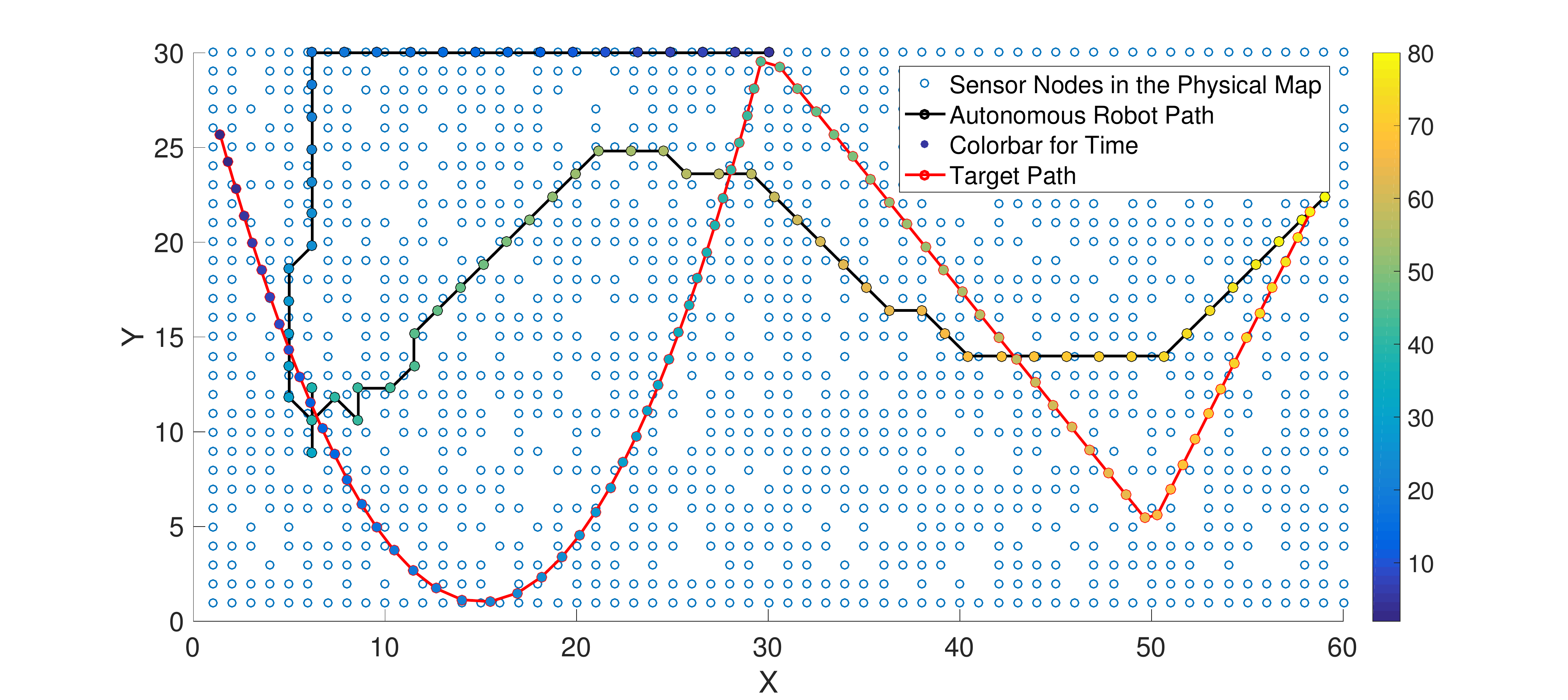}
   \label{fig::DeTarSK:rp}
   }
    \quad 
 \subfigure[Target search using P-G algorithm]{
  \includegraphics[width=0.7\textwidth]{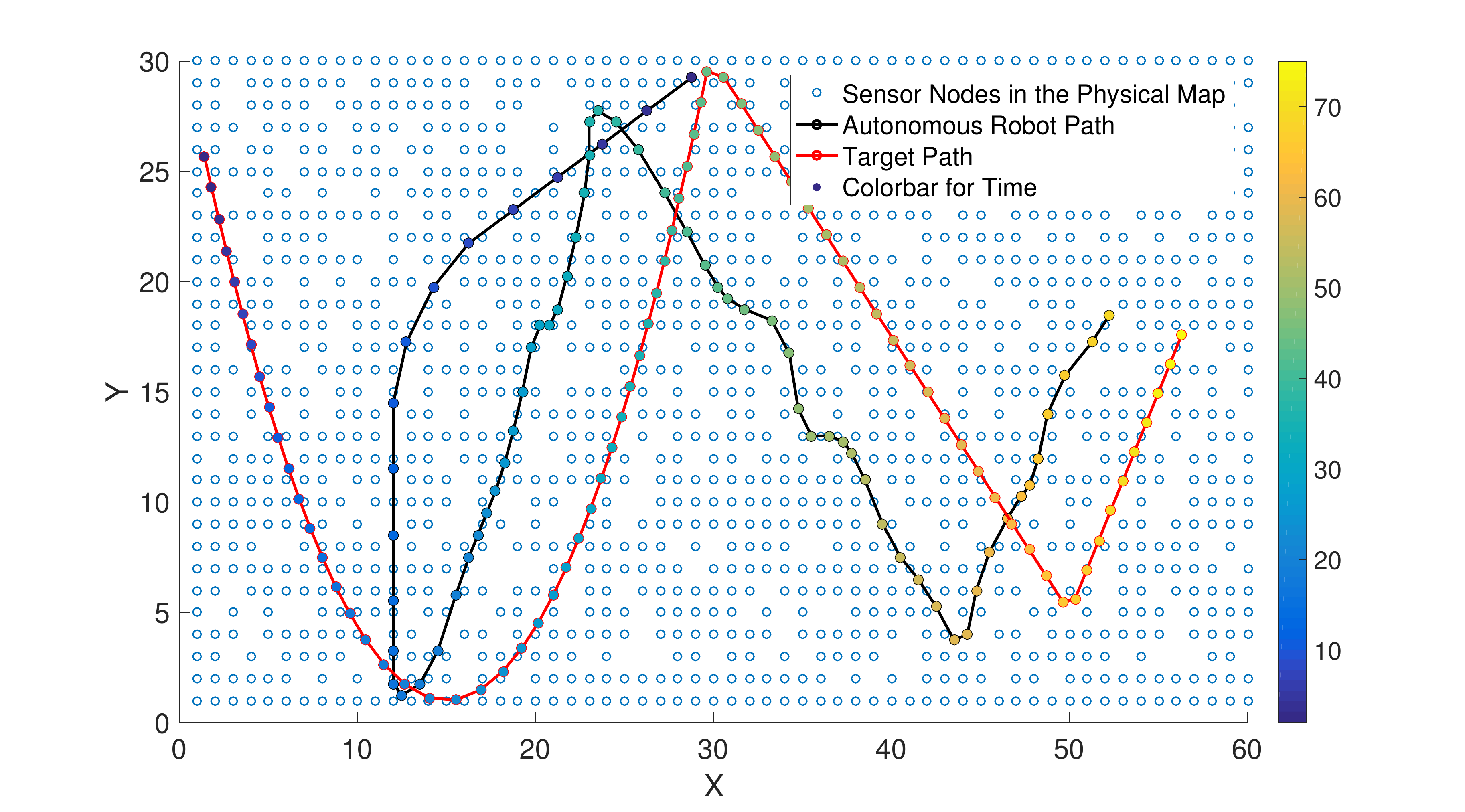}
   \label{fig::DeTarSK:rc}
   }
   \subfigure[Distance between target and robot ]{
  \includegraphics[width=0.7\textwidth]{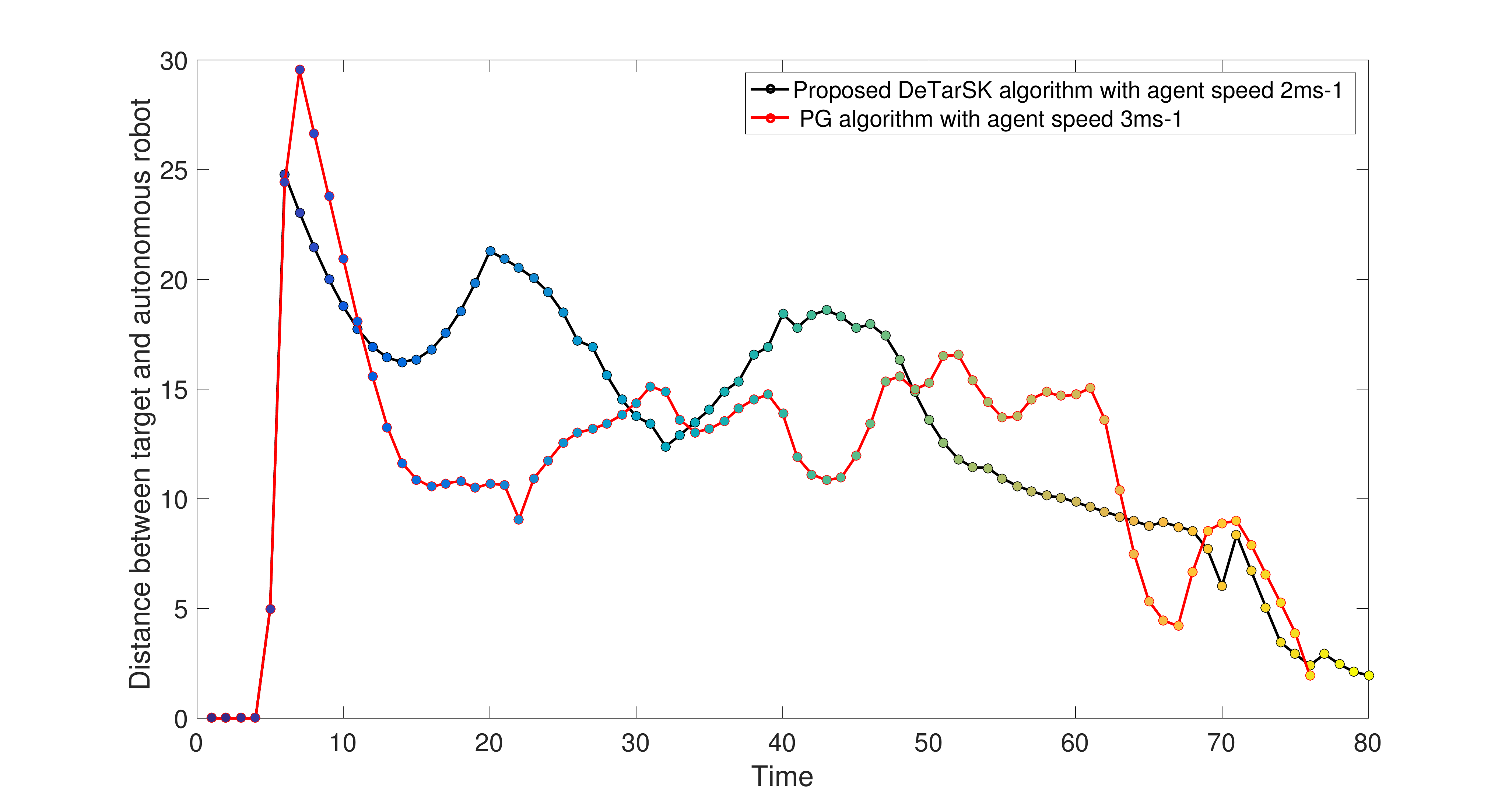}
   \label{fig::DeTarSK:rdis}
   }
   \caption{Performance comparison for Case 2} \label{fig::DeTarSK:sr1}
\end{figure*}
\begin{figure*}
 \centering
 \subfigure[Target search using DeTarSK]{
  \includegraphics[width=0.7\textwidth]{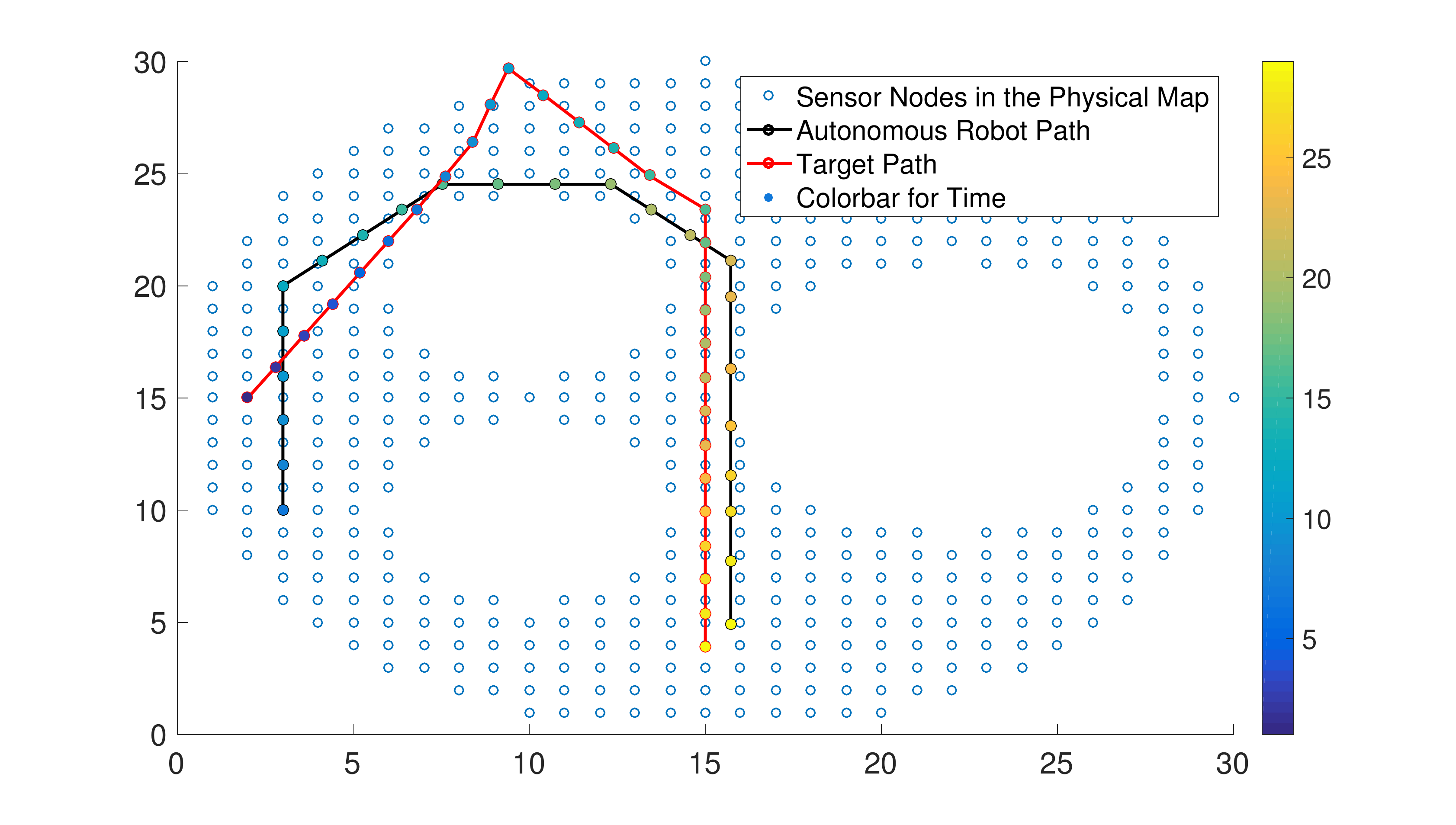}
   \label{fig::DeTarSK:cp}
   }
    \quad 
 \subfigure[Target search using P-G algorithm]{
  \includegraphics[width=0.7\textwidth]{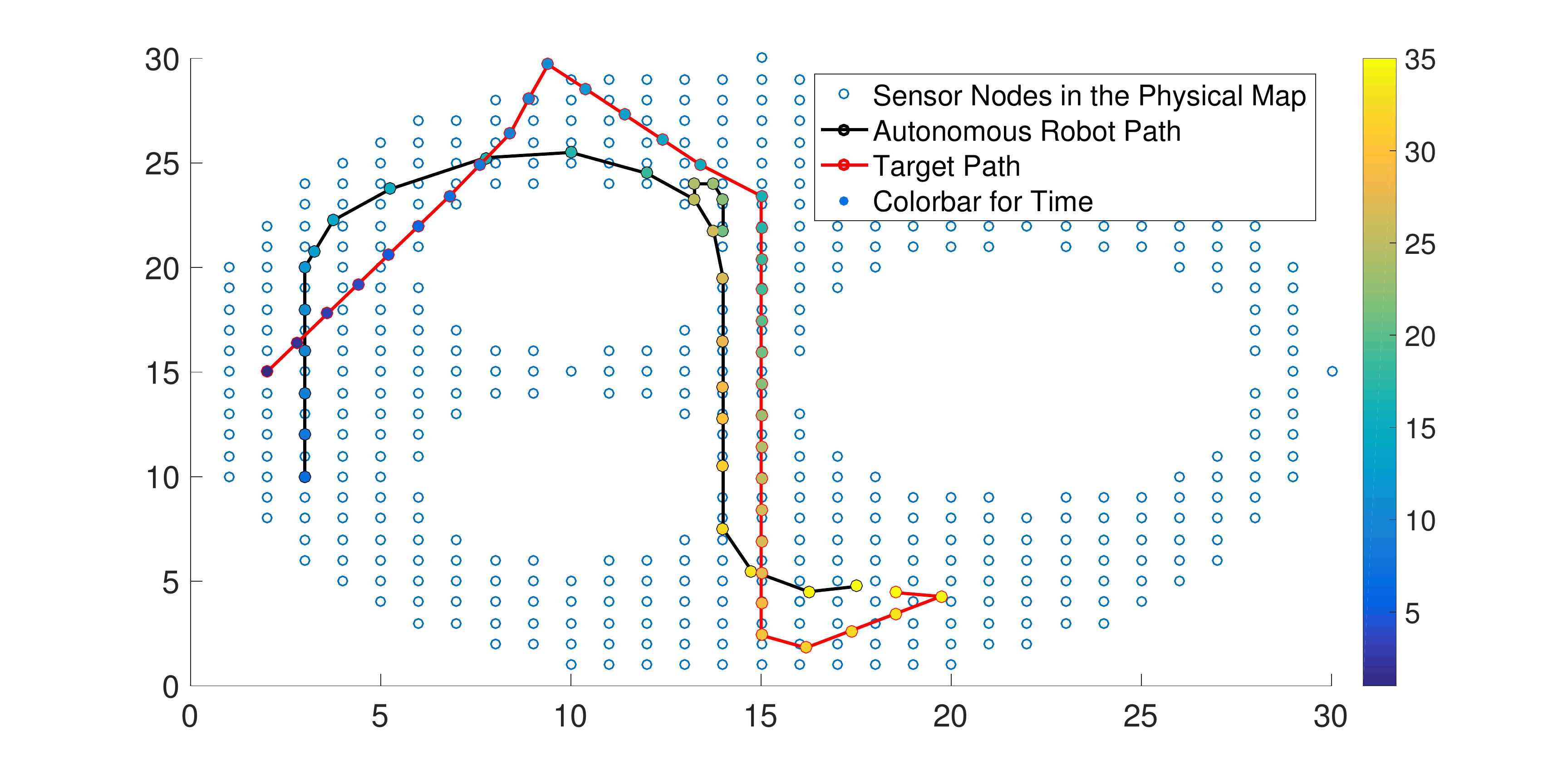}
   \label{fig::DeTarSK:cc}
   }
   \subfigure[Distance between target and robot ]{
  \includegraphics[width=0.7\textwidth]{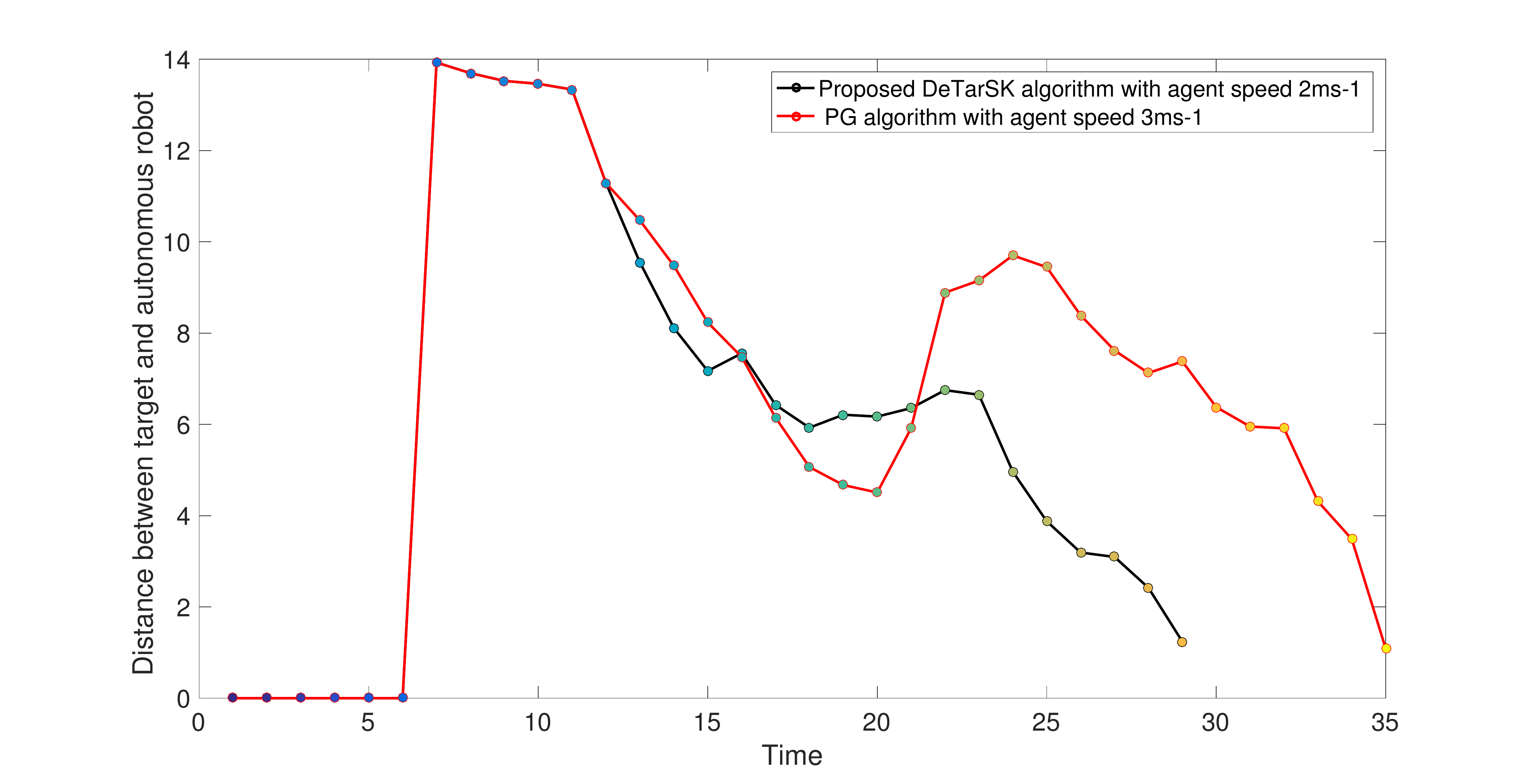}
   \label{fig::DeTarSK:cdis}
   }
   \caption{Performance comparison for Case 3} \label{fig::DeTarSK:cr1}
\end{figure*}

Since, DeTarSK algorithm is proposed for topology maps and P-G algorithms is proposed for physical map, it is needed to convert the results for one domain to perform a fair evaluation. Therefore, DeTarSK robot trajectory in ML-TM is mapped to physical map using Procrustes Analysis \cite{procrusters}. Let consider $X$ and $Y$ are the topological and physical coordinates matrices. Then the Procrustes transformation factors can be calculated as in equation (\ref{eqn::DeTarSK:Procr}).
\begin{equation}\label{eqn::DeTarSK:Procr}
Y = bXT + c;
\end{equation}
where $b$ is the scaling factor, $T$ is the rotation angle and $c$ is the shift value. In this Chapter, transformation factors are calculated locally. In other words, robot's each step coordinates are transformed to the physical map using its local neighbourhood sensor information.

Figure \ref{fig::DeTarSK:se1}-\ref{fig::DeTarSK:cr1} show the results of three cases. In those three figures, sub-figures (a) show the proposed algorithm’s autonomous robot search path in the physical map and sub-figures (b) show the robot search path using the P-G algorithm in the physical map. The distance between target and the autonomous robot in each time value is shown in sub-figures (c). The autonomous robot’s speed, and capturing time of the target in each case is stated in Table \ref{tab::DeTarSK:ttime}. In all three cases the target is moving with a 1.5$ms^{-1}$ speed. Also, the P-G algorithm was unable to capture the target with 2$ms{-1}$ robot speed as the target leaves the network as described in the three cases. The reason that P-G algorithm was unable to capture the target with a lower robot speed is, P-G algorithm does not have a prediction method to calculate the future behaviour of the target, thus, it takes time to adjust the moving direction of robot according to the target moving pattern. Therefore, the robot speed was increased up to 3$ms{-1}$ and find out the time required to capture the target. In case I, described by figure \ref{fig::DeTarSK:se1}, proposed DeTarSK algorithm detects the target in less time and lower speed than the P-G algorithm. In Case II also the proposed algorithm captures the target much faster than the P-G algorithm with a lower speed. The reason, as can be seen in Figure \ref{fig::DeTarSK:sr1} is  that DeTarSK autonomous robot path does not follow all the steps in the target trajectory, but it predicts the target’s future positions from past knowledge and makes a decision on its next movement. Hence, it is faster than the P-G algorithm. Case III result in Figure \ref{fig::DeTarSK:cr1} shows that proposed algorithm can follow a target and capture it in a network filled with obstacles. Even in this case DeTarSK was able to catch the target with lower speed and in less time compared to P-G algorithm. Figures (i.e Figure 6(c)-8(c)) show that when target changes its direction suddenly the distance increases. However, compared to P-G algorithm, proposed DeTarSK algorithm detected it and made required changes to reduce the distance within a few seconds. These results demonstrate that the DeTarSK algorithm can catch the target within less time and lower robot speed.
\begin{table}
\renewcommand{\arraystretch}{1.3}
\caption{\textsc{Time Required to Capture the Target}}\label{tab::DeTarSK:ttime} 
\center
\small
  \begin{tabular}{ |l|c|c|c|c| }
    \hline
    \multirow{2}{*}{Case} &
      \multicolumn{2}{c}{Robot Speed = 2$ms^{-1}$} &
      \multicolumn{2}{c}{Robot Speed = 3$ms^{-1}$} \\ 
      &DeTarSK & P-G    &DeTarSK & P-G    \\
    \hline
   1 & 70s&-\tnote{*}&47s&79s \\ \hline
2 & 80s&-\tnote{*}&52s&76s \\ \hline
3 & 29s&-\tnote{*}&18s&35s \\ 
    \hline
  \end{tabular}
\end{table}

Moreover, the DeTarSK consumes less energy in the WSN compared to the P-G algorithm. The reason is pseudo-gradient need to be calculated in each step of the robot movement. This required obtaining the hop count from target to each and every node in the network, which consumes more energy for packet transmission and reception. Also, it increases the traffic in the network. However, in the proposed method, the robot communicates with its local neighbourhood that requires only shorter length or one hop communication links. Thus the proposed scheme gives real-time result with less energy consumption.

\section{Conclusion}\label{sec::DeTarSK:sconclusion}
This chapter presented DeTarSK, a decentralized target search and prediction algorithm for sensor network based environments where it is not feasible to measure physical distances accurately using techniques such as RSSI. The algorithm uses the maximum likelihood topology coordinates instead of physical coordinates. Knowledge of the target location or direction is estimated from time stamps of observations of the target by nodes and conveyed to the autonomous robot that is searching for the target. DeTarSK is based on decentralized RKF and a non-linear least square method. It also removes the errors due to distortion of topology coordinate domain compared to physical domain. 

Even though, this method was considered in topology maps, the proposed algorithm can be also be used in geographical (physical) coordinate system. Other advantages include the real time decision making with less traffic and energy consumption in the network. To solve the drawbacks in decentralized algorithms, the future target locations are predicted using past behaviours of the target. 

DeTarSK shows better performance compared to the recently proposed P-G algorithm. Additionally, the proposed algorithm can be used in unknown environments containing obstacles in the search path. 

\chapter{Application 2: Sensor Network Based Navigation of a Mobile Robot for Extremum Seeking Using a Topology Map}\label{chapter:esa}
This chapter presents another WSN application that uses topology maps. Extremum or source seeking is an important task in emergency environments that has attracted the attention of researchers recently. To this end, a navigational algorithm for source seeking in a sensor network environment is presented in this chapter. The solution consists of a gradient-free approach and a ML-TMs of sensor networks. A robot is navigated using an angular velocity limited by maximum and minimum constants, and by measurements gathered by sensors close to the robot's current location. The location of the robot is calculated using sensor topology coordinates. However, actual physical distances are hidden in topology maps because of non-linear distortions of physical distances between nodes. As a result, the control law proposed does not depend on distance-based information. The performance of the algorithm is evaluated using a computer simulation and a real experimental setup. 

This chapter is structured as follows and the main results of the chapter were originally published in \cite{CCCesk, Jesk}. Section \ref{sec::esa:Introduction} offers an introduction and motivation for the research presented in this chapter. Section \ref{sec::esa:RW} reviews background research conducted in extremum seeking. Section \ref{sec::esa:algo} presents the proposed algorithm and is followed by the results in Section \ref{sec::esa:results}. Finally, Section \ref{sec::esa:conclusion} provides a conclusion to the chapter.

\section{Introduction}\label{sec::esa:Introduction}
In this chapter, the problem of navigating a mobile robot to the extrema of an environmental field based on information obtained from a WSN and a topology map is considered. The environmental field is an unknown scalar function, which represents a spatially distributed signal that decays away from the source at the maxima. This source may represent a fire \cite{fire}, a poisonous gas leak \cite{gas}, a chemical spill \cite{chemical} or a radioactively contaminated areas \cite{ref3}. As extremum seeking is an important task in emergency source seeking applications, target tracking applications based on signal decaying and environmental studies \cite{Dynamicenv, trackingcontrol}, it has attracted the attention of researchers recently \cite{ref1, ref2}. However, extremum seeking in emergency environments presents certain challenges; making real time decisions from available information and making correct decisions based on data corrupted by noise and fading \cite{ref4, noise}. 

Existing extremum seeking algorithms \cite{ref4, ref3, ref1, ref2} assume that the mobile robot is equipped with sensors to measure the environmental field while moving. The response time of the sensors varies from milliseconds to seconds depending on environmental characteristics and the sensor's components \cite{tempS1, tempS2}. Thus, measuring the field value by using a sensor attached to the robot may not be feasible when the robot is moving fast and the sensor's response time is high. On the other hand, if the robot stops at discrete time intervals to get a stable sensor reading, it may take longer to reach the desired location. To address these issues, this research proposes a robot navigation algorithm based on information gathered by a WSN. Sensor nodes are randomly distributed over the environment to measure field values and transmit these measurements to a robot. Since the sensors are static, the accuracy of the readings is higher and the time required to obtain a stable reading is less than that required in existing systems.

In the proposed algorithm, WSN performs two tasks: measuring field values and providing information to calculate the robot's location in the network map. To do this, WSN requires an accurate localization algorithm. Numerous localization algorithms based on RSSI, ToA, and AoA have been proposed in the literature \cite{fingerprintloc, IntelligentROM}. However, in varous complex and harsh environments, these range-based measurements are affected by noise, fading, interference, and multipath \cite{multipathref}. As a result, the accuracy of these algorithms decreases \cite{sensorbased3}. Under such circumstances, WSN topology maps \cite{VC, mltm, topologycontrol} play a vital role in these environments. The topology map used in this chapter is ML-TM, presented in Chapter \ref{chapter:mltm}, which is obtained using a mobile robot and a packet reception probability function sensitive to distance. A binary matrix is recorded by the robot. This matrix represents the packet reception of sensor nodes at the robot's various locations. The topology coordinates of sensors are extracted from the binary matrix using the packet reception probability function. Consequently, this method neither relies solely on connectivity \cite{VC, rangefree4} nor depends on range based parameters \cite{rssiprbabilistic}.

Existing research performed in extremum seeking can be grouped into two categories based on the approach used. The first involves a gradient-based approach that employs online estimation of the gradient at the robot's current location. The algorithms in this approach are complex and highly sensitive to measurement noises. The second category incorporates a gradient-free approach that directly uses the field value of the robot's current location. This chapter focuses on a gradient-free algorithm. The advantaged of this approach have been mentioned previously. In addition, it helps avoid distance distortion in WSN topology maps. The control law depends on sensor readings and an angular velocity limited by maximum and minimum constants. Also, the robot is modelled as a unicycle that travels at a constant speed. An unknown scalar field represents the strength of a spatially distributed signal in the environment, where the source of the distributed signal is at the extrema. Here, the distribution is arbitrary.

\section{Background on Extremum Seeking}\label{sec::esa:RW}
Prior work in the area of environmental boundary tracking can be grouped in to two categories, namely gradient-dependent and gradient free approach \cite{ref2}. Gradient-dependent algorithms measure a gradient of a field in the environment and navigate the robot along the gradients using a control law. For an example, a bio-inspired control model is proposed in \cite{gd1}, an underwater vehicle model is developed in \cite{gd2}, and recently, a model is proposed in \cite{gd3} to track dynamic plumes. In \cite{gd4}, several gradient dependent algorithms are demonstrated based on gradient-based contour estimation methods, extensions of the ‘snake’ algorithms in image segmentation, artificial potential approach, cooperative distribution of the sensors over the estimated contour, tracking a level curve by the centre of a rigid formation of multiple sensors based on collaborative estimation of the field gradient and Hessian. Rosero et al. proposed a gradient based source seeking algorithm in \cite{GBcomp}, which is based on formation control law to estimate the gradient direction. In \cite{GBcomp2}, they have extended the approach to linear time invariant models. A gradient based perturbation extremum seeking control scheme is presented in \cite{GBperturbation}, to decrease the fluctuation of convergence. Kebir et al. proposed an extremum seeking control algorithm \cite{GBneural} based on neural network model, which gives a real-time estimate of the optimal operating point based on the measurement of external disturbance. However, in practical, derivative information is unavailable for direct measurement and requires access to the field values at several nearby locations \cite{ref2}. Thus, these estimations contain some error factors due to measurement noises. 

The second type, gradient-free algorithm, uses directly the field value at the current location. In \cite{gf1,gf2}, the steering angle of the robot switches between alternatives based on comparison of current field value with the threshold of interest. Similarly, Barat et al. \cite{gf3}, proposed an approach with larger set of alternatives to an underwater vehicle. A control method based on segmentation of the infrared local images of the forest fire was proposed in \cite{gf4}. A linear PD controller was designed in \cite{gf5}, which ensures convergence of a unicycle-like vehicle to a level curve of a radial harmonic field. A sliding mode control method for tracking environmental level sets is offered in \cite{ref4} and an extremum seeking gradient-free algorithm based on numerical optimization method is proposed in \cite{GFoptimization}. Moreover, Zhang et al. proposed a feedback linearizable system in \cite{GFfeedback}, which is a trust region based extremum seeking control that do not require gradient information.  

\section{System Description and Problem Setup}\label{sec::esa:algo}
This section describes a navigation strategy that originates from \cite{navigationmodel} (see \cite{R1,R2} for details) and uses the topological coordinates instead of the actual physical coordinates where the real distance values are hidden. The proposed robot navigation based on sensor network information can be viewed as an example of networked control systems \cite{newA15, newA16, newA17, newA18, newA19, newA20, newA21, newA22, newA32, newA33}. Additionally, this navigation law does not employ a gradient estimate and it steers the robot to a point where the distribution assumes a pre-specified value. After approaching to the required position robot can either stops or switches to another guidance law.

Here, a planar mobile robot modelled as a unicycle \cite{ref2} which controlled by the time-varying angular velocity $\omega$ limited by a given constant $\omega_{max}$. The robot travels with a constant speed $V_r$ in the area supporting an unknown field distribution $D(r)$. Here $r$ is the vector of Cartesian coordinates $(x,y)$ in the plane ${\Bbb R^2}$. The position of the robot can be represented by a triplet $P_r = (X_r,Y_r,\theta_t)$, where $(X_r,Y_r)$ is the location of the robot in topology map and $\theta_r$ is the heading angle measured counter clockwise from x-axis in topology coordinate system.

The objective of this algorithm is to navigate the robot to the level curve $D(x,y)\:=\:d_0$. Sensor nodes in the network measure the field value at its location and store it in their memory. While robot is moving, it communicates with it's neighbour sensor nodes, i.e. nodes located in within robot's communication range, and calculates the distribution value $d(t)$ at it's current location. Let consider sensor node $s_i$ measures the distribution value at its physical location in the network $(x_i,y_i)$ as $sm_i\::=\:D(x_i,y_i)$. $N_t$ is the set of nodes located in robot neighbourhood at time $t$. Then, robot calculation for it's topological coordinates at time $t$ and field distribution $d(t)$ is shown in equation (\ref{eqn::esa:robottc}) and equation (\ref{eqn::esa:robotdf}) respectively. 

\begin{equation}\label{eqn::esa:robottc}
X_r(t)=\frac{\displaystyle\sum_{s_i\in N_t}w_i X_i}{\displaystyle\sum_{s_i\in N_t}w_i}\:\:\:\:\:and\:\:\;\;\:Y_r(t)=\frac{\displaystyle\sum_{s_i\in N_t}w_i Y_i}{\displaystyle\sum_{s_i\in N_t}w_i }
\end{equation}

\begin{equation}\label{eqn::esa:robotdf}
d(t)=\frac{\displaystyle\sum_{s_i\in N_t}w_i sm_i}{\displaystyle\sum_{s_i\in N_t}w_i}
\end{equation}

where, $w_i=\frac{P_{rx_i}}{P_{tx_i}}$ and $(X_i,Y_i)$ is sensor node $s_i$ topology coordinates. $P_{tx_i}$ is transmitting power of node $s_i$ and $P_{rx_i}$ is receiving power of the message sent by node $s_i$. The reason of introducing a weighting factor $w_i$ in equation (\ref{eqn::esa:robottc}) and equation (\ref{eqn::esa:robotdf}) is to give more weight for the values received from near by nodes and lesser weight for the values received from far away nodes. As the real distance between nodes and the robot is hidden, transmitting and receiving power is used to get an indication of the distance i.e. $d\propto \frac{P_{rx_i}}{P_{tx_i}}$. 

The kinematic model used in this chapter is given in equation (\ref{eqn::esa:kinematic}) and such a model describes planar motion of many ground robots, missiles, UAVs and underwater vehicles \cite{newA1, newA2, newA3, newA4, newA5, newA6, newA7}. 
\begin{eqnarray}\label{eqn::esa:kinematic}
&\dot{X}_r(t)=V_r cos(\theta _r(t))
 \nonumber \\
&\dot{Y}_r(t)=V_r sin(\theta _r(t))
 \nonumber \\
& \dot{\theta}_r(t)=\omega (t)
\end{eqnarray} 
with, $\omega (t) \in [-\omega _{max},\omega_{max}]$ and the initial conditions $[X_r(0)=X_0,\:Y_r(0)=Y_0,\:\theta _r(0)=\theta _0]$.

To design a controller that ensures the convergence $d(t)\rightarrow d_0$ as $t\rightarrow \infty $, the navigation law stated in equation \ref{eqn::esa:controle} is used \cite{ref2}. This navigation law belongs to the class of hybrid or sliding mode systems \cite{newA8, newA9, newA10, newA11, newA12, newA13, SAVKINnew2}.

\begin{equation}\label{eqn::esa:controle}
u(t)=sgn\{\dot{d}(t)+\Bbb X[d(t)-d_0]\}\omega _{max}
\end{equation}

where,
\[
 sgn(\alpha) =
  \begin{cases}
   1 & \text{if } \alpha >0 \\
   0     & \text{if } \alpha = 0\\
   -1 & \text{if } \alpha < 0
  \end{cases}
\]

\[
 \Bbb X(p) = 
  \begin{cases}
   \gamma p & \text{if } \vert p\vert \leq \delta  \\
   sgn(p)\gamma \delta & \text{otherwise} 
  \end{cases}
\]

In this equation gain coefficient ($\gamma $) and saturation threshold ($\delta $) are design parameters. Moreover, it is assumed that the function $D(.)$ is twice differentiable \cite{ref2}.

\section{Performance Evaluation}\label{sec::esa:results}
The performance of the proposed algorithm is evaluated in this section. First, a computer simulation is carried out to evaluate the performance of the algorithm and compared it with an existing algorithm. Then, an experimental setup is used to evaluate the performance of the proposed algorithm. Following sub sections describe the two evaluation methods in detail. 

\subsection{Computer Simulations}
Matlab simulation software is used to simulate the test environments. A $60\times 60$ sparse sensor network deployed in a suburban area with 2800 sensor nodes is considered as the simulation environment. In this section, two scenarios are considered as below.\\
\textbf{Scenario 1:} One emergency source is located at (38, 35) with a field distribution function $D(x,y)=10e^{-\{(x-38)^2+(y-35)^2\}/600}$.\\
\textbf{Scenario 2:} Two emergency sources are located at (10,18) and (40, 38) with a field distribution function $D(x,y)=10e^{-\{(x-40)^2+(y-38)^2\}/300} + 10e^{-\{(x-10)^2+(y-18)^2\}/200}$\\

The sensor nodes measure the field distribution in the deployed environment $D(.)$. These measurements are considered as point measurements. If the sensors are perfect, the measurements should be $sm_i\:=\:D(x_i,y_i)$ for sensor node $s_i$ located at $(x_i, y_i)$ in physical map. However, the sensor measurements encounter some errors due to aging, ambient humidity, gases etc. \cite{smmodel}. Hence, to emulate real sensor readings, following model is considered in this paper.
\begin{equation}\label{eqn::esa:smr}
\widehat{sm}_i=D(x_i,y_i)+e_i
\end{equation}
where, $\widehat{sm}_i$ is the sensor $s_i$ reading and $e_i$ is the sensor offset. In each simulation environment we define a SNR to calculate the $e_i$ value.
\begin{figure*}
 \centering
 \subfigure[Sensor measurements ]{
  \includegraphics[width=0.45\textwidth]{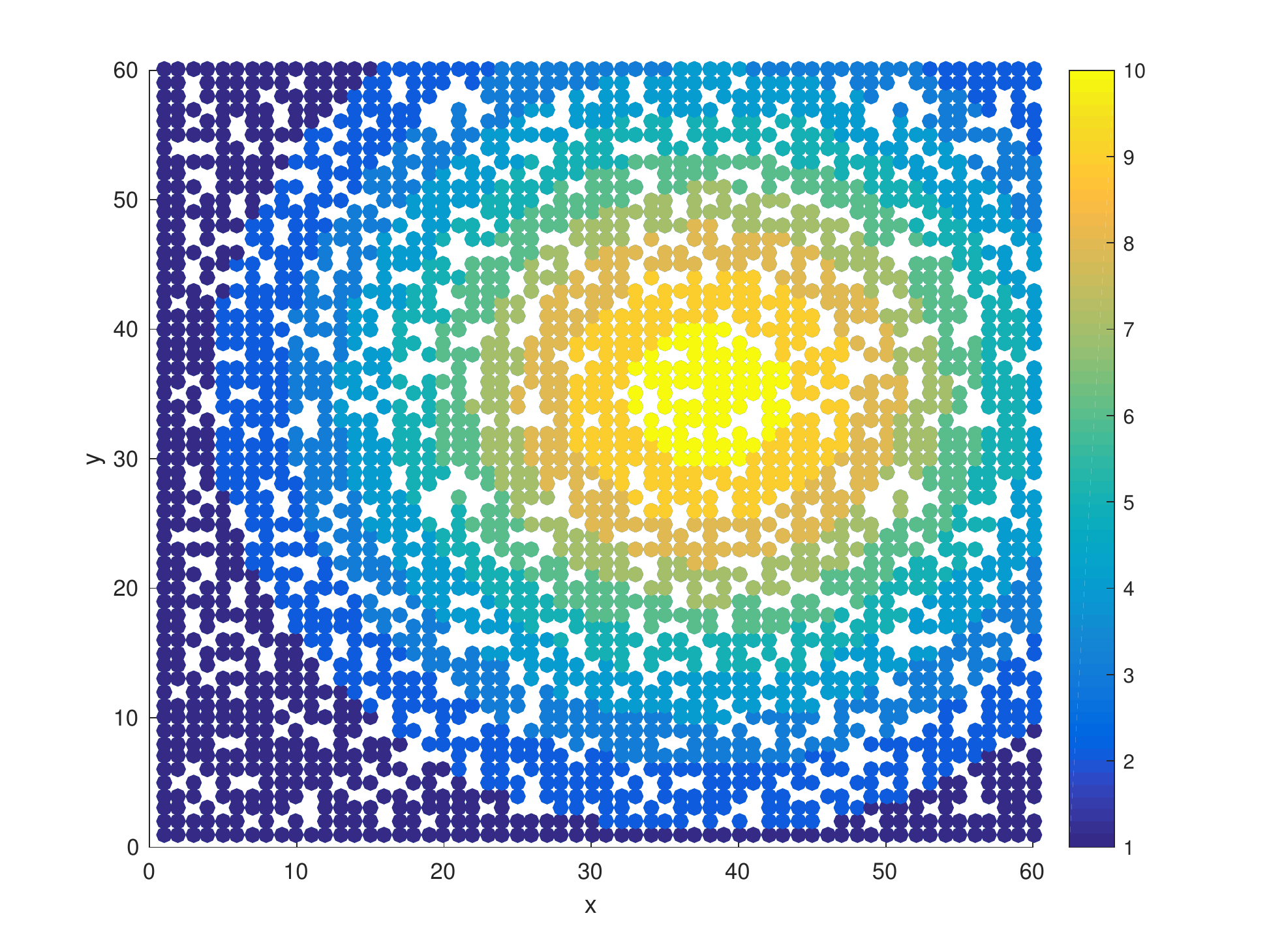}
   \label{fig::esa:sr0}
   }
 \subfigure[Field values over the network]{
  \includegraphics[width=0.45\textwidth]{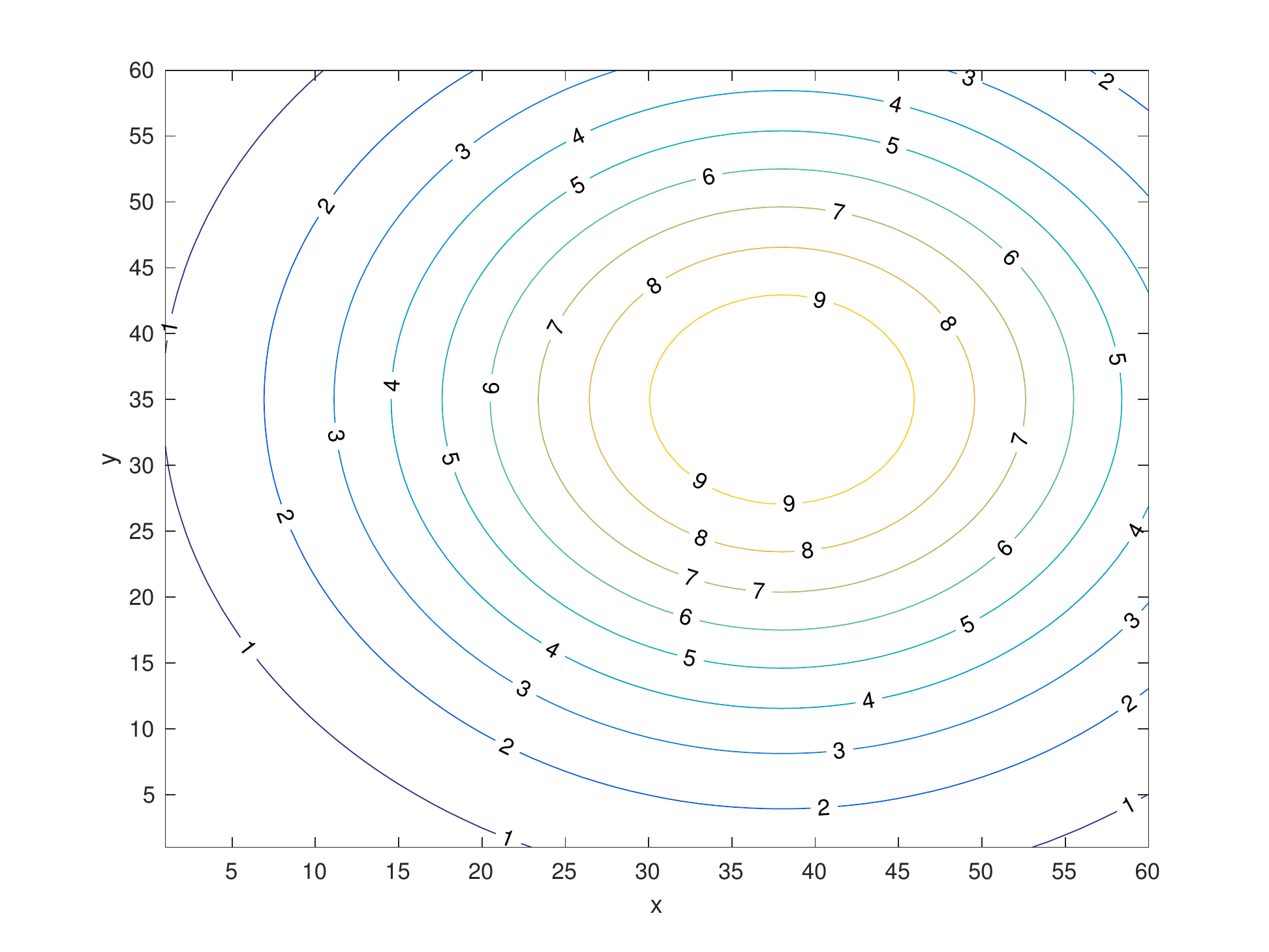}
   \label{fig::esa:2fd0}
   }  
  \subfigure[Field distribution over the network]{
  \includegraphics[width=0.45\textwidth]{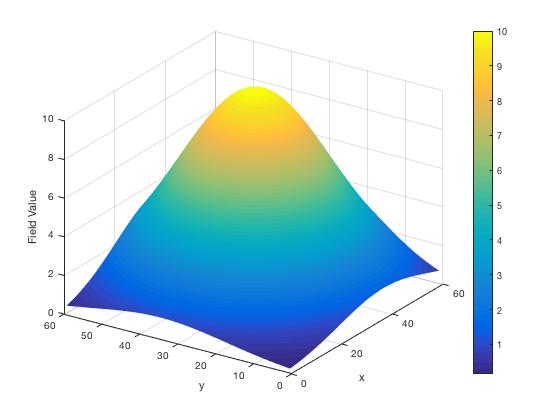}
   \label{fig::esa:3fd0}
   }  
  \caption{Field distribution of Scenario 1} \label{fig::esa:sc10}
\end{figure*}

\begin{figure}[ht!]
  \centering
    \includegraphics[width=0.7\textwidth]{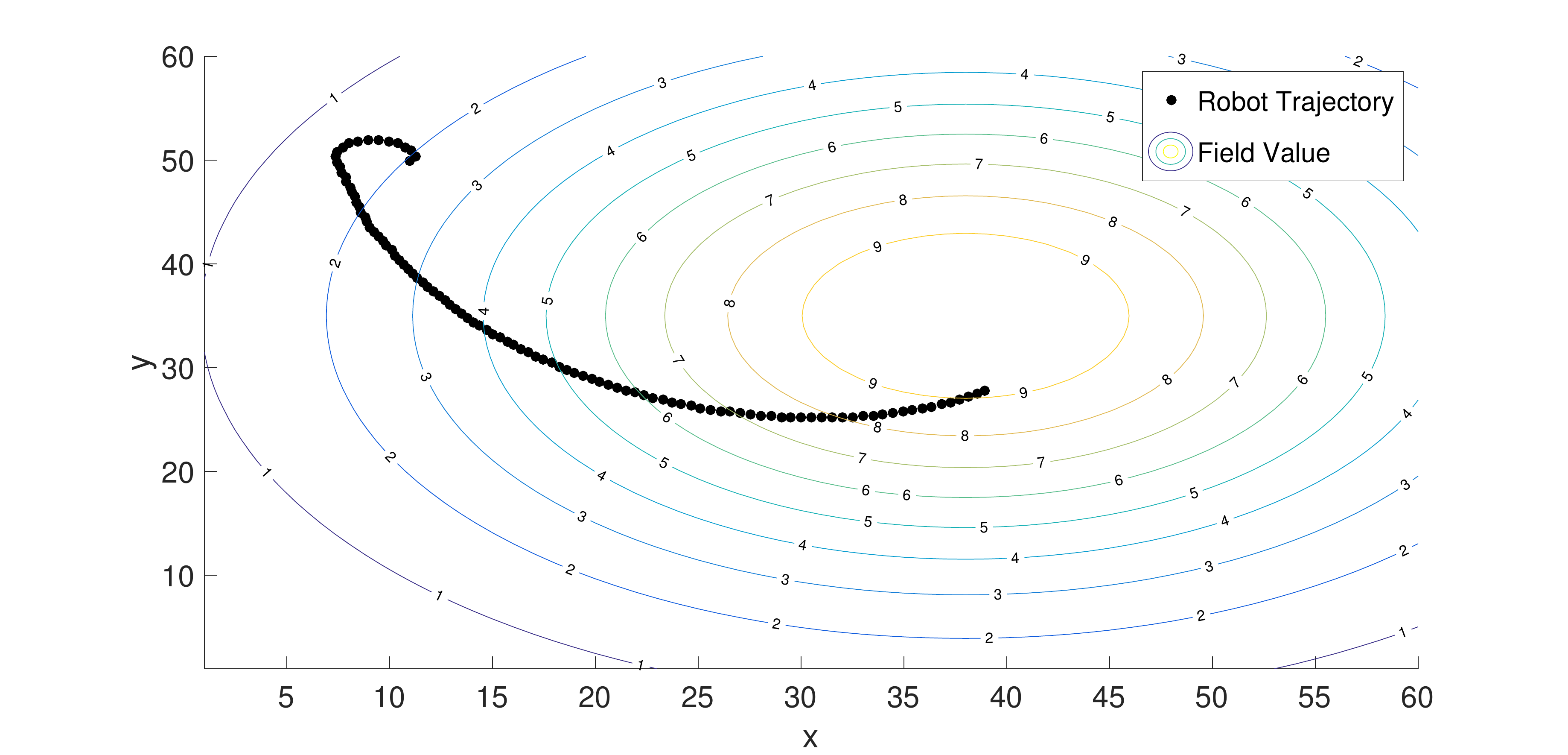}
    \caption{Robot trajectory when measurement error is zero}\label{fig::esa:1rt0}
\end{figure}
\begin{figure*}
 \centering
 \subfigure[Sensor measurements ]{
  \includegraphics[width=0.48\textwidth]{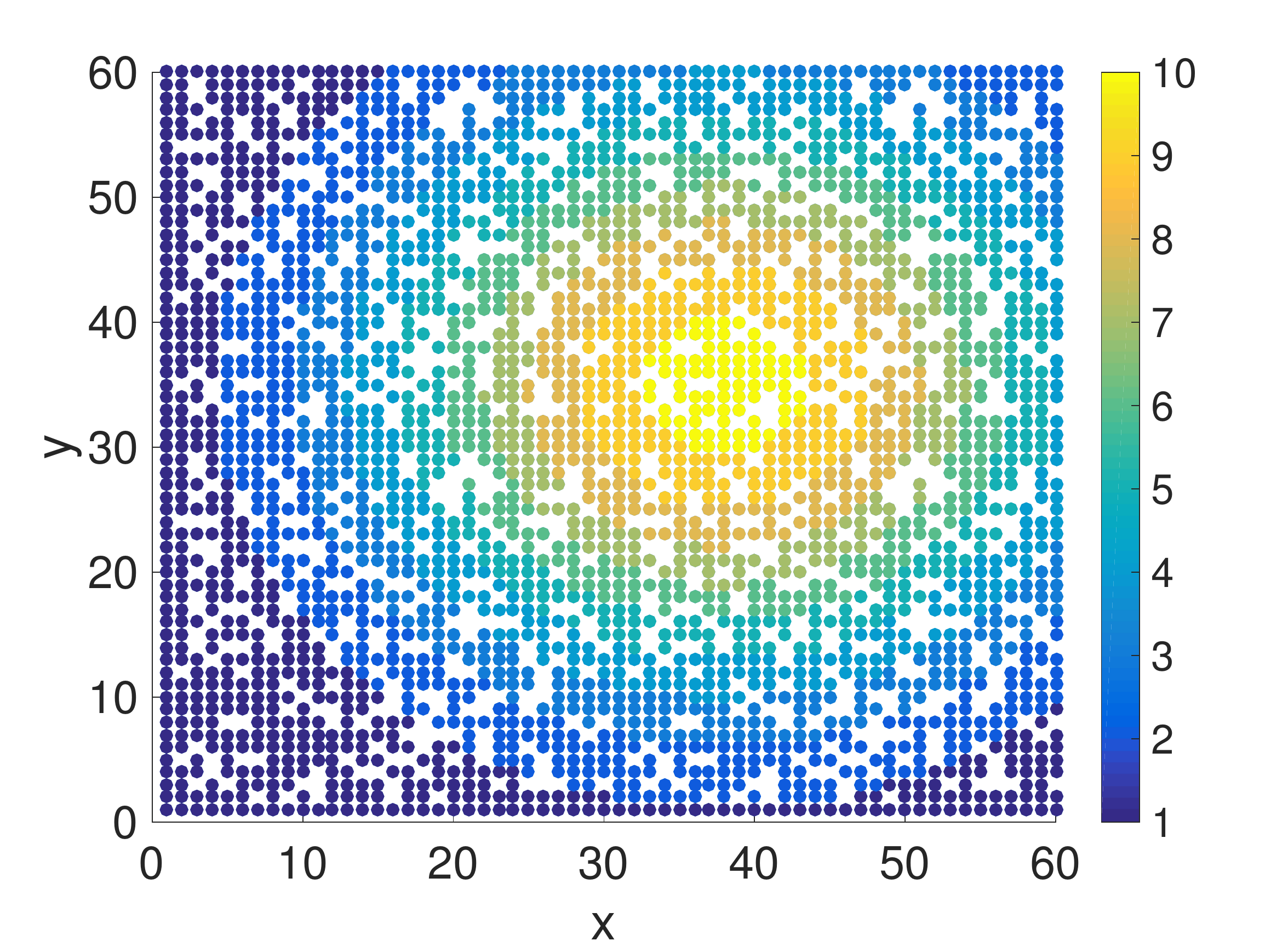}
   \label{fig::esa:sr30}
   }
 \subfigure[Field distribution values over the network ]{
  \includegraphics[width=0.48\textwidth]{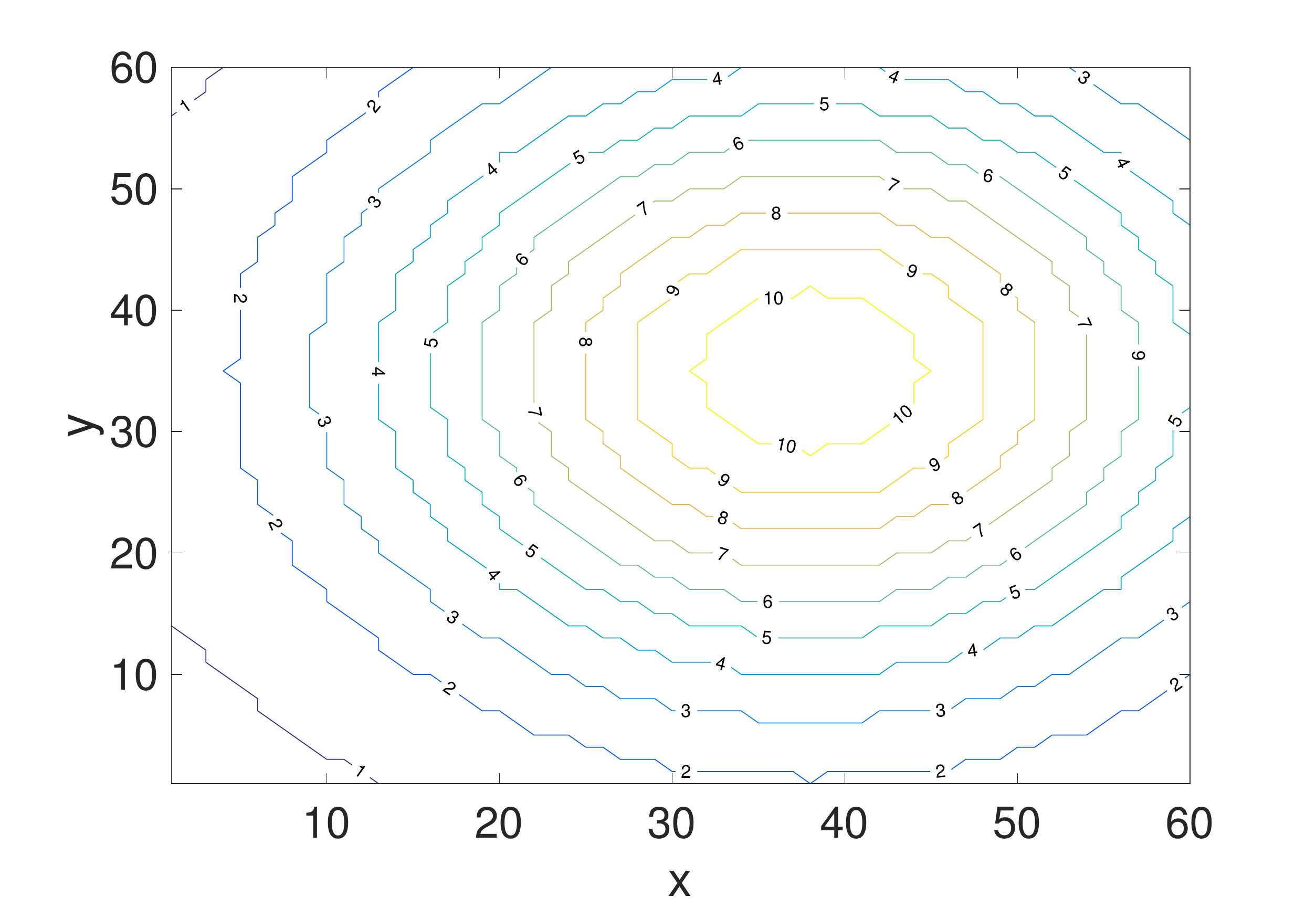}
   \label{fig::esa:fd30}
   }
  \subfigure[Robot trajectory]{
  \includegraphics[width=0.55\textwidth]{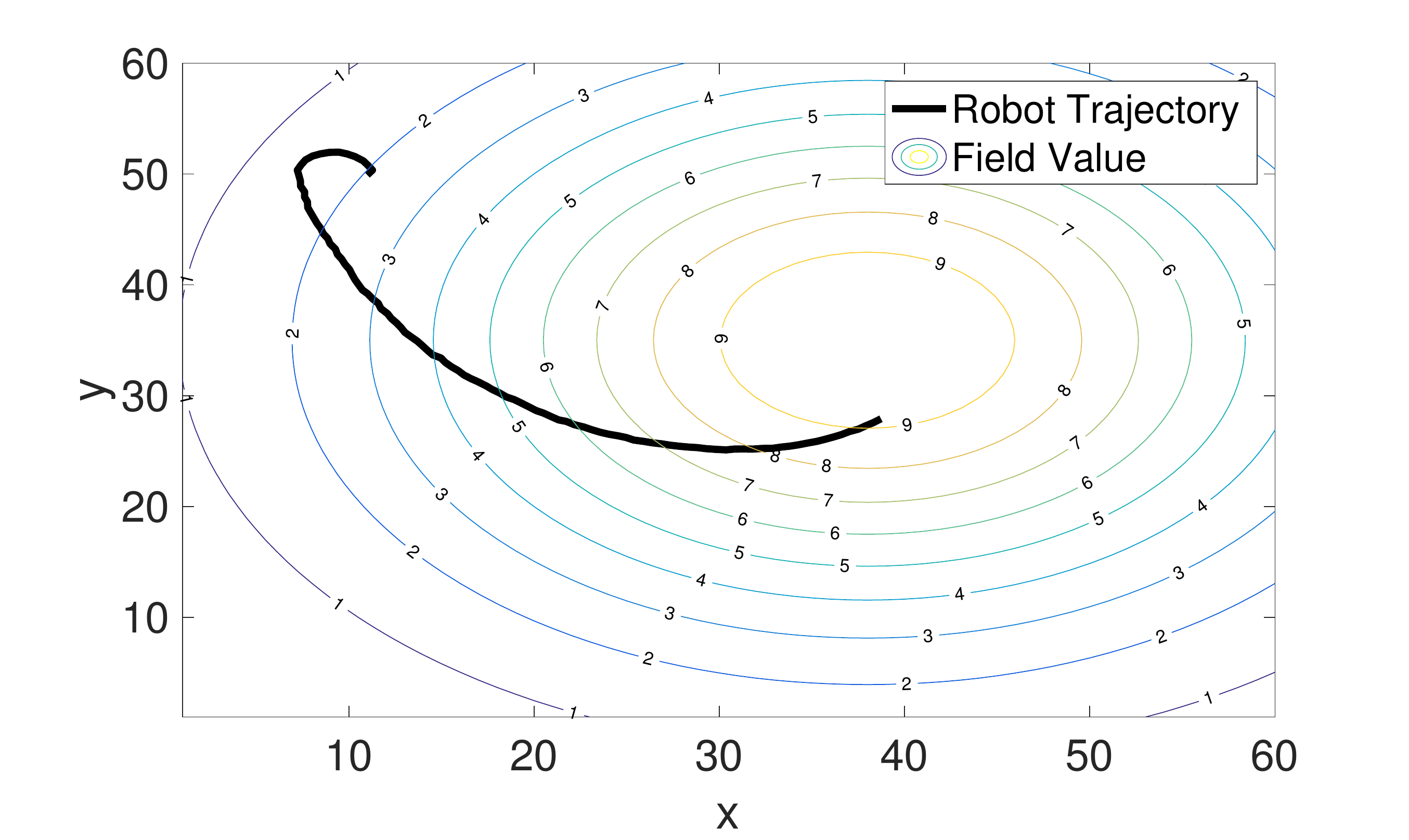}
   \label{fig::esa:rt30}
   }
  \caption{ When SNR = 30 dB in Scenario 1} \label{fig::esa:sc130}
\end{figure*}
Moreover, the propagation model proposed in Chapter \ref{chapter:mltm} is used to simulate the communication link between two sensors or sensor to robot. This model considers the path loss and shadowing as shown in equation \ref{eqn::esa:pm2}.
\begin{equation}\label{eqn::esa:pm2}
P_{rx_i} = P_{tx_j}-10\varepsilon log d_{ij} + X_{i,\sigma }                           
\end{equation}
where, the received signal strength at node $s_i$ is $P_{rx_i} $, the transmitted signal strength of the signal at node $s_j$ is $P_{tx_j}$, the path-loss exponent is $\varepsilon$, the distance between node $s_i$ and node $s_j$ is $d_{ij} $ and the logarithm of shadowing component with a $\sigma $ standard deviation on node $i$ at is $X_{i,\sigma }$. 

\begin{figure*}
 \centering
 \subfigure[Sensor measurements ]{
  \includegraphics[width=0.48\textwidth]{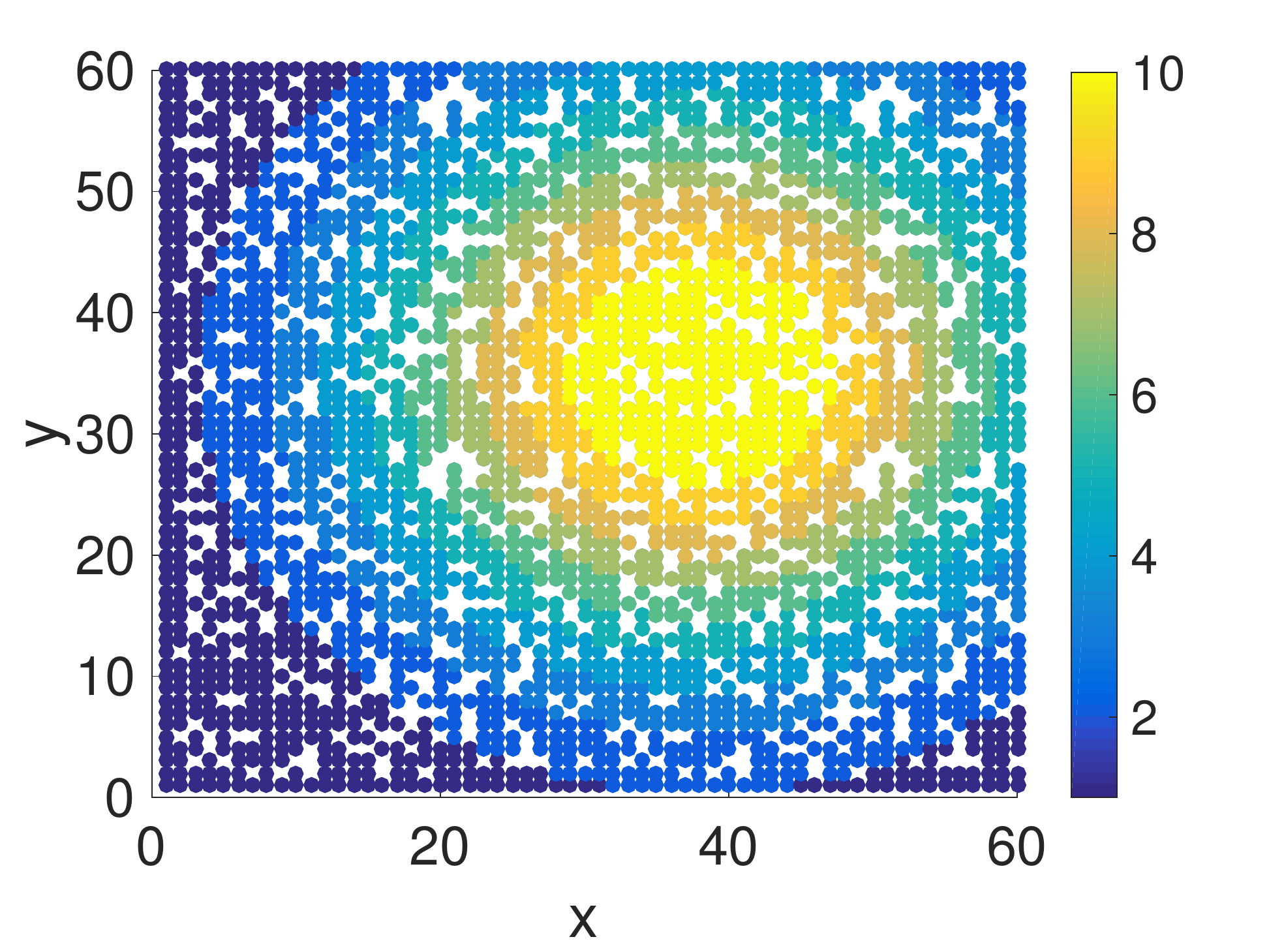}
   \label{fig::esa:sr20}
   }
 \subfigure[Field distribution values over the network]{
  \includegraphics[width=0.48\textwidth]{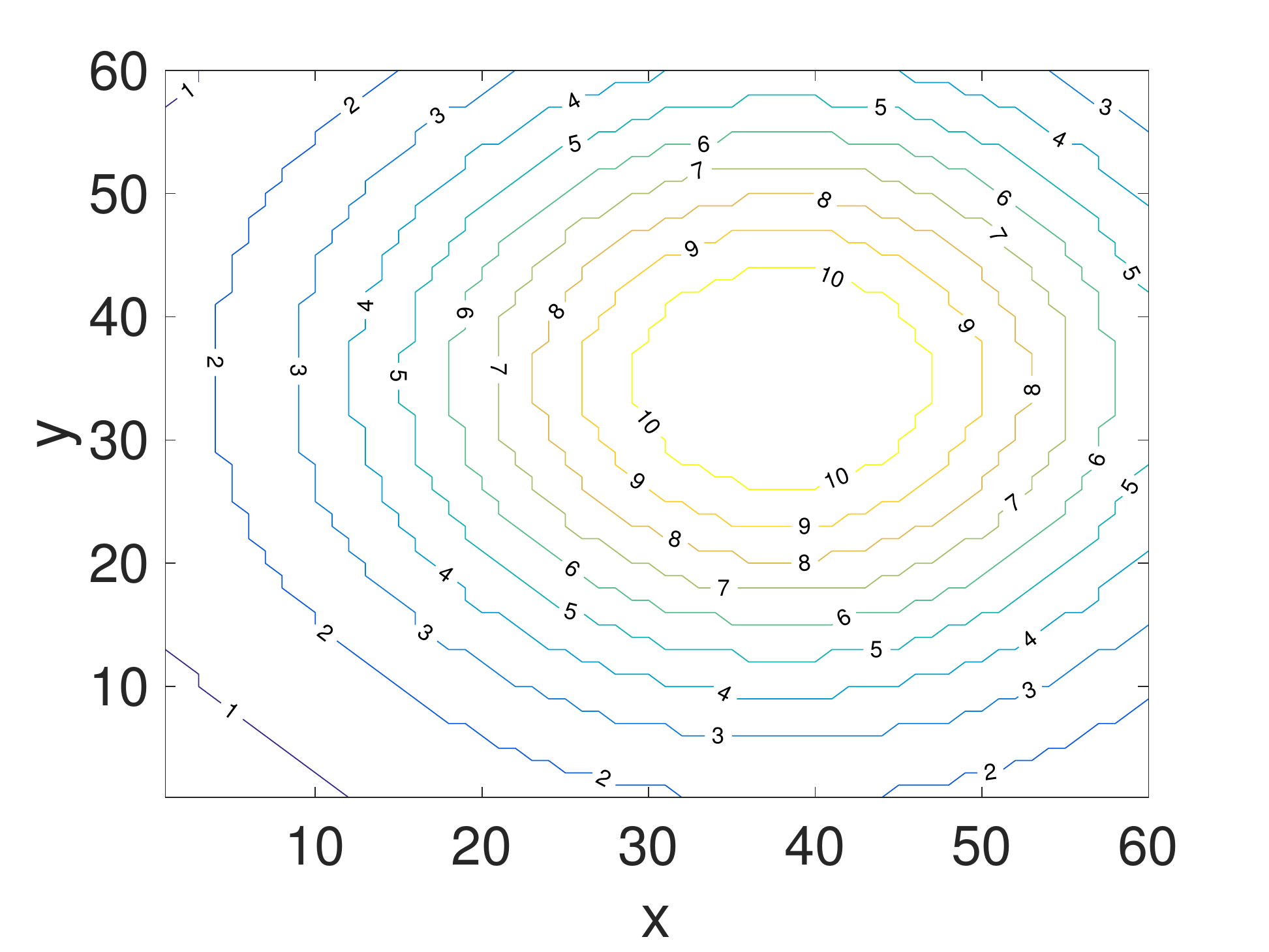}
   \label{fig::esa:fd20}
   }  
  \subfigure[Robot trajectory]{
  \includegraphics[width=0.55\textwidth]{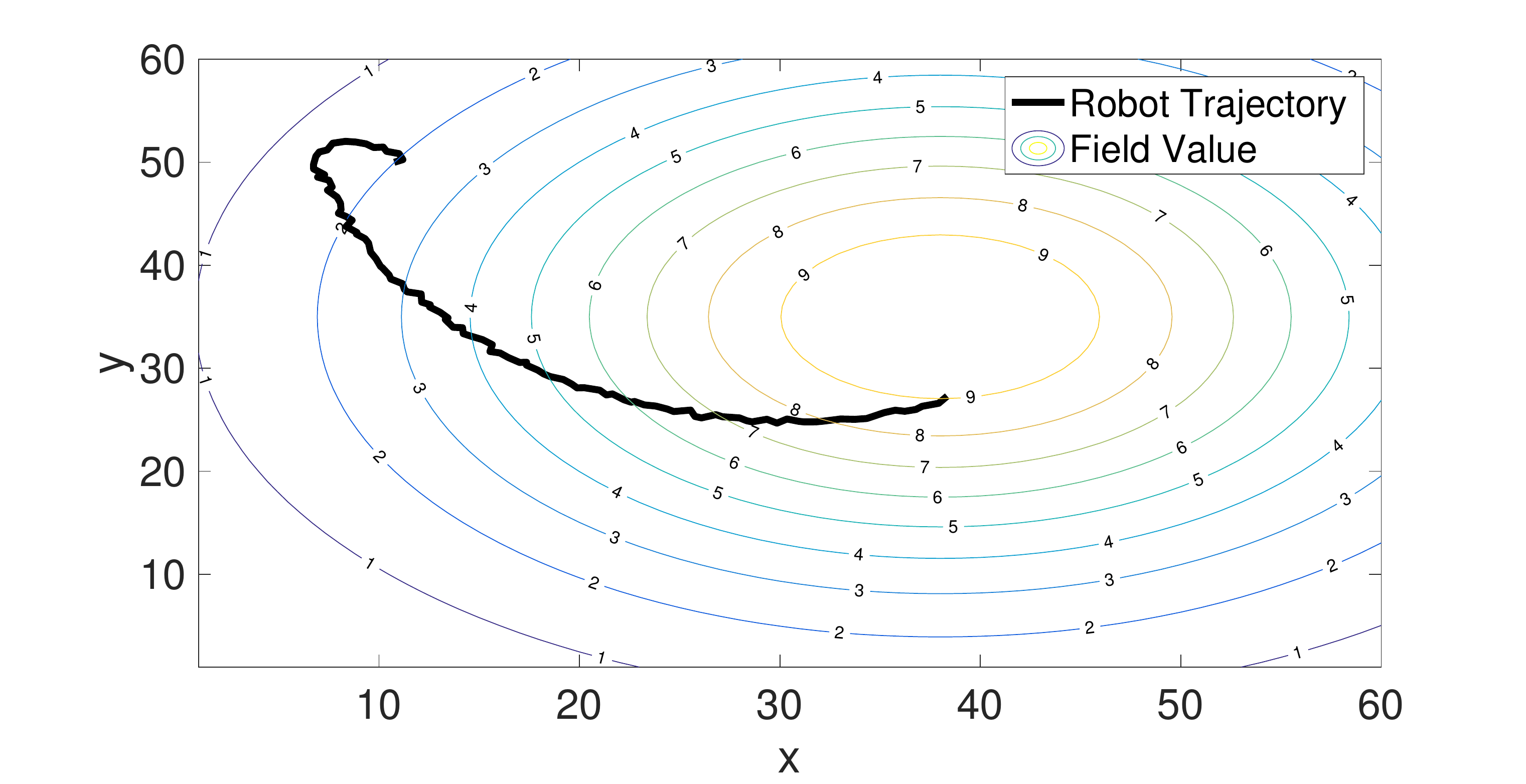}
   \label{fig::esa:rt20}
   }  
  \caption{ When SNR = 20 dB in Scenario 1} \label{fig::esa:sc120}
\end{figure*}

\begin{figure*}
 \centering
 \subfigure[Sensor measurements ]{
  \includegraphics[width=0.48\textwidth]{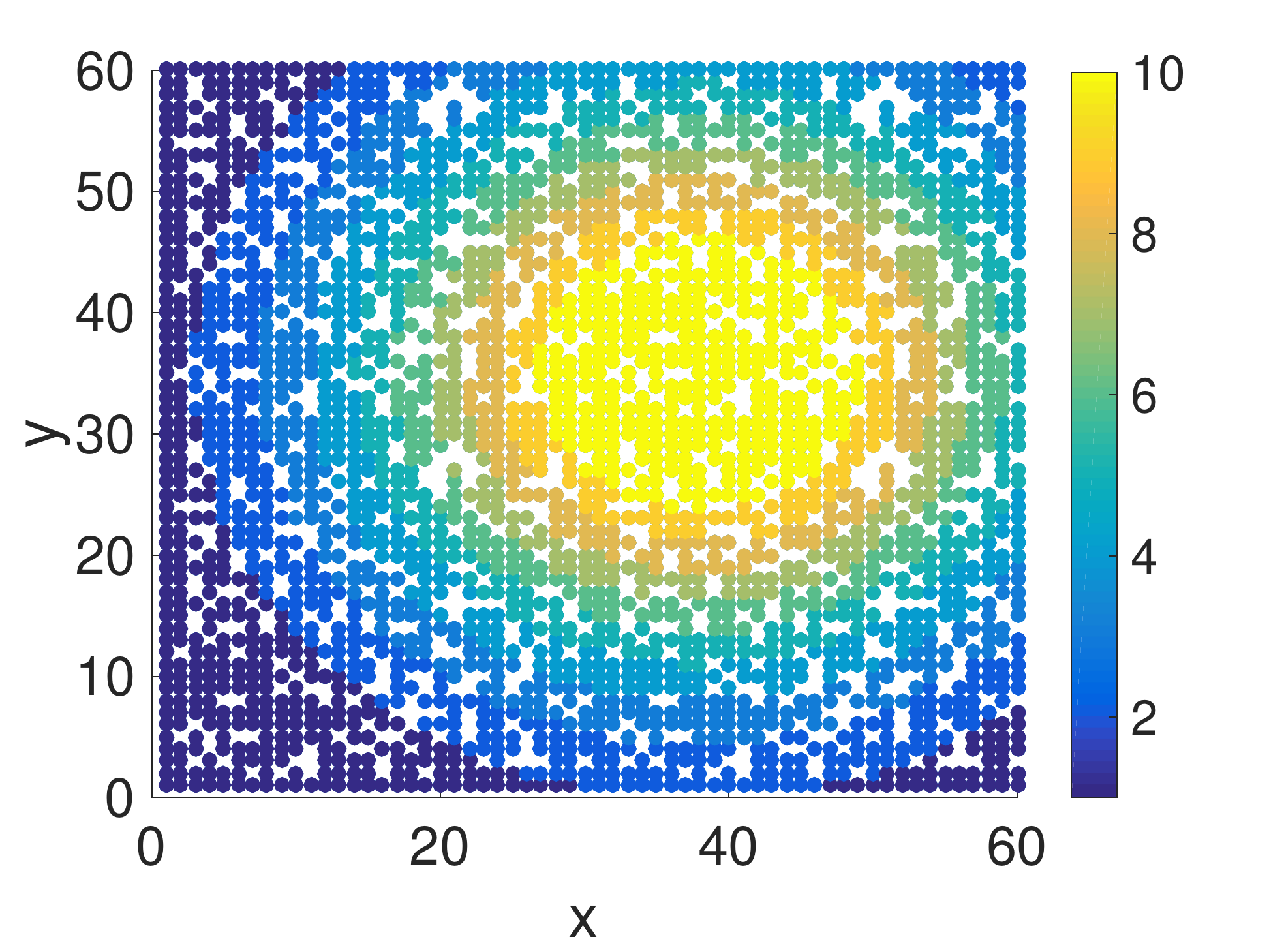}
   \label{fig::esa:sr10}
   }
 \subfigure[Field distribution values over the network]{
  \includegraphics[width=0.48\textwidth]{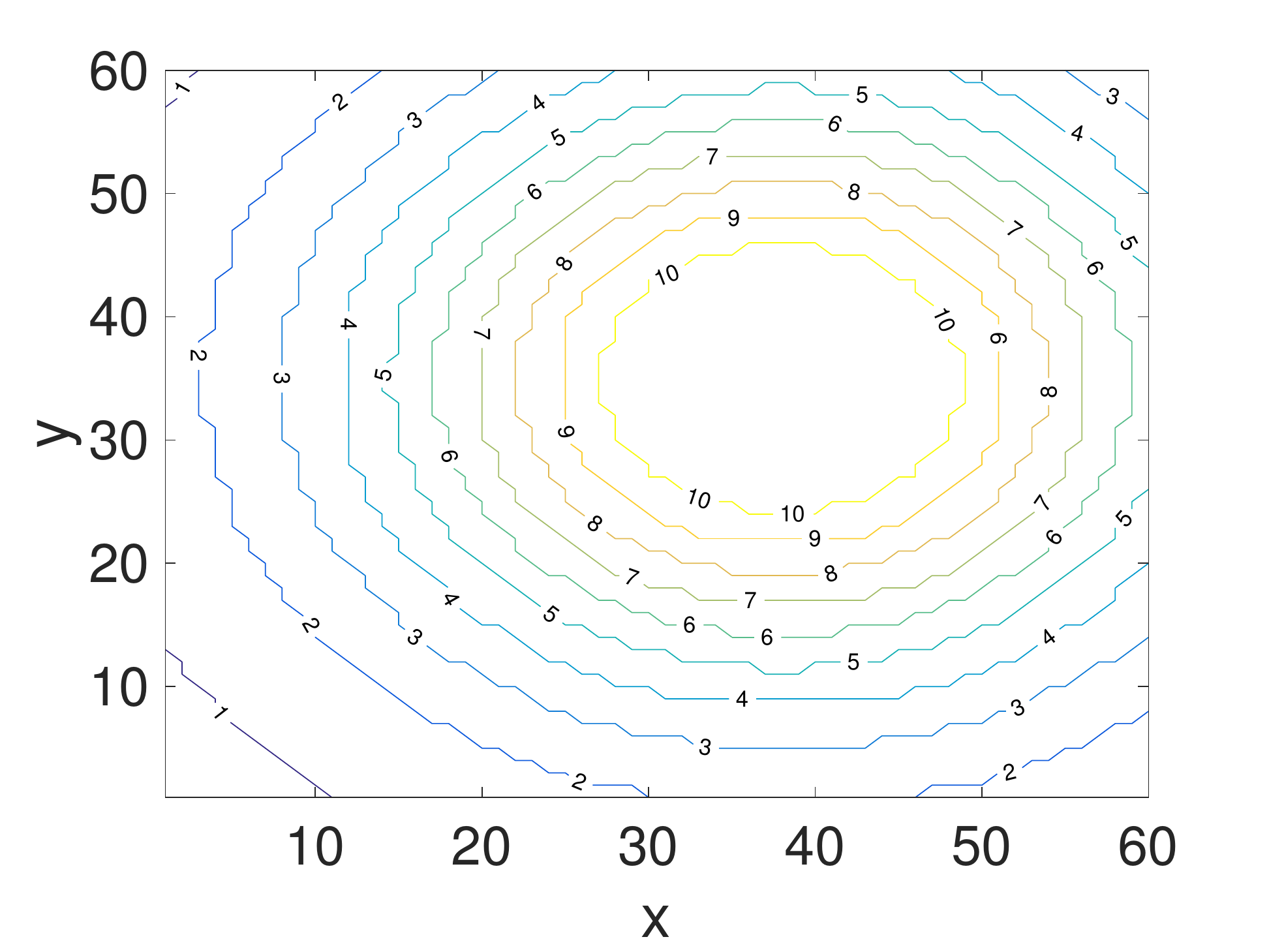}
   \label{fig::esa:fd10}
   }  
  \subfigure[Robot trajectory]{
  \includegraphics[width=0.55\textwidth]{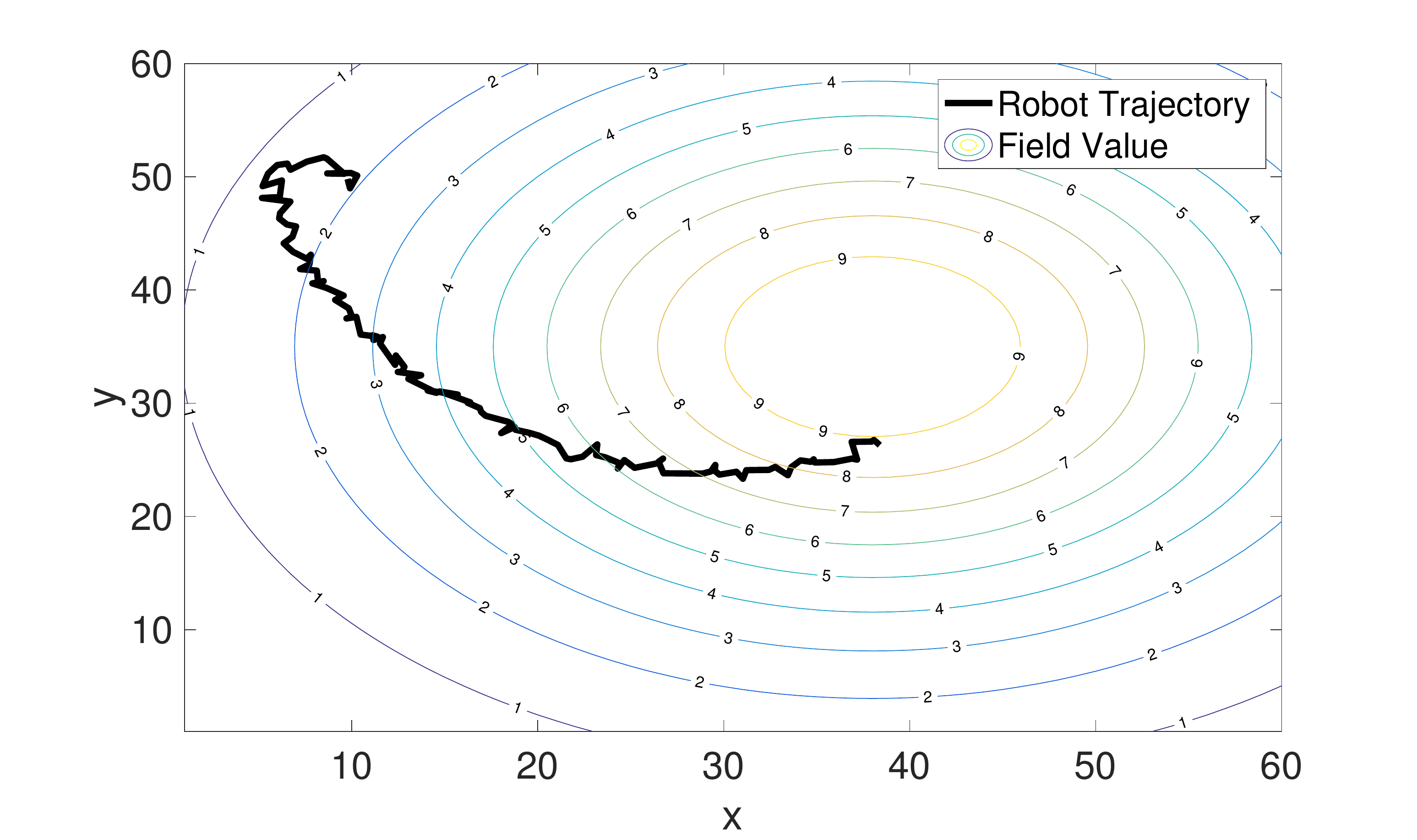}
   \label{fig::esa:rt10}
   }  
  \caption{ When SNR = 10 dB in Scenario 1} \label{fig::esa:sc110}
\end{figure*}

\begin{figure*}
 \centering
 \subfigure[Sensor measurements ]{
  \includegraphics[width=0.48\textwidth]{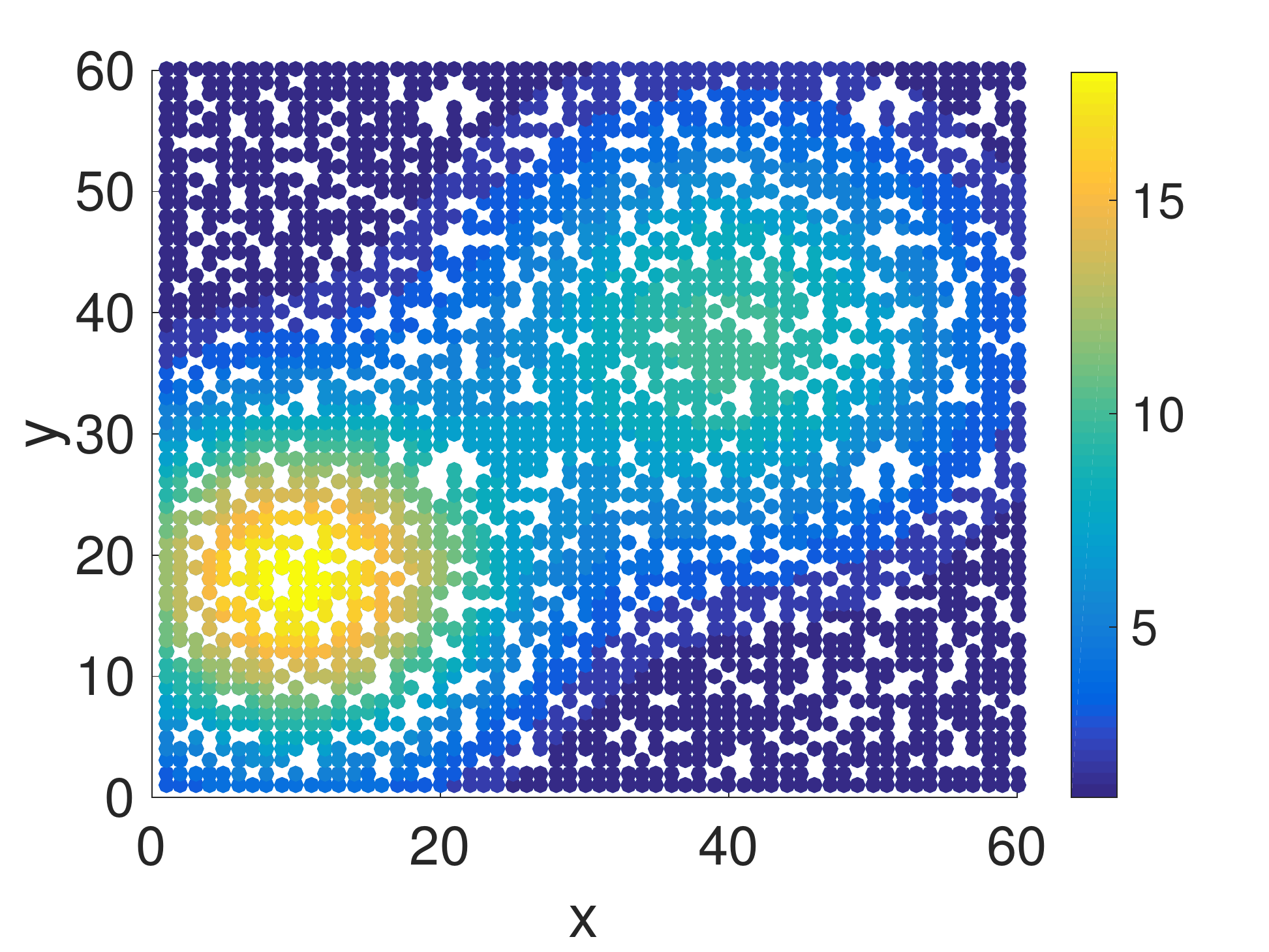}
   \label{fig::esa:2sr0}
   }
 \subfigure[Field values over the network]{
  \includegraphics[width=0.48\textwidth]{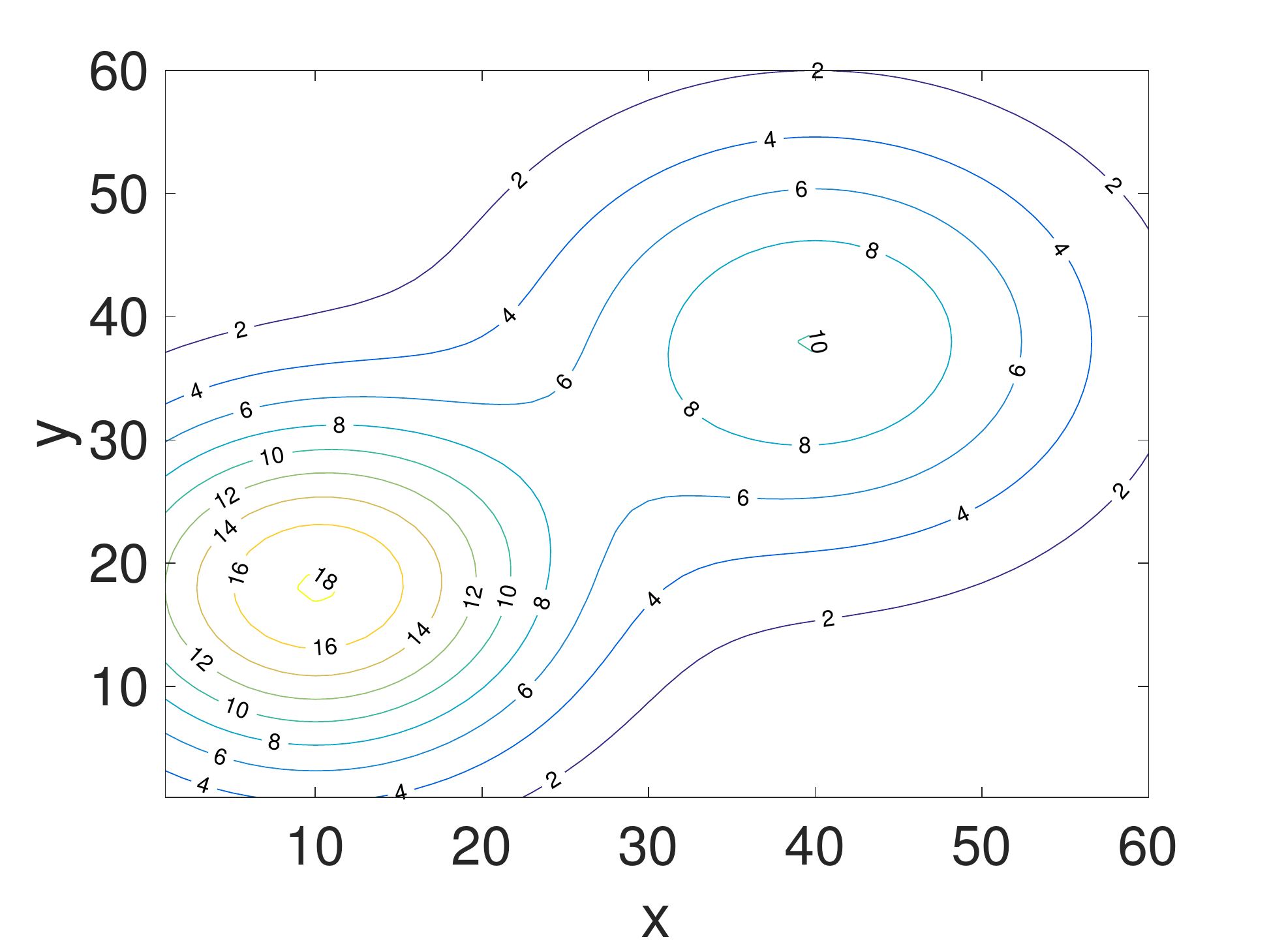}
   \label{fig::esa:22fd0}
   }  
  \subfigure[Field distribution over the network]{
  \includegraphics[width=0.55\textwidth]{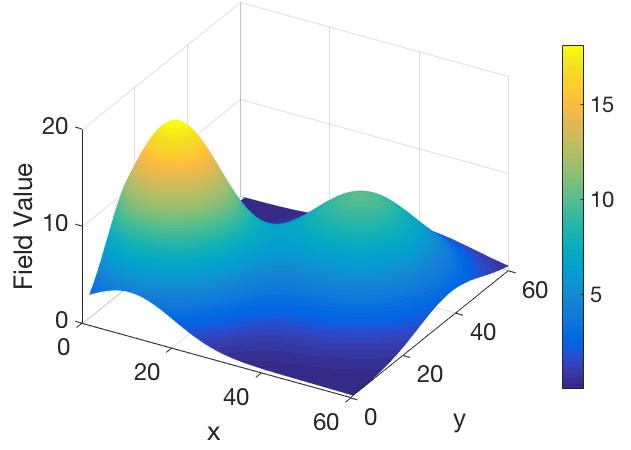}
   \label{fig::esa:23fd0}
   }  
  \caption{Field distribution of Scenario 2 } \label{fig::esa:sc20}
\end{figure*}

\begin{figure*}
 \centering
 \subfigure[Sensor measurements at t=1s]{
  \includegraphics[width=0.45\textwidth]{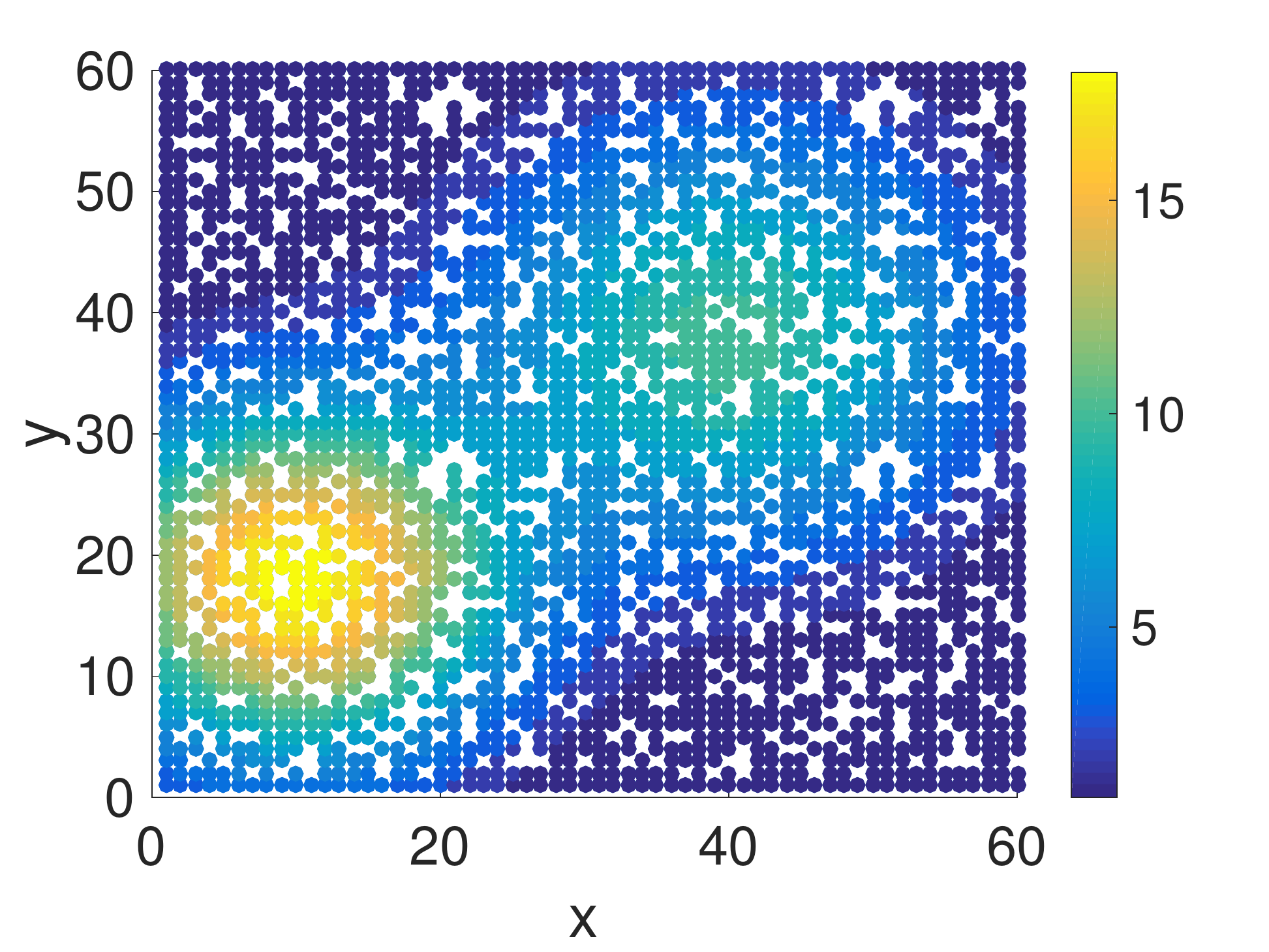}
   \label{fig::esa:2sr1}
   }
   \quad 
    \subfigure[Sensor measurements at t=77s]{
  \includegraphics[width=0.45\textwidth]{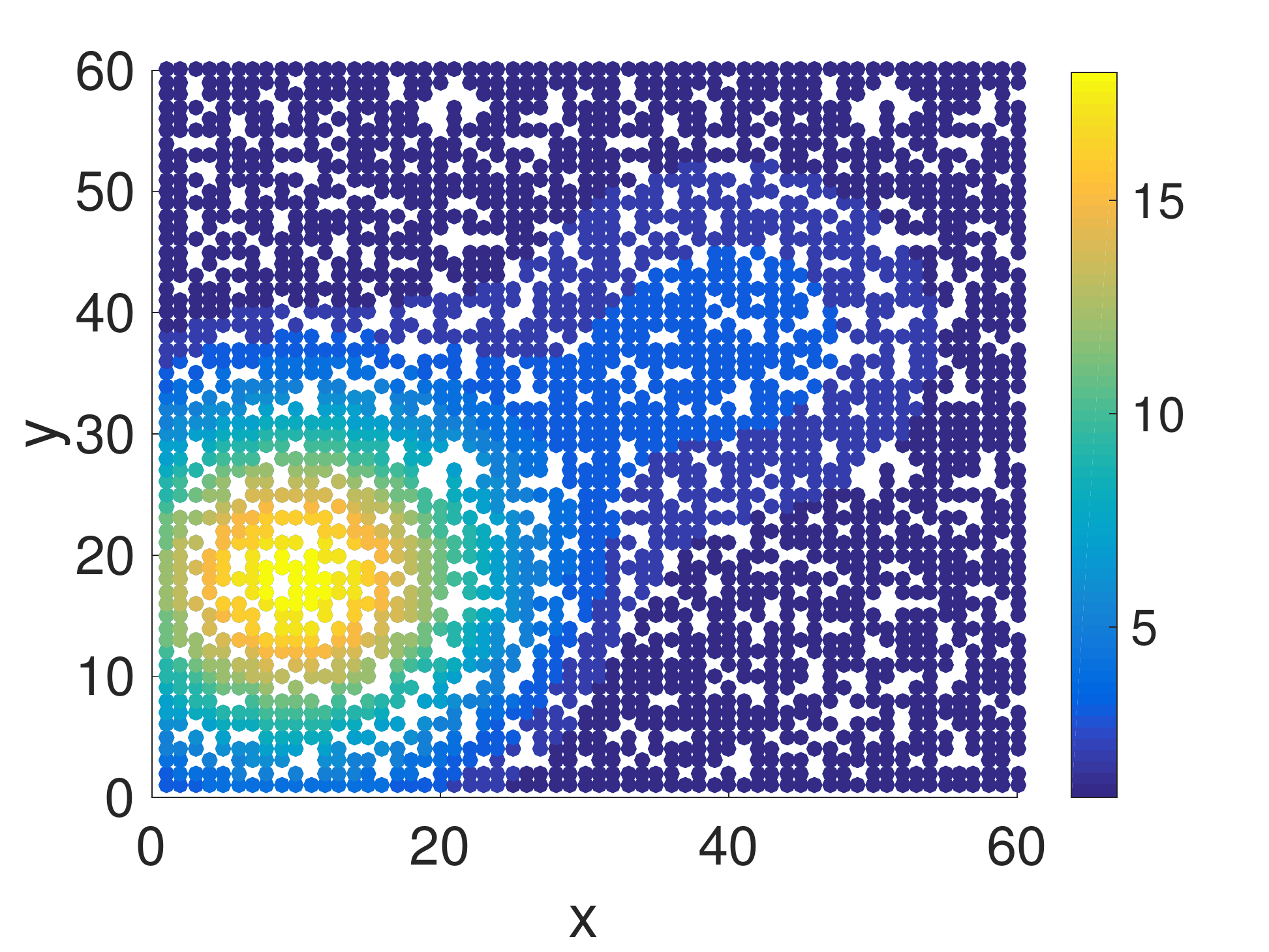}
   \label{fig::esa:2sr2}
   }  
 
   \subfigure[Sensor measurements at t=111s]{
  \includegraphics[width=0.45\textwidth]{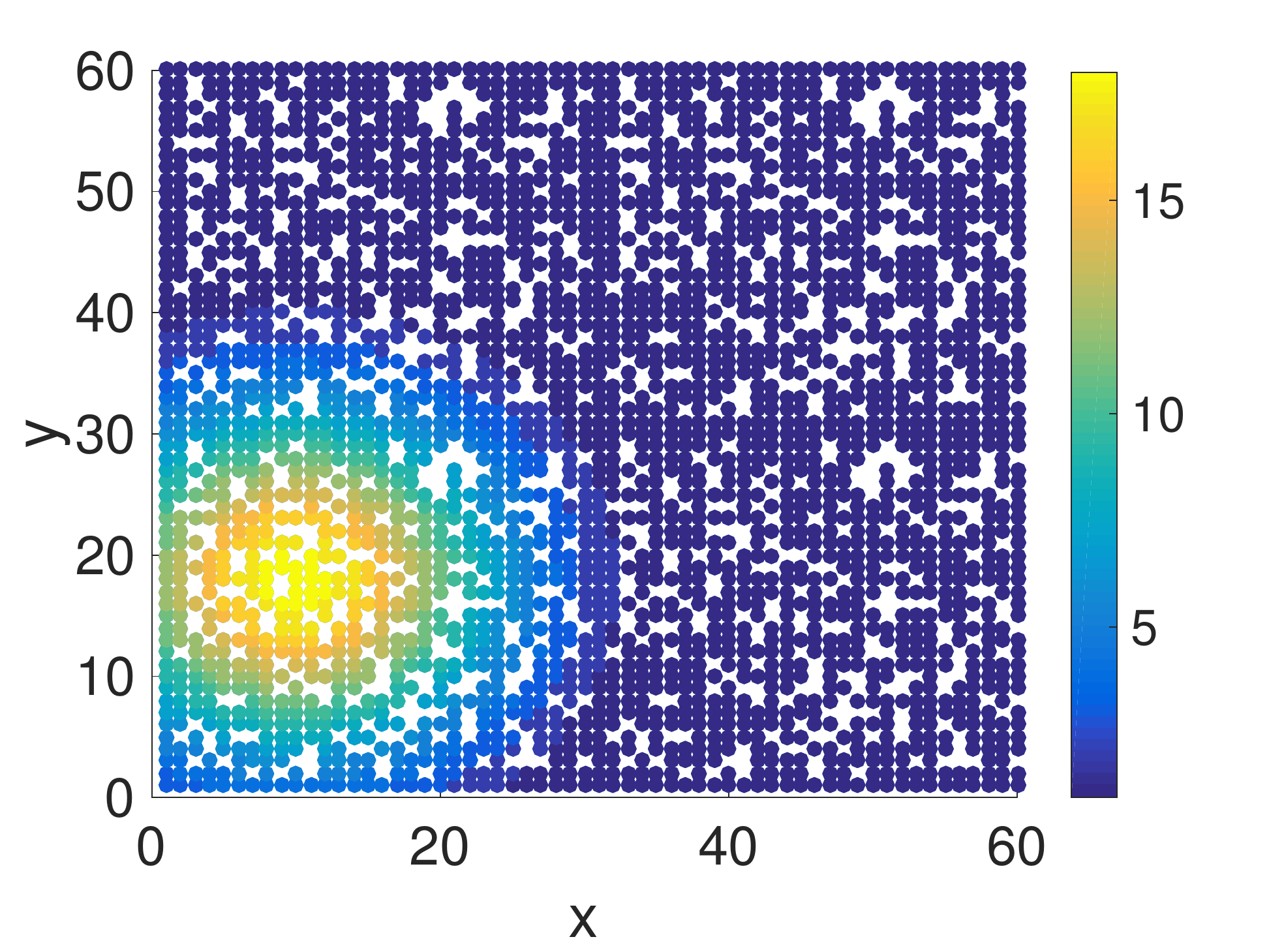}
   \label{fig::esa:2sr3}
   }  
  
  \caption{ Sensor measurements Scenario 2} \label{fig::esa:sc220_1}
\end{figure*}

\begin{figure*}
 \centering

    \subfigure[Robot trajectory upto t=76s]{
  \includegraphics[width=0.46\textwidth]{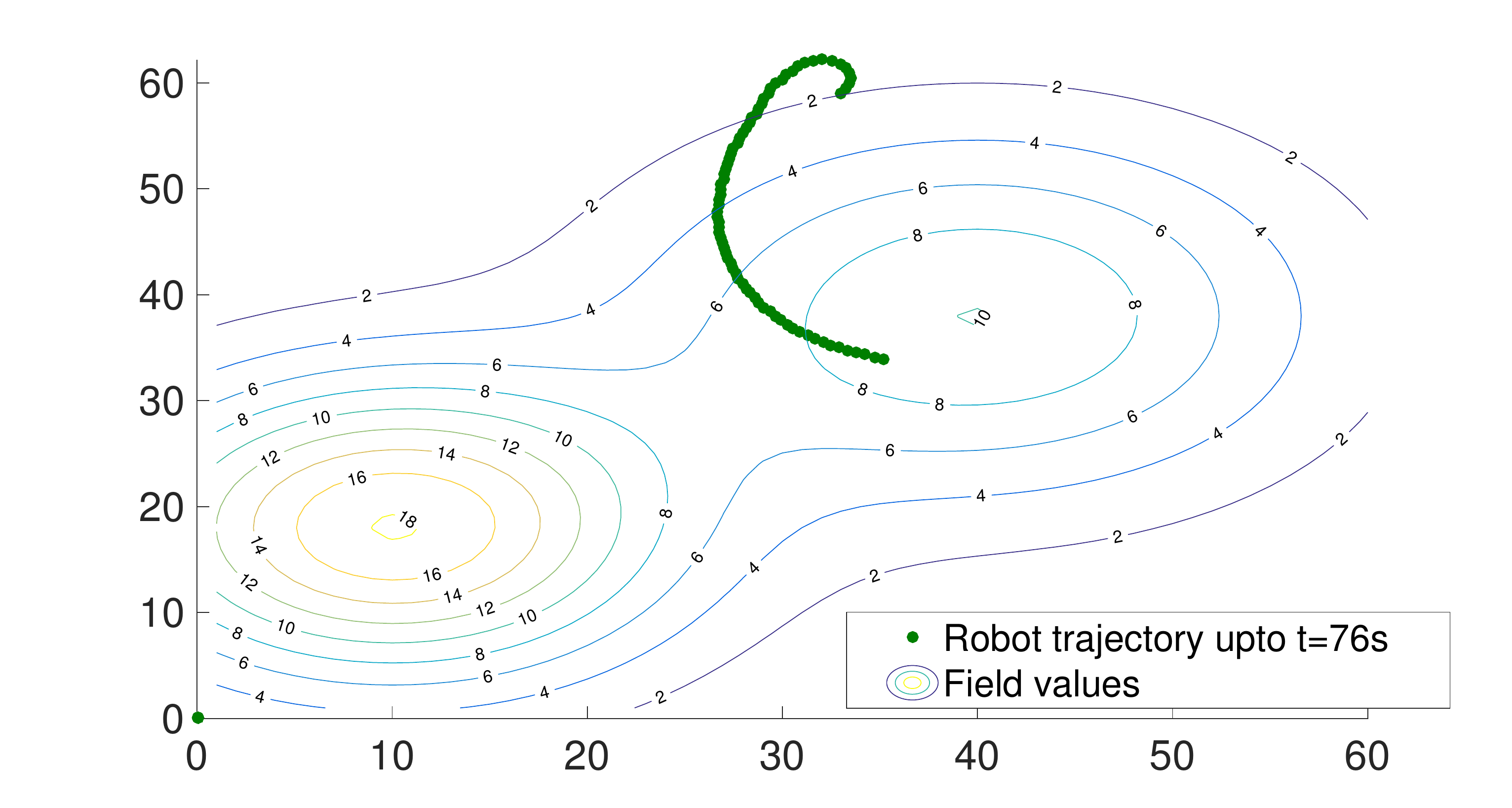}
   \label{fig::esa:2rt1}
   }
    \quad	 
 \subfigure[Robot trajectory upto t=110s]{
  \includegraphics[width=0.46\textwidth]{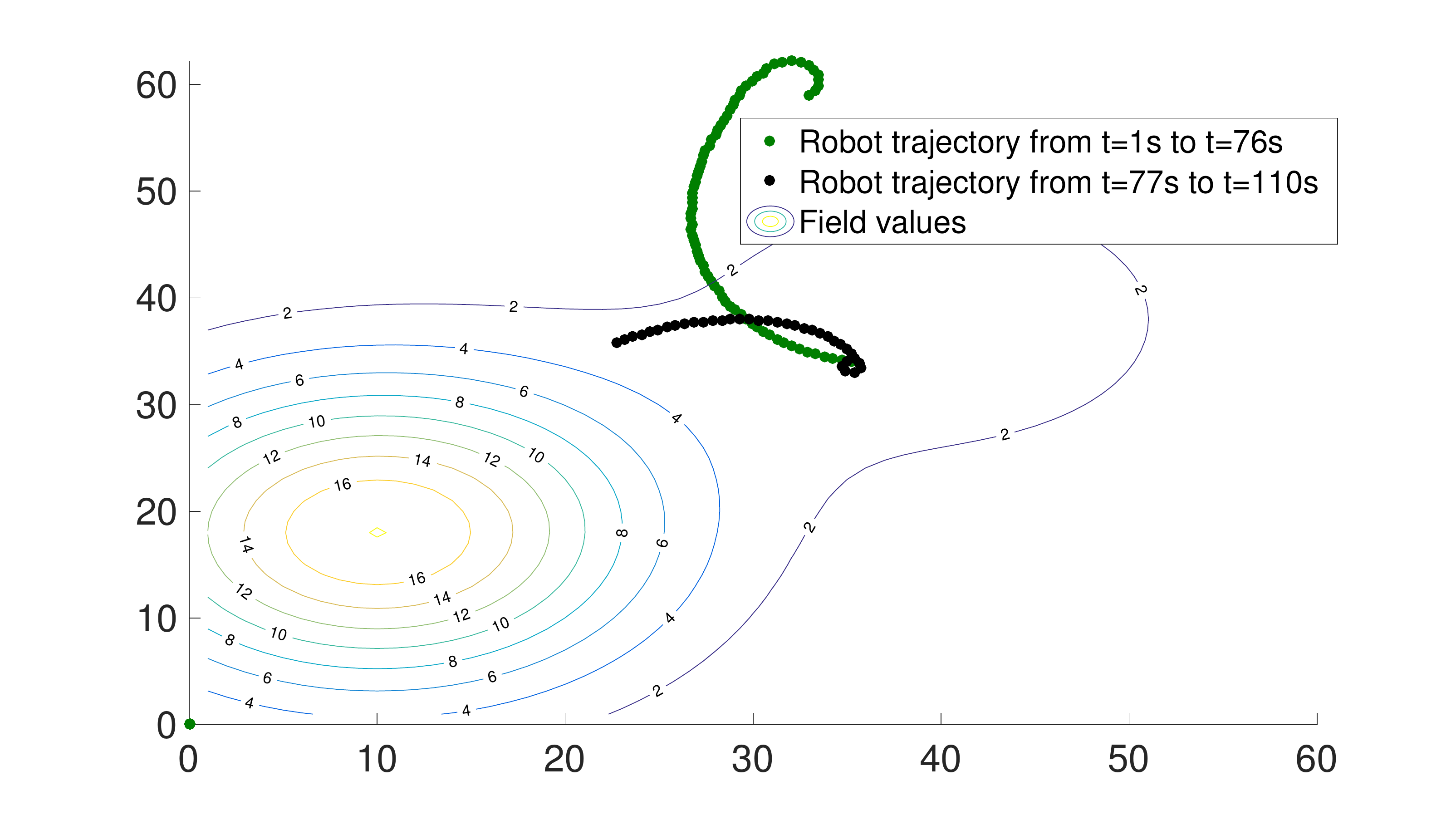}
   \label{fig::esa:2rt2}
   }  
    
  \subfigure[Robot trajectory upto t=160s]{
  \includegraphics[width=0.48\textwidth]{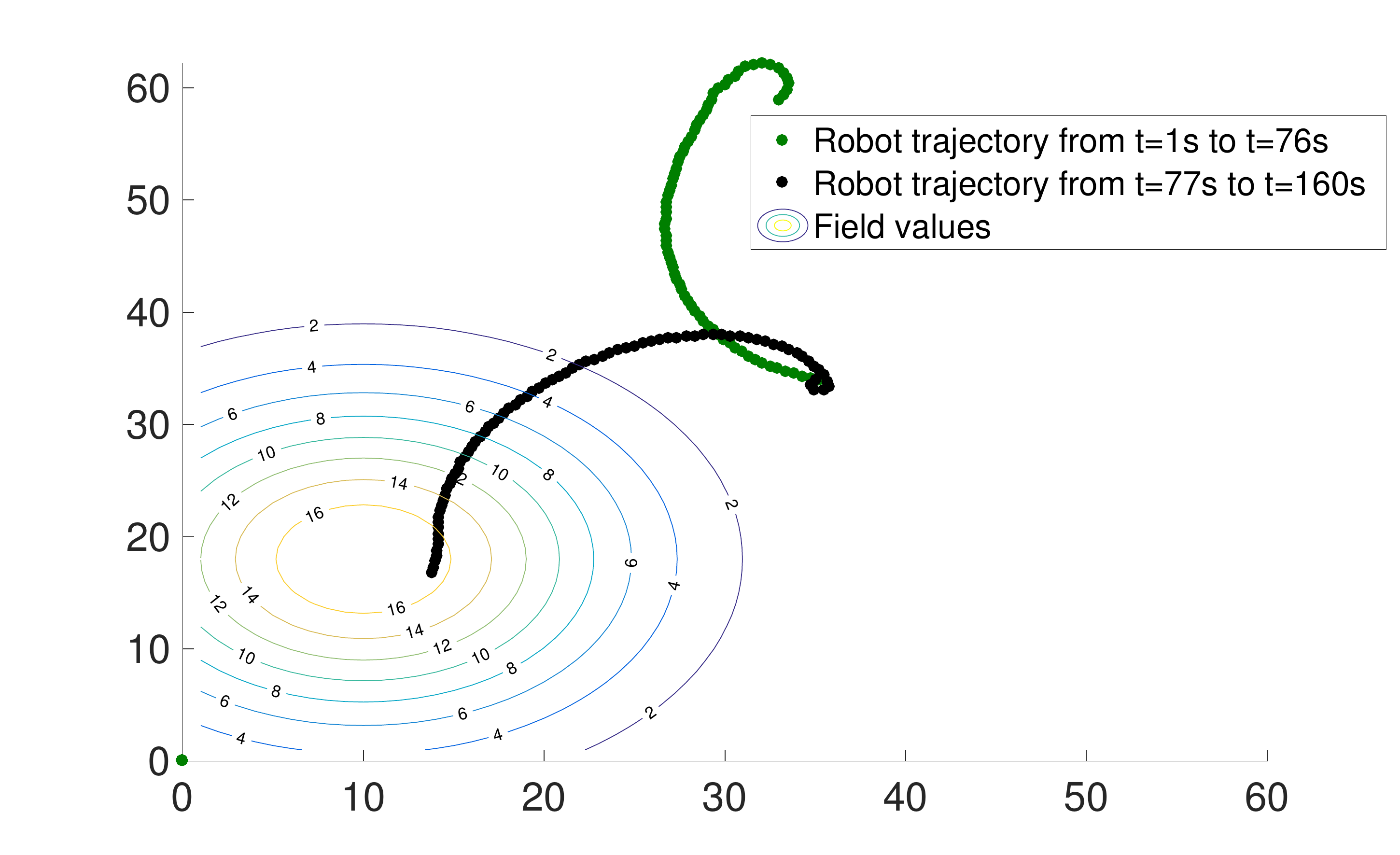}
   \label{fig::esa:2rt3}
   } 
  \caption{ Robot trajectories of Scenario 2} \label{fig::esa:sc220_2}
\end{figure*}

The field distribution for Scenario 1 is shown in Figure \ref{fig::esa:sc10}. The sensor readings at each sensor location when $e_i=0$ is shown in Figure \ref{fig::esa:sr0} and level curves are shown in Figure \ref{fig::esa:2fd0}. The field value distribution is shown in Figure \ref{fig::esa:3fd0}. First, the performance of the algorithm is evaluated when sensor readings are accurate. In other words, the noise is zero i.e. $e_i=0$. The goal of the robot is to approach the desired level $d_0=9$. The robots constant speed is $V_r = 0.5 m/s$ and time-varying angular velocity is limited by $\omega _{max} = 1 rad/s$. Moreover, $\gamma$ and $\delta$  is considered as 1 and 0.1 respectively. The robot starts the search at (10, 50) and navigation trajectory is shown in Figure \ref{fig::esa:1rt0}. From the figure, it can be seen that the robot was able to approach the desired level with the proposed guidance law and topology coordinates of sensors. 

Then the performance of the algorithm is evaluated against the measurement noise in sensor readings. Three cases with different SNR values are considered. Those are 30 dB, 20 dB and 10 dB. The results for each case is shown in Figure \ref{fig::esa:sc130}-\ref{fig::esa:sc110} respectively. Figure \ref{fig::esa:sr30}-\ref{fig::esa:sr10} show the sensor reading with measurement noise at each sensor location and Figure \ref{fig::esa:fd30}-\ref{fig::esa:fd10} show the level curves in the environment. The robot trajectories are shown in Figure \ref{fig::esa:rt30}-\ref{fig::esa:rt10}. From the trajectory plots, it can be seen that the smoothness/directness of the robot trajectory is less when measurement noise is high. For an example, in Figure \ref{fig::esa:rt30}, the robot trajectory is smooth and robot takes 53s to reach to the desired level. But in Figure \ref{fig::esa:rt10}, robot needs 68s to reach to the desired level and also the trajectory is not smooth as previous two cases. However, the control law satisfactorily guide the robot towards the desired field level with the sensor topology coordinates. 

In real world scenarios, more than one emergency source may exist in a single environment. Thus, two emergency sources located at the same environment as described in Scenario 2 is simulated. The sensor readings at each sensor location when $e_i=0$ and $t=1s$ is shown in Figure \ref{fig::esa:2sr0} and the corresponding level curves are shown in Figure \ref{fig::esa:22fd0}. The field value distribution for the same conditions is shown in Figure \ref{fig::esa:23fd0}. The goal of this simulation is to approach the same desired level $d_0=9$ in either source and remove it from the network. Then, navigate towards the desired level of remaining source and remove it from the network. In this case field function is changing from time to time and it is illustrated in Figure \ref{fig::esa:sc220_1}. Figure \ref{fig::esa:2sr1} shows the initial sensor readings at each sensor locations. This remains same until robot removes one source from the network. The robot navigation path when two sources are exist in the network is shown in Figure \ref{fig::esa:2rt1}. As in the figure, first robot reaches to the source located at (40, 38) and removes it from the network. Even though the source is removed, some particles emitted from that source may remain in the environment for a while and sensors can sense them. Thus, sensor readings at that time is shown in Figure \ref{fig::esa:2sr2} and robot trajectory in that time period is shown in Figure \ref{fig::esa:2rt2}. Then, it is assumed that all the particles emitted from the removed source is faded after t=110s and the sensor readings at that time is shown in Figure \ref{fig::esa:2sr3}. The robot navigation path towards the remaining source is shown in Figure \ref{fig::esa:2rt3}. Thus, the control law has been able to navigate the robot towards the desired levels and remove sources from the network even multiple sources are exist in the environment. 

\begin{figure*}
 \centering
 \subfigure[Proposed algorithm with zero noise]{
  \includegraphics[width=0.48\textwidth]{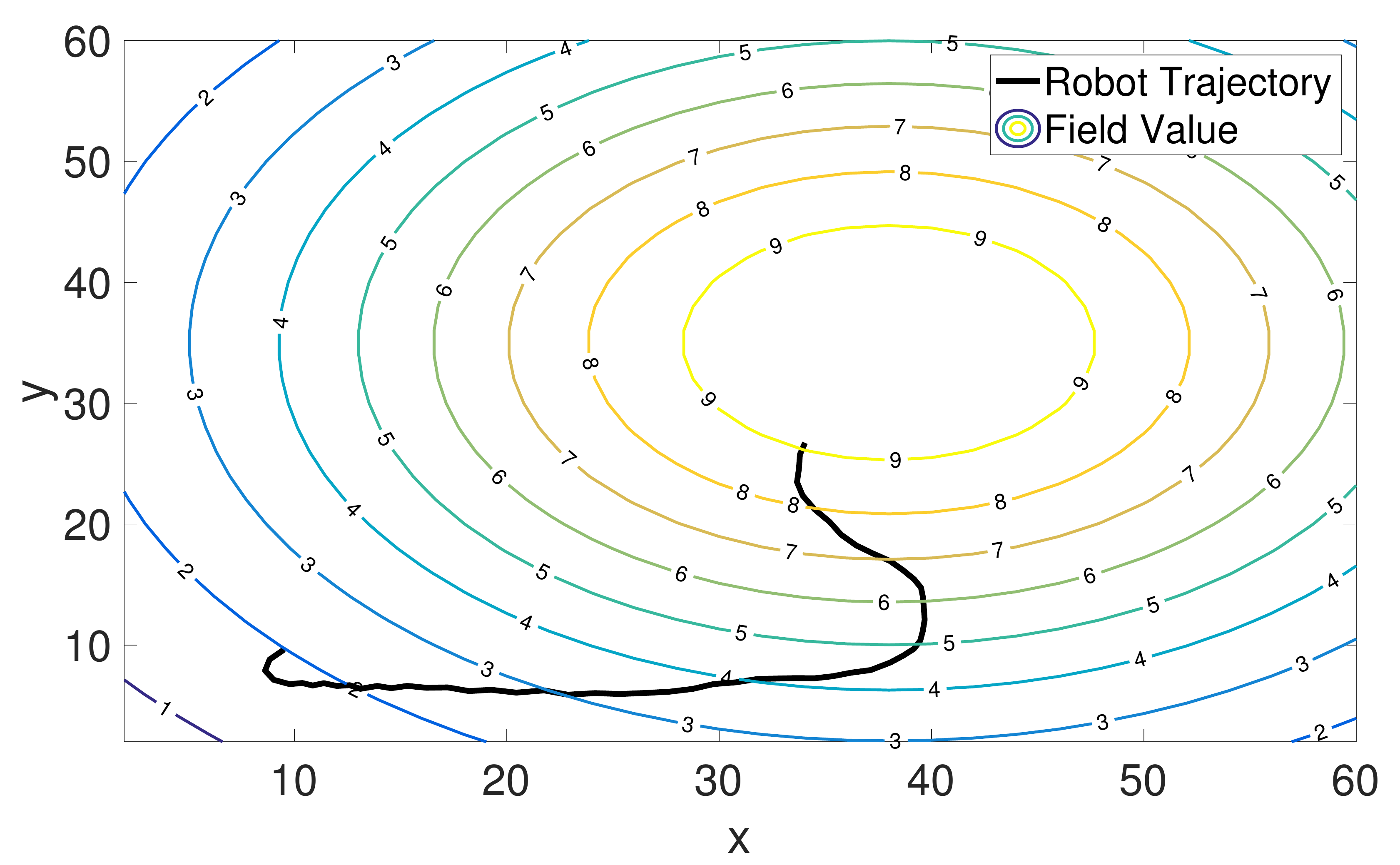}
   \label{fig::esa:Pr0}}
    \subfigure[Algorithm in \cite{GBcomp} with zero noise]{
  \includegraphics[width=0.48\textwidth]{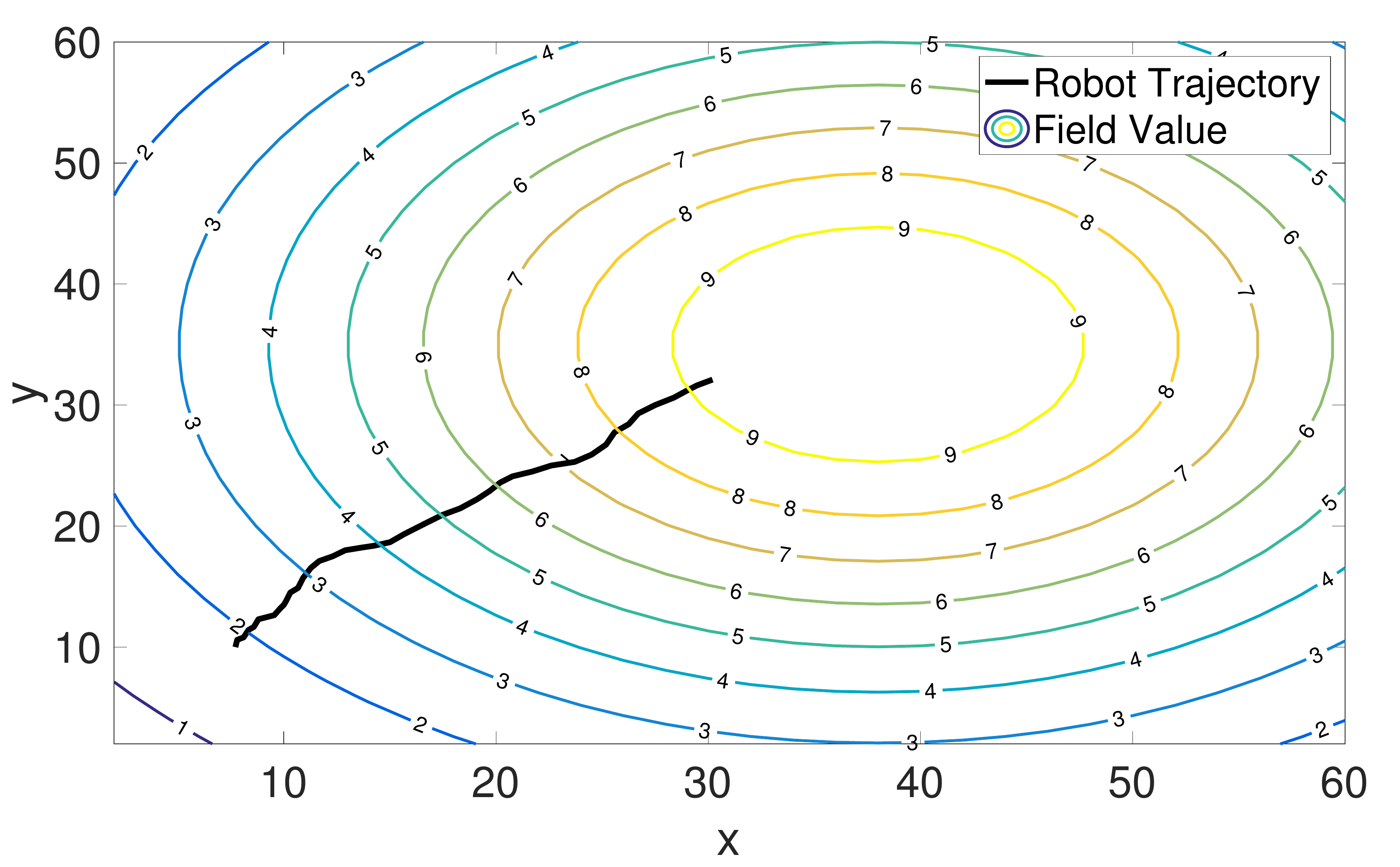}
   \label{fig::esa:Com0}}
   \subfigure[Proposed algorithm with SNR=20dB]{
  \includegraphics[width=0.48\textwidth]{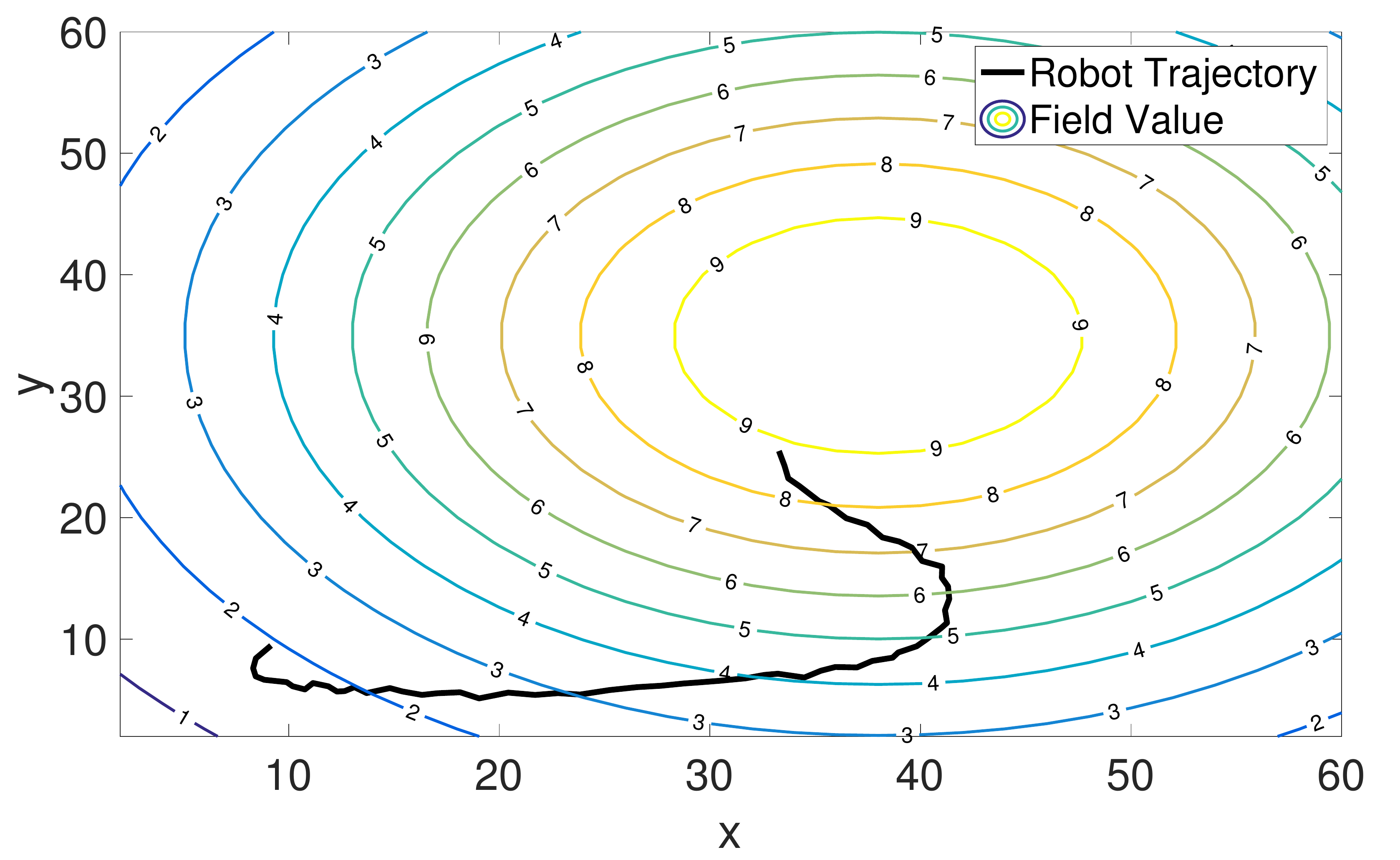}
     \label{fig::esa:Pr20}}
 \subfigure[Algorithm in \cite{GBcomp} with SNR=20dB]{
  \includegraphics[width=0.48\textwidth]{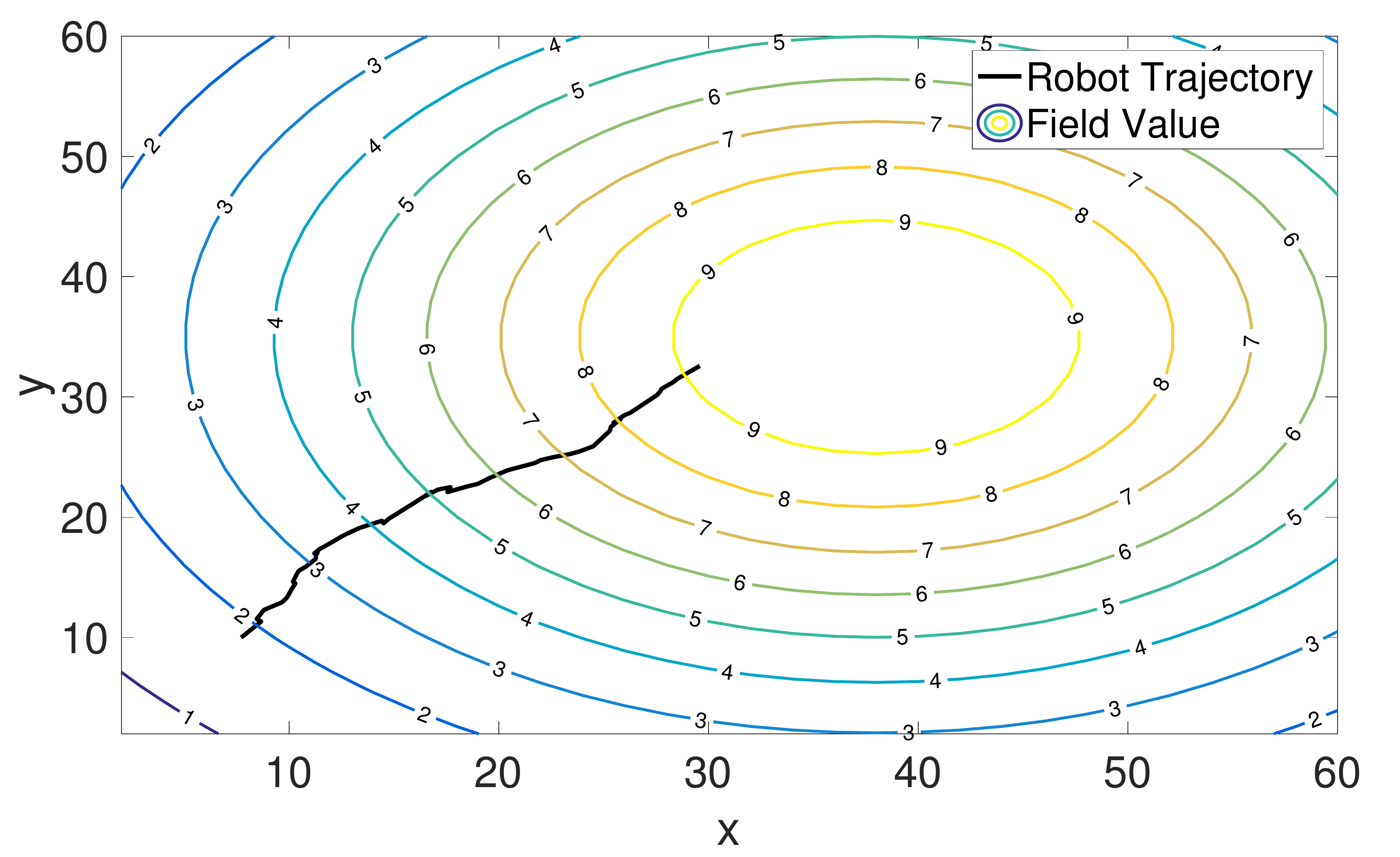}
     \label{fig::esa:Com20}}
     \subfigure[Proposed algorithm with SNR=10dB]{
  \includegraphics[width=0.48\textwidth]{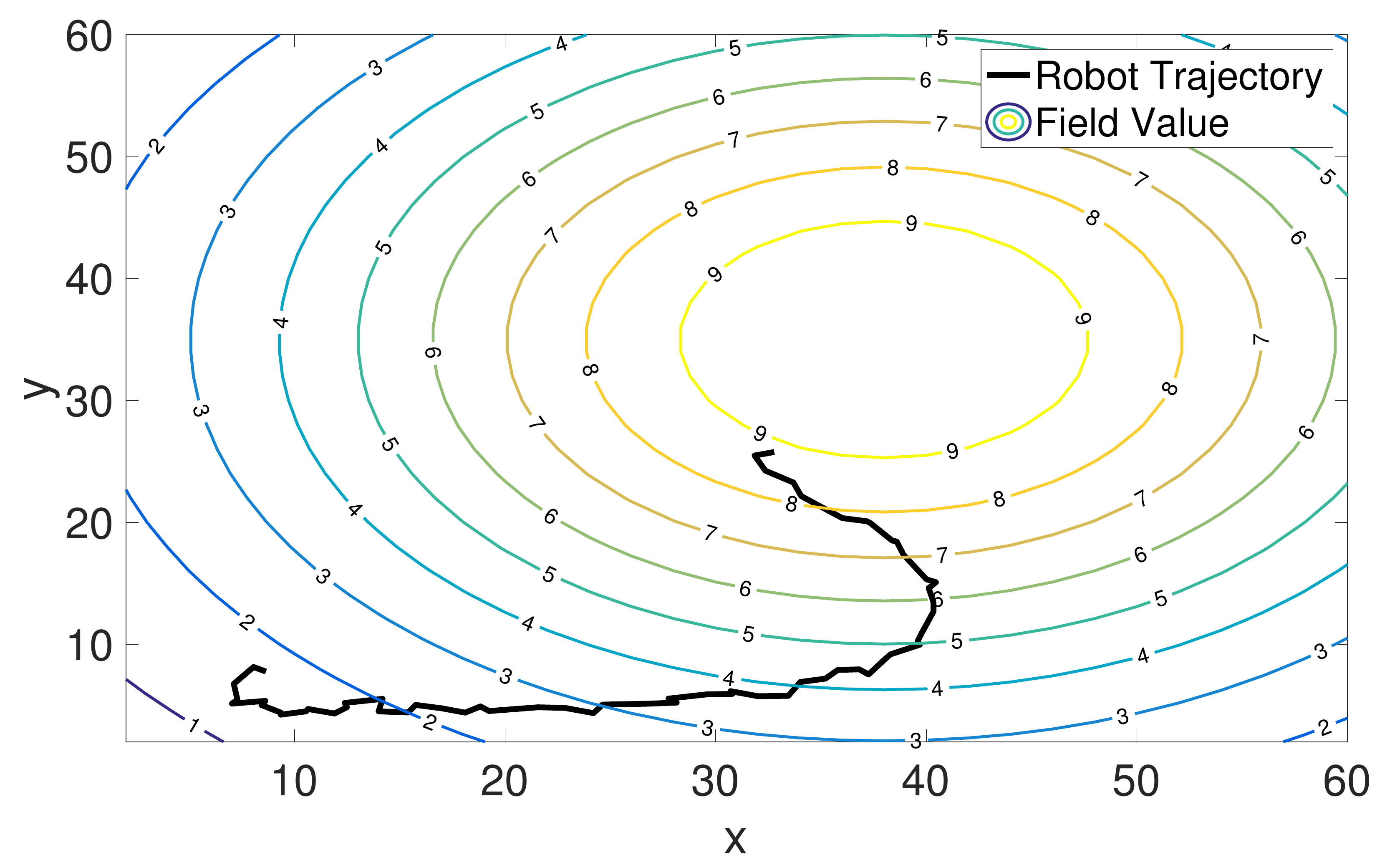}
     \label{fig::esa:Pr10}  }
  \subfigure[Algorithm in \cite{GBcomp} with SNR=10dB]{
  \includegraphics[width=0.48\textwidth]{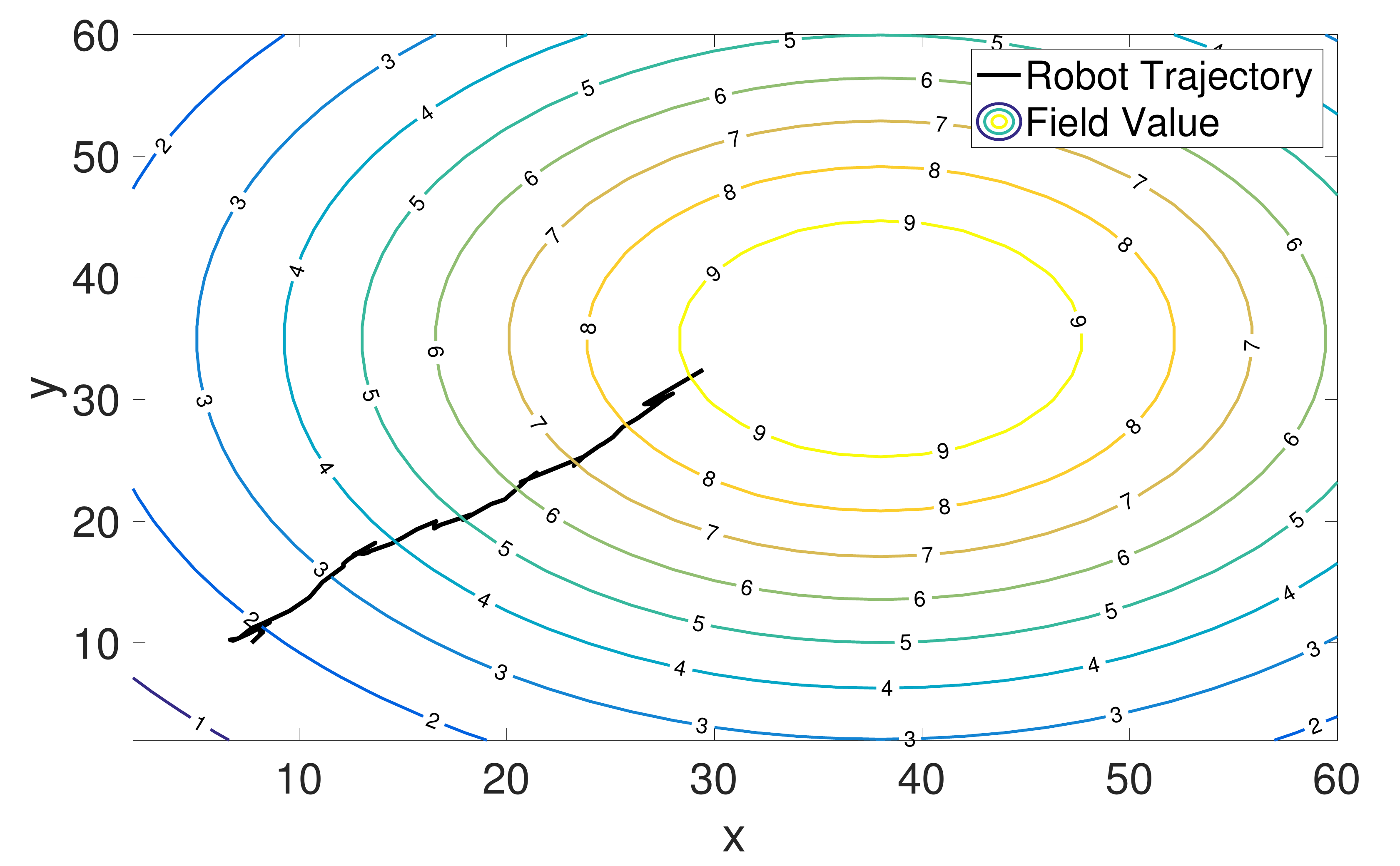}
   \label{fig::esa:Com10}}
  \caption{ Robot trajectory comparison for different noise levels} \label{fig::esa:comparsn}
\end{figure*}

\begin{figure}
 \centering
  \includegraphics[width=0.65\textwidth]{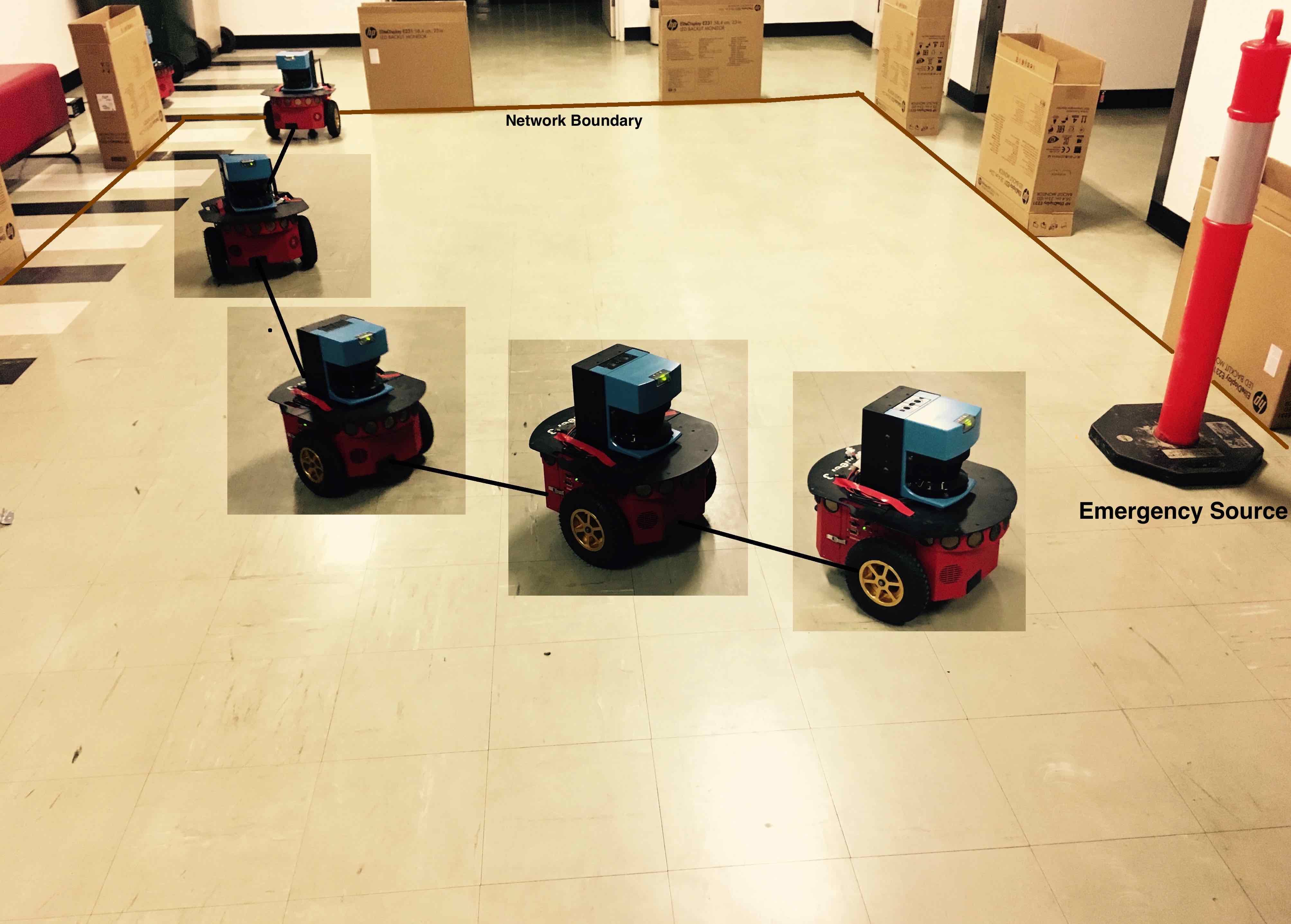}
  \caption{ Robot trajectory in experimental environment with zero measurement error} \label{fig::esa:practicl}
\end{figure}

Finally, the proposed algorithm is compared with an existing gradient based algorithm proposed in \cite{GBcomp}. In this algorithm, they have considered as each node is equipped with sensors measuring their relative signal strength and position. Then, it estimates the gradient directions using neighbours relative signal strength and position. In the comparison, first the algorithms are compared with accurate sensor readings i.e. $e_i=0$. Then, the SNR ratio increased upto 20 dB and 10 dB. The results of three cases are shown in Figure \ref{fig::esa:comparsn}. The field distribution function used for the comparison is described in Scenario 1. The robot speed is considered as $V_r = 1 m/s$ and starts the search at (9,10). Figure \ref{fig::esa:Pr0} and Figure \ref{fig::esa:Com0} show the robot trajectories of the proposed algorithm and gradient based algorithm proposed in \cite{GBcomp} when $e_i=0$. In this case, the proposed algorithm requires 55s to reach to the desired level, but gradient based algorithm requires 44s. However, compared to gradient based algorithm, the proposed algorithm has advantages such as no need of any special sensors to find position or relative signal strength, and less computation cost due to the gradient free approach. Then, as the second case, the SNR was increased to 20 dB. The results for the proposed algorithm and gradient based algorithm are shown in Figure \ref{fig::esa:Pr20} and Figure \ref{fig::esa:Com20} respectively. The proposed algorithm takes 59s to reach to the desired level and gradient based algorithm needs 54s, which is almost same as the proposed method. To consider a noisy environment, the SNR ratio was increased to 10 dB. Figure \ref{fig::esa:Pr10} and Figure \ref{fig::esa:Com10} show the robot trajectories of proposed algorithm and gradient based algorithm when SNR=10dB. The robot takes 63s to reach the desired level using the proposed control algorithm and 68s using gradient based algorithm. Hence, it can be seen that when noise is high in the environment, the performance of gradient based algorithm is less compared to our proposed algorithm. Thus it can be concluded that the proposed algorithm performs well in noisy environments and also it has above mentioned advantages over the gradient based approach. 
\begin{figure*}
 \centering
    \subfigure[Robot trajectory with zero noise]{
  \includegraphics[width=0.45\textwidth]{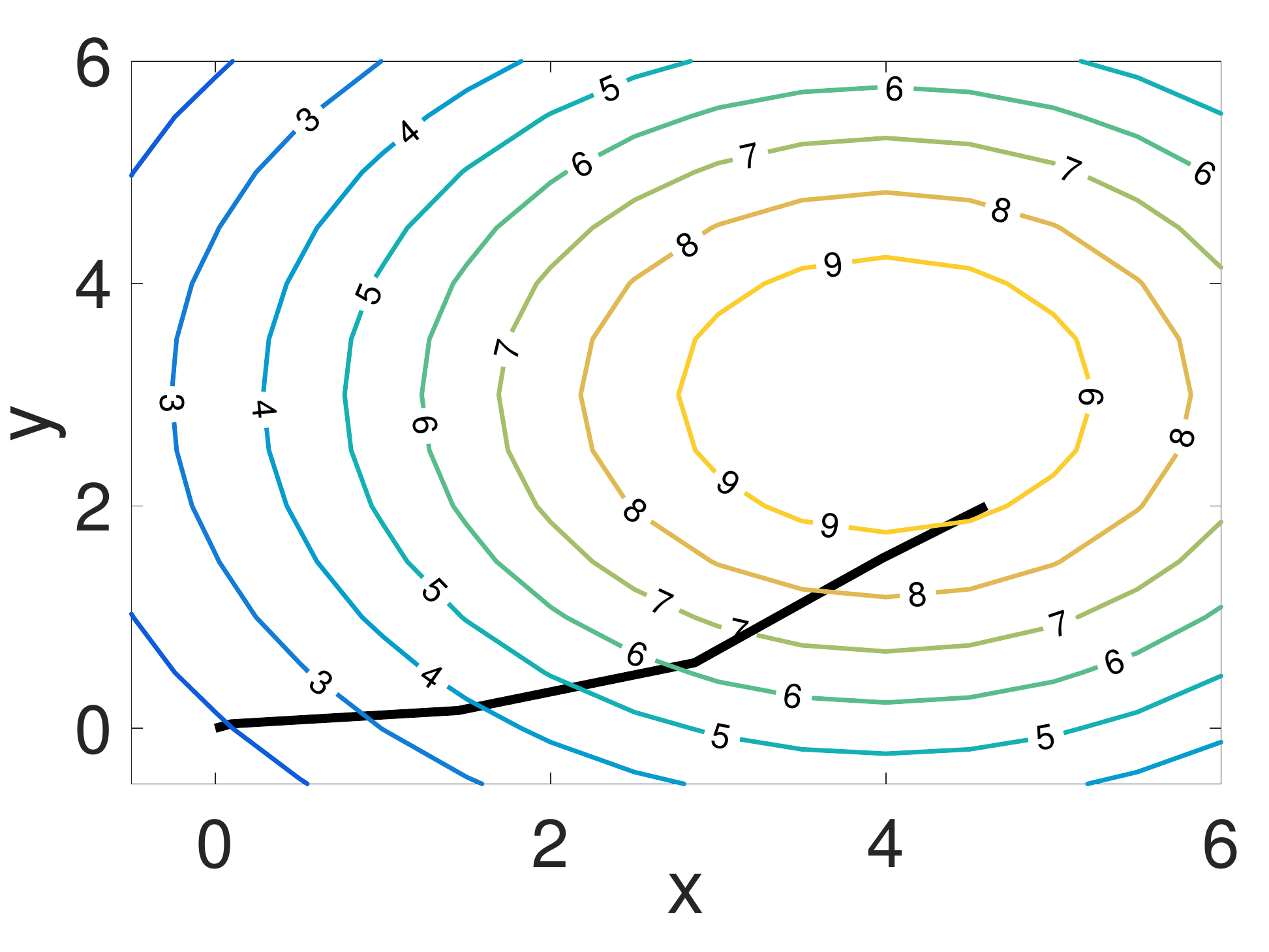}
   \label{fig::esa:Exp0}}
 \subfigure[Robot trajectory with SNR=20dB]{
  \includegraphics[width=0.45\textwidth]{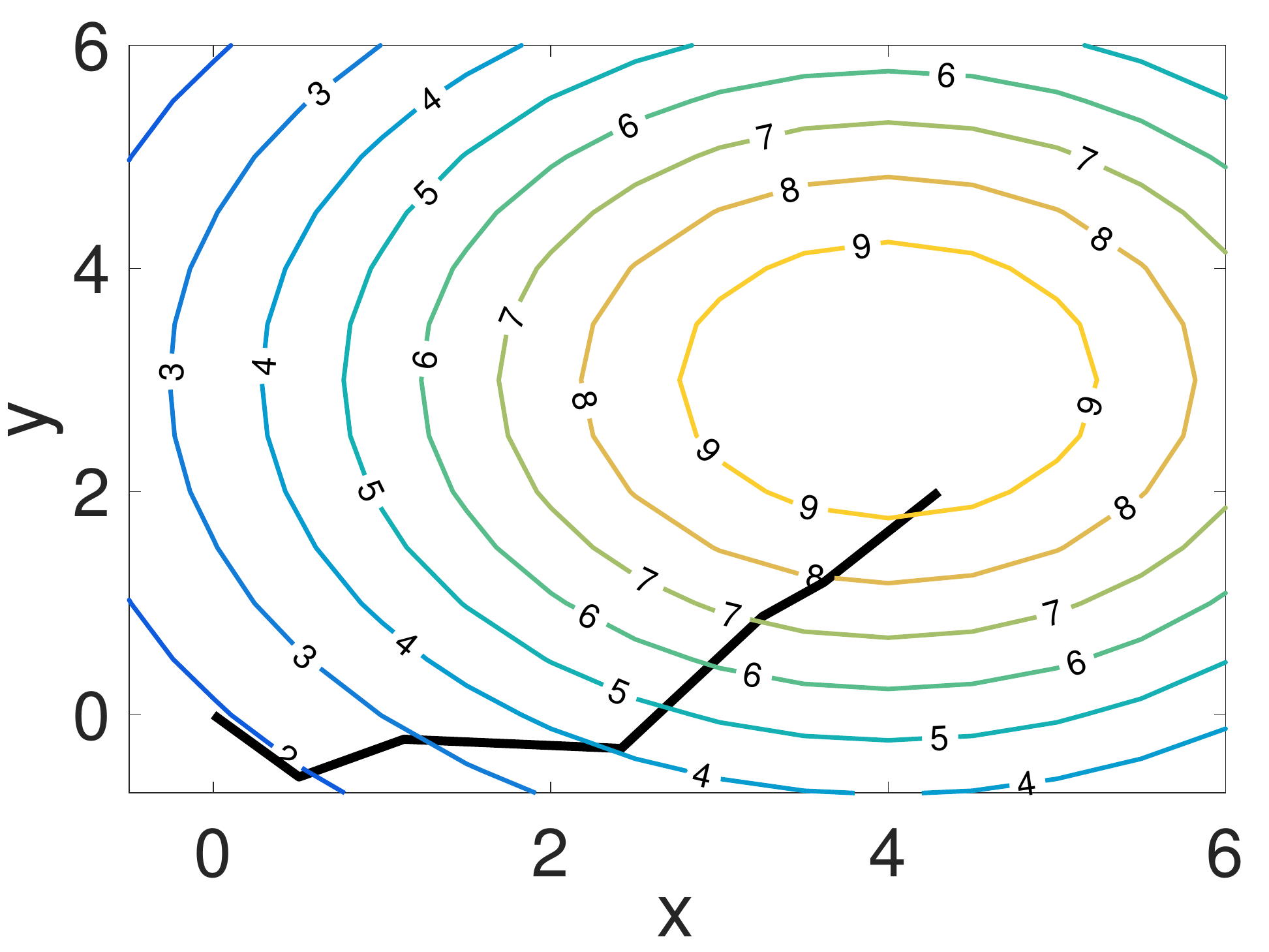}
     \label{fig::esa:Exp20}}
  \subfigure[Robot trajectory with SNR=10dB]{
  \includegraphics[width=0.45\textwidth]{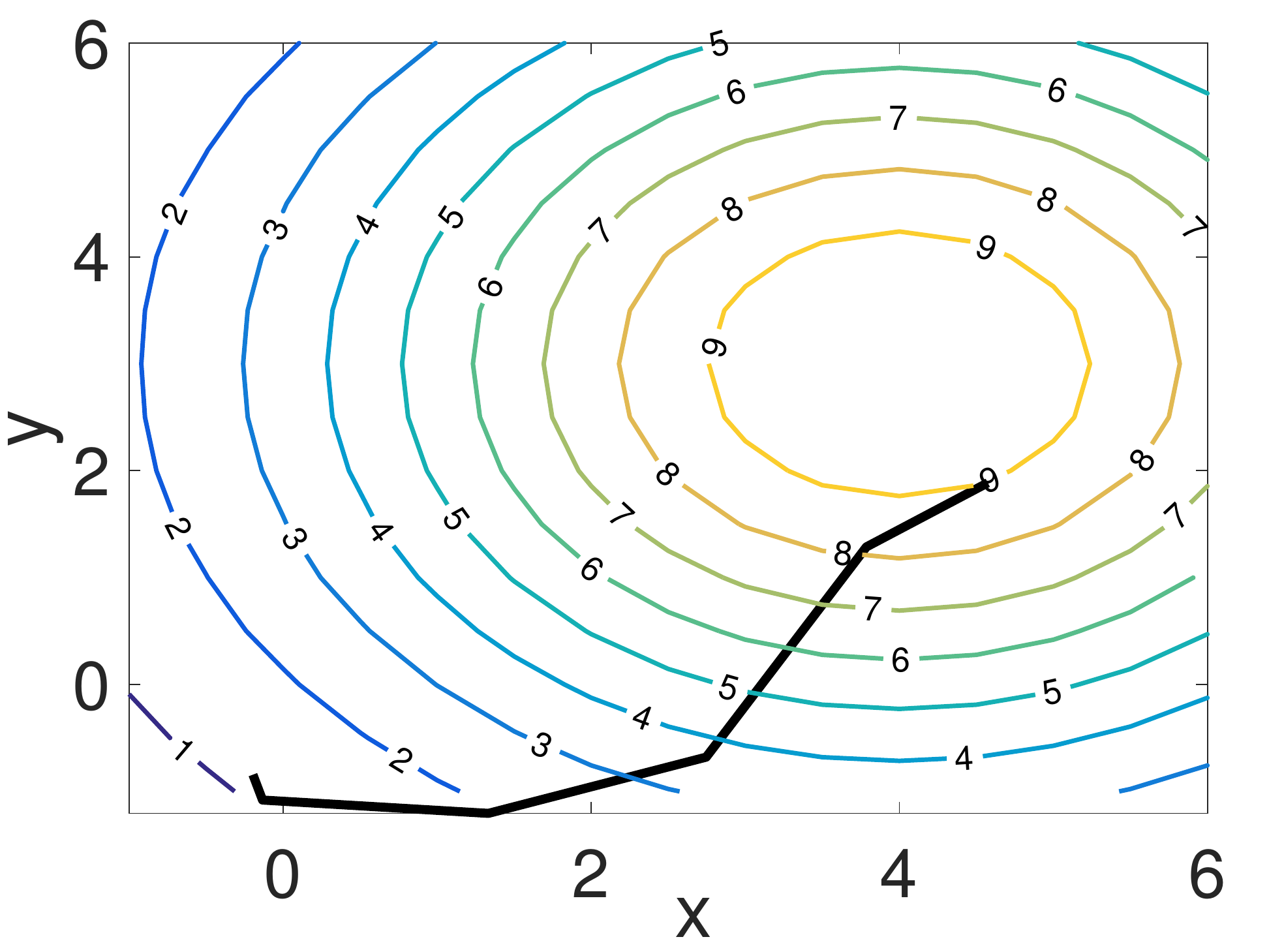}
    \label{fig::esa:Exp10}}
  \caption{ Robot trajectories in experimental environment} \label{fig::esa:Expcomparsn}
\end{figure*}
\subsection{Experimental Results}
To verify the validity and study the performance of the algorithm in real environment,the proposed navigational algorithm was evaluated using Pioneer 3-DX robot. An area of $8m\times 7m$ was used as the experimental network area and field distribution function was considered as $D(x,y)=10e^{-\{(x-4)^2+(y-3)^2\}/15}$. First, a situation with zero measurement error was considered and the robot positions at different time values in real environment are shown in Figure \ref{fig::esa:practicl}. The collected robot positions are then plotted in a graph using Matlab simulation software. This Matlab plot of robot trajectory with the field values are shown in Figure \ref{fig::esa:Exp0}. As the next step, measurement errors were added to the sensor readings and checked the robot navigation path in the real environment. The navigation paths when SNR = 20 dB and SNR = 10 dB is shown in Figure \ref{fig::esa:Exp20} and Figure \ref{fig::esa:Exp10} respectively. From the results it can be seen that the proposed navigational algorithm performs well in noisy environments and that the experiment results are similar to those of the computer simulation results.

\section{Conclusion}\label{sec::esa:conclusion}
This chapter presented a robot navigation algorithm for extremum seeking using WSN topology maps. The robot's control law is gradient-free, and the robot is controlled by a turning radius bounded by a maximum angular velocity. Notably, this control law does not depend on any gradient measurements that cannot directly be taken. In literature, navigational algorithms use physical coordinates of sensors for calculating the robot's coordinates or the gradient. This requires special hardware devices embedded in the sensors to obtain the range measurements such as RSSI and AoA. However, in emergency environments, due to the high level of noise and fading, it is not feasible to measure physical distances accurately using range measurements. Given these points, the proposed algorithm uses the maximum likelihood topology coordinates of sensor nodes that use a packet reception probability function and a packet reception binary matrix for coordinate calculation. 

The performance of the algorithm is evaluated using computer simulation and a real experimental setup. A sensor measurement model and a propagation model equivalent to the actual communication and measurement models are used in the computer simulation to obtain realistic data. The proposed algorithm is compared with a gradient based algorithm proposed in the literature. The results of the comparison reveal that the proposed algorithm performs better than gradient based algorithms in noisy environments. Moreover, the proposed algorithm has advantages over gradient based algorithms. These include the fact that no special sensors are required to find position or relative signal strength, and the lower computation cost of employing a gradient free approach. From the experimental results, it can be concluded that the proposed navigational algorithm performs well in noisy environments and that these results are similar to those of the computer simulation.

\chapter{Conclusion and Future Work}\label{chapter:Conclusion}
This chapter summarises the research discussed in previous chapters. It presents the contributions of this research and possible future extension. Section \ref{sec::Conclusion:summary} presents the summary and conclusions of the contributions made in this dissertation and Section \ref{sec::Conclusion:future} discusses the potential future extensions of the main contributions.

\section{Summary and Conclusion}\label{sec::Conclusion:summary}
This report primarily focuses on proposing a novel topology mapping algorithm to present physical layout information such as network shape and voids or obstacles, more accurately. As discussed in Chapter \ref{chapter:intro}, this is made challenging by wireless communication effects such as noise, fading, and interference, and cost constraints that prevent the incorporation of expensive hardware components such as GPS in large-scale deployments. As a result, a topology map that is fully isomorphic to the physical map of the network has not yet been proposed. To this end, this report presents a topology map that provides a more accurate physical representation of the network by using probability of packet reception and maximum likelihood estimation for coordinate calculation. In doing so, the proposed topology map does not depend on range based parameters that are unacceptable in complex environments containing obstacles and does not solely depend on connectivity information unlike previous topology mapping algorithms proposed in the literature. 

A novel ML-TM algorithm for RF WSNs based on maximum likelihood estimation is developed as the first contribution. The objective of this contribution is to come up with a topology map that is closer to the actual physical map of the network, but without requiring the expenses associated with localization based on actual physical distance measurements. ML-TM is a more accurate map to represent 2D and 3D physical layouts with voids and obstacles compared to existing alternative topology maps. It uses a packet reception probability function sensitive to distance along with a mobile robot that traverses the network to calculate the topology coordinates. This algorithm centrally calculates the topology coordinates, which reduces the computational complexity of the nodes and increases the accuracy of the generated map. Additionally, a one-hop connectivity error parameter is proposed to evaluate the accuracy of topology maps by considering the connectivity of a node in the neighbourhood. The ML-TM algorithm is then evaluated against the recently proposed RSSI localization algorithm and SVD based TPM algorithm. The results indicate that the error percentage is less than 7\% in ML-TM and that it outperformed the other algorithms. Furthermore, in Chapter \ref{chapter:mltm}, this method was demonstrated to be able to capture various network shapes with obstacles under different environmental conditions. Moreover, ML-TM scales seamlessly to 3D-WSNs, thus enabling its use in networks consisting of both 3D volumes and 2D surfaces. Therefore, it can be used as an alternative to geographical maps in the automation of sensor network protocol.

As the second contribution of this report, the proposed topology mapping algorithms is modified for a MmWave sensor network. Applications that demand WSN bandwidth, such as habitat monitoring, medical and smart city applications, are now moving from the RF frequency band to the MmWave frequency band. For this reason, it is necessary to develop a topology map for MmWave WSN by utilizing the MmWave communication characteristics. One of the major differences in MmWave WSN is the use of narrow beamwidth antenna arrays instead of omni antennas. The topology map of a MmWave sensor network needs to preserve the connectivity as well as the directionality of nodes. Accordingly, an MmTM algorithm is presented in Chapter \ref{chapter:mmtm} to estimate maximum likelihood topology coordinates for 3D MmWave WSNs. The coordinates are estimated by utilizing the sector antennas used in MmWave communication. As in the ML-TM calculation, an automated mobile robot is used to extract information from sensor nodes. The robot also keeps track of the packet reception from sensor nodes along with the IDs of the best sectors to communicate with. The collected information is then mapped to a coordinate system using a signal receiving probability function. In this portion of the research, three robot movements were simulated- namely a 2D robot path with VAA, a 2D robot path with VBS and a 3D robot path to generate an accurate topology map. As MmTM requires preserving the directivity information in the estimated topology map, a novel sector displacement metric is proposed. The MmTM algorithm is then compared with two existing algorithms, DR-MDS and NTLDV-HOP. The results indicate that the MmTM algorithm outperformed these algorithms in both connectivity and directivity. The CDF of distance error shows that the distance error of all nodes in MmTM is less than 5\% of the distance error in NTLDV-HOP and DR-MDS. Moreover, more than 35\% of the nodes have having zero sector displacement compared to the other two algorithms.

This centralized topology coordinate calculation is modified to a distributed coordinate calculation as the third contribution of this report. Although centralized coordinate calculation reduces the computation complexity in sensor nodes, it requires more time. The objective is to calculate the topology coordinates of the sensor nodes accurately in a distributed way using the same packet reception probability function as that used in centralized coordinate calculation. In Chapter \ref{chapter:dmmtm}, a DMmTM algorithm is presented, which distributively calculates a topology map, closely resembling the physical layout. The calculation of DMmTM was executed in two ways: DMmTM-SS for networks with static anchors, and DMmTM-HS for networks with  static and mobile anchors. Both static and mobile anchors are aware of their locations, but static anchors are unaware of their beam direction information. As the DMmTM algorithm calculates topology coordinates at each sensor, it is more efficient than the two proposed topology mapping algorithms. Furthermore, initially deployed anchors or mobile anchors in DMmTM do not require access to all the sensors in the network because they select new anchors from sensor nodes after initial coordinate calculation. Thus, the proposed algorithm does not require careful mobile anchor path planning or anchor distribution, which pose significant challenges. Finally, the performance of DMmTM was compared with three algorithms- MmTM, DR-MDS, and NTLDV-HOP. The performance results demonstrate that the proposed algorithms localize with distance errors and displacement errors similarly to MmTM. However, the two algorithms are significantly more efficient than MmTM, yet, unlike MmTM they do not require each sensor node to be directly accessible by an anchor. Furthermore, the two algorithms outperformed NTLDV-HOP and DR-MDS in both performance metrics. 

Next, as the fourth contribution, a target searching and prediction algorithm with a WSN topology map is proposed as an application of the proposed topology map. The existing research done in WSN target tracking is based on sensors' physical coordinates and distances. The topology coordinate space is a robust alternative to physical coordinates, and it contains significant non-linear distortions when compared to physical distances between nodes. Because of this, the objective is to come up with an algorithm that uses topology coordinates of sensors to search for a target moving in the network. Chapter \ref{chapter:DeTarSK} presents a novel algorithm, DeTarSK, to track and predict target locations in environments where it is not feasible to accurately estimate physical distances. In such environments, where measuring physical distances is not feasible, time stamps corresponding to target detection by sensor nodes are used to track the target. DeTarSK is based on decentralized robust Kalman filtering and a non-linear least square method. It also eliminates errors arising from distortion of the topology coordinate domain compared to the physical domain. Other advantages include real time decision making with lower traffic and energy consumption in the network. To address the drawbacks of decentralized algorithms, the target's future locations are predicted using its past behaviours. Moreover, DeTarSK demonstrated better performance than the recently proposed P-G algorithm, and it can be used in unknown environments that contain obstacles in the search path. 

As the final contribution of this report, a sensor based extremum seeking algorithm is proposed in Chapter \ref{chapter:esa}. This algorithm uses a topology map to navigate a mobile robot towards the extrema. Similarly to the target tracking algorithm, this algorithm does not depend on any distance-based information, given the non-linear distortions of distances in topology maps. The control law of the robot is gradient-free and controlled by a turning radius bounded by a maximum angular velocity. Additionally, the performance of the algorithm is evaluated using computer simulation and a real experimental setup with a Pioneer 3-DX ground robot. A sensor measurement model and a propagation model equivalent to the actual communication and measurement models are used in the computer simulation to obtain more realistic data. The proposed algorithm is compared with a gradient based algorithm proposed in the literature. The results of the comparison reveal that the proposed algorithm performed better than gradient based algorithms in noisy environments. The proposed algorithm has further advantages over gradient based algorithms. These include the fact that no special sensors are required to find position or relative signal strength, and the lower computation cost of employing a gradient free approach. From the experimental results, it can be concluded that the proposed navigational algorithm performed well in noisy environments and that these results are similar to that of the computer simulation.

\section{Future Work}\label{sec::Conclusion:future}

This section discusses important expansions to the research presented in this report. The topology map calculations presented in Chapter \ref{chapter:mltm} and Chapter \ref{chapter:mmtm} use a single robot to traverse the network to record information gathered by sensor nodes. When the network size increases, the time required to gather information from sensor nodes also increases as the robot needs to covers a larger network area. To reduce the time required to cover the network and to increase scalability, network partitioning can be implemented. Several methods have been proposed in the literature to partition a sensor network \cite{networkpartition, networkpartition1}. By using one of those methods, a sensor network can be partitioned into two or more partitions, with a single robot being allocated to each partition to gather information from sensor nodes. Calculating the topology map can then be executed in two ways. The first is a distributed topology map calculation, which calculates the topology map of each partition and then stitches the topology maps of the partitions together. The second way is a central topology calculation in which sensor information from all the partitions is collected in one location, after which the topology map calculation is executed. 

Another possible extension to increase the accuracy of topology mapping algorithms is to use more than one robot to traverse the network and scan the same area. Subsequently, the best possible packet reception matrix can be obtained using similarities and non-similarities in packet reception matrices recorded by all the robots.

In the topology map calculation for MmWave WSNs, the optimum sectors are chosen using the SLS phase of the IEEE 802.11ad standard protocol. However, in reality, this selection will be affected by multipath effects in wireless communication, which should to be addressed in future research \cite{ mmMultipath}. One way to address this is to use beam refinement protocols proposed in the IEEE 802.11ad protocol, which allow more accurate calculation of node directivity by eliminating the multipath effect \cite{11ad}. However, the problem that arises with using these complex beam training protocols is high energy consumption in sensor nodes as these refinement protocols require greater energy \cite{ nit14}. Therefore, optimizing the sensor transceiver to use such protocols should also be considered \cite{mmEnergyO}. 

Moreover, it is possible that some or all of the nodes in the network are mobile \cite{ mobilewsn}. Hence, making topology map calculation based on maximum likelihood estimation adaptive to such environments should be addressed in the future. Furthermore, as the DMmTM calculation proposed in Chapter \ref{chapter:dmmtm} uses anchor nodes to generate the topology map, an optimised protocol for anchor node distribution can be considered in future as an extension to this research \cite{ anchorplacement}. 

Moving further, topology map calculation for a smart dust sensor network can be considered \cite{ smartdust}. Smart dust sensor nodes can be mobile and low in power. Thus, the complexity of the algorithm need to be reduced. The current topology map calculation considers only the information gathered by the sensors. However, the accuracy of the topology map can be increased by considering network properties such as voids or obstacles, along with node connectivity. These physical properties can be recorded by the robot while moving through the network to gather node connectivity information from sensors.

Real world sensor network environments contain obstacles that can be steady (for example buildings and trees) or moving (for example humans and vehicles). Avoiding collisions with these obstacles is one of the key components in robot navigation. In this case, combining the extremum seeking algorithm proposed in Chapter \ref{chapter:esa} with obstacle avoidance techniques would be another direction for future research. Existing collision avoidance navigation approaches can be generally classified into two groups: global and local.  As global path planning algorithms use a priori information about the environment to find the safest path to the extrema, the problem of avoiding collisions with moving obstacles is harder to handle. On the other hand, local path-planning algorithms use real-time sensory data for optimum safest path calculation, which enables their use in time-changing environments. Thus, the proposed algorithm in Chapter \ref{chapter:esa} can be combined with one of the local obstacle avoidance techniques proposed in \cite{newA23, newA24, newA25}. 

Additionally, the sensor network based robot navigation algorithm for source seeking can be developed to be more efficient using the advanced communication and limited control methods of \cite{newA26, newA27}, as another direction for future research. Many real world robot navigation applications are in 3D space. However, moving from planar space to 3D space is complex and challenging. In contrast, future research can involve extending the algorithm to a 3D environment and obtaining the required results. In particular, the technique in \cite{ newA28} can be used to extend the proposed algorithm to an extremum seeking navigational algorithm for a 3D environmental.  

Another direction for future research can be to apply the approach of this report to various problems of coverage control. By using maximum likelihood coordinate systems for robotic sensor networks, the efficiency of coverage control in problems of barrier coverage, sweep coverage and blanket coverage studied in \cite{ newA29, newA30} can be improved. 

\addtocontents{toc}{\vspace{2em}} 

\appendix 

\chapter{Localization Techniques}\label{LT}
There are three most popular location calculation techniques used in sensor localization namely, trilateration, triangulation and maximum likelihood estimation. Those three techniques are discussed in following sub sections.

\section{Trilateration}
\begin{figure}[t]
  \centering
    \includegraphics[width=.7\textwidth]{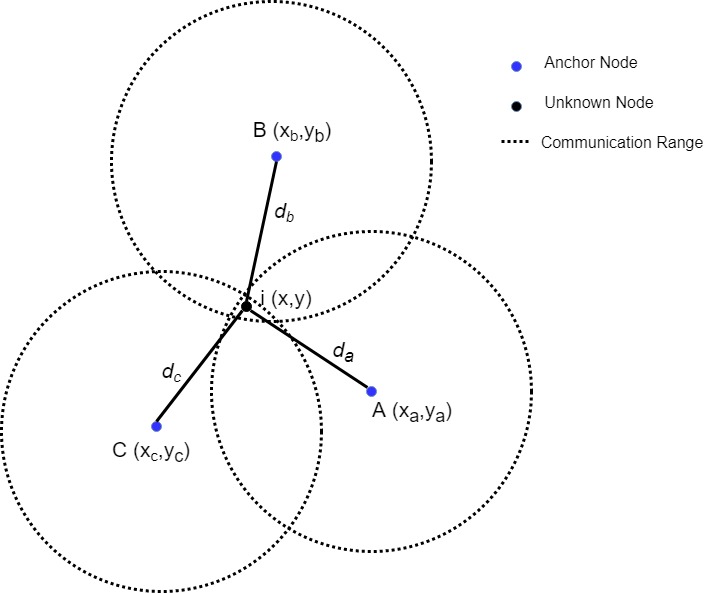} 
    \caption{An Illustration of Trilateration}
    \label{fig::LT:trilateration}
\end{figure}

Trilateration technique finds the location of a unknown node using the distance to anchor nodes located within its communication range. In 2D space, this technique requires at least three anchor nodes as shown in Figure \ref{fig::LT:trilateration} and four anchor nodes in 3D space. For the simplicity, a 2D scenario is considered to explain the calculation. One of the drawbacks in this technique is that it relies on an accurate distance measurement to determine the position of a sensor node. 

Let $A$, $B$ and $C$ are anchor nodes and $i$ is a location unknown node. The coordinates of $i$ is $(x,y)$. $(x_a,y_a)$,$(x_b,y_b)$ and $(x_c,y_c)$ are the coordinates of $A$, $B$, and $C$ respectively. The distance from node $i$ to $A$, $B$, $C$ are $d_a$, $d_b$ and $d_ac$. Then following geometric constraints can be obtained.
\begin{eqnarray}\label{eqn::LT:trilateration}
\sqrt{(x-x_a)^2+(y-y_a)^2}&=&d_a \nonumber \\ 
\sqrt{(x-x_b)^2+(y-y_b)^2}&=&d_b  \nonumber\\
\sqrt{(x-x_c)^2+(y-y_c)^2}&=&d_c 
\end{eqnarray}

By solving equation \ref{eqn::LT:trilateration}, the equation \ref{eqn::LT:trilateration2} can be obtained to calculate the coordinates of unknown node $i$.
\begin{equation}\label{eqn::LT:trilateration2}
PX=Q
\end{equation}
where, 
\[
X=
\begin{bmatrix}
    x \\
    y 
\end{bmatrix}
,P=2
\begin{bmatrix}
    (x_a-x_c)&(y_a-y_c)\\
    (x_b-x_c)&(y_b-y_c)
\end{bmatrix}
,Q=
\begin{bmatrix}
    {x_a}^2-{x_c}^2+{y_a}^2-{y_c}^2+{d_c}^2-{d_a}^2\\
    {x_b}^2-{x_c}^2+{y_b}^2-{y_c}^2+{d_c}^2-{d_b}^2
\end{bmatrix}
\]

\subsection{Multilateration}\label{ssec::LT:multitrilateration}
If the location of an unknown node is estimated using more than three anchor nodes in 2D space, it is called as multilateration. This produces better results than trilateration in the presence of erroneous distance measurements. In other words, when more than three anchors nodes are used, it results an over determined system of equations. Thus, by solving this linear system, the mean square error in the calculation can be minimized. This leads to produce a better result than trilateration.

Let, the coordinate of an unknown node $i$ is $(x,y)$ and $n$ number of anchor nodes are located within its neighbourhood. The coordinates of $n$ number of anchor nodes are $(x_1,y_1)$, $(x_2,y_2)$,..., $(x_n,y_n)$ and distance to those nodes are $d_1$, $d_2$,...,$d_n$ respectively. Then equation \ref{eqn::LT:trilateration} can be rewritten as,
\begin{eqnarray}\label{eqn::LT:multitrilateration}
\sqrt{(x-x_1)^2+(y-y_1)^2}&=&d_1 \nonumber \\ 
\sqrt{(x-x_2)^2+(y-y_2)^2}&=&d_2  \nonumber\\
\vdots\\
\sqrt{(x-x_n)^2+(y-y_n)^2}&=&d_n 
\end{eqnarray}

Then, $P$ and $Q$ matrices in equation \ref{eqn::LT:trilateration2} can be rewritten as,
\[
P=2
\begin{bmatrix}
    (x_1-x_n)&(y_1-y_n)\\
    (x_2-x_n)&(y_2-y_n)\\
    \vdots&\vdots\\
    (x_{n-1}-x_n)&(y_{n-1}-y_n)
\end{bmatrix}
\]
\[
Q=
\begin{bmatrix}
    {x_1}^2-{x_n}^2+{y_1}^2-{y_n}^2+{d_n}^2-{d_1}^2\\
    {x_2}^2-{x_n}^2+{y_2}^2-{y_n}^2+{d_n}^2-{d_2}^2\\
    \vdots\\
    {x_{n-1}}^2-{x_n}^2+{y_{n-1}}^2-{y_n}^2+{d_n}^2-{d_{n-1}}^2
\end{bmatrix}
\]

\section{Triangulation}
\begin{figure}
  \centering
    \includegraphics[width=.7\textwidth]{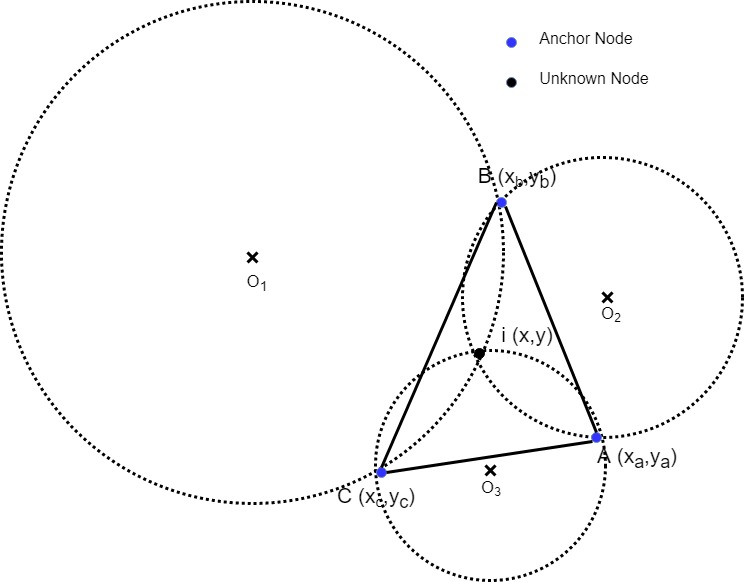} 
    \caption{An Illustration of Triangulation}
    \label{fig::LT:Triangulation}
\end{figure}

In triangulation, the position of a location unknown node is calculated based on the angular distance between three different pairs of anchors, measured from the unknown node.
Figure \ref{fig::LT:Triangulation} illustrates an example of triangulation calculation. Let $A$, $B$ and $C$ are anchor nodes and $i$ is an unknown node. The coordinates of $i$ is $(x,y)$. $(x_a,y_a)$,$(x_b,y_b)$ and $(x_c,y_c)$ are the coordinates of $A$, $B$, and $C$ respectively. The angles between the line segments connecting unknown and anchors are $\measuredangle AiB$, $\measuredangle AiC$ and $\measuredangle BiC$. Furthermore, unknown node $i$ is located at the intersection of the three (imaginary) circles centred at $O_1$,$O_2$ and $O_3$. If the angular distances are known, then centre of the circles can be obtained. 

First consider the anchor nodes $B$, $C$ and the angle $\measuredangle BiC$. If the arc $BC$ is within the scope of $\vartriangle ABC$, the circle centred at $O_1=(x_{o_1},y_{o_1})$ with a radius of $r_1$ can be uniquely identified. The $BC$ major arc subtends a central angle of 2$\measuredangle BiC$. Hence, the $BC$ minor arc subtends a central angle $\measuredangle BO_1C=2(\pi - \measuredangle BiC)$. Then the centre $O_1$ and the radius $r_1$ of the circle can be calculated as,
\begin{eqnarray}
\sqrt{(x_{o_1}-x_b)^2+(y_{o_1}-y_c)^2}&=&r_1 \nonumber \\ 
\sqrt{(x_{o_1}-x_c)^2+(y_{o_1}-y_c)^2}&=&r_1 \nonumber \\
(x_b-x_c)^2+(y_b-y_c)^2 &=&2r_1^2-2r_1^2cos(\measuredangle BO_1C)
\end{eqnarray}

Similarly, $O_2$, $r_2$ and $O_3$, $r_3$ can be calculated using angle measurements $\measuredangle BiA$ and $\measuredangle AiC$ respectively.

\section{Maximum Likelihood Estimation}
Maximum likelihood based methods finds the position of unknown nodes based on a likelihood function $\mathcal{L}(x,y)$. In other words, the value $(x,y)$ that maximize the likelihood function is selected as the coordinate of unknown sensor node. This likelihood function $\mathcal{L}(x,y)$ can be any function that defines by a localization algorithm. Furthermore, the maximum likelihood estimation assumes that measurement values from different anchor nodes are independent from each other. Thus, the likelihood function can be written as,
\begin{equation}
\mathcal{L}(x,y) = \prod _{i=1}^{n} f(d_i)
\end{equation}
where $n$ is the number of measurements and $f(d_i)$ is a function that depends on the measurement value $d_i$. By solving this equation to have a maximum value for $\mathcal{L}(x,y)$ the coordinates $(x,y)$ can be obtained. For an example, in multilateration discussed in Section \ref{ssec::LT:multitrilateration}, the maximum likelihood coordinate estimation for the unknown node can be obtained as $X=(P^TP)^{-1}P^TQ$. This likelihood function is derived using least square method to reduce the total distance error to the anchor nodes with respect to the calculated location of the unknown node. 
\chapter{Communication Protocols}\label{CP}
\section{IEEE 802.15.4 Protocol for RF Communication}
The IEEE 802.15.4 protocol is designed for applications that require low data throughout and have limited resources of power and computation capability. The main aim of this protocol is to overcome the drawbacks associated with the existing standards such as WiFi and Bluetooth \cite{rfcomprotocol2}. The IEEE 802.15.4 protocol defines the specification of the physical and MAC layers and supports the network topologies such as peer-to-peer and star topologies.

There are two types of devices in this protocol, namely, full-function device (FFD) and reduced function device (RFD). An FFD has the full functions of protocol stack. Thus it can initiate and manage the whole network by functioning as a coordinator. On the other hand, it can become a normal device as well. An RFD is a device, which has the basic functions of the stack to execute extremely simple tasks. Thus the objective of RFD is to regularly send sensor readings to the user.

There are four types of frames used in the IEEE 802.15.4 protocol. Those are
\begin{enumerate}
\item \textbf{Beacon frame:} Used by a coordinator to start a communication or synchronize with other devices.
\item \textbf{Acknowledgement frame:} Used for confirm successful frame reception.
\item \textbf{Data frame:} Used for all data communication.
\item \textbf{Command frame:} Used for handle all peer entity control transfers.
\end{enumerate}

Moreoever, there two data transmission types in this protocol. First is a Beacon-disable networks that FFD coordinator does not send beacons to synchronize with RFDs. These networks use unslotted CSMA-CA channel access mechanism. Second is a Beacon-enable networks that FFD coordinator sends beacons periodically to synchronize nodes that communicate with it and to define a super frame in which all transmissions must occur. The super frame is bounded by network beacons sent by the coordinator and is divided into 16 equally sized slots \cite{rfcomprotocol}. These networks use slotted CSMA-CA channel access mechanism, where the backoff slots are aligned with the start of the beacon transmission. 

\section{IEEE 802.11ad Protocol for MmWave Communication}
To reduce the majority of signal propagation issues, narrow beamwidth antenna arrays are used in MmWave communication. Therefore, adaptive beamforming has become an essential aspect of MmWave communication that determines the pair of antenna sectors with highest signal quality between the transmitter and receiver. In addition, practical feasibility of adaptive beamforming has been demonstrated through the recent indoor MmWave (60 Ghz) communication standard IEEE 802.11ad for WLAN. The beamforming protocol with regards to IEEE 802.11ad is explained in the following subsections.

\subsection{Overview of beamforming in IEEE 802.11ad}
\label{BF}

The complete process of beamforming contains three steps: i) Sector Level Sweep phase (SLS), ii) Beam Refinement Protocol phase (BRP), and iii) Beam Tracking phase (BT). SLS phase identifies course-grained sector pair and optional BRP and BT phases further refine the selection. Here the mandatory SLS phase is focused as it provides necessary background for the development of proposed topology mapping algorithm in this report.

\begin{figure}
\centering
\includegraphics [width=.7\textwidth]{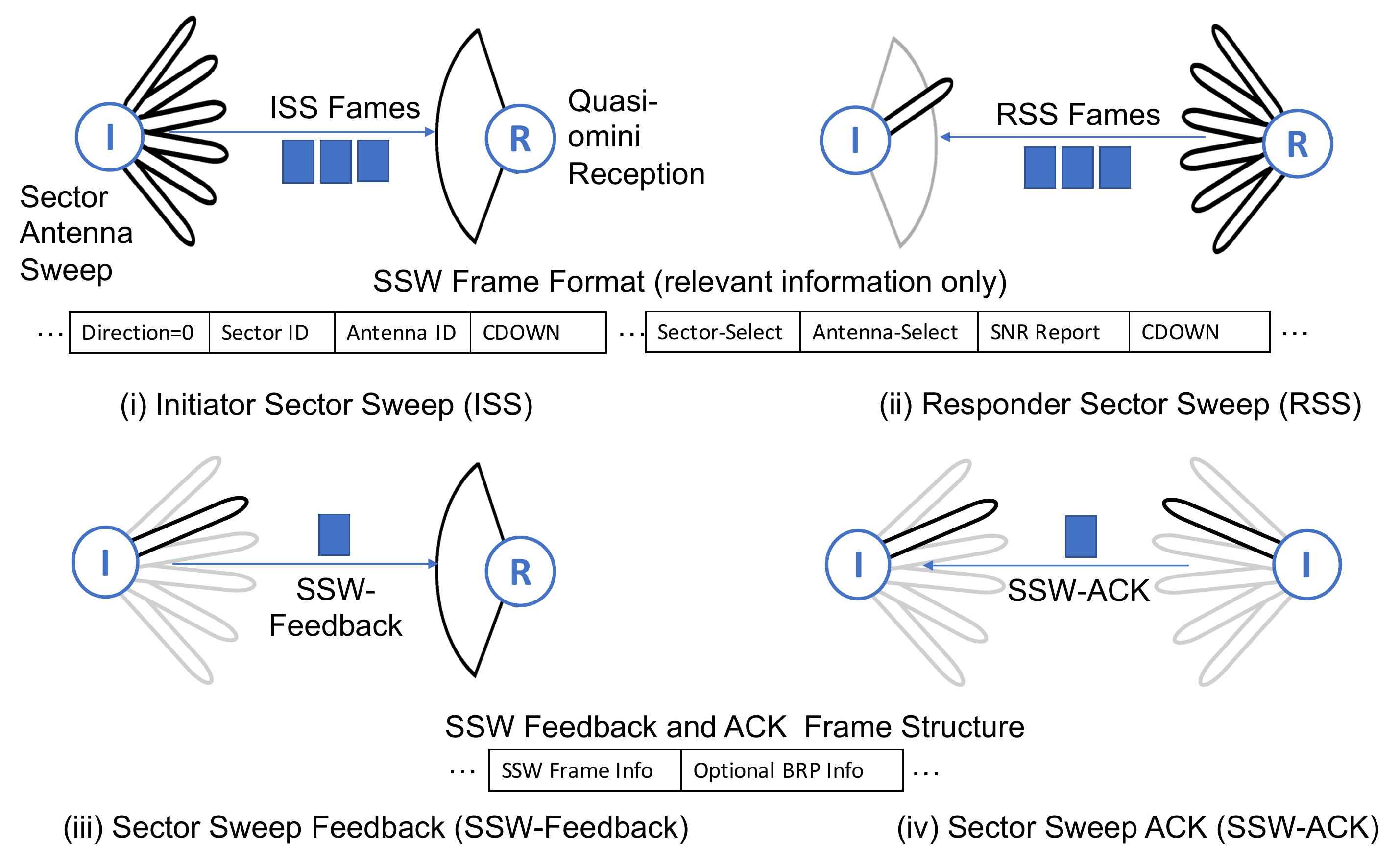} 
\caption{Sector Level Sweep (SLS) phase of beamforming}
\label{fig:bf}
\end{figure}

In SLS phase, a series of Sector Sweep (SSW) frames exchange between initiator (the node that initiates SSW frame transmission) and responder (the pairing node) to find the optimum sector pairs for data exchange. SLS includes four sub-phases that are schematically depicted by Figure \ref {fig:bf} and summarized below. 

 \noindent {\it i) Initiator Sector Sweep (ISS):}
The initiator transmits ISS frames over all its antenna sectors of the total number of antennas (if there is more than one transmit antenna). Each ISS message contains sector antenna IDs that the packet is being transmitted, the number of remaining frames (CDOWN) in the sector sweep and the SSW feedback filed does not contain any information. The responder listens on Quasi-omni pattern and record all received ISS frame information from all directions.

 \noindent  {\it ii) Responder Sector Sweep (RSS):} 
Similar to the initiator, the responder transmits a series of RSS frames sweeping through all sectors with similar information. In addition, from the ISS packet with highest received signal quality (SNR), the responder includes the best sector antenna IDs of the initiator in the SSW-Feedback filed of the RSS frame. In this sub-phase, the initiator listens to RSS packets on Quasi-omni antenna pattern.

 \noindent  {\it iii) Sector Sweep Feedback (SSW-Feedback):}
SSW-Feedback packet is transmitted by the initiator with the best antenna configuration (sector and antenna ID) set in the RSS frame by the responder. In addition, the initiator determines the best sector antenna IDs for the responder by evaluating SNR information of all received RSS frames and includes this information in the SSW-feedback field. Some fields in SSW-Feedback frame is set, only if BRP phase is desired to be followed. 

 \noindent {\it iv) Sector Sweep ACK (SSW-ACK):} 
It is the last frame transmitted by the responder for completing the beamforming process. SSW-ACK frame should be transmitted through the selected sector antenna IDs in the SSW-Feedback. The remaining fields shall set if the responder desires the transmitter to refine the selection as part of BRP phase.

This type of beamforming will be a desirable process for a reliable MmWave communication. On the other hand, this leads to additional collisions which handles by carrier sense multiple access with collision avoidance (CSMA/CA) mechanism in the IEEE 802.11ad protocol \cite{nit14, Niu2015}.

\chapter{Pseudo Codes}\label{PC}

\section{Maximum Likelihood Topology Map Algorithm}

\subsection{Information Gathering and Mapping}\label{sec::PC::infogathandmap}

\begin{algorithm}
$ N\leftarrow total\_number\_of\_nodes$\;
 $T\leftarrow Initializing\_time$\;
 
 \While{$t_i < T$}{
  sensors transmit signal\;
  robot update the M matrix\;
  }
  \For{each node $i$ $\in$ $N$ }{
  divide R-neighborhood of packet receiving points in robot trajectory into grids\;
  calculate $P_i(x_i,y_i)$ for vertices in the grid\;
  find maximum $P_i(l_i)$ \;
  $coordinate_i\leftarrow vertex $\: coordinate of maximum $P_i(l_i)$\;
  } 
\end{algorithm}

\newpage
\subsection{Robot Trajectory Automation}\label{sec::PC::rta}
 \begin{algorithm}
\SetAlgoLined
angle $\leftarrow $ 90  \tcc*[r]{moving direction w.r.t. x axis}
fwdObstacles $\leftarrow $ 0 \tcc*[r]{obstacle detected when angle=90}
bwdObstacles $\leftarrow $ 0 \tcc*[r]{obstacle detected when angle=-90}
\While{receive $N_p$ number of packets from all nodes }{
\uIf{Obstacle detected}{
obstacleRegistry(angle) \tcc*[r]{see Algorithm \ref{sec::PC::obstacleRegistry}}
storeAngle $\leftarrow $ angle\;
angle $\leftarrow $ 0; go forward \;
angle $\leftarrow $ -1$\times $ storeAngle\;
}
\uElseIf{Receive packets from nodes}{
go forward\;
}
\Else{
storeAngle $\leftarrow $ angle\;
angle $\leftarrow $ 0; go forward \;
angle$ \leftarrow $ -1$\times $ storeAngle\;
}
\uIf{fwdObstacles==1 and angle==90}{
coverObstacles(angle,fwdObstacleCoordinates)\tcc*[r]{see Algorithm \ref{sec::PC::coverObstacles}}
}
\uElseIf{bwdObstacles==1 and angle==-90}{
coverObstacles(angle,bwdObstacleCoordinates)
}
}
\end{algorithm}

\subsubsection{obstacleRegistry Function} \label{sec::PC::obstacleRegistry}
 \begin{algorithm}
 \SetAlgoLined
\Fn{obstacleRegistry(angle) }{
\uIf{angle==90 and fwdobstacles==0}{
fwdobstacles $\leftarrow $ 1\;
fwdObstacleCoordinates $\leftarrow (X_R,Y_R)$ \tcc*[r]{robot coordinates}
}
\uElseIf{bwdobstacles==0}{
bwdobstacles $\leftarrow $ 1\;
bwdObstacleCoordinates $\leftarrow (X_R,Y_R)$\; 
}
\Else{
return\;
}
} 
\label{FunRpathAlgo}
\end{algorithm}
\newpage
\subsubsection{coverObstacles Function}\label{sec::PC::coverObstacles}
 \begin{algorithm}
 \SetAlgoLined
\Fn{coverObstacles(angle,ObstacleCoordinate) }{
\uIf{No obstacle in moving direction 180}{
\uIf{angle==90 and $Y_R>$ObstacleCoordinate(1,2)}{
angle $\leftarrow$ 180\;
go to ObstacleCoordinate(1,1) or until obstacle detected\;
angle $\leftarrow$ 90; fwdobstacles $\leftarrow $ 0\;
}
\uElseIf{angle==-90 and $Y_R<$ObstacleCoordinate(1,2)}{
angle $\leftarrow$ 180\;
go to ObstacleCoordinate(1,1) or until obstacle detected\;
angle $\leftarrow$ -90; fwdobstacles $\leftarrow $ 0\;
}
\Else{
return\;
}
}
\Else{
return\;
}
} 
\end{algorithm}

\section{Millimetre Wave Topology Map Algorithm}

\subsection{Information Gathering and Mapping}\label{sec::PC::mmtminfogathandmap}

\begin{algorithm}
$ N\leftarrow total\_number\_of\_nodes$\;
 \While{robot receives $N_p$ number of packets from all nodes}{
  robot transmit ISS packets from all sectors\;
  sensors receiving those packets transmit RSS packets\;
  robot update the M and A matrices\;
  robot moves to the next step\;
  }
  \For{each node $i$ $\in$ $N$ }{
  divide R-neighborhood of packet receiving points in robot trajectory into grids\;
  calculate $P_i(x_i,y_i,z_i)$ for vertices in the grid\;
  find maximum $P_i(x_i,y_i,z_i)$ \;
  $coordinate_i\leftarrow vertex $\: coordinate of maximum $P_i(x_i,y_i,z_i)$\;
  } 
\end{algorithm}

\newpage

\section{Distributed Millimetre Wave Topology Map Algorithm}
\subsection{Distributed Millimeter Wave Topology Map for Static System (DMmTM-SS)}
\subsubsection{Direction Estimation of Anchor Sectors}\label{sec::PC::dmmtm1}

\begin{algorithm}
 $N_A \gets$ number of anchors\;
  \For{$i\gets1$ \KwTo $N_A$ }{
    anchor $a_i$ find it's set of neighbor anchors $\kappa _{a_i}$ and optimum sectors to communicate $\mathcal{H} _{a_i}$\;
    \For{$j\gets1$ \KwTo $\vert \kappa _{a_i}\vert $ }{
    calculate initial sector directions using equation (\ref{direction})\;
    }
    calculate optimum sector directions using equation (\ref{Fdirection})\;
    }
    \end{algorithm}

\subsubsection{Information Gathering}\label{sec::PC::dmmtm2}
\begin{algorithm}
 $N_A \gets$ number of anchors\;
 $N_S \gets$ number of sectors\;
  \For{$i\gets1$ \KwTo $N_A$ }{
  \For{$s\gets1$ \KwTo $N_S$ }{
    anchor $a_i$ broadcast a beacon packet from sector $s$\;
    sensors record anchor location and sector direction angle \;
    }
    }
\end{algorithm}
\newpage
\subsubsection{Information Mapping}\label{sec::PC::dmmtm3}
\begin{algorithm}
 $N \gets$ number of sensors\;
  \For{$j\gets1$ \KwTo $N$ }{
  sensor $s_j$ divide R-neighborhood of neighbor anchors into set of grids $G$\;
  \For{$l\gets1$ \KwTo $\vert G\vert $ }{
    calculate $P_j((x_l, y_l, z_l))$ using equation (\ref{probability})\;
    }
    choose maximum $P_j((x_l, y_l, z_l))$ value\;
    set that vertex as sensor $s_j$ topology coordinate\;
    }
\end{algorithm}

\subsubsection{Node Filtration}\label{sec::PC::dmmtm4}
\begin{algorithm}
 $N \gets$ number of sensors\;
 $N_{C_b} \gets N$\;
  $iterations \gets iT$\; 
 \While{$N_{C_b} >0$ or $iterations<iT$ }{
 $N_{C_b} \gets 0$\;
  \For{$j\gets1$ \KwTo $N$ }{
  get neighbor coordinates \;
  calculate $E_{N_S}^{s_j}$ using equation (\ref{Nsacttering})\;
  choose node category\;
  \If{category==$C_a$}{
    assign as a new anchor \;
    broadcast beacon packets\;
 	 }
  \ElseIf{category==$C_g$}{
    do nothing\;
  	}
  \Else{
    $N_{C_b} ++$ \;
    recalculate topology coordinates with new and old anchors using Algorithm 3\;
  	}
    }
   $ iterations ++$\;
    }
\end{algorithm}
\newpage
\subsection{Distributed Millimeter Wave Topology Map for Hybrid System (DMmTM-HS)}
\subsubsection{Information Gathering via Mobile Anchor}\label{sec::PC::dmmtm5}
\begin{algorithm}
 $T \gets$ time mobile anchor traverse in the network\;
  \For{$t\gets1$ \KwTo $T$ }{
  \For{$s\gets1$ \KwTo $N_S$ }{
    mobile anchor broadcast a beacon packet from sector $s$\;
    sensors record anchor location and sector direction angle \;
    }
    }
\end{algorithm}


\clearpage
\label{Bibliography}

\lhead{\emph{Bibliography}} 
\bibliographystyle{IEEEtran}

\bibliography{Report}

\begin{thebibliography}{100}
\providecommand{\url}[1]{#1}
\csname url@samestyle\endcsname
\providecommand{\newblock}{\relax}
\providecommand{\bibinfo}[2]{#2}
\providecommand{\BIBentrySTDinterwordspacing}{\spaceskip=0pt\relax}
\providecommand{\BIBentryALTinterwordstretchfactor}{4}
\providecommand{\BIBentryALTinterwordspacing}{\spaceskip=\fontdimen2\font plus
\BIBentryALTinterwordstretchfactor\fontdimen3\font minus
  \fontdimen4\font\relax}
\providecommand{\BIBforeignlanguage}[2]{{%
\expandafter\ifx\csname l@#1\endcsname\relax
\typeout{** WARNING: IEEEtran.bst: No hyphenation pattern has been}%
\typeout{** loaded for the language `#1'. Using the pattern for}%
\typeout{** the default language instead.}%
\else
\language=\csname l@#1\endcsname
\fi
#2}}
\providecommand{\BIBdecl}{\relax}
\BIBdecl

\bibitem{wsnarchi1}
O.~Diallo, J.~J. P.~C. Rodrigues, M.~Sene, and J.~Lloret, ``Distributed
  database management techniques for wireless sensor networks,'' in \emph{IEEE
  Transactions on Parallel and Distributed Systems}, vol.~26, no.~2, Feb 2015,
  pp. 604--620.

\bibitem{wsnarchi2}
V.~Potdar, A.~Sharif, and E.~Chang, ``Wireless sensor networks: A survey,'' in
  \emph{International Conference on Advanced Information Networking and
  Applications Workshops}, May 2009, pp. 636--641.

\bibitem{routingsurvey}
N.~Sabor, S.~Sasaki, M.~Abo-Zahhad, and S.~M. Ahmed, ``A comprehensive survey
  on hierarchical-based routing protocols for mobile wireless sensor networks:
  Review, taxonomy, and future directions,'' in \emph{Wireless Communications
  and Mobile Computing}, 2015.

\bibitem{routingsurvey2}
A.~Chatap and S.~Sirsikar, ``Review on various routing protocols for
  heterogeneous wireless sensor network,'' in \emph{International Conference on
  I-SMAC (IoT in Social, Mobile, Analytics and Cloud)}, Feb 2017, pp. 440--444.

\bibitem{wsnTypes}
J.~Yick, B.~Mukherjee, and D.~Ghosal, ``Wireless sensor network survey,'' in
  \emph{Computer Networks}, vol.~52, no.~12, 2008, pp. 2292 -- 2330.

\bibitem{vcthesis}
D.~C. Dhanapala, ``Anchor centric virtual coordinate systems in wireless sensor
  networks : from self-organization to network awareness,'' Ph.D. dissertation,
  Colorado State University, 2012.

\bibitem{introtoWSN}
K.~Sohraby, D.~Minoli, and T.~Znati, \emph{Wireless sensor networks,
  Technology, Protocols, and Applications}.\hskip 1em plus 0.5em minus
  0.4em\relax John Wiley, 2007.

\bibitem{wsnadvantages}
M.~Raza, N.~Aslam, H.~Le-Minh, S.~Hussain, Y.~Cao, and N.~M. Khan, ``A critical
  analysis of research potential, challenges and future directives in
  industrial wireless sensor networks,'' in \emph{IEEE Communications Surveys
  Tutorials}, vol.~20, no.~99, 2017, pp. 1--1.

\bibitem{wsnApplicationTypes}
M.~A. Rassam, A.~Zainal, and M.~A. Maarof, ``Advancements of data anomaly
  detection research in wireless sensor networks: A survey and open issues,''
  in \emph{Sensors}, vol.~13, no.~8, 2013, pp. 10\,087--10\,122.

\bibitem{volcano}
G.~Werner-Allen, K.~Lorincz, M.~Ruiz, O.~Marcillo, J.~Johnson, J.~Lees, and
  M.~Welsh, ``Deploying a wireless sensor network on an active volcano,'' in
  \emph{IEEE Internet Computing}, vol.~10, no.~2, March 2006, pp. 18--25.

\bibitem{duckisland}
A.~Mainwaring, D.~Culler, J.~Polastre, R.~Szewczyk, and J.~Anderson, ``Wireless
  sensor networks for habitat monitoring,'' in \emph{ACM International Workshop
  on Wireless Sensor Networks and Applications}.\hskip 1em plus 0.5em minus
  0.4em\relax ACM, 2002, pp. 88--97.

\bibitem{military}
M.~P. Durisic, Z.~Tafa, G.~Dimic, and V.~Milutinovic, ``A survey of military
  applications of wireless sensor networks,'' in \emph{Mediterranean Conference
  on Embedded Computing}, June 2012, pp. 196--199.

\bibitem{militarymonitoring}
H.~Deng and R.~Xu, ``Acoustic threatening sound detection and recognition using
  wireless sensor networks,'' Jan 2009, pp. 73\,050S--73\,050S.

\bibitem{trafficmonitoring}
M.~Mbodila, E.~Obeten, and I.~Bassey, ``Implementation of novel vehicles'
  traffic monitoring using wireless sensor network in south africa,'' in
  \emph{IEEE International Conference on Communication Software and Networks},
  June 2015, pp. 282--286.

\bibitem{industrymonitoing}
M.~Franceschinis, M.~A. Spirito, R.~Tomasi, G.~Ossini, and M.~Pidalà, ``Using
  {WSN} technology for industrial monitoring: A real case,'' in \emph{Second
  International Conference on Sensor Technologies and Applications}, Aug 2008,
  pp. 282--287.

\bibitem{codeblue}
D.~J. Malan, T.~R.~F. Fulford-Jones, M.~Welsh, and S.~Moulton, ``Codeblue: An
  ad hoc sensor network infrastructure for emergency medical care,'' in
  \emph{International Workshop on Wearable and Implantable Body Sensor
  Networks}, 2004.

\bibitem{encapsule}
T.~Aoyagi, K.~Takizawa, T.~Kobayashi, J.~I. Takada, and R.~Kohno, ``Development
  of a {WBAN} channel model for capsule endoscopy,'' in \emph{IEEE Antennas and
  Propagation Society International Symposium}, June 2009, pp. 1--4.

\bibitem{agreemonitoring}
Z.~Liqiang, Y.~Shouyi, L.~Leibo, Z.~Zhen, and W.~Shaojun., ``A crop monitoring
  system based on wireless sensor network,'' in \emph{Procedia Environmental
  Sciences}, vol.~11, 2011, pp. 558 -- 565.

\bibitem{smartapp}
H.~Ghayvat, S.~C. Mukhopadhyay, X.~Gui, and J.~Liu, ``Enhancement of {WSN}
  based smart home to a smart building for assisted living: Design issues,'' in
  \emph{Fifth International Conference on Communication Systems and Network
  Technologies}, April 2015, pp. 219--224.

\bibitem{zebranet}
T.~Liu, C.~M. Sadler, P.~Zhang, and M.~Martonosi, ``Implementing software on
  resource-constrained mobile sensors: Experiences with impala and zebranet,''
  in \emph{2nd International Conference on Mobile Systems, Applications, and
  Services}, ser. MobiSys.\hskip 1em plus 0.5em minus 0.4em\relax ACM, 2004,
  pp. 256--269.

\bibitem{militarytracking}
P.~Naz, S.~Hengy, and P.~Hamery, ``Soldier detection using unattended acoustic
  and seismic sensors,'' in \emph{SPIE}, vol. 8389, 2012, pp. 83\,890T--12.

\bibitem{vehicletracking}
R.~C. Jisha, M.~V. Ramesh, and G.~S. Lekshmi, ``Intruder tracking using
  wireless sensor network,'' in \emph{IEEE International Conference on
  Computational Intelligence and Computing Research}, Dec 2010, pp. 1--5.

\bibitem{schoolvehicle}
S.-e. Yoo, P.~K. Chong, and D.~Kim, ``S3: School zone safety system based on
  wireless sensor network,'' in \emph{Sensors}, vol.~9, no.~8, 2009, pp.
  5968--5988.

\bibitem{industrytracking}
A.~Mason, A.~I. Al-Shamma'a, and A.~Shaw, ``Wireless sensor network for
  intelligent inventory management for packaged gases,'' in \emph{Second
  International Conference on Developments in eSystems Engineering}, Dec 2009,
  pp. 413--417.

\bibitem{motecomparison}
R.~Narayanan, T.~Sarath, and V.~Vineeth, ``Survey on motes used in wireless
  sensor networks: performance \& parametric analysis,'' in \emph{Wireless
  Sensor Networks}, vol.~8, 2016, pp. 51--60.

\bibitem{Libelium}
\BIBentryALTinterwordspacing
Libelium. (2013) Libelium launches new generation of waspmote sensor nodes.
  [Online]. Available:
  \url{http://www.libelium.com/libelium-launches-new-generation-of-waspmote-sensor-nodes/}
\BIBentrySTDinterwordspacing

\bibitem{mmwaveUWB}
L.~Jin, C.~Loyez, N.~Rolland, and P.~A. Rolland, ``An {UWB} millimeter-wave
  transceiver architecture for wireless sensor networks applications,'' in
  \emph{The 40th European Microwave Conference}, Sept 2010, pp. 377--380.

\bibitem{smartdustintro}
\BIBentryALTinterwordspacing
A.~Robinson, F.~Hardisty, J.~A. Dutton, and G.~Chaplin. (2017) Technology
  trends - smart dust and sensor networks. [Online]. Available:
  \url{https://www.e-education.psu.edu/geog583/node/77}
\BIBentrySTDinterwordspacing

\bibitem{selforgsurvey}
K.~L. Mills, ``A brief survey of self-organization in wireless sensor networks:
  Research inproceedingss,'' in \emph{Wireless Communication and Mobile
  Computing}, vol.~7, no.~7.\hskip 1em plus 0.5em minus 0.4em\relax Chichester,
  UK: John Wiley and Sons Ltd., Sep. 2007, pp. 823--834.

\bibitem{rssiemperical}
X.~Wanga, S.~Yuanb, R.~Laura, and W.~Langb, ``Dynamic localization based on
  spatial reasoning with {RSSI} in wireless sensor networks for transport
  logistics,'' in \emph{Sensors and Actuators}, 2011, pp. 421--428.

\bibitem{VC}
D.~Dhanapala and A.~Jayasumana, ``Topology preserving maps ;extracting layout
  maps of wireless sensor networks from virtual coordinates,'' in
  \emph{IEEE/ACM Transactions on Networking}, vol.~22, no.~3, June 2014, pp.
  784--797.

\bibitem{802154}
\BIBentryALTinterwordspacing
I.~Poole. Ieee 802.15.4 air interface and frequency bands. [Online]. Available:
  \url{http://www.radio-electronics.com/info/wireless/ieee-802-15-4/bands-frequencies-radio-air-interface.php}
\BIBentrySTDinterwordspacing

\bibitem{80211ad}
\BIBentryALTinterwordspacing
I.~Poole. Ieee 802.11ad microwave wi-fi or wigig tutorial. [Online]. Available:
  \url{http://www.radio-electronics.com/info/wireless/wi-fi/ieee-802-11ad-microwave.php}
\BIBentrySTDinterwordspacing

\bibitem{locSurvey}
A.~K. Paul and T.~Sato, ``Localization in wireless sensor networks: A survey on
  algorithms, measurement techniques, applications and challenges,'' in
  \emph{booktitle of Sensor and Actuator Networks}, vol.~6, no.~4, 2017.

\bibitem{locSurvey2}
J.~Kuriakose, V.~Amruth, and N.~S. Nandhini, ``A survey on localization of
  wireless sensor nodes,'' in \emph{International Conference on Information
  Communication and Embedded Systems}, Feb 2014, pp. 1--6.

\bibitem{locSurvey3}
A.~Mesmoudi, M.~Feham, and N.~Labraoui, ``Wireless sensor networks localization
  algorithms: a comprehensive survey,'' in \emph{CoRR}, vol. 1312.4082, 2013.

\bibitem{locSurvey4}
N.~A. Alrajeh, M.~Bashir, and B.~Shams, ``Localization techniques in wireless
  sensor networks,'' in \emph{International booktitle of Distributed Sensor
  Networks}, vol.~9, no.~6, 2013.

\bibitem{locSurvey5}
E.~Zanaj, A.~Rista, B.~Zanaj, E.~Kotobelli, and E.~Elbasani, ``Wireless sensor
  networks localization algorithm,'' in \emph{12th International Workshop on
  Intelligent Solutions in Embedded Systems}, Oct 2015, pp. 63--66.

\bibitem{locSurvey6}
L.~Cheng, C.~Wu, Y.~Zhang, H.~Wu, M.~Li, and C.~Maple, ``A survey of
  localization in wireless sensor network,'' in \emph{International booktitle
  of Distributed Sensor Networks}, vol.~8, no.~12, 2012.

\bibitem{locSurvey7}
G.~Han, J.~Jiang, C.~Zhang, T.~Q. Duong, M.~Guizani, and G.~K. Karagiannidis,
  ``A survey on mobile anchor node assisted localization in wireless sensor
  networks,'' in \emph{IEEE Communications Surveys Tutorials}, vol.~18, no.~3,
  2016, pp. 2220--2243.

\bibitem{rssisurvey}
H.~P. Mistry and N.~H. Mistry, ``{RSSI} based localization scheme in wireless
  sensor networks: A survey,'' in \emph{Fifth International Conference on
  Advanced Computing Communication Technologies}, Feb 2015, pp. 647--652.

\bibitem{rssiadaptive}
H.~Zhang, J.~Zhang, and H.~Wu, ``An adaptive localization algorithm based on
  {RSSI} in wireless sensor networks,'' in \emph{IEEE 2nd International
  Conference on Cloud Computing and Intelligent Systems}, vol.~03, Oct 2012,
  pp. 1133--1136.

\bibitem{rssidirect}
B.~Mukhopadhyay, S.~Sarangi, and S.~Kar, ``Novel {RSSI} evaluation models for
  accurate indoor localization with sensor networks,'' in \emph{20th National
  Conference on Communications}, Feb 2014, pp. 1--6.

\bibitem{rssidirct2}
R.~Al~Alawi, ``{RSSI} based location estimation in wireless sensors networks,''
  in \emph{17th IEEE International Conference on Networks}, Dec 2011, pp.
  118--122.

\bibitem{rssiprbabilistic}
W.~Chengdong, C.~Shifeng, Z.~Yunzhou, C.~Long, and W.~Hao, ``A {RSSI}-based
  probabilistic distribution localization algorithm for wireless sensor
  network,'' in \emph{6th IEEE Joint International Information Technology and
  Artificial Intelligence Conference}, vol.~1, Aug 2011, pp. 333--337.

\bibitem{LSmethod}
J.~Xu, W.~Liu, F.~Lang, Y.~Zhang, and C.~Wang, ``Distance measurement model
  based on rssi in wsn,'' in \emph{Wireless Sensor Network (SciRes)}, vol.~2,
  no.~8, 2010, pp. 606--611.

\bibitem{GMMmethod}
N.~A. Dieng, M.~Charbit, C.~Chaudet, L.~Toutain, and T.~B. Meriem, ``Indoor
  localization in wireless networks based on a two-modes gaussian mixture
  model,'' in \emph{IEEE 78th Vehicular Technology Conference (VTC Fall)}, Sept
  2013, pp. 1--5.

\bibitem{empericalindoor}
O.~G. Adewumi, K.~Djouani, and A.~M. Kurien, ``{RSSI} based indoor and outdoor
  distance estimation for localization in wsn,'' in \emph{IEEE International
  Conference on Industrial Technology}, Feb 2013, pp. 1534--1539.

\bibitem{gaussianfilter}
L.~Xiao, Y.~Yin, X.~Wu, and J.~Wang, ``A large-scale rf-based indoor
  localization system using low-complexity gaussian filter and improved
  bayesian inference,'' in \emph{Radio engineering}, vol.~22, 2013, pp.
  371--380.

\bibitem{shadowingmitigate}
N.~Chuku, A.~Pal, and A.~Nasipuri, ``An {RSSI} based localization scheme for
  wireless sensor networks to mitigate shadowing effects,'' in \emph{IEEE
  Southeastcon}, April 2013, pp. 1--6.

\bibitem{mimolocalization}
S.~Hamdoun, A.~Rachedi, and A.~Benslimane, ``Comparative analysis of
  {RSSI}-based indoor localization when using multiple antennas in wireless
  sensor networks,'' in \emph{International Conference on Selected Topics in
  Mobile and Wireless Networking}, Aug 2013, pp. 146--151.

\bibitem{indoortri1}
Z.~M. Livinsa and S.~Jayashri, ``Performance analysis of diverse environment
  based on {RSSI} localization algorithms in wsns,'' in \emph{IEEE Conference
  on Information Communication Technologies}, April 2013, pp. 572--576.

\bibitem{indoortri2}
A.~Buchman and C.~Lung, ``Received signal strength based room level accuracy
  indoor localisation method,'' in \emph{IEEE 4th International Conference on
  Cognitive Infocommunications}, Dec 2013, pp. 103--108.

\bibitem{indoortri3}
B.~Mukhopadhyay, S.~Sarangi, and S.~Kar, ``Novel {RSSI} evaluation models for
  accurate indoor localization with sensor networks,'' in \emph{Twentieth
  National Conference on Communications}, Feb 2014, pp. 1--6.

\bibitem{indoorml1}
K.Vadivukkarasi and R.Kumar, ``A new approach for error reduction in
  localization for wireless sensor networks,'' in \emph{International booktitle
  on Recent Trends in Engineering and Technology}, vol.~8, no.~2, 2013.

\bibitem{indoorml2}
Y.~Chen, L.~Zhang, and J.~Wang, ``Localization indoor patient in wireless
  sensor networks,'' in \emph{First International Symposium on Future
  Information and Communication Technologies for Ubiquitous HealthCare}, July
  2013, pp. 1--3.

\bibitem{rssipowerdecay}
J.-Y. Wang, C.-P. Chen, T.-S. Lin, C.-L. Chuang, T.-Y. Lai, and J.-A. Jiang,
  ``High-precision {RSSI}-based indoor localization using a transmission power
  adjustment strategy for wireless sensor networks,'' in \emph{IEEE 14th
  International Conference on High Performance Computing and Communication},
  June 2012, pp. 1634--1638.

\bibitem{MLRSSI}
R.~I. Rusnac and A.~S. Gontean, ``Maximum likelihood estimation algorithm
  evaluation for wireless sensor networks,'' in \emph{12th International
  Symposium on Symbolic and Numeric Algorithms for Scientific Computing}, Sept
  2010, pp. 95--98.

\bibitem{convexRSSI}
M.~Beko, ``A complex convex relaxation for approximate maximum likelihood 2{D}
  energy-based source localization in sensor networks,'' in \emph{7th
  International Symposium on Wireless Communication Systems}, Sept 2010, pp.
  150--153.

\bibitem{CN2}
Y.~Yao and N.~Jiang, ``Distributed wireless sensor network localization based
  on weighted search,'' in \emph{Computer Networks}, vol.~86, 2015, pp. 57 --
  75.

\bibitem{rssimobile}
Y.~Zhu, B.~Zhang, F.~Yu, and S.~Ning, ``A {RSSI} based localization algorithm
  using a mobile anchor node for wireless sensor networks,'' in
  \emph{International Joint Conference on Computational Sciences and
  Optimization}, vol.~1, April 2009, pp. 123--126.

\bibitem{rssimobile2}
V.~Vivekanandan and V.~Wong, ``Concentric anchor beacon localization algorithm
  for wireless sensor networks,'' in \emph{IEEE Transactions on Vehicular
  Technology,}, vol.~56, no.~5, Sept 2007, pp. 2733--2744.

\bibitem{rssimobile3}
P.~Sahu, E.-K. Wu, and J.~Sahoo, ``Du{RT}: Dual {RSSI} trend based localization
  for wireless sensor networks,'' in \emph{IEEE Sensors booktitle,}, vol.~13,
  no.~8, Aug 2013, pp. 3115--3123.

\bibitem{mobileanchor1}
M.~Gai and A.~Azadmanesh, ``Sensor localization for indoor wireless sensor
  networks,'' in \emph{International Symposium on Performance Evaluation of
  Computer and Telecommunication Systems}, July 2014, pp. 536--541.

\bibitem{tdoasurvey}
S.~Ji, D.~Liu, and J.~Shen, ``Localization technology in wireless sensor
  networks using {RSSI} and {LQI}: A survey,'' in \emph{Advances in Computer
  Science and Ubiquitous Computing}, J.~J. J.~H. Park, Y.~Pan, G.~Yi, and
  V.~Loia, Eds., 2017, pp. 342--351.

\bibitem{tdoasurvey2}
K.~Stone and T.~Camp, ``A survey of distance‐based wireless sensor network
  localization techniques,'' in \emph{International booktitle of Pervasive
  Computing and Communications}, vol.~8, no.~2, 2012, pp. 158--183.

\bibitem{tdoa1}
C.~Savarese, J.~Rabaey, and K.~Langendoen, ``Robust positioning algorithms for
  distributed ad-hoc wireless sensor networks,'' in \emph{USENIX Annual
  Technical Conference}, 2002, pp. 317--327.

\bibitem{tdoa2}
X.~L. Luo, W.~Li, and J.~R. Lin, ``Geometric location based on {TDOA} for
  wireless sensor networks,'' in \emph{ISRN Applied Mathematics}, vol. 2012,
  2012.

\bibitem{tdoa3}
S.~Liu, Y.~Tang, C.~Zhang, and S.~Yue, ``Self-map building in wireless sensor
  network based on {TDOA} measurements,'' in \emph{IEEE Conference on
  Multisensor Fusion and Integration for Intelligent Systems}, Sept 2012, pp.
  150--155.

\bibitem{tdoatimesync}
S.~Kim and J.-W. Chong, ``An efficient tdoa-based localization algorithm
  without synchronization between base stations,'' in \emph{International
  booktitle of Distributed Sensor Networks}, vol.~11, no.~9, 2015.

\bibitem{tdoatimesync2}
R.~Kaune, ``Accuracy studies for {TDOA} and {TOA} localization,'' in \emph{2012
  15th International Conference on Information Fusion}, July 2012, pp.
  408--415.

\bibitem{tdoa4}
Y.~Wang and K.~Ho, ``{TDOA} source localization in the presence of
  synchronization clock bias and sensor position errors,'' in \emph{IEEE
  Transactions on Signal Processing}, vol.~61, no.~18, Sept 2013, pp.
  4532--4544.

\bibitem{tdoa5}
B.~Huang, L.~Xie, and Z.~Yang, ``{TDOA}-based source localization with
  distance-dependent noises,'' in \emph{IEEE Transactions on Wireless
  Communications,}, vol.~14, no.~1, Jan 2015, pp. 468--480.

\bibitem{toa1}
H.~Wymeersch, J.~Lien, and M.~Win, ``Cooperative localization in wireless
  networks,'' in \emph{Proceedings of the IEEE}, vol.~97, no.~2, Feb 2009, pp.
  427--450.

\bibitem{toa2}
M.~Gholami, S.~Gezici, and E.~Strom, ``Improved position estimation using
  hybrid {TW}-{TOA} and {TDOA} in cooperative networks,'' in \emph{IEEE
  Transactions on Signal Processing}, vol.~60, no.~7, July 2012, pp.
  3770--3785.

\bibitem{toa3}
P.~Oguz-Ekim, J.~Gomes, P.~Oliveira, M.~Reza~Gholami, and E.~Strom,
  ``{TW}-{TOA} based cooperative sensor network localization with unknown
  turn-around time,'' in \emph{IEEE International Conference on Acoustics,
  Speech and Signal Processing}, May 2013, pp. 6416--6420.

\bibitem{toasurvey}
I.~Guvenc and C.~C. Chong, ``A survey on {TOA} based wireless localization and
  {NLOS} mitigation techniques,'' in \emph{IEEE Communications Surveys
  Tutorials}, vol.~11, no.~3, rd 2009, pp. 107--124.

\bibitem{toas6}
S.~Gezici, Z.~Tian, G.~B. Giannakis, H.~Kobayashi, A.~F. Molisch, H.~V. Poor,
  and Z.~Sahinoglu, ``Localization via ultra-wideband radios: a look at
  positioning aspects for future sensor networks,'' in \emph{IEEE Signal
  Processing Magazine}, vol.~22, no.~4, July 2005, pp. 70--84.

\bibitem{toas3}
S.~Lanzisera, D.~T. Lin, and K.~S.~J. Pister, ``{RF} time of flight ranging for
  wireless sensor network localization,'' in \emph{International Workshop on
  Intelligent Solutions in Embedded Systems}, June 2006, pp. 1--12.

\bibitem{toa4}
A.~Yeredor, ``Decentralized {TOA}-based localization in non-synchronized
  wireless networks with partial, asymmetric connectivity,'' in \emph{IEEE 15th
  International Workshop on Signal Processing Advances in Wireless
  Communications}, June 2014, pp. 165--169.

\bibitem{toas1}
K.~H. Lee, C.~H. Yu, J.~W. Choi, and Y.~B. Seo, ``{ToA} based sensor
  localization in underwater wireless sensor networks,'' in \emph{SICE Annual
  Conference}, Aug 2008, pp. 1357--1361.

\bibitem{toas4}
N.~Patwari, J.~N. Ash, S.~Kyperountas, A.~O. Hero, R.~L. Moses, and N.~S.
  Correal, ``Locating the nodes: cooperative localization in wireless sensor
  networks,'' in \emph{IEEE Signal Processing Magazine}, vol.~22, no.~4, July
  2005, pp. 54--69.

\bibitem{toas5}
A.~Boukerche, H.~A. B.~F. Oliveira, E.~F. Nakamura, and A.~A.~F. Loureiro,
  ``Localization systems for wireless sensor networks,'' in \emph{IEEE Wireless
  Communications}, vol.~14, no.~6, December 2007, pp. 6--12.

\bibitem{toas2}
H.~Chen, B.~Liu, P.~Huang, J.~Liang, and Y.~Gu, ``Mobility-assisted node
  localization based on toa measurements without time synchronization in
  wireless sensor networks,'' in \emph{Mobile Networks and Applications},
  vol.~17, no.~1, Feb 2012, pp. 90--99.

\bibitem{aoa4}
J.~Xu, M.~Ma, and C.~L. Law, ``{AOA} cooperative position localization,'' in
  \emph{IEEE Global Telecommunications Conference}, Nov 2008, pp. 1--5.

\bibitem{aoa1}
D.~J. Torrieri, ``Statistical theory of passive location systems,'' in
  \emph{IEEE Transactions on Aerospace and Electronic Systems}, vol. AES-20,
  no.~2, March 1984, pp. 183--198.

\bibitem{aoa2}
A.~Pages-Zamora, J.~Vidal, and D.~H. Brooks, ``Closed-form solution for
  positioning based on angle of arrival measurements,'' in \emph{The 13th IEEE
  International Symposium on Personal, Indoor and Mobile Radio Communications},
  vol.~4, Sept 2002, pp. 1522--1526 vol.4.

\bibitem{aoa3}
A.~Urruela, A.~Pages-Zamora, and J.~Riba, ``Divide-and-conquer based
  closed-form position estimation for aoa and tdoa measurements,'' in
  \emph{IEEE International Conference on Acoustics Speech and Signal Processing
  Proceedings}, vol.~4, May 2006, pp. IV ­ 921--IV ­ 924.

\bibitem{aoa5}
D.~Niculescu and B.~Nath, ``Ad hoc positioning system ({APS}) using {AOA},'' in
  \emph{Twenty-second Annual Joint Conference of the IEEE Computer and
  Communications Societies}, vol.~3, March 2003, pp. 1734--1743 vol.3.

\bibitem{aoa7}
P.~Kulakowski, J.~Vales-Alonso, E.~Egea-Lopez, W.~Ludwin, and J.~G. Haro,
  ``Angle-of-arrival localization based on antenna arrays for wireless sensor
  networks,'' in \emph{Computers and Electrical Engineering}, vol.~36, no.~6,
  2010, pp. 1181 -- 1186.

\bibitem{aoa6}
I.~Amundson, J.~Sallai, X.~Koutsoukos, A.~Ledeczi, and M.~Maroti, ``{RF} angle
  of arrival-based node localisation,'' in \emph{IJSNet}, vol.~9, 05 2011, pp.
  209--224.

\bibitem{aoa8}
H.~J. Shao, X.~P. Zhang, and Z.~Wang, ``Efficient closed-form algorithms for
  {AOA} based self-localization of sensor nodes using auxiliary variables,'' in
  \emph{IEEE Transactions on Signal Processing}, vol.~62, no.~10, May 2014, pp.
  2580--2594.

\bibitem{aoa9}
Q.~Zhou and Z.~Duan, ``Weighted intersections of bearing lines for {AOA} based
  localization,'' in \emph{17th International Conference on Information
  Fusion}, July 2014, pp. 1--8.

\bibitem{aoa10}
A.~N. Bishop, B.~D.~O. Anderson, B.~Fidan, P.~N. Pathirana, and G.~Mao,
  ``Bearing-only localization using geometrically constrained optimization,''
  in \emph{IEEE Transactions on Aerospace and Electronic Systems}, vol.~45,
  no.~1, Jan 2009, pp. 308--320.

\bibitem{aoa11}
L.~Badriasl and K.~Dogancay, ``Three-dimensional target motion analysis using
  azimuth/elevation angles,'' in \emph{IEEE Transactions on Aerospace and
  Electronic Systems}, vol.~50, no.~4, October 2014, pp. 3178--3194.

\bibitem{aoa14}
Y.~Wang and K.~C. Ho, ``An asymptotically efficient estimator in closed-form
  for 3-d aoa localization using a sensor network,'' in \emph{IEEE Transactions
  on Wireless Communications}, vol.~14, no.~12, Dec 2015, pp. 6524--6535.

\bibitem{aoa12}
C.-Y. Tseng, D.~D. Feldman, and L.~J. Griffiths, ``Steering vector estimation
  in uncalibrated arrays,'' in \emph{IEEE Transactions on Signal Processing},
  vol.~43, no.~6, Jun 1995, pp. 1397--1412.

\bibitem{aoa13}
Y.-M. Chen, J.-H. Lee, and C.-C. Yeh, ``Two-dimensional angle-of-arrival
  estimation for uniform planar arrays with sensor position errors,'' in
  \emph{IEE Proceedings F - Radar and Signal Processing}, vol. 140, no.~1, Feb
  1993, pp. 37--42.

\bibitem{rangefree}
H.~Tian, C.~Huang, B.~M. Blum, J.~Stankovic, and T.~Abdelzaher, ``Range-free
  localization schemes for large scale sensor networks,'' in \emph{9th Annual
  International Conference on Mobile Computing and Networking}, ser. MobiCom
  '03, 2003, pp. 81--95.

\bibitem{dv}
D.~Niculescu and B.~Nath, ``{DV} based positioning in ad hoc networks,'' in
  \emph{Kluwer booktitle of Telecommunication Systems}, vol.~22, no.~1, Jan.
  2003, p. 267â280.

\bibitem{rangefree2}
C.~Liu, K.~Wu, and T.~He, ``Sensor localization with ring overlapping based on
  comparison of received signal strength indicator,'' in \emph{IEEE
  International Conference on Mobile Ad-hoc and Sensor Systems}, Oct 2004, pp.
  516--518.

\bibitem{rfa1}
W.~Yu and H.~Li, ``An improved dv-hop localization method in wireless sensor
  networks,'' in \emph{IEEE International Conference on Computer Science and
  Automation Engineering}, vol.~3, May 2012, pp. 199--202.

\bibitem{rangefree3}
Y.~Wang, X.~Wang, D.~Wang, and D.~Agrawal, ``Range-free localization using
  expected hop progress in wireless sensor networks,'' in \emph{IEEE
  Transactions on Parallel and Distributed Systems}, vol.~20, no.~10, Oct 2009,
  pp. 1540--1552.

\bibitem{rangefree4}
Q.~Xiao, B.~Xiao, J.~Cao, and J.~Wang, ``Multihop range-free localization in
  anisotropic wireless sensor networks: A pattern-driven scheme,'' in
  \emph{IEEE Transactions on Mobile Computing,}, vol.~9, no.~11, Nov 2010, pp.
  1592--1607.

\bibitem{CN3}
G.~Wu, S.~Wang, B.~Wang, Y.~Dong, and S.~Yan, ``A novel range-free localization
  based on regulated neighborhood distance for wireless ad hoc and sensor
  networks,'' in \emph{Computer Networks}, vol.~56, no.~16, 2012, pp. 3581 --
  3593.

\bibitem{rfa2}
J.~Z. Wang and H.~Jin, ``Improvement on apit localization algorithms for
  wireless sensor networks,'' in \emph{2009 International Conference on
  Networks Security, Wireless Communications and Trusted Computing}, vol.~1,
  April 2009, pp. 719--723.

\bibitem{rfa3}
J.~SHU, C.~YAN, and L.~lan LIU, ``Improved three-dimensional localization
  algorithm based on volume-test scan for wireless sensor networks,'' in
  \emph{The booktitle of China Universities of Posts and Telecommunications},
  vol.~19, 2012, pp. 1 -- 6.

\bibitem{rangefreemobile}
K.~F. Ssu, C.~H. Ou, and H.~Jiau, ``Localization with mobile anchor points in
  wireless sensor networks,'' in \emph{IEEE Transactions on Vehicular
  Technology,}, vol.~54, no.~3, May 2005, pp. 1187--1197.

\bibitem{rangefreemobile2}
M.~Sichitiu and V.~Ramadurai, ``Localization of wireless sensor networks with a
  mobile beacon,'' in \emph{IEEE International Conference on Mobile Ad-hoc and
  Sensor Systems}, Oct 2004, pp. 174--183.

\bibitem{rfa7}
M.~Singh and P.~M. Khilar, ``Mobile beacon based range free localization method
  for wireless sensor networks,'' in \emph{Wireless Networks}, vol.~23, no.~4,
  May 2017, pp. 1285--1300.

\bibitem{rangefreemobile3}
C.~H. Ou, ``Range-free node localization for mobile wireless sensor networks,''
  in \emph{3rd International Symposium on Wireless Pervasive Computing}, May
  2008, pp. 535--539.

\bibitem{rfa4}
E.~Guerrero, H.~G. Xiong, Q.~Gao, G.~Cova, R.~Ricardo, and J.~Estévez,
  ``{ADAL}: A distributed range-free localization algorithm based on a mobile
  beacon for wireless sensor networks,'' in \emph{International Conference on
  Ultra Modern Telecommunications Workshops}, Oct 2009, pp. 1--7.

\bibitem{rfa8}
S.~Lee, E.~Kim, C.~Kim, and K.~Kim, ``Localization with a mobile beacon based
  on geometric constraints in wireless sensor networks,'' in \emph{IEEE
  Transactions on Wireless Communications}, vol.~8, no.~12, December 2009, pp.
  5801--5805.

\bibitem{rfa9}
B.~Xiao, H.~Chen, and S.~Zhou, ``Distributed localization using a moving beacon
  in wireless sensor networks,'' in \emph{IEEE Transactions on Parallel and
  Distributed Systems}, vol.~19, no.~5, May 2008, pp. 587--600.

\bibitem{rfa5}
Q.~Xiao, B.~Xiao, J.~Cao, and J.~Wang, ``Multihop range-free localization in
  anisotropic wireless sensor networks: A pattern-driven scheme,'' in
  \emph{IEEE Transactions on Mobile Computing}, vol.~9, no.~11, Nov 2010, pp.
  1592--1607.

\bibitem{rfa6}
S.~Zaidi, A.~E. Assaf, S.~Affes, and N.~Kandil, ``Accurate range-free
  localization in multi-hop wireless sensor networks,'' in \emph{IEEE
  Transactions on Communications}, vol.~64, no.~9, Sept 2016, pp. 3886--3900.

\bibitem{thinplate}
A.~F. Buoud and A.~P. Jayasumana, ``Topology preserving map to physical map - a
  thin-plate spline based transform,'' in \emph{IEEE 41st Conference on Local
  Computer Networks}, Nov 2016, pp. 262--270.

\bibitem{allmap}
Y.~Bengio, J.~F. Paiement, and P.~Vincent, ``Out-of-sample extensions for
  {LLE}, {I}somap, {MDS}, {E}igenmaps, and {S}pectral clustering,'' in \emph{In
  Advances in Neural Information Processing Systems}.\hskip 1em plus 0.5em
  minus 0.4em\relax MIT Press, 2003, pp. 177--184.

\bibitem{mds}
Y.~Shang, W.~Rumi, Y.~Zhang, and M.~Fromherz, ``Localization from connectivity
  in sensor networks,'' in \emph{IEEE Transactions on Parallel and Distributed
  Systems,}, vol.~15, no.~11, Nov 2004, pp. 961--974.

\bibitem{isomap}
J.~Tenenbaum, V.~deSilva, and J.~Langford, ``A global geometric framework for
  nonlinear dimensionality reduction,'' in \emph{Science}, vol. 290, Dec. 2000,
  pp. 2319--2323.

\bibitem{mds2}
Y.~Shang, W.~Ruml, Y.~Zhang, and M.~P.~J. Fromherz, ``Localization from mere
  connectivity,'' in \emph{4th ACM International Symposium on Mobile Ad Hoc
  Networking \&Amp; Computing}, ser. MobiHoc '03, 2003, pp. 201--212.

\bibitem{mds3}
Y.~Shang, W.~Ruml, and M.~P. Fromherz, ``Positioning using local maps,'' in
  \emph{Ad Hoc Networks}, vol.~4, no.~2, 2006, pp. 240 -- 253.

\bibitem{mds4}
H.~Junfeng, C.~Jun, Z.~Yafeng, and M.~Xue, ``A {MDS}-based localization
  algorithm for large-scale wireless sensor network,'' in \emph{2010
  International Conference On Computer Design and Applications}, vol.~2, June
  2010, pp. V2--566--V2--570.

\bibitem{kernal1}
C.~Wang, J.~Chen, Y.~Sun, and X.~Shen, ``A graph embedding method for wireless
  sensor networks localization,'' in \emph{IEEE Global Telecommunications
  Conference}, Nov 2009, pp. 1--6.

\bibitem{kernal2}
C.~Wang, J.~Chen, Y.~Sun, and X.~Shen, ``Wireless sensor networks localization with isomap,'' in \emph{2009
  IEEE International Conference on Communications}, June 2009, pp. 1--5.

\bibitem{kernal3}
J.~Chen, C.~Wang, Y.~Sun, and X.~S. Shen, ``Semi-supervised laplacian
  regularized least squares algorithm for localization in wireless sensor
  networks,'' in \emph{Computer Networks}, vol.~55, no.~10, 2011, pp. 2481 --
  2491.

\bibitem{CN4}
H.~Xu, H.~Sun, Y.~Cheng, and H.~Liu, ``Wireless sensor networks localization
  based on graph embedding with polynomial mapping,'' in \emph{Computer
  Networks}, vol. 106, 2016, pp. 151 -- 160.

\bibitem{nit14}
T.~Nitsche, C.~Cordeiro, A.~B. Flores, E.~W. Knightly, E.~Perahia, and J.~C.
  Widmer, ``{IEEE} 802.11ad: directional 60 {GH}z communication for
  multi-{G}igabit-per-second {W}i-{F}i [invited paper],'' in \emph{IEEE
  Communications Magazine}, vol.~52, no.~12, December 2014, pp. 132--141.

\bibitem{mmwa6}
F.~Lemic, J.~Martin, C.~Yarp, D.~Chan, V.~Handziski, R.~Brodersen, G.~Fettweis,
  A.~Wolisz, and J.~Wawrzynek, ``Localization as a feature of mmwave
  communication,'' in \emph{International Wireless Communications and Mobile
  Computing Conference}, Sept 2016, pp. 1033--1038.

\bibitem{MMWlocalization}
H.~El-Sayed, G.~Athanasiou, and C.~Fischione, ``Evaluation of localization
  methods in millimeter-wave wireless systems,'' in \emph{IEEE 19th
  International Workshop on Computer Aided Modeling and Design of Communication
  Links and Networks}, Dec 2014, pp. 345--349.

\bibitem{mmwa7}
T.~Wei and X.~Zhang, ``m{T}rack: High-precision passive tracking using
  millimeter wave radios,'' in \emph{21st Annual International Conference on
  Mobile Computing and Networking}, ser. MobiCom '15, 2015, pp. 117--129.

\bibitem{mmwa8}
H.~Deng and A.~Sayeed, ``Mm-wave {MIMO} channel modeling and user localization
  using sparse beamspace signatures,'' in \emph{IEEE 15th International
  Workshop on Signal Processing Advances in Wireless Communications}, June
  2014, pp. 130--134.

\bibitem{mmLocalization1}
A.~Olivier, G.~Bielsa, I.~Tejado, M.~Zorzi, J.~Widmer, and P.~Casari,
  ``Lightweight indoor localization for 60-{GH}z millimeter wave systems,'' in
  \emph{13th Annual IEEE International Conference on Sensing, Communication,
  and Networking}, June 2016, pp. 1--9.

\bibitem{mmLocalization3}
J.~Palacios, P.~Casari, and J.~Widmer, ``Jade: {Z}eroknowledge device
  localization and environment mapping for millimeter wave systems,'' in
  \emph{{IEEE} Conference on Computer Communications}, May 2017.

\bibitem{mmLocalization2}
J.~Chen, D.~Steinmetzer, J.~Classen, E.~Knightly, and M.~Hollick, ``Pseudo
  lateration: Millimeter-wave localization using a single {RF} chain,'' in
  \emph{{IEEE} Wireless Communications and Networking Conference}, March 2017,
  pp. 1--6.

\bibitem{mmwa1}
M.~Bocquet, C.~Loyez, M.~Fryziel, and N.~Rolland, ``Millimeter-wave broadband
  positioning system for indoor applications,'' in \emph{IEEE/MTT-S
  International Microwave Symposium Digest}, June 2012, pp. 1--3.

\bibitem{mmwa4}
C.~Zhang, F.~Li, J.~Luo, and Y.~He, ``{iLocScan}: Harnessing multipath for
  simultaneous indoor source localization and space scanning,'' in \emph{12th
  ACM Conference on Embedded Network Sensor Systems}, ser. SenSys '14.\hskip
  1em plus 0.5em minus 0.4em\relax New York, NY, USA: ACM, 2014, pp. 91--104.

\bibitem{mmwa3}
K.~Kaemarungsi and P.~Krishnamurthy, ``Modeling of indoor positioning systems
  based on location fingerprinting,'' in \emph{IEEE INFOCOM 2004}, vol.~2,
  March 2004, pp. 1012--1022 vol.2.

\bibitem{mmwa5}
H.~Chu, P.~Xu, S.~Jiang, and X.~You, ``Joint design of axis alignment and
  positioning for nlos indoor mmwave wlans/wpans,'' in \emph{IEEE 80th
  Vehicular Technology Conference (VTC2014-Fall)}, Sept 2014, pp. 1--6.

\bibitem{mltm}
A.~Gunathillake, A.~V. Savkin, and A.~P. Jayasumana, ``Maximum likelihood
  topology maps for wireless sensor networks using an automated robot,'' in
  \emph{IEEE 41st Conference on Local Computer Networks}, Nov 2016, pp.
  339--347.

\bibitem{Jmltm}
A.~Gunathillake, A.~V. Savkin, and A.~P. Jayasumana, ``Topology mapping algorithm for 2{D} and 3{D} wireless sensor networks
  based on maximum likelihood estimation,'' in \emph{Computer Networks}, vol.
  130, 2018, pp. 1 -- 15.

\bibitem{SNadvantages}
D.~Bhattacharyya, T.~Kim, and S.~Pal, ``A comparative study of wireless sensor
  networks and their routing protocols,'' in \emph{Sensors}, November 2010, pp.
  10\,506--10\,523.

\bibitem{CN5}
C.~Hou, Y.~Hou, and Z.~Huang, ``A framework based on barycentric coordinates
  for localization in wireless sensor networks,'' in \emph{Computer Networks},
  vol.~57, no.~17, 2013, pp. 3701 -- 3712.

\bibitem{CN1}
Z.~Wang, Y.~Wang, M.~Ma, and J.~Wu, ``Efficient localization for mobile sensor
  networks based on constraint rules optimized monte carlo method,'' in
  \emph{Computer Networks}, vol.~57, no.~14, 2013, pp. 2788 -- 2801.

\bibitem{locationinwild}
J.~Zhao, W.~Xi, Y.~He, Y.~Liu, X.~Y. Li, L.~Mo, and Z.~Yang, ``Localization of
  wireless sensor networks in the wild: Pursuit of ranging quality,'' in
  \emph{IEEE/ACM Transactions on Networking,}, vol.~21, no.~1, Feb 2013, pp.
  311--323.

\bibitem{sensornets16}
A.~Gunathillake, A.~V. Savkin, A.~Jayasumana, and A.~Seneviratne, ``Sensor
  localization using signal receiving probability and procrustes analysis,'' in
  \emph{Proceedings of the 5th International Confererence on Sensor Networks},
  2016, pp. 113--120.

\bibitem{errorref}
S.~T. Roweis and L.~K. Saul, ``Nonlinear dimensionality reduction by locally
  linear embedding,'' in \emph{Science}, vol. 290, Dec. 2000, pp. 2323--2326.

\bibitem{robotpath}
K.~M. Hasan, A.~A. Nahid, and K.~J. Reza, ``Path planning algorithm development
  for autonomous vacuum cleaner robots,'' in \emph{International Conference on
  Informatics, Electronics Vision}, May 2014, pp. 1--6.

\bibitem{robotpath2d}
M.~A. Yakoubi and M.~T. Laskri, ``The path planning of cleaner robot for
  coverage region using genetic algorithms,'' in \emph{booktitle of Innovation
  in Digital Ecosystems}, vol.~3, no.~1, 2016, pp. 37 -- 43.

\bibitem{propagation}
P.~Pathirana, N.~Bulusu, A.~Savkin, and S.~Jha, ``Node localization using
  mobile robots in delay-tolerant sensor networks,'' in \emph{IEEE Transactions
  on Mobile Computing,}, vol.~4, no.~3, May 2005, pp. 285--296.

\bibitem{Pathirananew1}
P.~N. Pathirana, A.~V. Savkin, and S.~Jha, ``Mobility modelling and trajectory
  prediction for cellular networks with mobile base stations,'' in \emph{4th
  ACM International Symposium on Mobile Ad Hoc Networking \&Amp; Computing},
  ser. MobiHoc '03, 2003, pp. 213--221.

\bibitem{mwm}
M.~Lott and I.~Forkel, ``A multi-wall-and-floor model for indoor radio
  propagation,'' in \emph{53rd IEEE Vehicular Technology Conference}, vol.~1,
  2001, pp. 464--468 vol.1.

\bibitem{nconstant}
S.~Seidel, T.~Rappaport, S.~Jain, M.~Lord, and R.~Singh, ``Path loss,
  scattering and multipath delay statistics in four european cities for digital
  cellular and microcellular radiotelephone,'' in \emph{IEEE Transactions on
  Vehicular Technology,}, vol.~40, no.~4, Nov 1991, pp. 721--730.

\bibitem{nconstant2}
V.~Erceg, L.~Greenstein, S.~Tjandra, S.~Parkoff, A.~Gupta, B.~Kulic, A.~Julius,
  and R.~Bianchi, ``An empirically based path loss model for wireless channels
  in suburban environments,'' in \emph{IEEE booktitle on Selected Areas in
  Communications,}, vol.~17, no.~7, Jul 1999, pp. 1205--1211.

\bibitem{mmtm}
A.~Gunathillake, M.~Moradi, K.~Thilakarathna, A.~P. Jayasumana, and A.~V.
  Savkin, ``Topology maps for 3d millimeter wave sensor networks with
  directional antennas,'' in \emph{IEEE 42nd Conference on Local Computer
  Networks}, Oct 2017, pp. 453--461.

\bibitem{MMW}
P.~Wang, Y.~Li, L.~Song, and B.~Vucetic, ``Multi-gigabit millimeter wave
  wireless communications for 5{G}: from fixed access to cellular networks,''
  in \emph{IEEE Communications Magazine}, vol.~53, no.~1, January 2015, pp.
  168--178.

\bibitem{mmCapacity}
M.~R. Akdeniz, Y.~Liu, M.~K. Samimi, S.~Sun, S.~Rangan, T.~S. Rappaport, and
  E.~Erkip, ``Millimeter wave channel modeling and cellular capacity
  evaluation,'' in \emph{{IEEE} booktitle on Selected Areas in Communications},
  vol.~32, no.~6, June 2014, pp. 1164--1179.

\bibitem{MMWimagiingS}
M.~Seo, B.~Ananthasubramaniam, M.~Rodwell, and U.~Madhow, ``Millimeterwave
  imaging sensor nets: A scalable 60-{GH}z wireless sensor network,'' in
  \emph{IEEE/MTT-S International Microwave Symposium}, June 2007, pp. 563--566.

\bibitem{kutty16}
S.~Kutty and D.~Sen, ``Beamforming for millimeter wave communications: An
  inclusive survey,'' in \emph{IEEE Communications Surveys Tutorials}, vol.~18,
  no.~2, 2016, pp. 949--973.

\bibitem{omnivsdirec}
Cisco, ``Omni antenna vs. directional antenna,''
  http://www.cisco.com/c/en/us/support/docs/wireless-mobility/wireless-lan-wlan/82068-omni-vs-direct.html,
  2007, accessed:05-05-2017.

\bibitem{pathloss2}
T.~S. Rappaport, G.~R. MacCartney, M.~K. Samimi, and S.~Sun, ``Wideband
  millimeter-wave propagation measurements and channel models for future
  wireless communication system design,'' in \emph{IEEE Transactions on
  Communications}, vol.~63, no.~9, Sept 2015, pp. 3029--3056.

\bibitem{pathloss1}
G.~R. Maccartney, T.~S. Rappaport, S.~Sun, and S.~Deng, ``Indoor office
  wideband millimeter-wave propagation measurements and channel models at 28
  and 73 {GH}z for ultra-dense 5{G} wireless networks,'' in \emph{IEEE Access},
  vol.~3, 2015, pp. 2388--2424.

\bibitem{drmds}
L.~j.~Peng and W.~w.~Li, ``The improvement of 3{D} wireless sensor network
  nodes localization,'' in \emph{The 26th Chinese Control and Decision
  Conference}, May 2014, pp. 4873--4878.

\bibitem{ntldvhop}
M.~Chen, X.~Ding, X.~Wang, and X.~Xu, ``A novel three-dimensional localization
  algorithm based on {DV-HOP},'' in \emph{IEEE International Conference on
  Signal Processing, Communications and Computing}, Aug 2014, pp. 70--73.

\bibitem{pathplanning1}
M.~Biglarbegian and F.~Al-Turjman, ``Path planning for data collectors in
  precision agriculture wsns,'' in \emph{International Wireless Communications
  and Mobile Computing Conference}, Aug 2014, pp. 483--487.

\bibitem{pathplanning2}
F.~Al-Turjman, M.~Karakoc, and M.~Gunay, ``Path planning for mobile {DC}s in
  future cities,'' in \emph{Annals of Telecommunications}, vol.~72, no.~3, Apr
  2017, pp. 119--129.

\bibitem{rezazadeh14}
J.~Rezazadeh, M.~Moradi, A.~S. Ismail, and E.~Dutkiewicz, ``Superior path
  planning mechanism for mobile beacon-assisted localization in wireless sensor
  networks,'' in \emph{IEEE Sensors booktitle}, vol.~14, no.~9, Sept 2014.

\bibitem{dmmtm}
A.~Gunathillake, K.~Thilakarathna, A.~P. Jayasumana, and A.~V. Savkin,
  ``Topology maps for millimeter wave iot networks with narrow beamwidth
  antennas,'' in \emph{Submitted to IEEE Internet of Things Journal}, 2018.

\bibitem{Hosoya15}
K.~Hosoya, N.~Prasad, K.~Ramachandran, N.~Orihashi, S.~Kishimoto,
  S.~Rangarajan, and K.~Maruhashi, ``Multiple sector {ID} capture ({MIDC}): A
  novel beamforming technique for 60-ghz band multi-gbps wlan/pan systems,'' in
  \emph{IEEE Transactions on Antennas and Propagation}, vol.~63, no.~1, Jan
  2015, pp. 81--96.

\bibitem{11ad}
``{IEEE} standard for information technology-- {T}elecommunications and
  information exchange between systemslocal and metropolitan area networks--
  {S}pecific requirements--{P}art 11: Wireless lan medium access control
  ({MAC}) and physical layer ({PHY}) specifications--{A}mendment 4:
  {E}nhancements for very high throughput for operation in bands below 6
  {G}hz.'' in \emph{IEEE Std 802.11ac-2013 (Amendment to IEEE Std 802.11-2012,
  as amended by IEEE Std 802.11ae-2012, IEEE Std 802.11aa-2012, and IEEE Std
  802.11ad-2012)}, Dec 2013, pp. 1--425.

\bibitem{Niu2015}
Y.~Niu, Y.~Li, D.~Jin, L.~Su, and A.~V. Vasilakos, ``A survey of millimeter
  wave communications (mm{W}ave) for 5{G}: opportunities and challenges,'' in
  \emph{Wireless Networks}, vol.~21, no.~8, Nov 2015, pp. 2657--2676.

\bibitem{cdetarsk}
A.~Gunathillake and A.~V. Savkin, ``Decentralized target search in topology
  maps based on weighted least square method,'' in \emph{IEEE 85th Vehicular
  Technology Conference (VTC Spring)}, 2017, pp. 1--4.

\bibitem{detarsk}
A.~Gunathillake, A.~V. Savkin, and A.~P. Jayasumana, ``Robust {K}alman
  filter-based decentralised target search and prediction with topology maps,''
  in \emph{IET Wireless Sensor Systems}, vol.~8, no.~2, 2018, pp. 60--67.

\bibitem{tt1}
D.~A. Grundel, ``Searching for a moving target: optimal path planning,'' in
  \emph{In Proceedings of IEEE Networking, Sensing and Control}, 2005, pp.
  867--872.

\bibitem{tt8}
N.~Deshpande, E.~Grant, and T.~C. Henderson, ``Target localization and
  autonomous navigation using wireless sensor networks;a pseudogradient
  algorithm approach,'' in \emph{IEEE Systems booktitle}, vol.~8, no.~1, March
  2014, pp. 93--103.

\bibitem{emergency}
S.Gayan, D.~M.Weeraddana, and A.~Gunathillake, ``Sensor network based adaptable
  system architecture for emergency situations,'' in \emph{Lecture Notes on
  Information Theory}, vol.~2, no.~1, March 2014, pp. 85--91.

\bibitem{adhoc5}
Z.~Can and M.~Demirbas, ``A survey on in-network querying and tracking services
  for wireless sensor networks,'' in \emph{Ad Hoc Networks}, vol.~11, no.~1,
  2013, pp. 596 -- 610.

\bibitem{tt7}
S.~Beyme and C.~Leung, ``Rollout algorithms for wireless sensor
  network-assisted target search,'' in \emph{IEEE Sensors booktitle}, vol.~15,
  no.~7, July 2015, pp. 3835--3845.

\bibitem{tt4}
H.~Lau, S.~Huang, and G.~Dissanayake, ``Probabilistic search for a moving
  target in an indoor environment,'' in \emph{IEEE/RSJ International Conference
  on Intelligent Robots and Systems}, Oct 2006, pp. 3393--3398.

\bibitem{tt3}
E.~Kagan, G.~Goren, and I.~Ben-Gal, ``Probabilistic double-distance algorithm
  of search after static or moving target by autonomous mobile agent,'' in
  \emph{IEEE 26th Convention of Electrical and Electronics Engineers in
  Israel}, 2010.

\bibitem{adhoc8}
W.~Wang, H.~Ma, Y.~Wang, and M.~Fu, ``Performance analysis based on least
  squares and extended {K}alman filter for localization of static target in
  wireless sensor networks,'' in \emph{Ad Hoc Networks}, vol.~25, 2015, pp. 1
  -- 15.

\bibitem{propagation2}
P.~Pathirana, A.~Savkin, and S.~Jha, ``Location estimation and trajectory
  prediction for cellular networks with mobile base stations,'' in \emph{IEEE
  Transactions on Vehicular Technology}, vol.~53, no.~6, 2004, pp. 1903--1913.

\bibitem{NLS}
K.~Madsen, H.~Nielsen, and O.~Tingleff, \emph{Methods for non-linear least
  squares problem}.\hskip 1em plus 0.5em minus 0.4em\relax Technical University
  of Denmark, April 2004.

\bibitem{tt2}
N.~Lo, J.~Berger, and M.~Noel, ``Toward optimizing static target search path
  planning,'' in \emph{IEEE Symposium on Computational Intelligence for
  Security and Defence Applications}, July 2012, pp. 1--7.

\bibitem{treebased}
S.~Yoon, O.~Soysal, M.~Demirbas, and C.~Qiao, ``Coordinated locomotion of
  mobile sensor networks,'' in \emph{5th Annual IEEE Communications Society
  Conference on Sensor, Mesh and Ad Hoc Communications and Networks}, June
  2008, pp. 126--134.

\bibitem{hierar}
X.~Lu and M.~Demirbas, ``Writing on water, a lightweight soft-state tracking
  framework for dense mobile ad hoc networks,'' in \emph{5th IEEE International
  Conference on Mobile Ad Hoc and Sensor Systems}, Sept 2008, pp. 359--364.

\bibitem{geometrical}
S.~Yoon and C.~Qiao, ``A new search algorithm using autonomous and cooperative
  multiple sensor nodes,'' in \emph{26th IEEE International Conference on
  Computer Communications}, May 2007, pp. 937--945.

\bibitem{bashbased}
J.~Li, J.~Jannotti, D.~S.~J. De~Couto, D.~R. Karger, and R.~Morris, ``A
  scalable location service for geographic ad hoc routing,'' in \emph{6th
  Annual International Conference on Mobile Computing and Networking}, ser.
  MobiCom '00, 2000.

\bibitem{adhoc6}
E.~K. Chong and B.~E. Brewington, ``Decentralized rate control for tracking and
  surveillance networks,'' in \emph{Ad Hoc Networks}, vol.~5, no.~6, 2007, pp.
  910 -- 928.

\bibitem{ietR3}
S.~Bhatti, J.~Xu, and M.~Memon, ``Clustering and fault tolerance for target
  tracking using {WSNs},'' in \emph{IET Wireless Sensor Systems}, vol.~1,
  no.~2, June 2011, pp. 66--73.

\bibitem{adhoc4}
S.~Pino-Povedano and F.-J. Gonzalez-Serrano, ``Comparison of optimization
  algorithms in the sensor selection for predictive target tracking,'' in
  \emph{Ad Hoc Networks}, vol.~20, 2014, pp. 182 -- 192.

\bibitem{ietR1}
A.~N. Njoya, C.~Thron, J.~Barry, W.~Abdou, E.~Tonye, N.~S.~L. Konje, and
  A.~Dipanda, ``Efficient scalable sensor node placement algorithm for fixed
  target coverage applications of {WSNs},'' in \emph{IET Wireless Sensor
  Systems}, vol.~7, no.~2, 2017, pp. 44--54.

\bibitem{ietR5}
W.~Li and W.~Zhang, ``Sensor selection for improving accuracy of target
  localisation in wireless visual sensor networks,'' in \emph{IET Wireless
  Sensor Systems}, vol.~2, no.~4, December 2012, pp. 293--301.

\bibitem{adhoc1}
S.~Misra, A.~Singh, S.~Chatterjee, and A.~K. Mandal, ``Qo{S}-aware sensor
  allocation for target tracking in sensor-cloud,'' in \emph{Ad Hoc Networks},
  vol.~33, 2015, pp. 140 -- 153.

\bibitem{sensorJ1}
A.~Eryildirim and M.~B. Guldogan, ``A bernoulli filter for extended target
  tracking using random matrices in a {UWB} sensor network,'' in \emph{IEEE
  Sensors booktitle}, vol.~16, no.~11, June 2016, pp. 4362--4373.

\bibitem{newA31}
A.~N. Bishop, B.~Fidan, K.~Doğançay, B.~D. Anderson, and P.~N. Pathirana,
  ``Exploiting geometry for improved hybrid {AOA/TDOA}-based localization,'' in
  \emph{Signal Processing}, vol.~88, no.~7, 2008, pp. 1775 -- 1791.

\bibitem{LRTA}
X.~Sun, S.~Koenig, and W.~Yeoh, ``Real-time adaptive {A}*,'' in
  \emph{{I}nternational {J}oint {C}onference on {A}utonomous {A}gents and
  {M}ultiagent {S}ystems}, 2008, pp. 281--288.

\bibitem{MTS}
T.~Ishida and R.~E. Korf, ``Moving-target search: a real-time search for
  changing goals,'' in \emph{IEEE Transactions on Pattern Analysis and Machine
  Intelligence}, vol.~17, no.~6, Jun 1995, pp. 609--619.

\bibitem{tt6}
E.~Kagan and I.~Ben-Gal, ``Moving target search algorithm with informational
  distance measures,'' in \emph{The Open Applied Informatics booktitle},
  vol.~6, 2013, pp. 1--10.

\bibitem{Rokhlin}
V.~A. Rokhlin, ``Lectures on the entropy theory of measure-preserving
  transformations,'' in \emph{Russian Mathematical Surveys}, vol.~22, 1967, pp.
  1--52.

\bibitem{ietR4}
T.~Hazim, G.~K. Karagiannidis, and T.~A. Tsiftsis, ``Probability of early
  detection of ultra-wideband positioning sensor networks,'' in \emph{IET
  Wireless Sensor Systems}, vol.~1, no.~3, September 2011, pp. 123--128.

\bibitem{A1}
B.~Dong and X.~Wang, ``Adaptive mobile positioning in {WCDMA} networks,'' in
  \emph{EURASIP booktitle on Wireless Communications and Networking}, vol.
  2005, no.~3, Aug 2005.

\bibitem{sensorJ2}
S.~Mahfouz, F.~Mourad-Chehade, P.~Honeine, J.~Farah, and H.~Snoussi,
  ``Non-parametric and semi-parametric {RSSI}/distance modeling for target
  tracking in wireless sensor networks,'' in \emph{IEEE Sensors booktitle},
  vol.~16, no.~7, April 2016, pp. 2115--2126.

\bibitem{sensorJ3}
X.~Yang, W.~A. Zhang, L.~Yu, and K.~Xing, ``Multi-rate distributed fusion
  estimation for sensor network-based target tracking,'' in \emph{IEEE Sensors
  booktitle}, vol.~16, no.~5, March 2016, pp. 1233--1242.

\bibitem{adhoc2}
T.~Marian, O.~O. Mokryn, and Y.~Shavitt, ``Sensing clouds: A distributed
  cooperative target tracking with tiny binary noisy sensors,'' in \emph{Ad Hoc
  Networks}, vol.~11, no.~8, 2013, pp. 2356 -- 2366.

\bibitem{A2}
B.-H. Liu, M.-L. Chen, and M.-J. Tsai, ``Message-efficient location prediction
  for mobile objects in wireless sensor networks using a maximum likelihood
  technique,'' in \emph{IEEE Transactions on Computers}, vol.~60, 2010, pp.
  865--878.

\bibitem{A3}
Z.~R. Zaidi and B.~L. Mark, ``Real-time mobility tracking algorithms for
  cellular networks based on {K}alman filtering,'' in \emph{IEEE Transactions
  on Mobile Computing}, vol.~4, no.~2, March 2005, pp. 195--208.

\bibitem{ietR2}
S.~Pagano, S.~Peirani, and M.~Valle, ``Indoor ranging and localisation
  algorithm based on received signal strength indicator using statistic
  parameters for wireless sensor networks,'' in \emph{IET Wireless Sensor
  Systems}, vol.~5, no.~5, 2015, pp. 243--249.

\bibitem{kalmann}
I.R.Petersen and A.V.Savkin, in \emph{Robust {K}alman Filtering for Signals and
  Systems with Large Uncertainties}.\hskip 1em plus 0.5em minus 0.4em\relax
  Birkhauser, Boston, 1999.

\bibitem{newA14}
M.~R. James and I.~R. Petersen, ``Nonlinear state estimation for uncertain
  systems with an integral constraint,'' in \emph{IEEE Transactions on Signal
  Processing}, vol.~46, no.~11, Nov 1998, pp. 2926--2937.

\bibitem{tt9}
B.~M. ElHalawany, H.~M. Abdel-Kader, A.~TagEldeen, A.~E. Elsayed, and Z.~B.
  Nossair, ``Modified a* algorithm for safer mobile robot navigation,'' in
  \emph{International Conference on Modelling, Identification Control}, Aug
  2013, pp. 74--78.

\bibitem{mytarget}
A.~Gunathillake, A.~Savkin, and A.~P. Jayasumana, ``Decentralized time-based
  target searching algorithm using sensor network topology maps,'' in
  \emph{12th IEEE International Workshop on Performance and Management of
  Wireless and Mobile Networks}, 2016, pp. 173--180.

\bibitem{procrusters}
V.~D.~M. Nhat, N.~Vo, S.~Challa, and S.~Lee, ``Nonmetric mds for sensor
  localization,'' in \emph{Wireless Pervasive Computing, 3rd International
  Symposium on}, May 2008, pp. 396--400.

\bibitem{CCCesk}
A.~Gunathillake and A.~V. Savkin, ``Mobile robot navigation for emergency
  source seeking using sensor network topology maps,'' in \emph{36th Chinese
  Control Conference}, 2017, pp. 6027--6030.

\bibitem{Jesk}
A.~Gunathillake, H.~Huang, and A.~V. Savkin, ``Mobile robot navigation for
  emergency source seeking using sensor network topology maps,''
  \emph{Sibmitted to IEEE Transaction on Industrial Informatics}, 2018.

\bibitem{fire}
D.~W. Casbeer, D.~B. Kingston, R.~W. Beard, T.~W. McLain, S.~M. Li, and
  R.~Mehra, ``Cooperative forest fire surveillance using a team of small
  unmanned air vehicles,'' in \emph{International booktitle of Systems
  Science}, vol.~36, no.~6, 2006, pp. 351--360.

\bibitem{gas}
L.~M. Pettersson, D.~Durand, O.~M. Johannessen, and D.~Pozdnyakov, ``Monitoring
  of harmful algal blooms,'' in \emph{UK: Praxis Publishing}, 2012.

\bibitem{chemical}
J.~Clark and R.~Fierro, ``Mobile robotic sensors for perimeter detection and
  tracking,'' in \emph{International Society of Automation Transactions},
  vol.~46, 2007, pp. 3--13.

\bibitem{ref3}
A.~Matveev, M.~Hoy, K.~Ovchinnikov, A.~Anisimov, and A.~Savkin, ``Robot
  navigation for monitoring unsteady environmental boundaries without field
  gradient estimation,'' in \emph{Automatica}, vol.~62, 2015, pp. 227--235.

\bibitem{Dynamicenv}
A.~Matveev, M.~Hoy, and A.~Savkin, ``Proofs of the technical results justifying
  an algorithm of extremum seeking navigation in dynamic environmental
  fields,'' in \emph{arXiv preprint arXiv:1502.02224}, 2015.

\bibitem{trackingcontrol}
D.~Chen, Y.~Zhang, and S.~Li, ``Tracking control of robot manipulators with
  unknown models: A jacobian-matrix-adaption method,'' in \emph{IEEE
  Transactions on Industrial Informatics}, vol.~99, 2017.

\bibitem{ref1}
A.~Matveev, H.Teimoori, and A.~Savkin, ``Navigation of a unicycle-like mobile
  robot for environmental extremum seeking,'' in \emph{Automatica}, vol.~47,
  no.~1, 2011, pp. 85--91.

\bibitem{ref2}
A.~Matveev, H.Teimoori, and A.~Savkin, ``Method for tracking of environmental level sets by a unicycle-like
  vehicle,'' in \emph{Automatica}, vol.~48, no.~9, 2012, pp. 2252--2261.

\bibitem{ref4}
A.~Matveev, M.~Hoy, and A.~Savkin, ``Extremum seeking navigation without
  derivative estimation of a mobile robot in a dynamic environmental field,''
  in \emph{IEEE Transactions on Control Systems Technology}, 2016, pp. 1084
  --1091.

\bibitem{noise}
Z.~Can and M.~Demirbas, ``A survey on in-network querying and tracking services
  for wireless sensor networks,'' in \emph{Ad Hoc Networks}, vol.~11, no.~1,
  2013, p. 596 – 610.

\bibitem{tempS1}
\emph{PT TEMPERATURE SENSOR}, Sensor solution, Oct 2015.

\bibitem{tempS2}
\emph{LM35 Precision Centigrade Temperature Sensors}, Texas Instruments, Jan
  2015.

\bibitem{fingerprintloc}
K.~Liu, ``Towards low overhead fingerprint-based indoor localization via
  transfer learning: Design, implementation and evaluation,'' in \emph{IEEE
  Transactions on Industrial Informatics}, vol.~99, 2017.

\bibitem{IntelligentROM}
M.~S. Miah, J.~Knoll, and K.~Hevrdejs, ``Intelligent range-only mapping and
  navigation for mobile robots,'' in \emph{IEEE Transactions on Industrial
  Informatics}, vol.~99, 2017.

\bibitem{multipathref}
X.~Wanga, S.~Yuanb, R.~Laura, and W.~Langb, ``Dynamic localization based on
  spatial reasoning with {RSSI} in wireless sensor networks for transport
  logistics,'' in \emph{Sensors and Actuators}, 2011, p. 421–428.

\bibitem{sensorbased3}
S.~Gayan, D.~M. Weeraddana, and A.~Gunathillake, ``Sensor network based
  adaptable system architecture for emergency situations,'' in \emph{Lecture
  Notes on Information Theory}, vol.~2, no.~1, 2014, p. 85–91.

\bibitem{topologycontrol}
G.~Zhou, P.~Wang, Z.~Zhu, H.~Wang, and W.~Li, ``Topology control strategy for
  movable sensor networks in ultra-deep shafts,'' in \emph{IEEE Transactions on
  Industrial Informatics}, vol.~99, 2017.

\bibitem{gd1}
R.~Bachmayer and N.~E. Leonard, ``Vehicle networks for gradient descent in a
  sampled environment,,'' in \emph{41st IEEE Conference on Decision and
  Control}, 2002, pp. 112--117.

\bibitem{gd2}
F.~Zhang, E.~Fiorelli, and N.~E. Leonard, ``Exploring scalar fields using
  multiple sensor platforms: Tracking level curves,'' in \emph{46th IEEE
  Conference on Decision and Control}, 2007, pp. 3579--3584.

\bibitem{gd3}
S.~Li, Y.~Guo, and B.~Bingham, ``Multi-robot cooperative control for monitoring
  and tracking dynamic plumes,'' in \emph{IEEE International Conference on
  Robotics and Automation}, 2014, pp. 267--273.

\bibitem{gd4}
K.~Ovchinnikov, A.~Semakova, and A.~Matveev, ``Decentralized multi-agent
  tracking of unknown environmental level sets by a team of nonholonomic
  robots,'' in \emph{6th International Congress on Ultra Modern
  Telecommunications and Control Systems and Workshops}, 2014, pp. 352--359.

\bibitem{GBcomp}
E.~Rosero and H.~Werner, ``Cooperative source seeking via gradient estimation
  and formation control (part 1),'' in \emph{UKACC International Conference on
  Control, Loughborough}, 2014, pp. 628--633.

\bibitem{GBcomp2}
E.~Rosero and H.~Werner, ``Cooperative source seeking via gradient estimation and formation
  control (part 2),'' in \emph{UKACC International Conference on Control,
  Loughborough}, 2014, pp. 634--639.

\bibitem{GBperturbation}
Z.~He, S.~Chen, and J.~Wu, ``A gradient-based perturbation extremum seeking
  control scheme,'' in \emph{36th Chinese Control Conference}, 2017, pp.
  3425--3430.

\bibitem{GBneural}
A.~Kebir, L.~Woodward, and O.~Akhrif, ``Extremum-seeking control with adaptive
  excitation: Application to a photovoltaic system,'' in \emph{IEEE
  Transactions on Industrial Electronics}, vol.~99, 2017.

\bibitem{gf1}
A.~Joshi, T.~Ashley, Y.~R., Huang, and A.~L. Bertozzi, ``Experimental
  validation of cooperative environmental boundary tracking with on-board
  sensors,'' in \emph{American control conference}, 2009, pp. 2630--2635.

\bibitem{gf2}
Z.~Jin and A.~L. Bertozzi, ``Environmental boundary tracking and estimation
  using multiple autonomous vehicles,'' in \emph{46th IEEE Conference on
  Decision and Control}, 2007, pp. 4918--4923.

\bibitem{gf3}
C.~Barat and M.~J. Rendas, ``Benthic boundary tracking using a profiler
  sonar,'' in \emph{IEEE/RSJ International Conference on Intelligent Robots and
  Systems}, 2003, pp. 830--835.

\bibitem{gf4}
D.~W. Casbeer, S.~M. Li, R.~W. Beard, T.~W. McLain, and R.~K. Mehra, ``Forest
  fire monitoring using multiple small {UAV}s,'' in \emph{American control
  conference}, 2005, p. 3530–3535.

\bibitem{gf5}
D.~Baronov and J.~Baillieul, ``Reactive exploration through following isolines
  in a potential field,'' in \emph{American control conference}, 2007, p.
  2141–2146.

\bibitem{GFoptimization}
A.~O. Vweza and K.~T.~C. andD. J.~Lee, ``Gradient-free numerical
  optimization-based extremum seeking control for multiagent systems,'' in
  \emph{International booktitle of Control, Automation and Systems}, vol.~13,
  no.~4, 2015, p. 877–886.

\bibitem{GFfeedback}
C.~Zhang and R.~Ordonez, ``Non-gradient axtremum seeking control of feedback
  linearizable systems with application to abs design,'' in \emph{45th IEEE
  Conference on Decision and Control}, 2006, pp. 6666--6671.

\bibitem{navigationmodel}
H.~Teimoori and A.~V. Savkin, ``Equiangular navigation and guidance of a
  wheeled mobile robot based on range-only measurements,'' in \emph{Robotics
  and Autonomous Systems}, vol.~58, no.~2, 2010, pp. 203--215.

\bibitem{R1}
M.Hoy, A.S.Matveev, and A.V.Savkin, ``Algorithms for collision free navigation
  of mobile robots in complex cluttered environments: A survey,'' in
  \emph{Robotica}, vol.~33, no.~3, 2015, pp. 463--497.

\bibitem{R2}
A.S.Matveev, A.V.Savkin, M.Hoy, and C.Wang, ``Safe robot navigation among
  moving and steady obstacles,'' in \emph{Elsevier}, 2015.

\bibitem{newA15}
A.~V. Savkin, ``Analysis and synthesis of networked control systems:
  Topological entropy, observability, robustness and optimal control,'' in
  \emph{Automatica}, vol.~42, no.~1, 2006, pp. 51 -- 62.

\bibitem{newA16}
A.~V. Savkin and T.~M. Cheng, ``Detectability and output feedback
  stabilizability of nonlinear networked control systems,'' in \emph{IEEE
  Transactions on Automatic Control}, vol.~52, no.~4, April 2007, pp. 730--735.

\bibitem{newA17}
A.~S. Matveev and A.~V. Savkin, \emph{Estimation and Control over Communication
  Networks}.\hskip 1em plus 0.5em minus 0.4em\relax Birkhauser, Boston, 2009.

\bibitem{newA18}
G.~C. Goodwin, H.~Haimovich, D.~E. Quevedo, and J.~S. Welsh, ``A moving horizon
  approach to networked control system design,'' in \emph{IEEE Transactions on
  Automatic Control}, vol.~49, no.~9, Sept 2004, pp. 1427--1445.

\bibitem{newA19}
G.~C. Goodwin, D.~E. Quevedo, and E.~I. Silva, ``Architectures and coder design
  for networked control systems,'' in \emph{Automatica}, vol.~44, no.~1, 2008,
  pp. 248--257.

\bibitem{newA20}
D.~E. Quevedo, J.~Ostergaard, and D.~Nesic, ``Packetized predictive control of
  stochastic systems over bit-rate limited channels with packet loss,'' in
  \emph{IEEE Transactions on Automatic Control}, vol.~56, no.~12, Dec 2011, pp.
  2854--2868.

\bibitem{newA21}
D.~Nesic and A.~Teel, ``Input-to-state stability of networked control
  systems,'' in \emph{Automatica}, vol.~40, no.~12, 2004, pp. 2121 -- 2128.

\bibitem{newA22}
W.~P. M.~H. Heemels, A.~R. Teel, N.~V. de~Wouw, and D.~Nesic, ``Networked
  control systems with communication constraints: Tradeoffs between
  transmission intervals, delays and performance,'' in \emph{IEEE Transactions
  on Automatic Control}, vol.~55, no.~8, Aug 2010, pp. 1781--1796.

\bibitem{newA32}
K.~Liu, E.~Fridman, and K.~H. Johansson, \emph{Discrete-Time Networked Control
  Under Scheduling Protocols}.\hskip 1em plus 0.5em minus 0.4em\relax Springer
  International Publishing, 2016, pp. 151--165.

\bibitem{newA33}
D.~Freirich and E.~Fridman, ``Decentralized networked control of
  discrete‐time systems with local networks,'' in \emph{International Journal
  of Robust and Nonlinear Control}, vol.~28, no.~1, pp. 365--380.

\bibitem{newA1}
I.~R. Manchester and A.~V. Savkin, ``Circular-navigation-guidance law for
  precision missile/target engagements,'' in \emph{Journal of Guidance,
  Control, and Dynamics}, vol.~29, no.~2, 2006, pp. 314--320.

\bibitem{newA2}
A.~V. Savkin and C.~Wang, ``Seeking a path through the crowd: Robot navigation
  in unknown dynamic environments with moving obstacles based on an integrated
  environment representation,'' in \emph{Robotics and Autonomous Systems},
  vol.~62, no.~10, 2014, pp. 1568 -- 1580.

\bibitem{newA3}
D.~Baronov and J.~Baillieul, ``Autonomous vehicle control for
  ascending/descending along a potential field with two applications,'' in
  \emph{American Control Conference}, June 2008, pp. 678--683.

\bibitem{newA4}
J.~Cochran and M.~Krstic, ``Nonholonomic source seeking with tuning of angular
  velocity,'' in \emph{IEEE Transactions on Automatic Control}, vol.~54, no.~4,
  April 2009, pp. 717--731.

\bibitem{newA5}
S.-J. Liu and M.~Krstic, ``Stochastic source seeking for nonholonomic
  unicycle,'' in \emph{Automatica}, vol.~46, no.~9, 2010, pp. 1443 -- 1453.

\bibitem{newA6}
A.~R. Mesquita, J.~P. Hespanha, and K.~{\AA}str{\"o}m, ``Optimotaxis: A
  stochastic multi-agent optimization procedure with point measurements,'' in
  \emph{Hybrid Systems: Computation and Control}, M.~Egerstedt and B.~Mishra,
  Eds., 2008, pp. 358--371.

\bibitem{newA7}
C.~Zhang, D.~Arnold, N.~Ghods, A.~Siranosian, and M.~Krstic, ``Source seeking
  with nonholonomic unicycle without position measurement---part i: Tuning of
  forward velocity,'' in \emph{45th IEEE Conference on Decision and Control},
  Dec 2006, pp. 3040--3045.

\bibitem{newA8}
A.~Savkin and R.~Evans, \emph{Hybrid Dynamical Systems: Controller and Sensor
  Switching Problems}, ser. Control Engineering.\hskip 1em plus 0.5em minus
  0.4em\relax Birkh{\"a}user Boston, 2002.

\bibitem{newA9}
E.~Skafidas, R.~J. Evans, A.~V. Savkin, and I.~R. Petersen, ``Stability results
  for switched controller systems,'' in \emph{Automatica}, vol.~35, no.~4,
  1999, pp. 553 -- 564.

\bibitem{newA10}
A.~S. Matveev and A.~V. Savkin, \emph{Qualitative Theory of Hybrid Dynamical
  Systems}.\hskip 1em plus 0.5em minus 0.4em\relax Birkhauser Boston, 2000.

\bibitem{newA11}
V.~Utkin, J.~Guldner, and J.~Shi, \emph{Sliding Mode Control in
  Electro-Mechanical Systems}.\hskip 1em plus 0.5em minus 0.4em\relax Taylor
  and Francis: London, 1999.

\bibitem{newA12}
V.~I. Utkin, \emph{Sliding modes in control and optimization}.\hskip 1em plus
  0.5em minus 0.4em\relax Springer, 2013.

\bibitem{newA13}
V.~I. Utkin, \emph{Sliding modes and their application invariable structure
  systems}.\hskip 1em plus 0.5em minus 0.4em\relax Mir publisher, 1978.

\bibitem{SAVKINnew2}
A.~V. Savkin, E.~Skafidas,  R.~J. Evans, ``Robust output feedback stabilizability via controller switching,'' in
  \emph{Automatica}, vol.~35, no.~1, 1999, pp. 69 -- 74.

\bibitem{smmodel}
D.~Hasenfratz, O.~Saukh, and L.~Thiele, ``Model-driven accuracy bounds for
  noisy sensor readings,'' in \emph{IEEE International Conference on
  Distributed Computing in Sensor Systems}, 2013, pp. 165--174.

\bibitem{networkpartition}
Z.~Rehena, D.~Das, S.~Roy, and N.~Mukherjee, ``A comparative study of
  partitioning algorithms for wireless sensor networks,'' in \emph{Advances in
  Computer Science and Information Technology. Networks and Communications},
  N.~Meghanathan, N.~Chaki, and D.~Nagamalai, Eds.\hskip 1em plus 0.5em minus
  0.4em\relax Berlin, Heidelberg: Springer Berlin Heidelberg, 2012, pp.
  445--454.

\bibitem{networkpartition1}
L.~Wang and X.~Wu, ``Distributed prevention mechanism for network partitioning
  in wireless sensor networks,'' vol.~16, no.~6, Dec 2014, pp. 667--676.

\bibitem{mmMultipath}
C.~Gustafson, K.~Haneda, S.~Wyne, and F.~Tufvesson, ``On mm-wave multipath
  clustering and channel modeling,'' vol.~62, no.~3, March 2014, pp.
  1445--1455.

\bibitem{mmEnergyO}
A.~Pizzo and L.~Sanguinetti, ``Optimal design of energy-efficient millimeter
  wave hybrid transceivers for wireless backhaul,'' in \emph{15th International
  Symposium on Modeling and Optimization in Mobile, Ad Hoc, and Wireless
  Networks}, May 2017, pp. 1--8.

\bibitem{mobilewsn}
G.~Song, Y.~Zhou, F.~Ding, and A.~Song, ``A mobile sensor network system for
  monitoring of unfriendly environments,'' in \emph{Sensors}, vol.~8, no.~11,
  2008, pp. 7259--7274.

\bibitem{anchorplacement}
L.~Huang, F.~Wang, C.~Ma, and W.~Duan, ``The analysis of anchor placement for
  self-localization algorithm in wireless sensor networks,'' in \emph{Advances
  in Wireless Sensor Networks}, R.~Wang and F.~Xiao, Eds.\hskip 1em plus 0.5em
  minus 0.4em\relax Berlin, Heidelberg: Springer Berlin Heidelberg, 2013, pp.
  117--126.

\bibitem{smartdust}
M.~Ilyas and I.~Mahgoub, \emph{Smart Dust: Sensor Network Applications,
  Architecture and Design}.\hskip 1em plus 0.5em minus 0.4em\relax CRC Press,
  2016.

\bibitem{newA23}
A.~S. Matveev, H.~Teimoori, and A.~V. Savkin, ``A method for guidance and
  control of an autonomous vehicle in problems of border patrolling and
  obstacle avoidance,'' in \emph{Automatica}, vol.~47, no.~3, 2011, pp. 515 --
  524.

\bibitem{newA24}
A.~S. Matveev, C.~Wang, and A.~V. Savkin, ``Real-time navigation of mobile
  robots in problems of border patrolling and avoiding collisions with moving
  and deforming obstacles,'' in \emph{Robotics and Autonomous Systems},
  vol.~60, no.~6, 2012, pp. 769 -- 788.

\bibitem{newA25}
A.~V. Savkin and C.~Wang, ``A simple biologically inspired algorithm for
  collision-free navigation of a unicycle-like robot in dynamic environments
  with moving obstacles,'' in \emph{Robotica}, vol.~31, no.~6, 2013, p.
  993–1001.

\bibitem{newA26}
A.~S. Matveev and A.~V. Savkin, ``The problem of {LOG} optimal control via a
  limited capacity communication channel,'' in \emph{Systems and Control
  Letters}, vol.~53, no.~1, 2004, pp. 51--64.

\bibitem{newA27}
A.~S. Matveev and A.~V. Savkin, ``An analogue of {S}hannon information theory for detection and
  stabilization via noisy discrete communication channels,'' in \emph{SIAM
  Journal on Control and Optimization}, vol.~46, no.~4, 2007, pp. 1323--1367.

\bibitem{newA28}
A.~S. Matveev, M.~C. Hoy, and A.~V. Savkin, ``3{D} environmental extremum
  seeking navigation of a nonholonomic mobile robot,'' in \emph{Automatica},
  vol.~50, no.~7, 2014, pp. 1802 -- 1815.

\bibitem{newA29}
T.~M. Cheng and A.~V. Savkin, ``A distributed self-deployment algorithm for the
  coverage of mobile wireless sensor networks,'' in \emph{IEEE Communications
  Letters}, vol.~13, no.~11, November 2009, pp. 877--879.

\bibitem{newA30}
A.~V. Savkin, T.~M. Cheng, Z.~Xi, F.~Javed, A.~S. Matveev, and H.~Nguyen,
  \emph{Decentralized Coverage Control Problems For Mobile Robotic Sensor and
  Actuator Networks}.\hskip 1em plus 0.5em minus 0.4em\relax IEEE Press-Wiley,
  2015.

\bibitem{rfcomprotocol2}
\emph{Wireless Sensor Networks: Principles, Design and Applications}.\hskip 1em
  plus 0.5em minus 0.4em\relax Springer-Verlag London, 2014.

\bibitem{rfcomprotocol}
L.~C. Feng, H.~Y. Chen, T.~H. Li, J.~J. Chiou, and C.~C. Shen, ``Design and
  implementation of an {IEEE} 802.15.4 protocol stack in embedded linux
  kernel,'' in \emph{IEEE 24th International Conference on Advanced Information
  Networking and Applications Workshops}, April 2010, pp. 251--256.

\end{thebibliography}

\end{document}